\documentclass{pasj00}

\def\commenta{$^*$}
\def\commentb{$^\dagger$}
\def\commentc{$^\ddagger$}
\def\commentd{$^\S$}
\def\commente{$^\|$}

\def\submitted{submitted}
\def\inpress{in press}

\def\arxiv#1{ (arXiv astro-ph/#1)}

\DeclareAbbreviation\AAHam{Astron. Abh. Hamburg. Sternw.}
\DeclareAbbreviation\AARv{Astron. Astrophys. Rev.}
\DeclareAbbreviation\AAS{American Astron. Soc. Meeting Abstracts}
\DeclareAbbreviation\AcA{Acta Astron.}
\DeclareAbbreviation\actaa{Acta Astron.}
\DeclareAbbreviation\Afz{Astrofizika}
\DeclareAbbreviation\AGAb{Astronomische Gesellschaft Abstract Ser.}
\DeclareAbbreviation\an{Astron. Nachr.}
\DeclareAbbreviation\AnAp{Annales d'Astrophysique}
\DeclareAbbreviation\AnTok{Tokyo Astron. Obs. Annals, Sec. Ser.}
\DeclareAbbreviation\Ap{Astrophysics}
\DeclareAbbreviation\ARep{Astron. Rep.}
\DeclareAbbreviation\AstBu{Astrophys. Bull.}
\DeclareAbbreviation\ATel{Astron. Telegram}
\DeclareAbbreviation\ATsir{Astron. Tsirk.}
\DeclareAbbreviation\AcApS{Acta Astrophys. Sinica}
\DeclareAbbreviation\AstL{Astron. Lett.}
\DeclareAbbreviation\BaltA{Baltic Astron.}
\DeclareAbbreviation\BANS{Bull. of the Astron. Institutes of the Netherlands Suppl. Ser.}
\DeclareAbbreviation\BASI{Bull. Astron. Soc. India}
\DeclareAbbreviation\BeSN{Be Newslett.}
\DeclareAbbreviation\BHarO{Harvard Coll. Obs. Bull.}
\DeclareAbbreviation\CBET{Cent. Bur. Electron. Telegrams}
\DeclareAbbreviation\ChJAA{Chinese J. of Astron. and Astrophys.}
\DeclareAbbreviation\caa{Chinese J. of Astron. and Astrophys.}
\DeclareAbbreviation\CoAsi{Asiago Contr.}
\DeclareAbbreviation\CoSka{Contributions of the Astronomical Observatory Skalnat\'e Pleso}
\DeclareAbbreviation\GCN{GRB Coord. Netw. Circ.}
\DeclareAbbreviation\ErgAN{Erg. Astron. Nachr.}
\DeclareAbbreviation\ibvs{IBVS}
\DeclareAbbreviation\IEEEP{IEEE Proc.}
\DeclareAbbreviation\JAD{J. Astron. Data}
\DeclareAbbreviation\JAVSO{J. American Assoc. Variable Star Obs.}
\DeclareAbbreviation\JBAA{J. Br. Astron. Assoc.}
\DeclareAbbreviation\JPhCS{J. of Physics Conference Series}
\DeclareAbbreviation\JPSJ{J. Phys. Soc. Japan}
\DeclareAbbreviation\JSARA{J. of the Southeastern Assoc. for Research in Astron.}
\DeclareAbbreviation\LowOB{Lowell Obs. Bull.}
\DeclareAbbreviation\MitAG{Mitteil. der Astronom. Gesell. Hamburg}
\DeclareAbbreviation\MitVS{Mitteil. Ver\"{a}nderl. Sterne}
\DeclareAbbreviation\MmSAI{Mem. Soc. Astron. Ital.}
\DeclareAbbreviation\memsai{Mem. Soc. Astron. Ital.}
\DeclareAbbreviation\Msngr{Messenger}
\DeclareAbbreviation\NewA{New Astron.}
\DeclareAbbreviation\na{New Astron.}
\DeclareAbbreviation\NewAR{New Astron. Rev.}
\DeclareAbbreviation\nar{New Astron. Rev.}
\DeclareAbbreviation\NInfo{Nauchnye Informatsii}
\DeclareAbbreviation\OAP{Odessa Astron. Publ.}
\DeclareAbbreviation\Obs{Observatory}
\DeclareAbbreviation\OEJV{Open Eur. J. on Variable Stars}
\DeclareAbbreviation\PASA{Publ. Astron. Soc. Australia}
\DeclareAbbreviation\PASAu{Publ. Astron. Soc. Australia}
\DeclareAbbreviation\PAZh{Pis'ma AZh}
\DeclareAbbreviation\POBeo{Publ. de l'Observatoire Astronomique de Beograd}
\DeclareAbbreviation\PCCP{Phys. Chem. Chem. Phys.}
\DeclareAbbreviation\PhR{Phys. Rep.}
\DeclareAbbreviation\PVSS{Publ. Variable Stars Sect. R. Astron. Soc. New Zealand}
\DeclareAbbreviation\PZ{Perem. Zvezdy}
\DeclareAbbreviation\PZP{Perem. Zvezdy, Prilozh.}
\DeclareAbbreviation\QJRAS{QJRAS}
\DeclareAbbreviation\RA{Ricerche Astronomiche}
\DeclareAbbreviation\RMxAA{Rev. Mexicana Astron. Astrof.}
\DeclareAbbreviation\RvMA{Reviews of Modern Astron.}
\DeclareAbbreviation\SASS{Society for Astronom. Sciences Ann. Symp.}
\DeclareAbbreviation\Sci{Science}
\DeclareAbbreviation\SPIE{SPIE Proc.}
\DeclareAbbreviation\SvA{Soviet Astronomy}
\DeclareAbbreviation\SvAL{Soviet Astronomy Letters}
\DeclareAbbreviation\VeSon{Ver\"{o}ff. Sternw. Sonneberg}
\DeclareAbbreviation\VSOLJBul{VSOLJ Variable Star Bull.}
\DeclareAbbreviation\yCat{VizieR Online Data Catalog}
\DeclareAbbreviation\ZA{Z. Astrophys.}

\def\ASPConf#1#2{ASP Conf. Ser. #1, #2}

\def\PublisherCambridge{Cambridge: Cambridge University Press}

\def\PublisherASP{San Francisco: ASP}
\def\PublisherAIP{Maryland: AIP}

\newcounter{author}
\def\authorcount#1#2{\refstepcounter{author}\label{#1}
                     \altaffiltext{\ref{#1}}{#2}}

\begin{document}
\SetRunningHead{T. Kato et al.}{Period Variations in SU UMa-Type Dwarf Novae IV}

\Received{201X/XX/XX}
\Accepted{201X/XX/XX}

\title{Survey of Period Variations of Superhumps in SU UMa-Type Dwarf Novae.
    IV: The Fourth Year (2011--2012)}

\author{Taichi~\textsc{Kato},\altaffilmark{\ref{affil:Kyoto}*}
        Franz-Josef~\textsc{Hambsch},\altaffilmark{\ref{affil:GEOS}}$^,$\altaffilmark{\ref{affil:BAV}}$^,$\altaffilmark{\ref{affil:Hambsch}}
        Hiroyuki~\textsc{Maehara},\altaffilmark{\ref{affil:HidaKwasan}}
        Gianluca~\textsc{Masi},\altaffilmark{\ref{affil:Masi}} 
        Ian~\textsc{Miller},\altaffilmark{\ref{affil:Miller}}
        Ryo~\textsc{Noguchi},\altaffilmark{\ref{affil:OKU}}
        Chihiro~\textsc{Akasaka},\altaffilmark{\ref{affil:OKU}}
        Tomoya~\textsc{Aoki},\altaffilmark{\ref{affil:OKU}}
        Hiroshi~\textsc{Kobayashi},\altaffilmark{\ref{affil:OKU}}
        Katsura~\textsc{Matsumoto},\altaffilmark{\ref{affil:OKU}}
        Shinichi~\textsc{Nakagawa},\altaffilmark{\ref{affil:OKU}}
        Takuma~\textsc{Nakazato},\altaffilmark{\ref{affil:OKU}}
        Takashi~\textsc{Nomoto},\altaffilmark{\ref{affil:OKU}}
        Kazuyuki~\textsc{Ogura},\altaffilmark{\ref{affil:OKU}}
        Rikako~\textsc{Ono},\altaffilmark{\ref{affil:OKU}}
        Keisuke~\textsc{Taniuchi},\altaffilmark{\ref{affil:OKU}}
        William~\textsc{Stein},\altaffilmark{\ref{affil:Stein}}
        Arne~\textsc{Henden},\altaffilmark{\ref{affil:AAVSO}}
        Enrique~de~\textsc{Miguel},\altaffilmark{\ref{affil:Miguel}}$^,$\altaffilmark{\ref{affil:Miguel2}}
        Seiichiro~\textsc{Kiyota},\altaffilmark{\ref{affil:Kis}}
        Pavol~A.~\textsc{Dubovsky},\altaffilmark{\ref{affil:Dubovsky}}
        Igor~\textsc{Kudzej},\altaffilmark{\ref{affil:Dubovsky}}
        Kazuyoshi~\textsc{Imamura},\altaffilmark{\ref{affil:OUS}}
        Hidehiko~\textsc{Akazawa},\altaffilmark{\ref{affil:OUS}}
        Ryosuke~\textsc{Takagi},\altaffilmark{\ref{affil:OUS}}
        Yuya~\textsc{Wakabayashi},\altaffilmark{\ref{affil:OUS}}
        Minako~\textsc{Ogi},\altaffilmark{\ref{affil:OUS}}
        Kenji~\textsc{Tanabe},\altaffilmark{\ref{affil:OUS}}
        Joseph~\textsc{Ulowetz},\altaffilmark{\ref{affil:Ulowetz}}
        Etienne~\textsc{Morelle},\altaffilmark{\ref{affil:Morelle}}
        Roger~D.~\textsc{Pickard},\altaffilmark{\ref{affil:BAAVSS}}$^,$\altaffilmark{\ref{affil:Pickard}}
        Tomohito~\textsc{Ohshima},\altaffilmark{\ref{affil:Kyoto}}
        Kiyoshi~\textsc{Kasai},\altaffilmark{\ref{affil:Kai}}
        Elena~P.~\textsc{Pavlenko},\altaffilmark{\ref{affil:CrAO}}
        Oksana~I.~\textsc{Antonyuk},\altaffilmark{\ref{affil:CrAO}}
        Aleksei~V.~\textsc{Baklanov},\altaffilmark{\ref{affil:CrAO}}
        Kirill~\textsc{Antonyuk},\altaffilmark{\ref{affil:CrAO}}
        Denis~\textsc{Samsonov},\altaffilmark{\ref{affil:CrAO}}
        Nikolaj~\textsc{Pit},\altaffilmark{\ref{affil:CrAO}}
        Aleksei~\textsc{Sosnovskij},\altaffilmark{\ref{affil:CrAO}}
        Colin~\textsc{Littlefield},\altaffilmark{\ref{affil:LCO}}
        Richard~\textsc{Sabo},\altaffilmark{\ref{affil:Sabo}}
        Javier~\textsc{Ruiz},\altaffilmark{\ref{affil:Ruiz1}}$^,$\altaffilmark{\ref{affil:Ruiz2}}
        Thomas~\textsc{Krajci},\altaffilmark{\ref{affil:Krajci}}
        Shawn~\textsc{Dvorak},\altaffilmark{\ref{affil:Dvorak}}
        Arto~\textsc{Oksanen},\altaffilmark{\ref{affil:Nyrola}}
        Kenji~\textsc{Hirosawa},\altaffilmark{\ref{affil:Hsk}}
        William~N.~\textsc{Goff},\altaffilmark{\ref{affil:Goff}}
        Berto~\textsc{Monard},\altaffilmark{\ref{affil:Monard}}
        Jeremy~\textsc{Shears},\altaffilmark{\ref{affil:Shears}}
        David~\textsc{Boyd},\altaffilmark{\ref{affil:DavidBoyd}} 
        Irina~B.~\textsc{Voloshina},\altaffilmark{\ref{affil:Sternberg}}
        Sergey~Yu.~\textsc{Shugarov},\altaffilmark{\ref{affil:Sternberg}}$^,$\altaffilmark{\ref{affil:Slovak}}
        Drahomir~\textsc{Chochol},\altaffilmark{\ref{affil:Slovak}}
        Atsushi~\textsc{Miyashita},\altaffilmark{\ref{affil:Seikei}}
        Jochen~\textsc{Pietz},\altaffilmark{\ref{affil:Pietz}}
        Natalia~\textsc{Katysheva},\altaffilmark{\ref{affil:Sternberg}}
        Hiroshi~\textsc{Itoh},\altaffilmark{\ref{affil:Ioh}}
        Greg~\textsc{Bolt},\altaffilmark{\ref{affil:Bolt}}
        Maksim~V.~\textsc{Andreev},\altaffilmark{\ref{affil:Terskol}}$^,$\altaffilmark{\ref{affil:ICUkraine}}
        Nikolai~\textsc{Parakhin},\altaffilmark{\ref{affil:Terskol}}
        Viktor~\textsc{Malanushenko},\altaffilmark{\ref{affil:APO}}
        Fabio~\textsc{Martinelli},\altaffilmark{\ref{affil:Montecatini}}
        Denis~\textsc{Denisenko},\altaffilmark{\ref{affil:Denisenko}}
        Chris~\textsc{Stockdale},\altaffilmark{\ref{affil:Stockdale}}
        Peter~\textsc{Starr},\altaffilmark{\ref{affil:Starr}}
        Mike~\textsc{Simonsen},\altaffilmark{\ref{affil:AAVSO}}
        Paul.~J.~\textsc{Tristram},\altaffilmark{\ref{affil:MtJohn}}
        Akihiko~\textsc{Fukui},\altaffilmark{\ref{affil:OAO}}
        Tamas~\textsc{Tordai},\altaffilmark{\ref{affil:Polaris}}
        Robert~\textsc{Fidrich},\altaffilmark{\ref{affil:Polaris}}
        Kevin~B.~\textsc{Paxson},\altaffilmark{\ref{affil:Paxson}}
        Koh-ichi~\textsc{Itagaki},\altaffilmark{\ref{affil:Itagaki}}
        Youichirou~\textsc{Nakashima},\altaffilmark{\ref{affil:Nakashima}}
        Seiichi~\textsc{Yoshida},\altaffilmark{\ref{affil:MISAO}}
        Hideo~\textsc{Nishimura},\altaffilmark{\ref{affil:Nmh}}
        Timur~V.~\textsc{Kryachko},\altaffilmark{\ref{affil:Astrotel}}
        Andrey~V.~\textsc{Samokhvalov},\altaffilmark{\ref{affil:Astrotel}}
        Stanislav~A.~\textsc{Korotkiy},\altaffilmark{\ref{affil:Astrotel}}
        Boris~L.~\textsc{Satovski},\altaffilmark{\ref{affil:Astrotel}}
        Rod~\textsc{Stubbings},\altaffilmark{\ref{affil:Stubbings}}
        Gary~\textsc{Poyner},\altaffilmark{\ref{affil:Poyner}}
        Eddy~\textsc{Muyllaert},\altaffilmark{\ref{affil:VVSBelgium}}
        Vladimir~\textsc{Gerke},\altaffilmark{\ref{affil:KaDar}}
        Walter~\textsc{MacDonald}~II,\altaffilmark{\ref{affil:Winchester}}
        Michael~\textsc{Linnolt},\altaffilmark{\ref{affil:AAVSO}}
        Yutaka~\textsc{Maeda},\altaffilmark{\ref{affil:Mdy}}
        Hubert~\textsc{Hautecler},\altaffilmark{\ref{affil:VVSBelgium}}
}

\authorcount{affil:Kyoto}{
     Department of Astronomy, Kyoto University, Kyoto 606-8502}
\email{$^*$tkato@kusastro.kyoto-u.ac.jp}

\authorcount{affil:GEOS}{
     Groupe Europ\'een d'Observations Stellaires (GEOS),
     23 Parc de Levesville, 28300 Bailleau l'Ev\^eque, France}

\authorcount{affil:BAV}{
     Bundesdeutsche Arbeitsgemeinschaft f\"ur Ver\"anderliche Sterne
     (BAV), Munsterdamm 90, 12169 Berlin, Germany}

\authorcount{affil:Hambsch}{
     Vereniging Voor Sterrenkunde (VVS), Oude Bleken 12, 2400 Mol, Belgium}

\authorcount{affil:HidaKwasan}{
     Kwasan and Hida Observatories, Kyoto University, Yamashina,
     Kyoto 607-8471}

\authorcount{affil:Masi}{
     The Virtual Telescope Project, Via Madonna del Loco 47, 03023
     Ceccano (FR), Italy}

\authorcount{affil:Miller}{
     Furzehill House, Ilston, Swansea, SA2 7LE, UK}

\authorcount{affil:OKU}{
     Osaka Kyoiku University, 4-698-1 Asahigaoka, Osaka 582-8582}

\authorcount{affil:Stein}{
     6025 Calle Paraiso, Las Cruces, New Mexico 88012, USA}

\authorcount{affil:AAVSO}{
     American Association of Variable Star Observers, 49 Bay State Rd.,
     Cambridge, MA 02138, USA}

\authorcount{affil:Miguel}{
     Departamento de F\'isica Aplicada, Facultad de Ciencias
     Experimentales, Universidad de Huelva,
     21071 Huelva, Spain}

\authorcount{affil:Miguel2}{
     Center for Backyard Astrophysics, Observatorio del CIECEM,
     Parque Dunar, Matalasca\~nas, 21760 Almonte, Huelva, Spain}

\authorcount{affil:Kis}{
     Variable Star Observers League in Japan (VSOLJ), 405-1003 Matsushiro,
     Tsukuba, Ibaraki 305-0035}

\authorcount{affil:Dubovsky}{
     Vihorlat Observatory, Mierova 4, Humenne, Slovakia}

\authorcount{affil:OUS}{
     Department of Biosphere-Geosphere System Science, Faculty of Informatics,
     Okayama University of Science, 1-1 Ridai-cho, Okayama, Okayama 700-0005}

\authorcount{affil:Ulowetz}{
     Center for Backyard Astrophysics Illinois,
     Northbrook Meadow Observatory, 855 Fair Ln, Northbrook,
     Illinois 60062, USA}

\authorcount{affil:Morelle}{
     9 rue Vasco de GAMA, 59553 Lauwin Planque, France}

\authorcount{affil:BAAVSS}{
     The British Astronomical Association, Variable Star Section (BAA VSS),
     Burlington House, Piccadilly, London, W1J 0DU, UK}

\authorcount{affil:Pickard}{
     3 The Birches, Shobdon, Leominster, Herefordshire, HR6 9NG, UK}

\authorcount{affil:Kai}{
     Baselstrasse 133D, CH-4132 Muttenz, Switzerland}

\authorcount{affil:CrAO}{
     Crimean Astrophysical Observatory, 98409, Nauchny, Crimea, Ukraine}

\authorcount{affil:LCO}{
     Department of Physics, University of Notre Dame, Notre Dame,
     Indiana 46556, USA}

\authorcount{affil:Sabo}{
     2336 Trailcrest Dr., Bozeman, Montana 59718, USA}

\authorcount{affil:Ruiz1}{
     Observatorio de C\'antabria, Ctra. de Rocamundo s/n, Valderredible, 
     Cantabria, Spain}

\authorcount{affil:Ruiz2}{
     Agrupaci\'on Astron\'omica C\'antabra, Apartado 573,
     39080-Santander, Spain}

\authorcount{affil:Krajci}{
     Astrokolkhoz Observatory,
     Center for Backyard Astrophysics New Mexico, PO Box 1351 Cloudcroft,
     New Mexico 83117, USA}

\authorcount{affil:Dvorak}{
     Rolling Hills Observatory, 1643 Nightfall Drive,
     Clermont, Florida 34711, USA}

\authorcount{affil:Nyrola}{
     Nyrola observatory, Jyvaskylan Sirius ry, Vertaalantie
     419, FI-40270 Palokka, Finland}

\authorcount{affil:Hsk}{
     216-4 Maeda, Inazawa-cho, Inazawa-shi, Aichi 492-8217}

\authorcount{affil:Goff}{
     13508 Monitor Ln., Sutter Creek, California 95685, USA}

\authorcount{affil:Monard}{
     Bronberg Observatory, Center for Backyard Astronomy Pretoria,
     PO Box 11426, Tiegerpoort 0056, South Africa}

\authorcount{affil:Shears}{
     ``Pemberton'', School Lane, Bunbury, Tarporley, Cheshire, CW6 9NR, UK}

\authorcount{affil:DavidBoyd}{
     Silver Lane, West Challow, Wantage, OX12 9TX, UK}

\authorcount{affil:Sternberg}{
     Sternberg Astronomical Institute, Lomonosov Moscow University, 
     Universitetsky Ave., 13, Moscow 119992, Russia}

\authorcount{affil:Slovak}{
     Astronomical Institute of the Slovak Academy of Sciences, 05960,
     Tatranska Lomnica, the Slovak Republic}

\authorcount{affil:Seikei}{
     Seikei Meteorological Observatory, Seikei High School,
     3-3-1, Kichijoji-Kitamachi, Musashino-shi, Tokyo 180-8633}

\authorcount{affil:Pietz}{
     Nollenweg 6, 65510 Idstein, Germany}

\authorcount{affil:Ioh}{
     VSOLJ, 1001-105 Nishiterakata, Hachioji, Tokyo 192-0153}

\authorcount{affil:Bolt}{
     Camberwarra Drive, Craigie, Western Australia 6025, Australia}

\authorcount{affil:Terskol}{
     Institute of Astronomy, Russian Academy of Sciences, 361605 Peak Terskol,
     Kabardino-Balkaria, Russia}

\authorcount{affil:ICUkraine}{
     International Center for Astronomical, Medical and Ecological Research
     of NASU, Ukraine 27 Akademika Zabolotnoho Str. 03680 Kyiv,
     Ukraine}

\authorcount{affil:APO}{
     Apache Point Observatory, New Mexico State University,
     2001 Apache Point Road, P.O. Box 59, Sunspot, New Mexico 88349-0059, USA}

\authorcount{affil:Montecatini}{
     Palareta, 18-56040 Montecatini, Val Di Cecina, Italy}

\authorcount{affil:Denisenko}{
     Space Research Institute (IKI), Russian Academy of Sciences, Moscow,
     Russia}

\authorcount{affil:Stockdale}{
     8 Matta Drive, Churchill, Victoria  3842, Australia}

\authorcount{affil:Starr}{
     Warrumbungle Observatory, Tenby, 841 Timor Rd,
     Coonabarabran NSW 2357, Australia}

\authorcount{affil:MtJohn}{
     Mt. John Observatory, P.O. Box 56, Lake Tekapo 8770, New Zealand}

\authorcount{affil:OAO}{
     Okayama Astrophysical Observatory, National Astronomical Observatory 
     of Japan, Asakuchi, Okayama 719-0232}

\authorcount{affil:Polaris}{
     Polaris Observatory, Hungarian Astronomical Association,
     Laborc utca 2/c, 1037 Budapest, Hungary}

\authorcount{affil:Paxson}{
     20219 Eden Pines, Spring, Texas 77379, USA}

\authorcount{affil:Itagaki}{
     Itagaki Astronomical Observatory, Teppo-cho, Yamagata 990-2492}

\authorcount{affil:Nakashima}{
     968-4 Yamadanoshou, Oku-cho, Setouchi-City, Okayama 701-4246}

\authorcount{affil:MISAO}{
     2-4-10-708 Tsunashima-nishi, Kohoku-ku, Yokohama-City, Kanagawa 223-0053}

\authorcount{affil:Nmh}{
     Miyawaki 302-6, Kakegawa, Shizuoka 436-0086}

\authorcount{affil:Astrotel}{
     Astrotel-Caucasus Observatory, 41 Lenin Street, village Zelenchukskaya
     Karachay-Cherkessiya, 369140 Russia}

\authorcount{affil:Stubbings}{
     Tetoora Observatory, Tetoora Road, Victoria, Australia}

\authorcount{affil:Poyner}{
     BAA Variable Star Section, 67 Ellerton Road, Kingstanding,
     Birmingham B44 0QE, UK}

\authorcount{affil:VVSBelgium}{
     Vereniging Voor Sterrenkunde (VVS),  Moffelstraat 13 3370
     Boutersem, Belgium}

\authorcount{affil:KaDar}{
     39-28 Razvilka, Moscow region, 142717 Russia}

\authorcount{affil:Winchester}{
     Winchester Observatory, P. O. Box 142, Winchester, ON K0C 2K0, Canada}

\authorcount{affil:Mdy}{
     Kaminishiyamamachi 12-14, Nagasaki, Nagasaki 850-0006}



\maketitle

\begin{abstract}
   Continuing the project described by \citet{Pdot}, we collected
times of superhump maxima for 86 SU UMa-type dwarf novae 
mainly observed during the 2011--2012 season.
We confirmed the general trends recorded in our previous studies, 
such as the relation between period derivatives and orbital periods.  
There are some systems showing positive period derivatives despite 
the long orbital periods.
We observed the 2011 outburst of the WZ Sge-type dwarf nova
BW Scl, and recorded an $O-C$ diagram similar to those of previously known
WZ Sge-type dwarf novae.  The WZ Sge-type dwarf nova OT J184228.1$+$483742
showed an unusual pattern of double outbursts composed of an outburst
with early superhumps and one with ordinary superhumps.  We propose
an interpretation that a very small growth rate of the 3:1 resonance 
due to an extremely low mass-ratio led to a quenching of the superoutburst
before the ordinary superhumps appeared.
We systematically studied ER UMa-type dwarf novae and found
that V1159 Ori showed positive superhumps similar to ER UMa in the 1990s.
The recently recognized ER UMa-type object BK Lyn dominantly showed
negative superhumps, and its behavior was very similar to the present-day
state of ER UMa.
The pattern of period variations in AM CVn-type objects
were very similar to short-period hydrogen-rich SU UMa-type dwarf novae,
making them helium analogue of hydrogen-rich SU UMa-type dwarf novae.
SBS 1108$+$574, a peculiar hydrogen-rich dwarf nova below
the period minimum, showed a very similar 
pattern of period variations to those of short-period SU UMa-type 
dwarf novae.  The mass-ratio derived from the detected orbital period
suggests that this secondary is a somewhat evolved
star whose hydrogen envelope was mostly stripped during 
the mass-exchange.  CC Scl, MASTER OT J072948.66$+$593824.4 and 
OT J173516.9$+$154708 showed only low-amplitude 
superhumps with complex profiles.  These superhumps are likely
a combination of closely separated two periods.
\end{abstract}

\section{Introduction}

   In papers \citet{Pdot}, \citet{Pdot2} and \citet{Pdot3},
we systematically surveyed period variations of superhumps in 
SU UMa-type dwarf novae (for general information of SU UMa-type 
dwarf novae and superhumps, see e.g. \cite{war95book}).
The period variation of superhumps in many SU UMa-type dwarf novae
is generally composed of three distinct stages: early evolutionary
stage with a longer superhump period ($P_{\rm SH}$) (stage A), 
middle stage with systematically varying periods (stage B), 
final stage with a shorter, stable superhump period (stage C).  
These stages are most distinct in objects with short orbital periods 
($P_{\rm orb}$).  Objects with longer orbital periods tend to
show more gradual changes around the transition from stage B to C.
It was also shown that the period derivatives ($P_{\rm dot} = \dot{P}/P$)
during stage B is correlated with $P_{\rm SH}$,
or binary mass-ratios ($q = M_2/M_1$).
In \citet{Pdot3}, we also studied global trends in the amplitudes
of superhumps, and found that the amplitudes of superhumps are 
strongly correlated with orbital periods, and the dependence on 
the inclination is weak in non-eclipsing systems.

   In the present study, we extended the survey to newly recorded objects
and superoutbursts since the publication of \citet{Pdot3}.

\section{Observation and Analysis}\label{sec:obs}

   The data were obtained under campaigns led by the VSNET Collaboration
\citep{VSNET}.  In some objects, we used the public data from 
the AAVSO International Database\footnote{
$<$http://www.aavso.org/data-download$>$.
}.
The majority of the data were acquired
by time-resolved CCD photometry by using 30 cm-class telescopes, whose
observational details on individual objects will be presented in
future papers dealing with analysis and discussion on individual objects.
The list of outbursts and observers is summarized in table \ref{tab:outobs}.
The data analysis was performed just in the same way described
in \citet{Pdot} and \citet{Pdot3}.  We particularly refer to
Phase Dispersion Minimization (PDM; \cite{PDM}).
We also used the Least Absolute 
Shrinkage and Selection Operator (Lasso) method \citep{kat12perlasso}
for separating closely spaced periods.
The times of all observations are expressed in Barycentric Julian Dates (BJD).
We also use the same abbreviations: $P_{\rm orb}$ for
the orbital period and $\epsilon = P_{\rm SH}/P_{\rm orb}-1$ for 
the fractional superhump excess.

   The derived $P_{\rm SH}$, $P_{\rm dot}$ and other parameters
are listed in table \ref{tab:perlist} in same format as in
\citet{Pdot}.  The definitions of parameters $P_1, P_2, E_1, E_2$
and $P_{\rm dot}$ are the same as in \citet{Pdot}.
We also present comparisons of $O-C$ diagrams between different
superoutbursts since this has been one of the motivations of
these surveys (cf. \cite{uem05tvcrv}).

   We use the same terminology of superhumps summarized in
\citet{Pdot3}.  We especially call reader's attention to
the term ``late superhumps''.  We only use ``traditional''
late superhumps when an $\sim$0.5 phase shift is confirmed.
Early superhumps are superhumps seen during the early stages
of WZ Sge-type dwarf novae, and have period close to the orbital
periods (\cite{kat96alcom}; \cite{kat02wzsgeESH}).

\begin{table*}
\caption{List of Superoutbursts.}\label{tab:outobs}
\begin{center}
\begin{tabular}{ccccl}
\hline
Subsection & Object & Year & Observers or references\commenta & ID\commentb \\
\hline
\ref{obj:v725aql}  & V725 Aql     & 2012 & AKz & \\
\ref{obj:egaqr}    & EG Aqr       & 2011 & LCO, Kis, KU & \\
\ref{obj:svari}    & SV Ari       & 2011 & KU, HaC, OKU, MEV, Mhh, OUS, & \\
                   &              &      & Mas, DPV, SRI, AAVSO, Kis, PKV & \\
\ref{obj:ttboo}    & TT Boo       & 2012 & IMi, OUS, Mhh, PXR & \\
\ref{obj:crboo}    & CR Boo       & 2012 & AAVSO, UJH, Nyr, MEV, DPV, GFB, SRI, HMB & \\
                   & CR Boo       & 2012b & UJH, SWI, AAVSO, MEV, Nyr, HMB, DKS & \\
\ref{obj:nncam}    & NN Cam       & 2011 & OKU, Mhh, SWI, IMi & \\
\ref{obj:sycap}    & SY Cap       & 2011 & Mhh, OUS & \\
\ref{obj:gzcet}    & GZ Cet       & 2011 & Mhh, IMi, Hsk & \\
\ref{obj:akcnc}    & AK Cnc       & 2012 & OUS & \\
\ref{obj:cccnc}    & CC Cnc       & 2011 & SWI, OKU, KU, Mhh & \\
\ref{obj:gocom}    & GO Com       & 2012 & DPV, OKU, Mhh, PXR, IMi, Pol & \\
\ref{obj:tucrt}    & TU Crt       & 2011 & Kis & \\
\ref{obj:v503cyg}  & V503 Cyg     & 2011 & Ter, LCO, KU, CRI, OKU, DPV, IMi, HMB & \\
                   & V503 Cyg     & 2011b & CRI & \\
\ref{obj:v1454cyg} & V1454 Cyg    & 2012 & IMi & \\
\ref{obj:aqeri}    & AQ Eri       & 2011 & HMB, SWI & \\
\ref{obj:uvgem}    & UV Gem       & 2011 & MEV, AAVSO & \\
\ref{obj:nyher}    & NY Her       & 2011 & AKz & \\
\ref{obj:prher}    & PR Her       & 2011 & Kai, OUS, OKU, JSh, DPV, deM, SXN, IMi, Ioh, SAc, PXR & \\
\ref{obj:v611her}  & V611 Her     & 2012 & Mas & \\
\ref{obj:v844her}  & V844 Her     & 2012 & OUS, Vol, DPV, PXR, HMB, Hsk & \\
\ref{obj:mmhya}    & MM Hya       & 2012 & HMB, Mhh, IMi, AAVSO & \\
\ref{obj:vwhyi}    & VW Hyi       & 2011 & HaC, AAVSO & \\
\ref{obj:rzlmi}    & RZ LMi       & 2012 & MEV, HMB & \\
                   & RZ LMi       & 2012b & HMB, DKS, AAVSO & \\
                   & RZ LMi       & 2012c & HMB, AAVSO & \\
\ref{obj:bklyn}    & BK Lyn       & 2012 & HMB, AAVSO, MEV, DKS, Boy, UJH, Mhh, GFB, Kai, SRI & \\
                   & BK Lyn       & 2012b & UJH, Nyr, AAVSO, DKS, Boy, SRI & \\
\ref{obj:v585lyr}  & V585 Lyr     & 2012 & Mhh & \\
\ref{obj:fqmon}    & FQ Mon       & 2011 & Kis & \\
\ref{obj:v1032oph} & V1032 Oph    & 2012 & Kai, Mhh & \\
\ref{obj:v2051oph} & V2051 Oph    & 2012 & Mhh & \\
\ref{obj:v1159ori} & V1159 Ori    & 2012 & UJH, SWI & \\
\ref{obj:arpic}    & AR Pic       & 2011 & HaC & \\
\ref{obj:gvpsc}    & GV Psc       & 2011 & SWI, IMi, Mas, OKU & \\
\ref{obj:bwscl}    & BW Scl       & 2011 & HaC, MLF, Mhh, SPE, Kis, Sto, DKS, KU, MOA, Hsk, Nyr, AAVSO & \\
\ref{obj:ccscl}    & CC Scl       & 2011 & HaC & \\
\ref{obj:v1208tau} & V1208 Tau    & 2011 & SWI, IMi, OKU & \\
\ref{obj:v1212tau} & V1212 Tau    & 2011b & MEV, IMi & \\
\ref{obj:diuma}    & DI UMa       & 2007 & \citet{rut09diuma} & \\
                   & DI UMa       & 2007b & \citet{rut09diuma} & \\
\ref{obj:iyuma}    & IY UMa       & 2011 & OUS & \\
\ref{obj:ksuma}    & KS UMa       & 2012 & OUS & \\
\ref{obj:mruma}    & MR UMa       & 2012 & DPV & \\
\ref{obj:puuma}    & PU UMa       & 2012 & IMi, Kai, LCO, CRI, Mhh, PXR, OKU, JSh & \\
\ref{obj:ssumi}    & SS UMi       & 2012 & HMB, AKz, AAVSO, Kai, UJH & \\
\hline
  \multicolumn{5}{l}{\parbox{530pt}{\commenta Key to observers:
AKz (Astrokolkhoz Obs.),
APO (Apache Point Obs.),
Boy\commentc (D. Boyd),
CRI (Crimean Astrophys. Obs.),
deM (E. de Miguel),
DKS\commentc (S. Dvorak),
DPV (P. Dubovsky),
GBo (G. Bolt),
GFB\commentc (W. Goff),
HaC (F.-J. Hambsch, remote obs. in Chile),
HMB (F.-J. Hambsch),
Hsk (K. Hirosawa),
IMi\commentc (I. Miller),
Ioh (H. Itoh),
JSh\commentc (J. Shears),
Kai (K. Kasai),
Kis (S. Kiyota),
Kra (T. Krajci),
KU (Kyoto U., campus obs.),
LCO\commentc (C. Littlefield),
Mas (G. Masi),
MEV\commentc (E. Morelle),
Mhh (H. Maehara),
MLF\commentc (B. Monard),
MOA (MOA team),
Mtc (Montecatini Obs.),
NKa (N. Katysheva),
Nyr (Nyrola and Hankasalmi Obs.),
OKU (Osaya Kyoiku U.),
OUS (Okayama U. of Science),
PIE (J. Pietz),
PKV\commentc (K. Paxson),
Pol (Polaris Obs.),
PXR\commentc (R. Pickard),
Rui (J. Ruiz),
SAc (Seikei High School),
Shu (S. Shugarov),
SPE\commentc (P. Starr),
SRI\commentc (R. Sabo),
Sto (C. Stockdale),
SWI\commentc (W. Stein),
SXN\commentc (M. Simonsen),
Ter (Terskol Obs.),
UJH\commentc (J. Ulowetz),
Vol (I. Voloshina)
}} \\
  \multicolumn{5}{l}{\commentb Original identifications or discoverers.} \\
  \multicolumn{5}{l}{\commentc Inclusive of observations from the AAVSO database.} \\
\end{tabular}
\end{center}
\end{table*}

\addtocounter{table}{-1}
\begin{table*}
\caption{List of Superoutbursts (continued).}
\begin{center}
\begin{tabular}{ccccl}
\hline
Subsection & Object & Year & Observers or references\commenta & ID\commentb \\
\hline
\ref{obj:j2319}    & 1RXS J231935 & 2011 & MEV, Rui, PIE, OKU, PXR, & \\
                   &              &      & deM, Mhh, AAVSO, Mtc & \\
\ref{obj:asas2243} & ASAS J224349 & 2011 & IMi & \\
\ref{obj:dde19}    & DDE 19       & 2011 & SWI & \\
\ref{obj:j0729}    & MASTER J072948 & 2012 & deM, SWI, Shu, IMi, Mhh & \citet{bal12j0729atl3935} \\
\ref{obj:j1743}    & MASTER J174305 & 2012 & Kra & \citet{bal12j1743atel4022} \\
\ref{obj:j1822}    & MASTER J182201 & 2012 & Mas & \citet{bal12j1822atel4084} \\
\ref{obj:misv1446} & MisV 1446    & 2012 & GBo, MLF, Kis, Kai, KU, deM, HaC & \\
\ref{obj:sbs1108}  & SBS 1108     & 2012 & Kai, deM, Vol, LCO, & \\
                   &              &      & APO, GFB, Mhh, NKa, & \\
                   &              &      & CRI, OKU, Kis, Shu & \\
\ref{obj:j0732}    & SDSS J073208 & 2012 & SRI, PXR & \citet{wil10newCVs} \\
\ref{obj:j0803}    & SDSS J080303 & 2011 & deM, OKU, Rui, IMi & \\
\ref{obj:j1653}    & SDSS J165359 & 2012 & IMi, Mhh, PXR, OKU, deM & \\
\ref{obj:j1702}    & SDSS J170213 & 2011 & MEV, OKU, IMi, DPV, & \\
                   &              &      & Mas, LCO, Boy, HMB & \\
\ref{obj:j1721}    & SDSS J172102 & 2012 & GFB, Mas & \citet{rau10HeDN} \\
\ref{obj:j2104}    & SDSS J210449 & 2011 & IMi & \\
\ref{obj:j2205}    & SDSS J220553 & 2011 & SWI, Mhh, NKa & \\
\ref{obj:j0019}    & OT J001952   & 2012 & deM & CSS120131:001952$+$433901 \\
\ref{obj:j0115}    & OT J011516   & 2012 & Mas & CSS101008:011517$+$245530 \\
\ref{obj:j0507}    & OT J050716   & 2012 & Mas & CSS081221:050716$+$125314 \\
\ref{obj:j0557}    & OT J055721   & 2011 & HaC, Mhh & SSS111229:055722$-$363055 \\
\ref{obj:j0646}    & OT J064608   & 2011 & SWI, Mas, Rui & CSS080512:064608$+$403305 \\
\ref{obj:j0811}    & OT J081117   & 2011 & Rui, Mhh & CSS111030:081117$+$152003 \\
\ref{obj:j0841}    & OT J084127   & 2012 & Mas, OKU, PXR & CSS090525:084127$+$210054 \\
\ref{obj:j0948}    & OT J094854   & 2012 & HMB, SWI, Mas & CSS120315:094854$+$014911 \\
\ref{obj:j1028}    & OT J102842   & 2012 & OKU, Kis, UJH, SWI, deM, HMB & CSS090331:102843$-$081927 \\
\ref{obj:j1051}    & OT J105122   & 2012 & SWI, CRI & CSS120101:105123$+$672528 \\
\ref{obj:j1259}    & OT J125905   & 2012 & Mas & CSS120424:125906$+$242634 \\
\ref{obj:j1316}    & OT J131625   & 2012 & Mas & CSS080427:131626$-$151313 \\
\ref{obj:j1425}    & OT J142548   & 2011 & Mas & CSS110628:142548$+$151502 \\
\ref{obj:j1442}    & OT J144252   & 2012 & MLF, HaC, LCO & CSS120417:144252$-$225040 \\
\ref{obj:j1444}    & OT J144453   & 2012 & Mhh, HaC & CSS120424:144453$-$131118 \\
\ref{obj:j1459}    & OT J145921   & 2011 & Kra, Mas, PIE & CSS110613:145922$+$354806 \\
\ref{obj:j1556}    & OT J155631   & 2012 & GBo, HMB & CSS090321:155631$-$080440 \\
\ref{obj:j1604}    & OT J160410   & 2012 & Mas & CSS120326:160411$+$145618 \\
\ref{obj:j1628}    & OT J162806   & 2011 & Mas, Mhh & CSS110611:162806$+$065316 \\
\ref{obj:j1639}    & OT J163942   & 2012 & IMi & CSS080131:163943$+$122414 \\
\ref{obj:j1706}    & OT J170609   & 2011 & Mas & CSS090205:170610$+$143452 \\
\ref{obj:j1735}    & OT J173516   & 2011 & OKU, Mas, Mhh, DPV, KU, HMB, Kis & CSS110623:173517$+$154708 \\
\ref{obj:j1842}    & OT J184228   & 2011 & Mas, Mhh, OKU, DPV, & \\
                   &              &      & OUS, Ioh, deM, SRI, & \\
                   &              &      & UJH, KU, AAVSO, HMB, & \\
                   &              &      & LCO, CRI, Hsk, IMi & Nishimura \citep{nak11j1842cbet2818} \\
\ref{obj:j2109}    & OT J210950   & 2011 & DKS, Rui, DPV, OUS, & \\
                   &              &      & Kis, SRI, IMi, LCO, & \\
                   &              &      & Mhh, AAVSO & Itagaki \citep{yam11j2109cbet2731} \\
\ref{obj:j2147}    & OT J214738   & 2011 & Mas, SWI, deM, HMB, OKU, & \\
                   &              &      & Nyr, UJH, CRI & CSS111004:214738$+$244554 \\
\ref{obj:j2158}    & OT J215818   & 2011 & SWI, Rui, JSh, deM, & \\
                   &              &      & OKU, SRI, IMi, UJH, & \\
                   &              &      & Mas, MEV & PNV J21581852$+$2419246 \\
\ref{obj:j2212}    & OT J221232   & 2011 & SWI, Kai, Mas, CRI, SAc & CSS 090911:221232$+$160140 \\
\ref{obj:j2247}    & OT J224736   & 2012 & Mas & CSS120616:224736$+$250436 \\
\ref{obj:j0846}    & TCP J084616  & 2012 & deM, Mas & TCP J08461690$+$3115554 \\
\ref{obj:j2313}    & TCP J231308  & 2011 & Rui, Mas, Mhh, Kra, Kis & TCP J23130812$+$2337018 \\
\hline
\end{tabular}
\end{center}
\end{table*}

\begin{table*}
\caption{Superhump Periods and Period Derivatives}\label{tab:perlist}
\begin{center}
\begin{tabular}{cccccccccccccc}
\hline
Object & Year & $P_1$ (d) & err & \multicolumn{2}{c}{$E_1$\commenta} & $P_{\rm dot}$\commentb & err\commentb & $P_2$ (d) & err & \multicolumn{2}{c}{$E_2$\commenta} & $P_{\rm orb}$ (d) & Q\commentc \\
\hline
EG Aqr & 2011 & 0.078577 & 0.000055 & 0 & 93 & $-$17.6 & 7.2 & -- & -- & -- & -- & -- & CGM \\
SV Ari & 2011 & 0.055524 & 0.000014 & 19 & 311 & 4.0 & 0.2 & 0.055350 & 0.000052 & 307 & 366 & -- & A \\
TT Boo & 2012 & 0.078083 & 0.000015 & 0 & 113 & 1.6 & 0.8 & -- & -- & -- & -- & -- & C \\
CR Boo & 2012 & 0.017265 & 0.000002 & 0 & 247 & 2.0 & 0.2 & 0.017193 & 0.000006 & 237 & 395 & 0.017029 & B \\
CR Boo & 2012b & 0.017257 & 0.000002 & 0 & 245 & 1.9 & 0.2 & -- & -- & -- & -- & 0.017029 & B \\
NN Cam & 2011 & 0.074197 & 0.000023 & 0 & 57 & 7.1 & 3.8 & 0.073843 & 0.000013 & 54 & 109 & 0.0717 & B \\
SY Cap & 2011 & 0.063750 & 0.000026 & 0 & 31 & -- & -- & -- & -- & -- & -- & -- & CG \\
AK Cnc & 2012 & 0.067239 & 0.000123 & 0 & 46 & -- & -- & -- & -- & -- & -- & 0.0651 & C \\
CC Cnc & 2011 & 0.075887 & 0.000001 & 0 & 27 & -- & -- & 0.075456 & 0.000028 & 42 & 103 & 0.07352 & C \\
GO Com & 2012 & 0.063016 & 0.000019 & 0 & 128 & 4.8 & 1.5 & 0.062492 & 0.000150 & 127 & 144 & -- & B \\
TU Crt & 2011 & -- & -- & -- & -- & -- & -- & 0.084962 & 0.000043 & 0 & 82 & 0.08209 & C \\
V503 Cyg & 2011 & 0.081309 & 0.000062 & 0 & 25 & -- & -- & 0.081046 & 0.000048 & 35 & 78 & 0.07777 & B \\
V503 Cyg & 2011b & 0.081241 & 0.000057 & 0 & 87 & $-$11.6 & 3.4 & -- & -- & -- & -- & 0.07777 & CGM \\
V1454 Cyg & 2012 & 0.057494 & 0.000015 & 0 & 18 & -- & -- & -- & -- & -- & -- & -- & C \\
AQ Eri & 2011 & -- & -- & -- & -- & -- & -- & 0.061648 & 0.000247 & 143 & 161 & 0.06094 & CG \\
UV Gem & 2011 & 0.092822 & 0.000094 & 0 & 13 & -- & -- & -- & -- & -- & -- & -- & C \\
NY Her & 2011 & 0.075802 & 0.000121 & 0 & 37 & -- & -- & -- & -- & -- & -- & -- & CG \\
PR Her & 2011 & 0.055022 & 0.000026 & 0 & 92 & 8.8 & 3.7 & -- & -- & -- & -- & 0.05422 & CE \\
V844 Her & 2012 & 0.055901 & 0.000021 & 22 & 124 & 12.4 & 1.5 & 0.055873 & 0.000031 & 124 & 183 & 0.054643 & B \\
MM Hya & 2012 & 0.058872 & 0.000026 & 0 & 122 & -- & -- & 0.058625 & 0.000049 & 119 & 201 & 0.057590 & C \\
VW Hyi & 2011 & 0.076914 & 0.000026 & 25 & 68 & 8.2 & 5.8 & 0.076540 & 0.000019 & 77 & 146 & 0.074271 & A \\
RZ LMi & 2012 & 0.059441 & 0.000021 & 0 & 126 & 2.4 & 1.5 & -- & -- & -- & -- & -- & C \\
RZ LMi & 2012b & 0.059472 & 0.000026 & 0 & 84 & 4.5 & 3.6 & -- & -- & -- & -- & -- & C \\
RZ LMi & 2012c & 0.059408 & 0.000011 & 0 & 133 & 2.9 & 0.4 & -- & -- & -- & -- & -- & B \\
BK Lyn & 2012b & 0.078510 & 0.000028 & 25 & 127 & 3.2 & 2.7 & -- & -- & -- & -- & 0.07498 & B \\
V585 Lyr & 2012 & 0.060350 & 0.000038 & 0 & 19 & -- & -- & -- & -- & -- & -- & -- & C \\
FQ Mon & 2011 & -- & -- & -- & -- & -- & -- & 0.072718 & 0.000180 & 0 & 14 & -- & C \\
V1032 Oph & 2012 & 0.085965 & 0.000288 & 0 & 47 & -- & -- & -- & -- & -- & -- & 0.081055 & C \\
AR Pic & 2011 & -- & -- & -- & -- & -- & -- & 0.083154 & 0.000149 & 0 & 50 & 0.0801 & CP \\
GV Psc & 2011 & 0.094313 & 0.000018 & 0 & 62 & $-$3.1 & 2.3 & -- & -- & -- & -- & -- & C2 \\
BW Scl & 2011 & 0.055000 & 0.000008 & 25 & 210 & 4.4 & 0.3 & -- & -- & -- & -- & 0.054323 & A \\
CC Scl & 2011 & -- & -- & -- & -- & -- & -- & 0.060012 & 0.000028 & 0 & 152 & 0.05858 & C \\
V1208 Tau & 2011 & -- & -- & -- & -- & -- & -- & 0.070481 & 0.000066 & 0 & 49 & -- & B \\
V1212 Tau & 2011b & 0.069692 & 0.000055 & 0 & 18 & -- & -- & -- & -- & -- & -- & -- & C2 \\
DI UMa & 2007 & 0.055306 & 0.000015 & 18 & 182 & 4.1 & 0.8 & -- & -- & -- & -- & 0.054566 & B \\
DI UMa & 2007b & 0.055340 & 0.000040 & 0 & 126 & 9.3 & 4.3 & -- & -- & -- & -- & 0.054566 & B \\
MR UMa & 2012 & -- & -- & -- & -- & -- & -- & 0.064746 & 0.000021 & 0 & 48 & -- & C \\
PU UMa & 2012 & 0.081090 & 0.000048 & 11 & 84 & $-$14.3 & 2.6 & 0.080724 & 0.000100 & 84 & 121 & 0.077881 & B \\
SS UMi & 2012 & 0.070358 & 0.000128 & 0 & 33 & -- & -- & -- & -- & -- & -- & 0.06778 & C \\
1RXS J231935 & 2011 & 0.065989 & 0.000019 & 0 & 79 & 11.6 & 1.7 & 0.065528 & 0.000014 & 75 & 159 & -- & B \\
DDE 19 & 2011 & -- & -- & -- & -- & -- & -- & 0.091210 & 0.000043 & 0 & 35 & -- & C \\
MisV 1446 & 2012 & 0.078072 & 0.000088 & 0 & 35 & -- & -- & 0.077304 & 0.000098 & 35 & 69 & -- & C \\
SBS 1108 & 2012 & 0.039118 & 0.000003 & 0 & 403 & 1.2 & 0.1 & 0.038869 & 0.000004 & 399 & 876 & 0.038449 & CP \\
SDSS J073208 & 2012 & 0.079571 & 0.000021 & 0 & 72 & -- & -- & -- & -- & -- & -- & -- & CG \\
SDSS J080303 & 2011 & 0.091949 & 0.000119 & 17 & 31 & -- & -- & 0.090393 & 0.000022 & 27 & 88 & -- & C \\
SDSS J165359 & 2012 & -- & -- & -- & -- & -- & -- & 0.065105 & 0.000150 & 91 & 121 & -- & C \\
SDSS J170213 & 2011 & 0.105005 & 0.000056 & 32 & 117 & 17.0 & 2.8 & -- & -- & -- & -- & 0.100082 & B \\
SDSS J172102 & 2012 & -- & -- & -- & -- & -- & -- & 0.026673 & 0.000008 & 0 & 463 & -- & C \\
SDSS J210449 & 2011 & 0.075315 & 0.000045 & 0 & 27 & -- & -- & -- & -- & -- & -- & -- & C \\
SDSS J220553 & 2011 & 0.058151 & 0.000021 & 0 & 99 & 7.7 & 0.9 & -- & -- & -- & -- & 0.05752 & B \\
\hline
  \multicolumn{13}{l}{\commenta Interval used for calculating the period (corresponding to $E$ in section \ref{sec:individual}).} \\
  \multicolumn{13}{l}{\commentb Unit $10^{-5}$.} \\
  \multicolumn{13}{l}{\commentc Data quality and comments. A: excellent, B: partial coverage or slightly low quality, C: insufficient coverage or}\\
  \multicolumn{13}{l}{\phantom{\commentc} observations with large scatter, G: $P_{\rm dot}$ denotes global $P_{\rm dot}$, M: observational gap in middle stage,}\\
  \multicolumn{13}{l}{\phantom{\commentc} 2: late-stage coverage, the listed period may refer to $P_2$, E: $P_{\rm orb}$ refers to the period of early superhumps.} \\
  \multicolumn{13}{l}{\phantom{\commentc} P: $P_{\rm orb}$ refers to a shorter stable periodicity recorded in outburst.} \\
\end{tabular}
\end{center}
\end{table*}

\addtocounter{table}{-1}
\begin{table*}
\caption{Superhump Periods and Period Derivatives (continued)}
\begin{center}
\begin{tabular}{cccccccccccccc}
\hline
Object & Year & $P_1$ & err & \multicolumn{2}{c}{$E_1$} & $P_{\rm dot}$ & err & $P_2$ & err & \multicolumn{2}{c}{$E_2$} & $P_{\rm orb}$ & Q \\
\hline
OT J001952 & 2012 & 0.056770 & 0.000039 & 0 & 18 & -- & -- & -- & -- & -- & -- & -- & C \\
OT J050716 & 2012 & 0.065916 & 0.000080 & 0 & 15 & -- & -- & -- & -- & -- & -- & -- & C \\
OT J055721 & 2011 & 0.059756 & 0.000017 & 0 & 153 & 4.6 & 0.9 & -- & -- & -- & -- & -- & B \\
OT J064608 & 2011 & 0.061105 & 0.000023 & 0 & 82 & 11.1 & 2.6 & -- & -- & -- & -- & -- & B \\
OT J081117 & 2011 & 0.058035 & 0.000027 & 0 & 63 & -- & -- & -- & -- & -- & -- & -- & C \\
OT J084127 & 2012 & 0.087686 & 0.000252 & 0 & 4 & -- & -- & -- & -- & -- & -- & -- & C \\
OT J094854 & 2012 & 0.057499 & 0.000021 & 0 & 77 & 8.3 & 2.8 & -- & -- & -- & -- & -- & C \\
OT J102842 & 2012 & 0.038168 & 0.000008 & 70 & 151 & -- & -- & -- & -- & -- & -- & -- & C \\
OT J105122 & 2012 & 0.061054 & 0.000109 & 0 & 30 & -- & -- & -- & -- & -- & -- & 0.0596 & C2 \\
OT J144252 & 2012 & 0.065126 & 0.000028 & 0 & 59 & 13.6 & 4.3 & 0.064639 & 0.000054 & 59 & 107 & -- & B \\
OT J144453 & 2012 & -- & -- & -- & -- & -- & -- & 0.082289 & 0.000060 & 0 & 58 & -- & C \\
OT J145921 & 2011 & 0.085114 & 0.000059 & 0 & 74 & 10.9 & 7.2 & -- & -- & -- & -- & -- & C \\
OT J155631 & 2012 & 0.089309 & 0.000053 & 0 & 41 & $-$21.3 & 5.8 & -- & -- & -- & -- & -- & CG \\
OT J162806 & 2011 & 0.068847 & 0.000008 & 0 & 140 & -- & -- & -- & -- & -- & -- & -- & CGM \\
OT J163942 & 2012 & 0.088585 & 0.000052 & 0 & 23 & -- & -- & -- & -- & -- & -- & -- & C \\
OT J170609 & 2011 & 0.059460 & 0.000076 & 0 & 16 & -- & -- & -- & -- & -- & -- & -- & C \\
OT J184228 & 2011 & 0.072342 & 0.000018 & 64 & 206 & $-$0.9 & 1.5 & -- & -- & -- & -- & 0.07168 & BE \\
OT J210950 & 2011 & 0.060045 & 0.000026 & 34 & 188 & 8.5 & 0.6 & 0.059742 & 0.000022 & 187 & 289 & 0.05865 & BP \\
OT J214738 & 2011 & 0.097147 & 0.000021 & 21 & 107 & 8.8 & 1.0 & -- & -- & -- & -- & 0.09273 & BP \\
OT J215818 & 2011 & 0.067397 & 0.000027 & 0 & 56 & 6.9 & 4.5 & 0.066852 & 0.000020 & 50 & 127 & -- & B \\
OT J221232 & 2011 & 0.090322 & 0.000097 & 0 & 29 & -- & -- & 0.090051 & 0.000028 & 29 & 106 & -- & B \\
OT J224736 & 2012 & 0.056673 & 0.000020 & 0 & 37 & -- & -- & -- & -- & -- & -- & -- & C \\
TCP J084616 & 2012 & 0.096333 & 0.000106 & 0 & 12 & -- & -- & -- & -- & -- & -- & 0.09139 & C \\
TCP J231308 & 2011 & 0.071364 & 0.000044 & 0 & 24 & -- & -- & 0.071016 & 0.000033 & 28 & 85 & -- & C \\
\hline
\end{tabular}
\end{center}
\end{table*}

\section{Individual Objects}\label{sec:individual}

\subsection{V725 Aquilae}\label{obj:v725aql}

   Y. Nakashima detected an outburst of this object on 2012 April 16
(vsnet-alert 14450).  Subsequent observations confirmed that
it is indeed a superoutburst (vsnet-alert 14460).  Due to the short
visibility in the morning, observations only on two nights were
obtained.  A PDM analysis yielded a period of 0.09047(5)~d.
We obtained a single superhump maximum of BJD 2456036.9734(8) ($N=137$).
It is noticeable that a likely superoutburst occurred in 2011 May
(vsnet-alert 14460), and the interval between the superoutburst
was only $\sim$340~d, which is much shorter than previously
considered \citep{uem01v725aql}.  The object faded quickly
(unfiltered CCD magnitude 17.6 on April 24) and we probably observed
the final stage of the superoutburst.  There was a visual
detection at a magnitude of 14.6 on April 27.  The object may have shown
a rebrightening as in the 1999 one \citep{uem01v725aql}.

\subsection{EG Aquarii}\label{obj:egaqr}

\begin{figure}
  \begin{center}
    \FigureFile(88mm,70mm){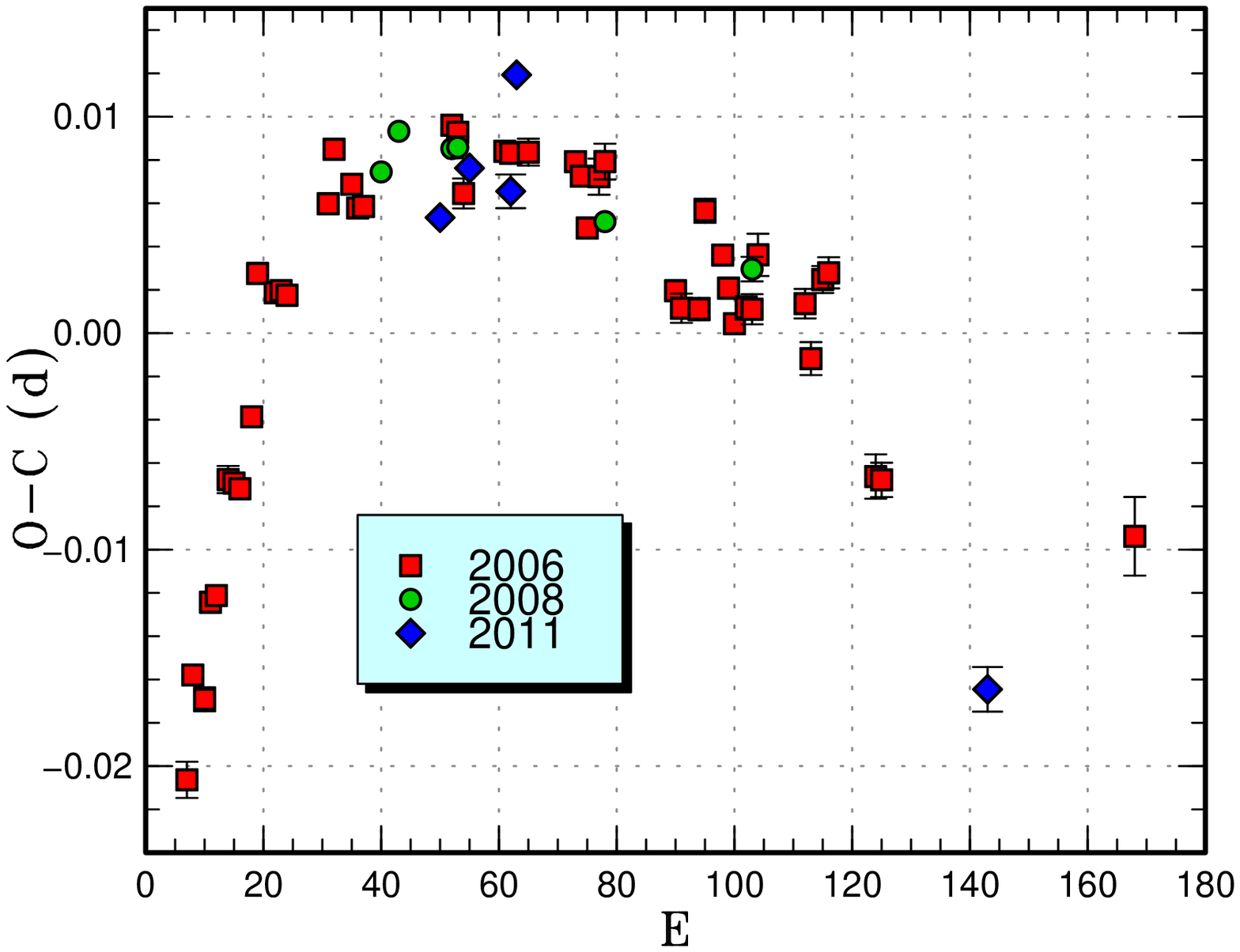}
  \end{center}
  \caption{Comparison of $O-C$ diagrams of EG Aqr between different
  superoutbursts.  A period of 0.07885~d was used to draw this figure.
  Approximate cycle counts ($E$) after the start of the superoutburst
  were used.  Since the starts of the 2008 and 2011 superoutbursts
  were not well constrained, we shifted the $O-C$ diagrams
  to best fit the best-recorded 2006 one.
  }
  \label{fig:egaqrcomp}
\end{figure}

   The 2011 June superoutburst of this object was detected by
R. Stubbings at a visual magnitude of 12.5 (vsnet-alert 13460).
The object was rather unfavorably located and the observations
were limited than in the past studies (\cite{ima08egaqr}; \cite{Pdot}).
The times of superhump maxima are listed in table
\ref{tab:egaqroc2011}.  Although there was likely a stage B--C
transition between $E=13$ and $E=93$, the epoch of this transition
was not covered by observations.  The $P_{\rm dot}$ listed in
table \ref{tab:perlist} refers to the global $P_{\rm dot}$.
A comparison of $O-C$ diagrams of EG Aqr between different
superoutbursts is shown in figure \ref{fig:egaqrcomp}.

\begin{table}
\caption{Superhump maxima of EG Aqr (2011).}\label{tab:egaqroc2011}
\begin{center}
\begin{tabular}{ccccc}
\hline
$E$ & max\commenta & error & $O-C$\commentb & $N$\commentc \\
\hline
0 & 55740.8707 & 0.0002 & $-$0.0044 & 221 \\
5 & 55741.2672 & 0.0004 & $-$0.0007 & 119 \\
12 & 55741.8181 & 0.0008 & 0.0001 & 88 \\
13 & 55741.9023 & 0.0007 & 0.0058 & 78 \\
93 & 55748.1819 & 0.0010 & $-$0.0008 & 146 \\
\hline
  \multicolumn{5}{l}{\commenta BJD$-$2400000.} \\
  \multicolumn{5}{l}{\commentb Against max $= 2455740.8750 + 0.078577 E$.} \\
  \multicolumn{5}{l}{\commentc Number of points used to determine the maximum.} \\
\end{tabular}
\end{center}
\end{table}

\subsection{SV Arietis}\label{obj:svari}

\begin{figure}
  \begin{center}
    \FigureFile(88mm,110mm){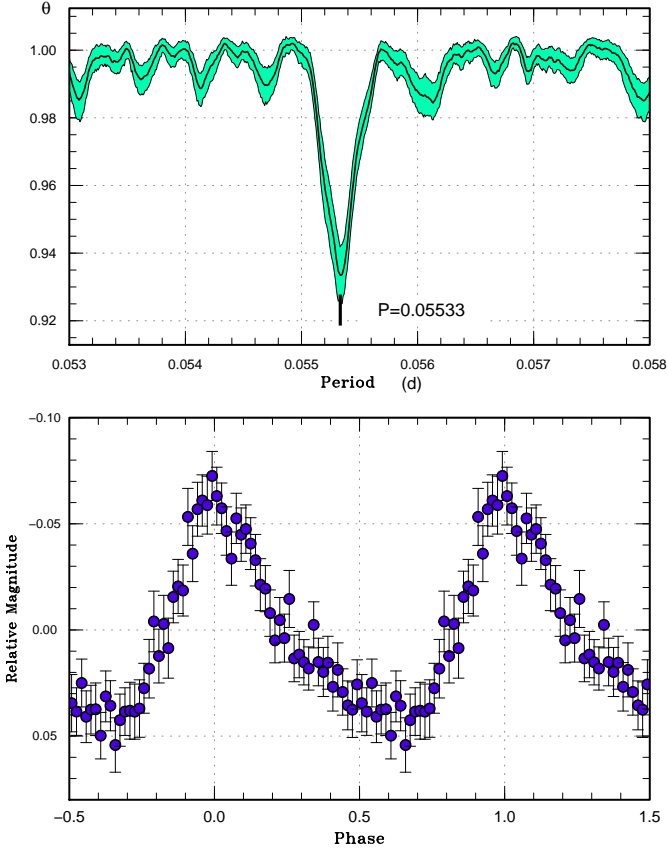}
  \end{center}
  \caption{Superhumps in SV Ari (2011). (Upper): PDM analysis.
     (Lower): Phase-averaged profile.}
  \label{fig:svarishpdm}
\end{figure}

   SV Ari was discovered by \citet{wol05svari} who recorded the object
at magnitude 12 on three Heidelberg plates taken on 1905 November 6.
The object was not detected on November 1, and it quickly faded
to magnitude 13.5 on November 21.  According to \citet{due87novaatlas},
there was a possible detection of a brightening to magnitude 15.7
on 1943 September 2 by Himpel and Jansch.  Although \citet{due87novaatlas}
even suggested an intergalactic nova, many observers, mostly amateur 
observers, intensively monitored the object suspecting that it is 
a dwarf nova.  \citet{rob98oldnovaproc} identified a $B=22.1$ mag
quiescent counterpart [see also \citet{rob00oldnova}; 
\citet{due87novaatlas} had also proposed the same 22-nd mag counterpart].  
After a long period of unsuccessful detection of an outburst,
R. Stubbings finally detected an outburst at a visual magnitude of 15.0
on 2011 August 2 (vsnet-outburst 13091).  The outburst was immediately
confirmed by T. Tordai and G. Masi who detected superhumps
(vsnet-alert 13541, 13552; figure \ref{fig:svarishpdm}).

   The times of superhump maxima are listed in table \ref{tab:svarioc2011}
The early to middle portion of the $O-C$ diagram shows clear
stages of A and B.  During the period of BJD 2455789--2477912, there
were sometimes two hump maxima, and humps with phases different
from main humps ($E=240, 253, 272$) were also included in the table.
There was some indication of a stage C around the terminal stage
($E \ge 364$).  The values given in table \ref{tab:perlist} were
determined after rejecting humps at $E=240, 253, 272$.
The resultant $P_{\rm dot}$ for stage B superhumps was
$+4.0(0.2) \times 10^{-5}$, comparable to those of extreme
WZ Sge-type dwarf novae.

   The 2011 outburst was much fainter than the 1905 outburst.
This difference can be understood as a combination of two effects:
(1) the magnitude scale in \citet{wol05svari} was about 2 mag
brighter than the present scale, which is confirmed from a comparison
of the magnitudes of the comparison stars, and (2) the brightness
maximum of the 2011 outburst was missed because there were
no observations in that season before the Stubbings' detection.
The lack of a stage of early superhumps, which is expected
for such a WZ Sge-type dwarf nova, can also be understood
for the same reason.  No post-superoutburst rebrightening was
recorded.

\begin{table}
\caption{Superhump maxima of SV Ari (2011).}\label{tab:svarioc2011}
\begin{center}
\begin{tabular}{ccccc}
\hline
$E$ & max\commenta & error & $O-C$\commentb & $N$\commentc \\
\hline
0 & 55776.5824 & 0.0026 & 0.0059 & 66 \\
5 & 55776.8591 & 0.0010 & 0.0052 & 47 \\
7 & 55776.9722 & 0.0011 & 0.0072 & 99 \\
11 & 55777.1956 & 0.0009 & 0.0086 & 250 \\
12 & 55777.2464 & 0.0007 & 0.0039 & 358 \\
13 & 55777.3083 & 0.0014 & 0.0103 & 154 \\
16 & 55777.4748 & 0.0013 & 0.0104 & 41 \\
17 & 55777.5259 & 0.0015 & 0.0059 & 30 \\
18 & 55777.5860 & 0.0006 & 0.0106 & 49 \\
19 & 55777.6419 & 0.0003 & 0.0109 & 37 \\
23 & 55777.8629 & 0.0003 & 0.0099 & 53 \\
24 & 55777.9167 & 0.0004 & 0.0082 & 38 \\
30 & 55778.2477 & 0.0010 & 0.0062 & 84 \\
34 & 55778.4729 & 0.0014 & 0.0095 & 32 \\
35 & 55778.5252 & 0.0009 & 0.0063 & 31 \\
36 & 55778.5792 & 0.0003 & 0.0048 & 91 \\
37 & 55778.6356 & 0.0004 & 0.0056 & 50 \\
41 & 55778.8563 & 0.0006 & 0.0043 & 34 \\
43 & 55778.9637 & 0.0010 & 0.0008 & 28 \\
54 & 55779.5730 & 0.0003 & $-$0.0004 & 38 \\
59 & 55779.8498 & 0.0005 & $-$0.0012 & 36 \\
60 & 55779.9032 & 0.0005 & $-$0.0033 & 28 \\
66 & 55780.2370 & 0.0028 & $-$0.0024 & 186 \\
67 & 55780.2891 & 0.0016 & $-$0.0059 & 167 \\
77 & 55780.8425 & 0.0010 & $-$0.0074 & 29 \\
78 & 55780.9012 & 0.0010 & $-$0.0043 & 75 \\
79 & 55780.9551 & 0.0008 & $-$0.0059 & 70 \\
83 & 55781.1839 & 0.0031 & 0.0009 & 100 \\
84 & 55781.2324 & 0.0012 & $-$0.0060 & 224 \\
85 & 55781.2897 & 0.0008 & $-$0.0043 & 252 \\
95 & 55781.8415 & 0.0005 & $-$0.0074 & 36 \\
96 & 55781.8981 & 0.0005 & $-$0.0064 & 32 \\
102 & 55782.2312 & 0.0010 & $-$0.0063 & 238 \\
103 & 55782.2838 & 0.0006 & $-$0.0092 & 288 \\
109 & 55782.6183 & 0.0006 & $-$0.0077 & 44 \\
113 & 55782.8402 & 0.0005 & $-$0.0078 & 35 \\
114 & 55782.8950 & 0.0006 & $-$0.0085 & 35 \\
127 & 55783.6165 & 0.0005 & $-$0.0085 & 51 \\
131 & 55783.8384 & 0.0008 & $-$0.0086 & 35 \\
132 & 55783.8941 & 0.0007 & $-$0.0083 & 35 \\
145 & 55784.6161 & 0.0007 & $-$0.0079 & 51 \\
150 & 55784.8916 & 0.0013 & $-$0.0099 & 65 \\
162 & 55785.5544 & 0.0037 & $-$0.0131 & 26 \\
163 & 55785.6194 & 0.0010 & $-$0.0036 & 49 \\
167 & 55785.8358 & 0.0019 & $-$0.0092 & 30 \\
168 & 55785.8944 & 0.0010 & $-$0.0061 & 30 \\
185 & 55786.8424 & 0.0035 & $-$0.0016 & 28 \\
186 & 55786.8891 & 0.0035 & $-$0.0104 & 19 \\
192 & 55787.2249 & 0.0039 & $-$0.0076 & 105 \\
203 & 55787.8394 & 0.0019 & $-$0.0036 & 28 \\
204 & 55787.8962 & 0.0024 & $-$0.0023 & 28 \\
\hline
  \multicolumn{5}{l}{\commenta BJD$-$2400000.} \\
  \multicolumn{5}{l}{\commentb Against max $= 2455776.5764 + 0.055500 E$.} \\
  \multicolumn{5}{l}{\commentc Number of points used to determine the maximum.} \\
\end{tabular}
\end{center}
\end{table}

\addtocounter{table}{-1}
\begin{table}
\caption{Superhump maxima of SV Ari (2011) (continued).}
\begin{center}
\begin{tabular}{ccccc}
\hline
$E$ & max\commenta & error & $O-C$\commentb & $N$\commentc \\
\hline
222 & 55788.8967 & 0.0043 & $-$0.0008 & 29 \\
240 & 55789.8813 & 0.0061 & $-$0.0152 & 20 \\
252 & 55790.5667 & 0.0027 & 0.0042 & 51 \\
253 & 55790.5904 & 0.0023 & $-$0.0276 & 51 \\
253 & 55790.6199 & 0.0007 & 0.0020 & 37 \\
257 & 55790.8503 & 0.0093 & 0.0103 & 19 \\
270 & 55791.5773 & 0.0049 & 0.0158 & 51 \\
272 & 55791.6554 & 0.0045 & $-$0.0171 & 29 \\
307 & 55793.6346 & 0.0020 & 0.0196 & 22 \\
310 & 55793.7952 & 0.0024 & 0.0137 & 30 \\
311 & 55793.8603 & 0.0009 & 0.0233 & 40 \\
364 & 55796.7895 & 0.0039 & 0.0110 & 20 \\
365 & 55796.8452 & 0.0034 & 0.0112 & 20 \\
366 & 55796.8987 & 0.0016 & 0.0092 & 16 \\
\hline
  \multicolumn{5}{l}{\commenta BJD$-$2400000.} \\
  \multicolumn{5}{l}{\commentb Against max $= 2455776.5764 + 0.055500 E$.} \\
  \multicolumn{5}{l}{\commentc Number of points used to determine the maximum.} \\
\end{tabular}
\end{center}
\end{table}

\subsection{TT Bootis}\label{obj:ttboo}

\begin{figure}
  \begin{center}
    \FigureFile(88mm,70mm){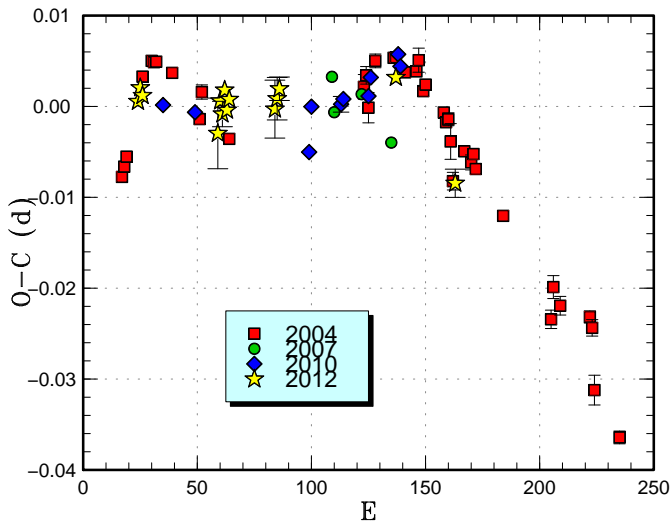}
  \end{center}
  \caption{Comparison of $O-C$ diagrams of TT Boo between different
  superoutbursts.  A period of 0.07807~d was used to draw this figure.
  Approximate cycle counts ($E$) after the start of the superoutburst
  were used.
  }
  \label{fig:ttboocomp2}
\end{figure}

   We observed the early part of the 2012 superoutburst.
The times of superhump maxima are listed in table \ref{tab:ttboooc2012}.
There were no detectable superhumps 0.8~d prior to the initial
epoch of superhump maximum.  The resultant $P_{\rm dot}$ for stage B
was smaller than in 2004 and 2010, and it is probably a result
of the limited observation of stage B and possibly from contamination 
of stage A or C superhumps (figure \ref{fig:ttboocomp2}).

\begin{table}
\caption{Superhump maxima of TT Boo (2012).}\label{tab:ttboooc2012}
\begin{center}
\begin{tabular}{ccccc}
\hline
$E$ & max\commenta & error & $O-C$\commentb & $N$\commentc \\
\hline
0 & 56016.4443 & 0.0002 & $-$0.0010 & 81 \\
1 & 56016.5239 & 0.0002 & 0.0005 & 83 \\
2 & 56016.6011 & 0.0002 & $-$0.0003 & 83 \\
35 & 56019.1732 & 0.0039 & $-$0.0034 & 30 \\
36 & 56019.2549 & 0.0011 & 0.0002 & 56 \\
37 & 56019.3315 & 0.0014 & $-$0.0012 & 42 \\
38 & 56019.4122 & 0.0005 & 0.0015 & 73 \\
39 & 56019.4881 & 0.0005 & $-$0.0007 & 78 \\
40 & 56019.5673 & 0.0005 & 0.0005 & 82 \\
60 & 56021.1277 & 0.0032 & 0.0001 & 44 \\
61 & 56021.2069 & 0.0023 & 0.0013 & 58 \\
62 & 56021.2861 & 0.0013 & 0.0024 & 54 \\
113 & 56025.2689 & 0.0007 & 0.0054 & 56 \\
139 & 56027.2870 & 0.0016 & $-$0.0054 & 55 \\
\hline
  \multicolumn{5}{l}{\commenta BJD$-$2400000.} \\
  \multicolumn{5}{l}{\commentb Against max $= 2456016.4454 + 0.078036 E$.} \\
  \multicolumn{5}{l}{\commentc Number of points used to determine the maximum.} \\
\end{tabular}
\end{center}
\end{table}

\subsection{CR Bootis}\label{obj:crboo}

   CR Boo is one of the prototypical ``helium dwarf novae''
[\citet{pat97crboo}; \citet{pro97crboo}; \citet{kat00crboo};
for representative theoretical analyses, see \citet{tsu97amcvn}, 
\citet{kot12amcvnoutburst}; for recent observational reviews 
of AM CVn stars, see \citet{sol10amcvnreview}, \citet{ram11amcvn},
\citet{ram12amcvnLC}].
Although superhumps in this object was
well established in the past, the published observations
were either obtained in an anomalous state \citep{pat97crboo}
or not very ideally sampled \citep{kat00crboo}.  The object
was in a state of regular pattern of outbursts (cf. \cite{kat00crboo})
with a supercycle of $\sim$50~d in 2011--2012 and is
ideal to study the behavior of superhumps in helium dwarf novae.

   We present here an analysis of a superoutburst in 2012 March
mainly using the AAVSO observations.
The superoutburst was first detected by G. Gualdoni on March 3
at $V=13.61$ (AAVSO data).  The existence of superhumps was
soon recognized (vsnet-alert 14305).  Although the object stayed
in its plateau phase for six days, it started oscillations
with a quasi-period of 1.0~d similar to \citet{pat97crboo},
and this state lasted for six days.  The object apparently
entered a more stable state, and finally started fading rapidly
on March 25.  Although the overall behavior of the superoutburst
was similar to those of hydrogen-rich SU UMa-type dwarf novae,
the presence of oscillatory state is different.  The relatively
large scatter in the supercycle-phase-folded light curve
(figure 4 of \cite{kat00crboo}) may have been a result of
these oscillations.

\begin{figure}
  \begin{center}
    \FigureFile(88mm,110mm){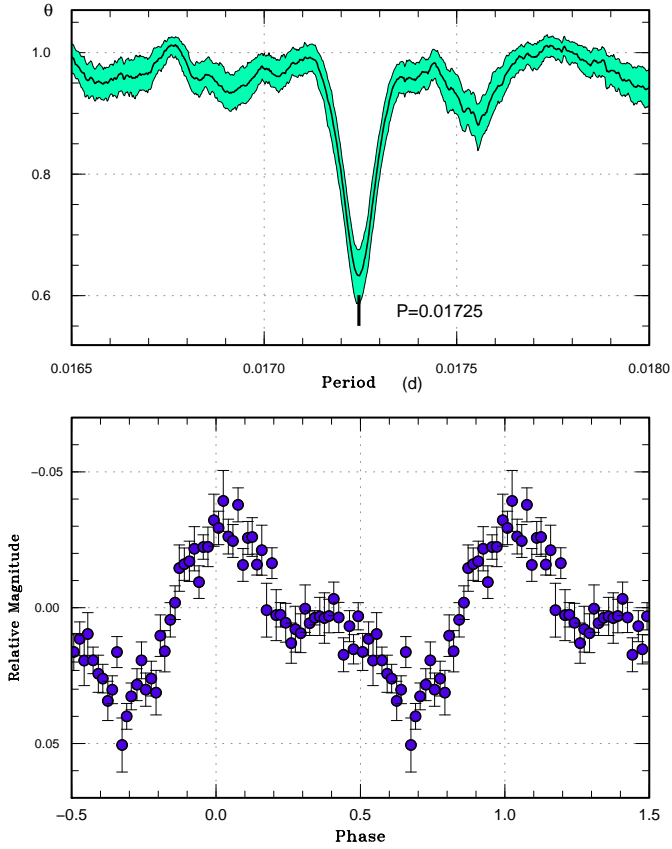}
  \end{center}
  \caption{Superhumps in CR Boo (2012 March) before the oscillatory phase. 
     (Upper): PDM analysis.
     (Lower): Phase-averaged profile.}
  \label{fig:crbooshpdm1}
\end{figure}

\begin{figure}
  \begin{center}
    \FigureFile(88mm,110mm){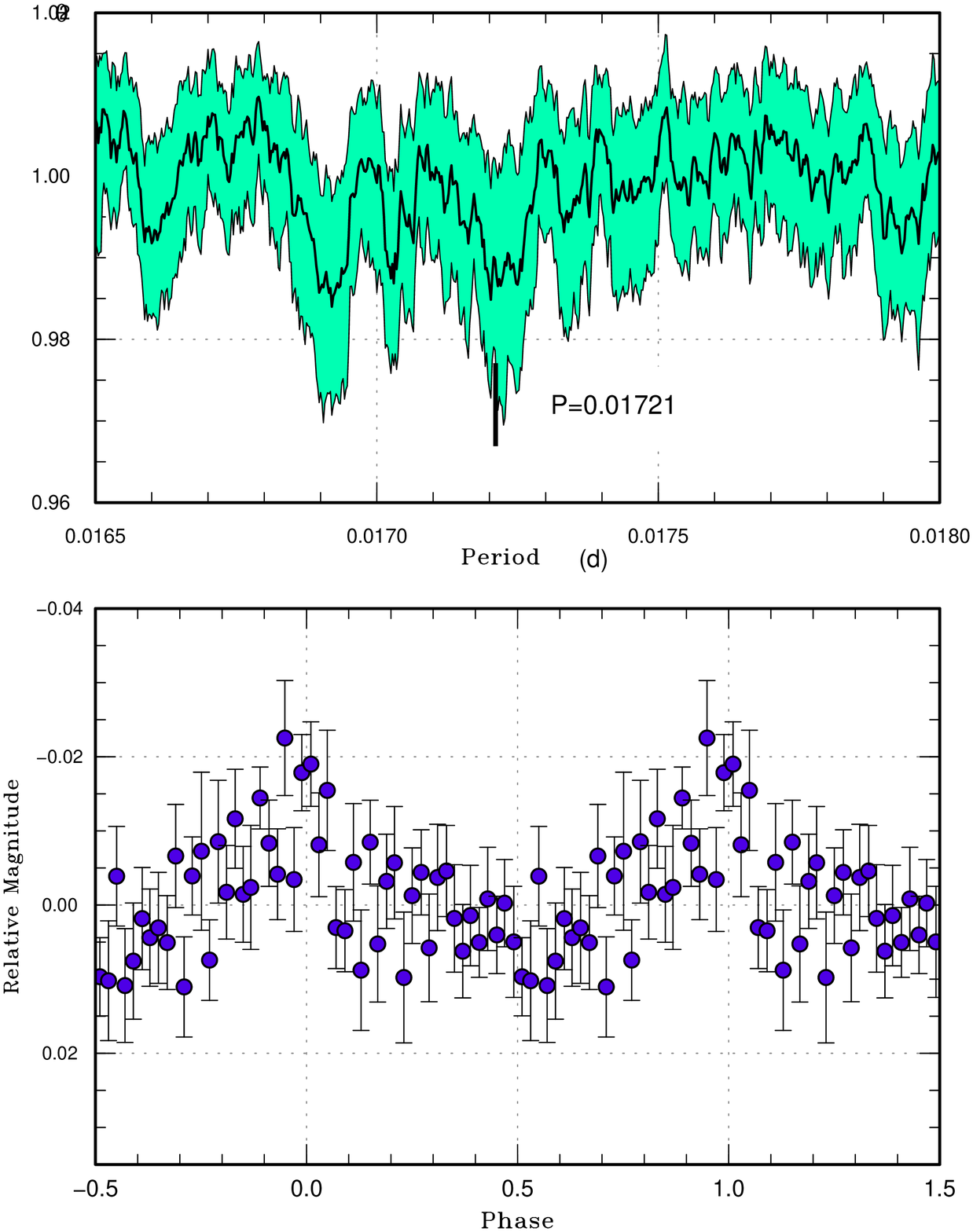}
  \end{center}
  \caption{Superhumps in CR Boo (2012 March) during the oscillatory phase. 
     (Upper): PDM analysis.
     (Lower): Phase-averaged profile.}
  \label{fig:crbooshpdm2}
\end{figure}

\begin{figure}
  \begin{center}
    \FigureFile(88mm,100mm){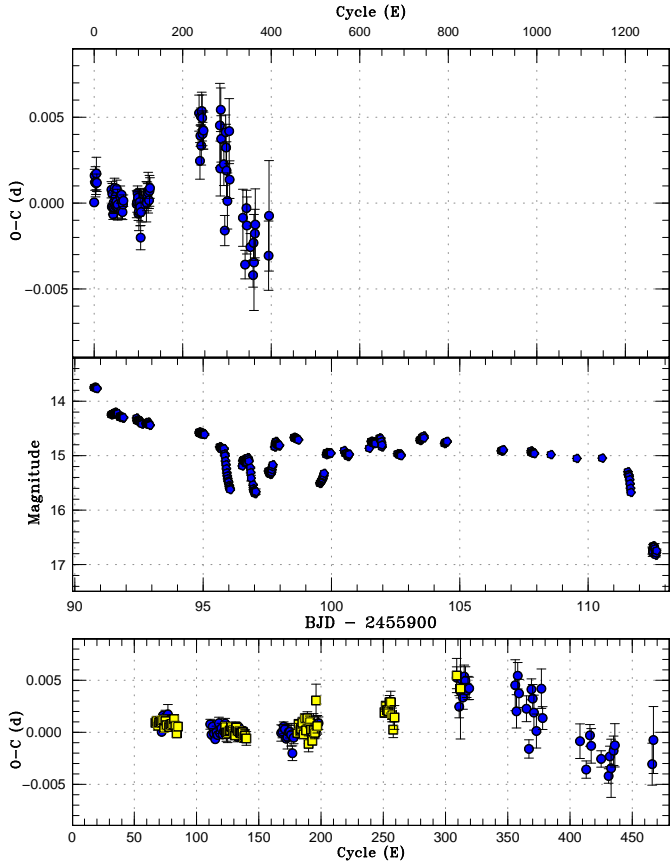}
  \end{center}
  \caption{$O-C$ diagram of superhumps in CR Boo.
     (Upper:) $O-C$ for the 2012 March superoutburst.
     We used a period of 0.017249~d for calculating the $O-C$ residuals.
     (Middle:) Light curve for the 2012 March superoutburst.
     (Lower:) Comparison of $O-C$ diagrams between two superoutbursts
     in 2012 March (filled circles) and April (filled squares).
     Approximate cycle counts ($E$) after the start of the superoutburst
     were used.
  }
  \label{fig:crboohumpall}
\end{figure}

   The times of superhump maxima until the early stage of
the oscillatory state are shown in table \ref{tab:crboooc2012}.
The $O-C$ diagram shows a pattern very similar to stages B and C
in hydrogen-rich SU UMa-type dwarf novae.  The $P_{\rm dot}$
for stage B was $+2.0(0.2) \times 10^{-5}$ and the $\epsilon$
for stage B and C superhumps (figures \ref{fig:crbooshpdm1}
and \ref{fig:crbooshpdm2}, respectively) were 1.39(1)\%, and 0.97(4)\%,
similar to those of WZ Sge-type dwarf novae, but are larger
than what is expected only from the mass-ratio.
The stage B--C transition occurred when the oscillation started
(figure \ref{fig:crboohumpall}).
This may be analogous to WZ Sge-type dwarf novae, which usually do not
show stage C superhumps by the end of plateau phase
(\cite{Pdot}; \cite{Pdot2}; \cite{Pdot3}).  The oscillatory
phase in CR Boo may correspond to post-superoutburst stage
in WZ Sge-type dwarf novae, when these objects tend to show
various kinds of rebrightenings (cf. \cite{Pdot}).
We might recall the past examples of V803 Cen \citep{kat04v803cen}
and V406 Hya \citep{nog04v406hya}, both of which showed rebrightenings
similar to WZ Sge-type dwarf novae.  \citet{ram11amcvn} also
noted the presence of a ``dip'' during superoutbursts of
short-$P_{\rm orb}$ AM CVn-type objects (see also \cite{lev11j0719}; 
\cite{kot12amcvnoutburst}).  Such phenomena may be more
prevalent than had been thought.

   Although the superhumps in the later stages were not readily
recognizable, we could detect the period with the PDM method:
0.017183(5)~d for BJD 2455997.8--2456002.0 (oscillatory phase)
and 0.017265(3)~d for BJD 2456002.0--2456009.0 (second stable
plateau).  These periods indicate the persistence of superhumps
until the end of the superoutburst.

   After one supercycle, the object underwent another superoutburst
in 2012 April.  The times of superhump maxima are listed in
table \ref{tab:crboooc2012b}.  Although the object started oscillatory
behavior as in the March superoutburst, the later part of the
superoutburst was not as well observed as in the March one.  The resultant
period and period derivatives were quite similar to those
of the March superoutburst.  The $O-C$ diagram of the stage B
very well reproduced that of the March superoutburst
(figure \ref{fig:crboohumpall}, lower panel).

\begin{table}
\caption{Superhump maxima of CR Boo (2012 March).}\label{tab:crboooc2012}
\begin{center}
\begin{tabular}{ccccc}
\hline
$E$ & max\commenta & error & $O-C$\commentb & $N$\commentc \\
\hline
0 & 55990.7536 & 0.0001 & $-$0.0006 & 5 \\
1 & 55990.7724 & 0.0004 & 0.0010 & 6 \\
2 & 55990.7893 & 0.0005 & 0.0006 & 6 \\
3 & 55990.8066 & 0.0009 & 0.0006 & 5 \\
4 & 55990.8237 & 0.0007 & 0.0006 & 6 \\
5 & 55990.8416 & 0.0009 & 0.0011 & 6 \\
6 & 55990.8583 & 0.0005 & 0.0006 & 8 \\
39 & 55991.4270 & 0.0005 & 0.0001 & 14 \\
40 & 55991.4433 & 0.0003 & $-$0.0008 & 17 \\
41 & 55991.4613 & 0.0006 & $-$0.0001 & 16 \\
42 & 55991.4779 & 0.0005 & $-$0.0008 & 16 \\
43 & 55991.4946 & 0.0003 & $-$0.0013 & 16 \\
44 & 55991.5125 & 0.0006 & $-$0.0006 & 12 \\
45 & 55991.5295 & 0.0003 & $-$0.0009 & 14 \\
46 & 55991.5479 & 0.0006 & 0.0003 & 13 \\
47 & 55991.5641 & 0.0004 & $-$0.0007 & 8 \\
48 & 55991.5819 & 0.0005 & $-$0.0003 & 12 \\
49 & 55991.5985 & 0.0003 & $-$0.0009 & 15 \\
50 & 55991.6161 & 0.0004 & $-$0.0006 & 16 \\
51 & 55991.6341 & 0.0005 & 0.0002 & 15 \\
52 & 55991.6503 & 0.0007 & $-$0.0008 & 17 \\
53 & 55991.6677 & 0.0009 & $-$0.0007 & 14 \\
59 & 55991.7716 & 0.0006 & $-$0.0003 & 7 \\
60 & 55991.7888 & 0.0003 & $-$0.0003 & 6 \\
61 & 55991.8053 & 0.0005 & $-$0.0011 & 6 \\
62 & 55991.8235 & 0.0002 & $-$0.0002 & 5 \\
63 & 55991.8403 & 0.0002 & $-$0.0006 & 6 \\
64 & 55991.8569 & 0.0004 & $-$0.0012 & 7 \\
65 & 55991.8746 & 0.0002 & $-$0.0007 & 7 \\
66 & 55991.8921 & 0.0003 & $-$0.0005 & 6 \\
96 & 55992.4093 & 0.0008 & $-$0.0007 & 28 \\
97 & 55992.4267 & 0.0006 & $-$0.0006 & 33 \\
98 & 55992.4445 & 0.0004 & $-$0.0001 & 30 \\
99 & 55992.4615 & 0.0004 & $-$0.0004 & 30 \\
100 & 55992.4778 & 0.0004 & $-$0.0013 & 22 \\
101 & 55992.4952 & 0.0011 & $-$0.0012 & 21 \\
102 & 55992.5129 & 0.0008 & $-$0.0007 & 20 \\
103 & 55992.5302 & 0.0006 & $-$0.0006 & 24 \\
104 & 55992.5472 & 0.0011 & $-$0.0009 & 19 \\
105 & 55992.5626 & 0.0007 & $-$0.0027 & 17 \\
106 & 55992.5814 & 0.0006 & $-$0.0012 & 18 \\
118 & 55992.7889 & 0.0012 & $-$0.0006 & 15 \\
119 & 55992.8061 & 0.0004 & $-$0.0007 & 13 \\
120 & 55992.8238 & 0.0008 & $-$0.0002 & 15 \\
121 & 55992.8413 & 0.0004 & 0.0000 & 16 \\
122 & 55992.8585 & 0.0012 & $-$0.0001 & 15 \\
124 & 55992.8925 & 0.0005 & $-$0.0005 & 15 \\
125 & 55992.9103 & 0.0009 & 0.0000 & 15 \\
126 & 55992.9277 & 0.0006 & 0.0002 & 12 \\
237 & 55994.8466 & 0.0010 & 0.0045 & 16 \\
239 & 55994.8783 & 0.0011 & 0.0017 & 15 \\
240 & 55994.8970 & 0.0006 & 0.0031 & 16 \\
\hline
  \multicolumn{5}{l}{\commenta BJD$-$2400000.} \\
  \multicolumn{5}{l}{\commentb Against max $= 2455990.7542 + 0.017249 E$.} \\
  \multicolumn{5}{l}{\commentc Number of points used to determine the maximum.} \\
\end{tabular}
\end{center}
\end{table}

\addtocounter{table}{-1}
\begin{table}
\caption{Superhump maxima of CR Boo (2012 March) (continued).}
\begin{center}
\begin{tabular}{ccccc}
\hline
$E$ & max\commenta & error & $O-C$\commentb & $N$\commentc \\
\hline
241 & 55994.9154 & 0.0005 & 0.0043 & 16 \\
242 & 55994.9310 & 0.0011 & 0.0026 & 14 \\
243 & 55994.9502 & 0.0011 & 0.0046 & 25 \\
244 & 55994.9671 & 0.0013 & 0.0042 & 24 \\
245 & 55994.9834 & 0.0007 & 0.0032 & 25 \\
246 & 55995.0009 & 0.0011 & 0.0035 & 25 \\
247 & 55995.0181 & 0.0011 & 0.0034 & 13 \\
284 & 55995.6566 & 0.0024 & 0.0037 & 21 \\
285 & 55995.6713 & 0.0016 & 0.0012 & 21 \\
286 & 55995.6920 & 0.0013 & 0.0046 & 21 \\
287 & 55995.7075 & 0.0014 & 0.0029 & 22 \\
293 & 55995.8095 & 0.0012 & 0.0015 & 6 \\
295 & 55995.8402 & 0.0009 & $-$0.0024 & 11 \\
297 & 55995.8804 & 0.0010 & 0.0033 & 14 \\
298 & 55995.8967 & 0.0014 & 0.0024 & 15 \\
299 & 55995.9126 & 0.0017 & 0.0011 & 14 \\
301 & 55995.9454 & 0.0016 & $-$0.0007 & 24 \\
305 & 55996.0184 & 0.0019 & 0.0034 & 9 \\
306 & 55996.0329 & 0.0011 & 0.0005 & 9 \\
336 & 55996.5481 & 0.0017 & $-$0.0017 & 22 \\
341 & 55996.6316 & 0.0008 & $-$0.0044 & 17 \\
344 & 55996.6866 & 0.0010 & $-$0.0011 & 37 \\
345 & 55996.7029 & 0.0017 & $-$0.0022 & 36 \\
353 & 55996.8396 & 0.0008 & $-$0.0034 & 10 \\
359 & 55996.9414 & 0.0007 & $-$0.0051 & 12 \\
360 & 55996.9606 & 0.0010 & $-$0.0032 & 7 \\
361 & 55996.9767 & 0.0028 & $-$0.0043 & 15 \\
363 & 55997.0128 & 0.0014 & $-$0.0026 & 9 \\
364 & 55997.0306 & 0.0021 & $-$0.0021 & 8 \\
394 & 55997.5462 & 0.0020 & $-$0.0039 & 16 \\
395 & 55997.5658 & 0.0032 & $-$0.0016 & 18 \\
\hline
  \multicolumn{5}{l}{\commenta BJD$-$2400000.} \\
  \multicolumn{5}{l}{\commentb Against max $= 2455990.7542 + 0.017249 E$.} \\
  \multicolumn{5}{l}{\commentc Number of points used to determine the maximum.} \\
\end{tabular}
\end{center}
\end{table}

\begin{table}
\caption{Superhump maxima of CR Boo (2012 April).}\label{tab:crboooc2012b}
\begin{center}
\begin{tabular}{ccccc}
\hline
$E$ & max\commenta & error & $O-C$\commentb & $N$\commentc \\
\hline
0 & 56039.6483 & 0.0002 & 0.0009 & 21 \\
1 & 56039.6657 & 0.0002 & 0.0011 & 33 \\
2 & 56039.6825 & 0.0001 & 0.0006 & 33 \\
3 & 56039.7001 & 0.0003 & 0.0009 & 33 \\
4 & 56039.7175 & 0.0001 & 0.0010 & 34 \\
5 & 56039.7346 & 0.0002 & 0.0009 & 34 \\
6 & 56039.7521 & 0.0002 & 0.0011 & 34 \\
7 & 56039.7685 & 0.0002 & 0.0003 & 34 \\
8 & 56039.7865 & 0.0001 & 0.0010 & 33 \\
9 & 56039.8033 & 0.0002 & 0.0006 & 34 \\
11 & 56039.8376 & 0.0002 & 0.0004 & 24 \\
12 & 56039.8551 & 0.0002 & 0.0006 & 32 \\
13 & 56039.8724 & 0.0002 & 0.0006 & 33 \\
14 & 56039.8896 & 0.0003 & 0.0006 & 33 \\
15 & 56039.9074 & 0.0002 & 0.0011 & 33 \\
16 & 56039.9240 & 0.0002 & 0.0005 & 33 \\
17 & 56039.9405 & 0.0003 & $-$0.0003 & 33 \\
18 & 56039.9584 & 0.0002 & 0.0004 & 33 \\
56 & 56040.6138 & 0.0005 & 0.0000 & 15 \\
57 & 56040.6305 & 0.0004 & $-$0.0005 & 16 \\
58 & 56040.6477 & 0.0003 & $-$0.0006 & 16 \\
59 & 56040.6652 & 0.0004 & $-$0.0004 & 15 \\
60 & 56040.6825 & 0.0003 & $-$0.0004 & 14 \\
61 & 56040.7001 & 0.0004 & $-$0.0000 & 16 \\
62 & 56040.7170 & 0.0003 & $-$0.0003 & 16 \\
63 & 56040.7340 & 0.0005 & $-$0.0006 & 14 \\
64 & 56040.7510 & 0.0003 & $-$0.0009 & 16 \\
65 & 56040.7690 & 0.0005 & $-$0.0001 & 13 \\
66 & 56040.7859 & 0.0004 & $-$0.0005 & 14 \\
67 & 56040.8031 & 0.0003 & $-$0.0005 & 15 \\
68 & 56040.8202 & 0.0003 & $-$0.0006 & 15 \\
69 & 56040.8374 & 0.0003 & $-$0.0007 & 14 \\
70 & 56040.8548 & 0.0005 & $-$0.0006 & 15 \\
71 & 56040.8715 & 0.0003 & $-$0.0011 & 13 \\
72 & 56040.8890 & 0.0006 & $-$0.0009 & 13 \\
73 & 56040.9059 & 0.0007 & $-$0.0012 & 15 \\
114 & 56041.6136 & 0.0004 & $-$0.0011 & 16 \\
115 & 56041.6317 & 0.0005 & $-$0.0003 & 16 \\
116 & 56041.6483 & 0.0007 & $-$0.0009 & 16 \\
117 & 56041.6657 & 0.0009 & $-$0.0007 & 13 \\
118 & 56041.6836 & 0.0005 & $-$0.0001 & 15 \\
119 & 56041.6997 & 0.0004 & $-$0.0013 & 15 \\
120 & 56041.7185 & 0.0004 & 0.0003 & 15 \\
121 & 56041.7346 & 0.0003 & $-$0.0009 & 11 \\
122 & 56041.7530 & 0.0006 & 0.0003 & 15 \\
123 & 56041.7678 & 0.0008 & $-$0.0022 & 14 \\
124 & 56041.7872 & 0.0006 & $-$0.0001 & 16 \\
125 & 56041.8037 & 0.0006 & $-$0.0008 & 15 \\
126 & 56041.8198 & 0.0004 & $-$0.0019 & 12 \\
127 & 56041.8381 & 0.0005 & $-$0.0009 & 16 \\
128 & 56041.8551 & 0.0006 & $-$0.0012 & 15 \\
129 & 56041.8754 & 0.0016 & 0.0019 & 14 \\
\hline
  \multicolumn{5}{l}{\commenta BJD$-$2400000.} \\
  \multicolumn{5}{l}{\commentb Against max $= 2456039.6474 + 0.017257 E$.} \\
  \multicolumn{5}{l}{\commentc Number of points used to determine the maximum.} \\
\end{tabular}
\end{center}
\end{table}

\addtocounter{table}{-1}
\begin{table}
\caption{Superhump maxima of CR Boo (2012 April) (continued).}
\begin{center}
\begin{tabular}{ccccc}
\hline
$E$ & max\commenta & error & $O-C$\commentb & $N$\commentc \\
\hline
130 & 56041.8902 & 0.0007 & $-$0.0006 & 15 \\
184 & 56042.8229 & 0.0005 & 0.0003 & 15 \\
185 & 56042.8408 & 0.0005 & 0.0009 & 17 \\
186 & 56042.8577 & 0.0004 & 0.0005 & 18 \\
187 & 56042.8748 & 0.0003 & 0.0004 & 17 \\
188 & 56042.8929 & 0.0006 & 0.0012 & 18 \\
189 & 56042.9102 & 0.0010 & 0.0012 & 17 \\
190 & 56042.9262 & 0.0007 & 0.0000 & 18 \\
191 & 56042.9420 & 0.0008 & $-$0.0014 & 18 \\
192 & 56042.9604 & 0.0011 & $-$0.0003 & 18 \\
242 & 56043.8269 & 0.0016 & 0.0033 & 16 \\
245 & 56043.8774 & 0.0048 & 0.0021 & 15 \\
\hline
  \multicolumn{5}{l}{\commenta BJD$-$2400000.} \\
  \multicolumn{5}{l}{\commentb Against max $= 2456039.6474 + 0.017257 E$.} \\
  \multicolumn{5}{l}{\commentc Number of points used to determine the maximum.} \\
\end{tabular}
\end{center}
\end{table}

\subsection{NN Camelopardalis}\label{obj:nncam}

   We observed a superoutburst in 2011 December.  The times of
superhump maxima are listed in table \ref{tab:nncamoc2011}.
Although stage A and early part of stage B were missed, a clear 
pattern of stage B--C superhumps was detected.

\begin{figure}
  \begin{center}
    \FigureFile(88mm,70mm){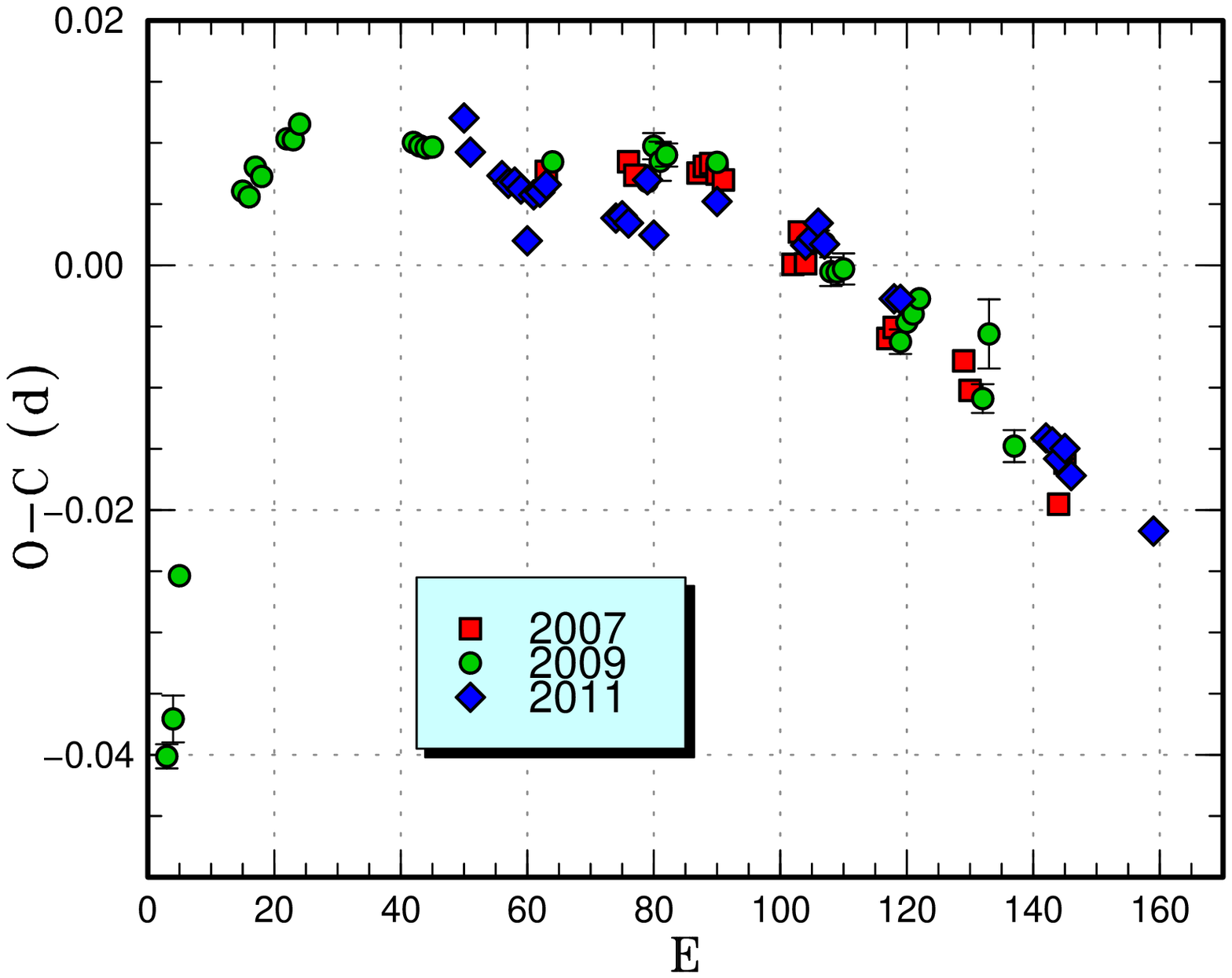}
  \end{center}
  \caption{Comparison of $O-C$ diagrams of NN Cam between different
  superoutbursts.  A period of 0.0743~d was used to draw this figure.
  Approximate cycle counts ($E$) after the start of the superoutburst
  were used.  The 2007 superoutburst was shifted by 63 cycles to 
  best match the others.
  }
  \label{fig:nncamcomp}
\end{figure}

   A comparison of $O-C$ diagrams between different superoutbursts 
is shown in figure \ref{fig:nncamcomp}.  The 2007 superoutburst, 
whose start of the main superoutburst was not observed, was shifted 
by 63 cycles to best match the others.  This cycle count placed 
the initial epoch of superhump evolution around BJD 2454358.9, 
shortly after the precursor outburst.  It was likely that superhumps 
started to grow just following the precursor outburst, and it is likely
the true start of the main superoutburst was missed.

\begin{table}
\caption{Superhump maxima of NN Cam (2011).}\label{tab:nncamoc2011}
\begin{center}
\begin{tabular}{ccccc}
\hline
$E$ & max\commenta & error & $O-C$\commentb & $N$\commentc \\
\hline
0 & 55904.9897 & 0.0002 & 0.0015 & 254 \\
1 & 55905.0612 & 0.0001 & $-$0.0010 & 241 \\
6 & 55905.4308 & 0.0005 & $-$0.0017 & 75 \\
7 & 55905.5044 & 0.0005 & $-$0.0021 & 63 \\
8 & 55905.5788 & 0.0004 & $-$0.0017 & 68 \\
9 & 55905.6526 & 0.0005 & $-$0.0020 & 56 \\
10 & 55905.7226 & 0.0010 & $-$0.0060 & 42 \\
11 & 55905.8007 & 0.0004 & $-$0.0020 & 78 \\
12 & 55905.8752 & 0.0004 & $-$0.0016 & 78 \\
13 & 55905.9501 & 0.0004 & $-$0.0007 & 76 \\
24 & 55906.7647 & 0.0005 & $-$0.0007 & 77 \\
25 & 55906.8392 & 0.0004 & $-$0.0003 & 77 \\
26 & 55906.9129 & 0.0005 & $-$0.0006 & 78 \\
29 & 55907.1393 & 0.0002 & 0.0037 & 294 \\
30 & 55907.2091 & 0.0003 & $-$0.0006 & 294 \\
40 & 55907.9548 & 0.0003 & 0.0046 & 261 \\
54 & 55908.9915 & 0.0002 & 0.0045 & 279 \\
55 & 55909.0664 & 0.0003 & 0.0053 & 433 \\
56 & 55909.1419 & 0.0005 & 0.0068 & 216 \\
57 & 55909.2145 & 0.0004 & 0.0053 & 156 \\
68 & 55910.0273 & 0.0004 & 0.0036 & 156 \\
69 & 55910.1016 & 0.0005 & 0.0038 & 101 \\
92 & 55911.7991 & 0.0005 & $-$0.0019 & 79 \\
93 & 55911.8731 & 0.0007 & $-$0.0020 & 78 \\
94 & 55911.9460 & 0.0003 & $-$0.0031 & 307 \\
95 & 55912.0212 & 0.0004 & $-$0.0020 & 236 \\
96 & 55912.0932 & 0.0006 & $-$0.0040 & 159 \\
109 & 55913.0546 & 0.0007 & $-$0.0053 & 135 \\
\hline
  \multicolumn{5}{l}{\commenta BJD$-$2400000.} \\
  \multicolumn{5}{l}{\commentb Against max $= 2455904.9881 + 0.074053 E$.} \\
  \multicolumn{5}{l}{\commentc Number of points used to determine the maximum.} \\
\end{tabular}
\end{center}
\end{table}

\subsection{SY Capricorni}\label{obj:sycap}

   We observed a superoutburst in 2011 August--September.
The times of superhump are listed in table \ref{tab:sycapoc2011}.
Since only a limited fragment of observation was obtained,
we adopted a period with the PDM analysis in table \ref{tab:perlist}.

\begin{table}
\caption{Superhump maxima of SY Cap (2011).}\label{tab:sycapoc2011}
\begin{center}
\begin{tabular}{ccccc}
\hline
$E$ & max\commenta & error & $O-C$\commentb & $N$\commentc \\
\hline
0 & 55803.0789 & 0.0005 & 0.0012 & 172 \\
1 & 55803.1409 & 0.0006 & $-$0.0005 & 154 \\
16 & 55804.0964 & 0.0017 & $-$0.0014 & 43 \\
31 & 55805.0550 & 0.0007 & 0.0007 & 50 \\
\hline
  \multicolumn{5}{l}{\commenta BJD$-$2400000.} \\
  \multicolumn{5}{l}{\commentb Against max $= 2455803.0777 + 0.063761 E$.} \\
  \multicolumn{5}{l}{\commentc Number of points used to determine the maximum.} \\
\end{tabular}
\end{center}
\end{table}

\subsection{GZ Ceti}\label{obj:gzcet}

\begin{figure}
  \begin{center}
    \FigureFile(88mm,70mm){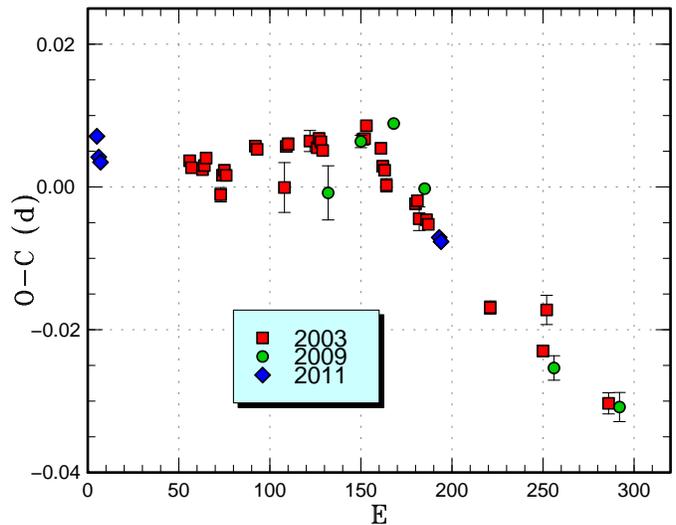}
  \end{center}
  \caption{Comparison of $O-C$ diagrams of GZ Cet between different
  superoutbursts.  A period of 0.05672~d was used to draw this figure.
  Approximate cycle counts ($E$) after the start of the superoutburst
  were used.  We assumed that the 2011 superoutburst was caught around
  its peak based on the brightness and evolution of superhumps,
  and assumed it to be the start of the superoutburst.
  }
  \label{fig:gzcetcomp}
\end{figure}

   This object (=SDSS J013701.06$-$091234.9) is an unusual
short-$P_{\rm orb}$ dwarf nova with a massive secondary
(\cite{ima06j0137}; \cite{ish07CVIR}).  We observed the 2011
superoutburst.  We only observed the initial and final parts
of the outburst.  The times of superhump maxima are listed in
table \ref{tab:gzcetoc2011}.  On BJD 2455923, the amplitudes of
superhumps were still less than 0.1 mag, and we must have caught
the initial stage of the outburst.
A comparison of $O-C$ diagrams between different outbursts
is shown in figure \ref{fig:gzcetcomp}.  Despite its unusual
properties, the $O-C$ curve is composed of stages B and C
similar to those of ordinary SU UMa-type dwarf novae.
The $P_{\rm dot}$ during stage B appears to be smaller than
those of ordinary SU UMa-type dwarf novae with similar $P_{\rm SH}$,
consistent with the result in \citet{Pdot}.

\begin{table}
\caption{Superhump maxima of GZ Cet (2011).}\label{tab:gzcetoc2011}
\begin{center}
\begin{tabular}{ccccc}
\hline
$E$ & max\commenta & error & $O-C$\commentb & $N$\commentc \\
\hline
0 & 55924.2865 & 0.0011 & 0.0021 & 31 \\
1 & 55924.3403 & 0.0003 & $-$0.0007 & 54 \\
2 & 55924.3963 & 0.0003 & $-$0.0014 & 46 \\
188 & 55934.9357 & 0.0003 & 0.0003 & 164 \\
189 & 55934.9918 & 0.0002 & $-$0.0003 & 165 \\
\hline
  \multicolumn{5}{l}{\commenta BJD$-$2400000.} \\
  \multicolumn{5}{l}{\commentb Against max $= 2455924.2844 + 0.056654 E$.} \\
  \multicolumn{5}{l}{\commentc Number of points used to determine the maximum.} \\
\end{tabular}
\end{center}
\end{table}

\subsection{AK Cancri}\label{obj:akcnc}

   We observed a superoutburst in 2012 January.  Due to the short
duration of the observation, the recorded superhumps were limited
(table \ref{tab:akcncoc2012}).  The resultant period suggests that
we observed stage B superhumps.

\begin{table}
\caption{Superhump maxima of AK Cnc (2012).}\label{tab:akcncoc2012}
\begin{center}
\begin{tabular}{ccccc}
\hline
$E$ & max\commenta & error & $O-C$\commentb & $N$\commentc \\
\hline
0 & 55952.0678 & 0.0047 & $-$0.0045 & 42 \\
1 & 55952.1424 & 0.0007 & 0.0028 & 74 \\
2 & 55952.2086 & 0.0007 & 0.0018 & 54 \\
45 & 55955.1043 & 0.0016 & 0.0062 & 74 \\
46 & 55955.1591 & 0.0011 & $-$0.0062 & 69 \\
\hline
  \multicolumn{5}{l}{\commenta BJD$-$2400000.} \\
  \multicolumn{5}{l}{\commentb Against max $= 2455952.0723 + 0.067239 E$.} \\
  \multicolumn{5}{l}{\commentc Number of points used to determine the maximum.} \\
\end{tabular}
\end{center}
\end{table}

\subsection{CC Cancri}\label{obj:cccnc}

   We observed a superoutburst in 2011 December.  The times of
superhump maxima are listed in table \ref{tab:cccncoc2011}.
Although the data were rather sparse, stages B and C were recorded.
The obtained periods were similar to those in 2001 \citep{Pdot}.

\begin{figure}
  \begin{center}
    \FigureFile(88mm,70mm){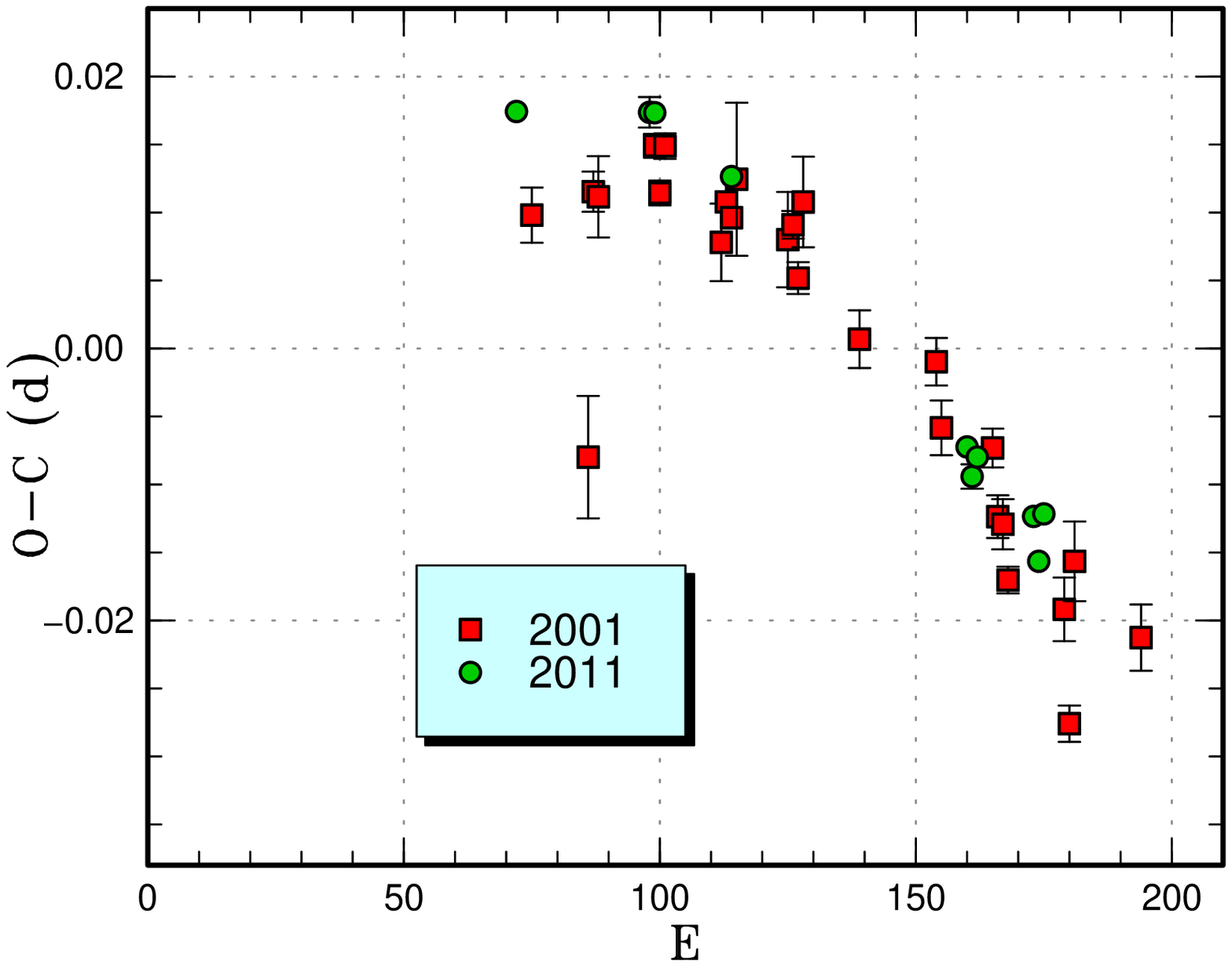}
  \end{center}
  \caption{Comparison of $O-C$ diagrams of CC Cnc between different
  superoutbursts.  A period of 0.07589~d was used to draw this figure.
  Approximate cycle counts ($E$) after the start of the superoutburst
  were used.  Since the start of the 2001 superoutburst was not
  well constrained, we shifted the $O-C$ diagrams
  to best fit the best-recorded 2011 one.
  }
  \label{fig:cccnccomp}
\end{figure}

   A comparison of $O-C$ diagrams between different superoutburst
is shown in figure \ref{fig:cccnccomp}.  Early stage observations
are still lacking for this object.

\begin{table}
\caption{Superhump maxima of CC Cnc (2011).}\label{tab:cccncoc2011}
\begin{center}
\begin{tabular}{ccccc}
\hline
$E$ & max\commenta & error & $O-C$\commentb & $N$\commentc \\
\hline
0 & 55910.1871 & 0.0006 & $-$0.0065 & 155 \\
26 & 55912.1601 & 0.0011 & 0.0028 & 91 \\
27 & 55912.2360 & 0.0006 & 0.0031 & 154 \\
42 & 55913.3697 & 0.0002 & 0.0038 & 200 \\
88 & 55916.8407 & 0.0004 & 0.0003 & 77 \\
89 & 55916.9144 & 0.0009 & $-$0.0015 & 37 \\
90 & 55916.9918 & 0.0007 & 0.0003 & 73 \\
101 & 55917.8222 & 0.0006 & $-$0.0001 & 79 \\
102 & 55917.8948 & 0.0008 & $-$0.0031 & 77 \\
103 & 55917.9741 & 0.0005 & 0.0008 & 79 \\
\hline
  \multicolumn{5}{l}{\commenta BJD$-$2400000.} \\
  \multicolumn{5}{l}{\commentb Against max $= 2455910.1935 + 0.075532 E$.} \\
  \multicolumn{5}{l}{\commentc Number of points used to determine the maximum.} \\
\end{tabular}
\end{center}
\end{table}

\subsection{GO Comae Berenices}\label{obj:gocom}

\begin{figure}
  \begin{center}
    \FigureFile(88mm,70mm){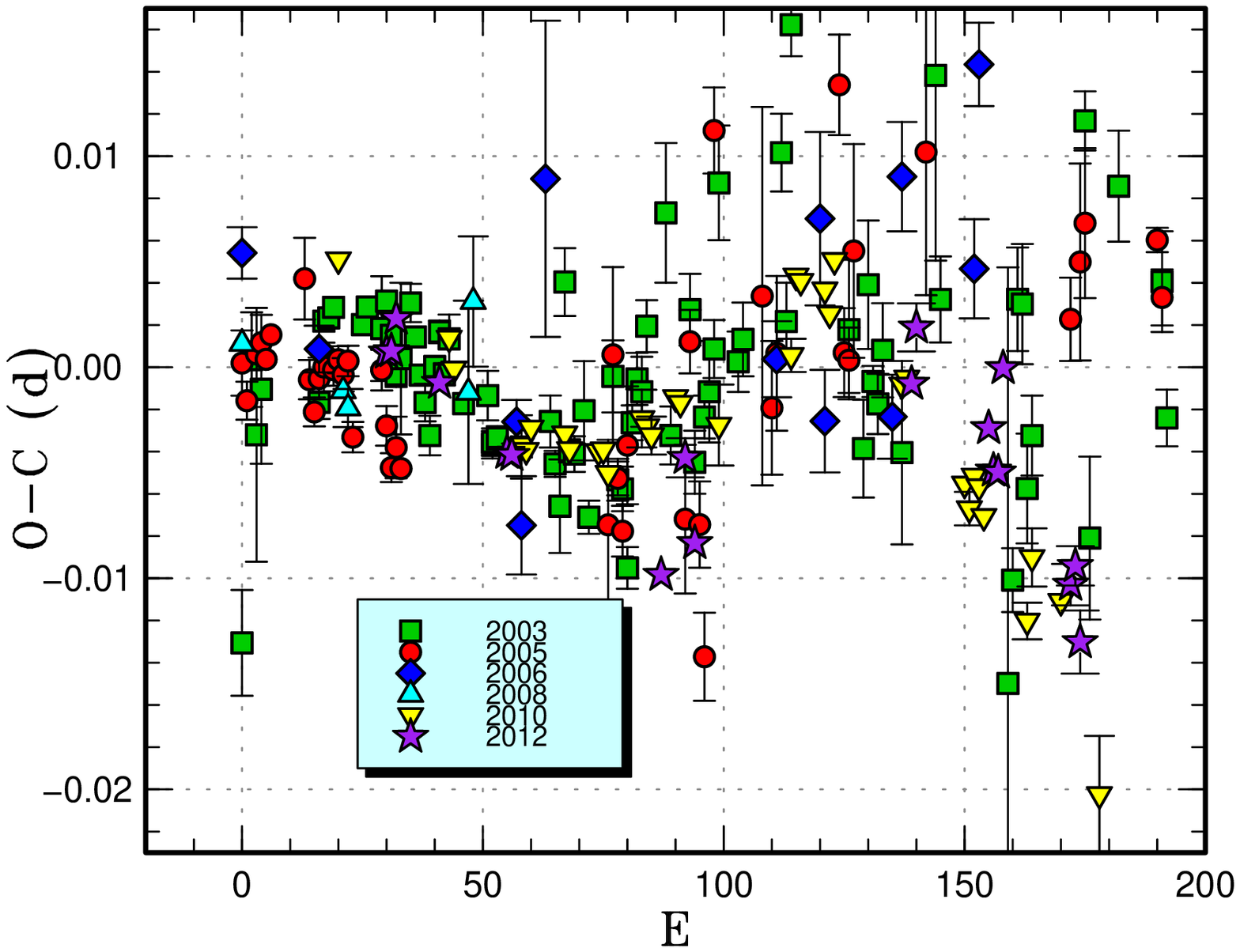}
  \end{center}
  \caption{Comparison of $O-C$ diagrams of GO Com between different
  superoutbursts.  A period of 0.06303~d was used to draw this figure.
  Approximate cycle counts ($E$) after the start of the superoutburst
  were used.
  }
  \label{fig:gocomcomp3}
\end{figure}

   We observed the 2012 superoutburst of this object.  The times
of superhump maxima are listed in table \ref{tab:gocomoc2012}.
Both typical stages B and C can be clearly identified.
The $O-C$ variation during this outburst was similar to those
in previous outbursts (figure \ref{fig:gocomcomp3}).

\begin{table}
\caption{Superhump maxima of GO Com (2012).}\label{tab:gocomoc2012}
\begin{center}
\begin{tabular}{ccccc}
\hline
$E$ & max\commenta & error & $O-C$\commentb & $N$\commentc \\
\hline
0 & 55983.5944 & 0.0002 & 0.0015 & 63 \\
1 & 55983.6575 & 0.0003 & 0.0017 & 61 \\
2 & 55983.7221 & 0.0003 & 0.0032 & 65 \\
11 & 55984.2863 & 0.0003 & 0.0006 & 107 \\
25 & 55985.1654 & 0.0002 & $-$0.0022 & 130 \\
26 & 55985.2284 & 0.0003 & $-$0.0023 & 128 \\
57 & 55987.1766 & 0.0003 & $-$0.0067 & 145 \\
62 & 55987.4973 & 0.0009 & $-$0.0009 & 54 \\
64 & 55987.6193 & 0.0008 & $-$0.0049 & 53 \\
109 & 55990.4632 & 0.0005 & 0.0045 & 129 \\
110 & 55990.5289 & 0.0011 & 0.0072 & 92 \\
125 & 55991.4696 & 0.0004 & 0.0030 & 128 \\
126 & 55991.5306 & 0.0005 & 0.0011 & 131 \\
127 & 55991.5936 & 0.0005 & 0.0010 & 122 \\
128 & 55991.6615 & 0.0008 & 0.0060 & 71 \\
142 & 55992.5337 & 0.0010 & $-$0.0037 & 55 \\
143 & 55992.5976 & 0.0009 & $-$0.0028 & 48 \\
144 & 55992.6570 & 0.0015 & $-$0.0063 & 45 \\
\hline
  \multicolumn{5}{l}{\commenta BJD$-$2400000.} \\
  \multicolumn{5}{l}{\commentb Against max $= 2455983.5929 + 0.062990 E$.} \\
  \multicolumn{5}{l}{\commentc Number of points used to determine the maximum.} \\
\end{tabular}
\end{center}
\end{table}

\subsection{TU Crateris}\label{obj:tucrt}

\begin{figure}
  \begin{center}
    \FigureFile(88mm,70mm){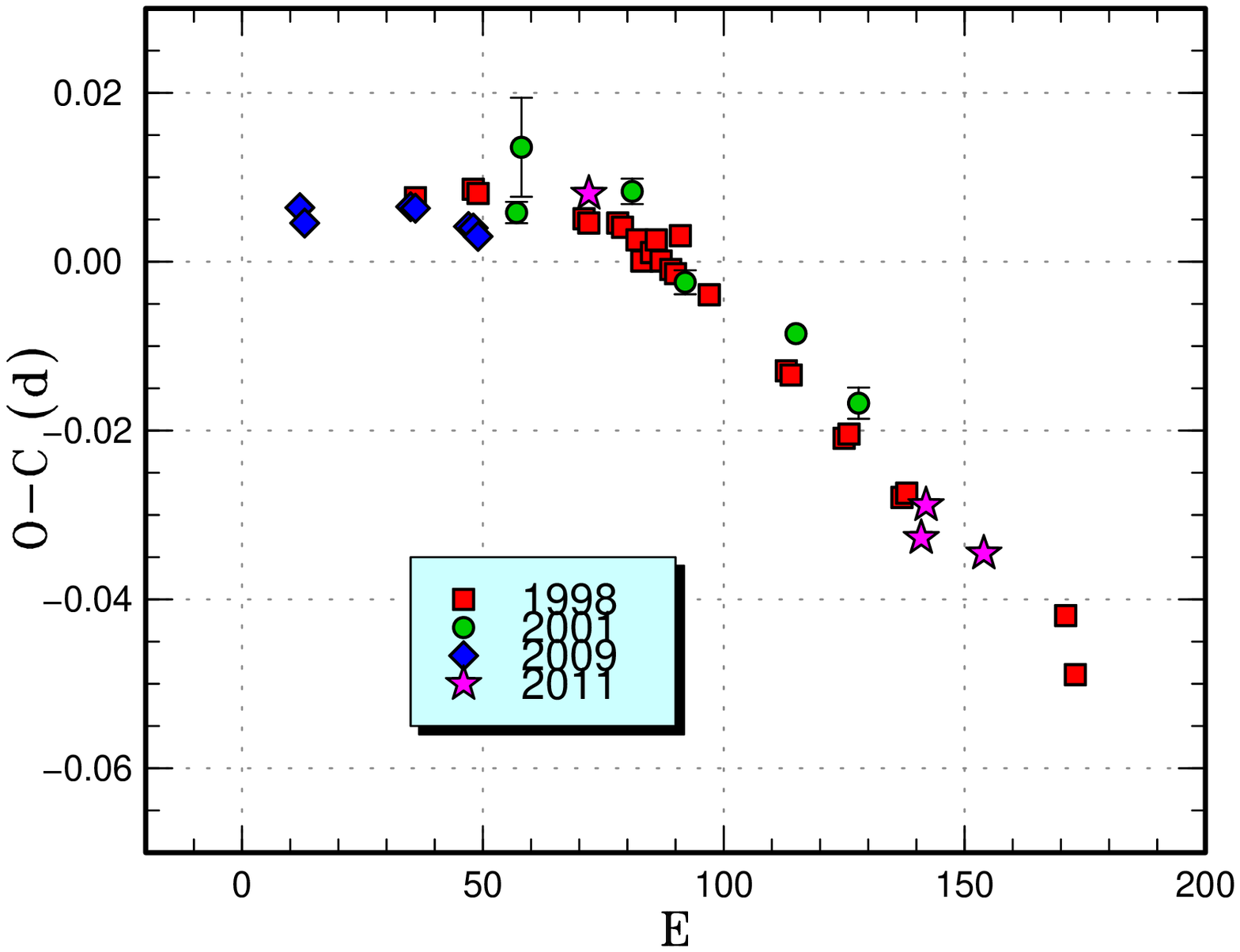}
  \end{center}
  \caption{Comparison of $O-C$ diagrams of TU Crt between different
  superoutbursts.  A period of 0.08550~d was used to draw this figure.
  Approximate cycle counts ($E$) after the start of the superoutburst
  were used.
  }
  \label{fig:tucrtcomp2}
\end{figure}

   We observed the late stage of the 2011 superoutburst of TU Crt.
The times of superhump maxima are listed in table \ref{tab:tucrtoc2011}.
We most likely observed only stage C superhumps.  The measured
period is in good agreement with that of stage C superhumps recorded
in 1998 \citep{men99tucrt} and analyzed in \citet{Pdot}.
A comparison of $O-C$ diagrams between different superoutburst
is shown in figure \ref{fig:tucrtcomp2}.

\begin{table}
\caption{Superhump maxima of TU Crt (2011).}\label{tab:tucrtoc2011}
\begin{center}
\begin{tabular}{ccccc}
\hline
$E$ & max\commenta & error & $O-C$\commentb & $N$\commentc \\
\hline
0 & 55925.2955 & 0.0003 & 0.0004 & 151 \\
69 & 55931.1543 & 0.0017 & $-$0.0033 & 99 \\
70 & 55931.2437 & 0.0005 & 0.0011 & 151 \\
82 & 55932.2639 & 0.0007 & 0.0018 & 151 \\
\hline
  \multicolumn{5}{l}{\commenta BJD$-$2400000.} \\
  \multicolumn{5}{l}{\commentb Against max $= 2455925.2952 + 0.084962 E$.} \\
  \multicolumn{5}{l}{\commentc Number of points used to determine the maximum.} \\
\end{tabular}
\end{center}
\end{table}

\subsection{V503 Cygni}\label{obj:v503cyg}

   \citet{har95v503cyg} established the SU UMa-type nature of this
object and reported a mean $P_{\rm SH}$ of 0.08101(4)~d.
They also detected negative superhumps in quiescence.
Although there may have been some evidence of a hump
corresponding to negative superhumps during superoutburst,
its presence was not well established.

   We observed the 2011 July superoutburst, subsequent phase with
normal outbursts and 2011 October superoutburst.
The times of superhump maxima during the July
superoutburst are listed in table \ref{tab:v503cygoc2011}.
There was some hint of a break in the $O-C$ diagram for
the superhumps during the superoutburst between $E=25$ and $E=35$,
and we attributed this to be a stage B--C transition.
A global $P_{\rm dot}$ corresponded to $-3.8(2.6) \times 10^{-5}$.

   The signals of the ordinary superhumps already became difficult to
trace even before the rapid fading (BJD 2455751).
A PDM analysis, however, to the data for the interval
BJD 2455751--2455754 yielded a period of 0.0814(1)~d, suggesting
that the ordinary superhumps were still the dominant signal,
rather than negative superhumps.

   After BJD 2455754, large-amplitude modulations appeared again.
The times of maxima were not on a smooth extension of the times
of superhump maxima during the superoutburst plateau.
These new signals appear to correspond to the traditional
late superhumps (e.g. \cite{vog83lateSH}), rather than ``stage C
superhumps'' in our designation (table \ref{tab:v503cygoc2011late}).

   The times of superhump maxima during the October
superoutburst are listed in table \ref{tab:v503cygoc2011b}.
Although the epoch $E=114$ is possibly a late superhump as in
the 2011 July superoutburst, the lack of subsequent observations
made the identification unclear.  We listed a global $P_{\rm orb}$
and $P_{\rm dot}$ in table \ref{tab:perlist}.  A period derived
from $E \le 27$ (stage B) was 0.08151(8)~d.

   We were not able to detect a signal of negative superhumps 
during the fading stage and subsequent quiescence, and the signal 
was dominated by positive superhumps.  The situation was thus 
different from ER UMa \citep{ohs12eruma}.  The mean period of 
(traditional late) superhumps during the post-superoutburst 
stage was 0.08032(3)~d (PDM method), 3.4\% longer than $P_{\rm orb}$,
and was significantly shorter than that of ordinary superhumps.
Although the superhump signal persisted during the quiescent
state following the superoutburst, the signal became dominated
by $P_{\rm orb}$ after the next normal outburst.
The $P_{\rm orb}$ determined from all the observations between
BJD 2455744--2455802 was 0.077773(2)~d.  This period is in agreement
with an analysis of the data set restricted to the phase when 
the object did not show superhumps within respective errors.
The period is also in good agreement with the period obtained
from the 2010 observations \citep{pav12v503cyg}.
We used this refined $P_{\rm orb}$ in table \ref{tab:perlist}.

   The lack of negative superhumps during these observations
made a clear contrast to the observation by \citet{har95v503cyg}.
V503 Cyg is known to display highly variable number of normal outbursts
between superoutbursts \citep{kat02v503cyg}, and normal outbursts were 
very infrequent (every $\sim$30~d) during the observation
by \citet{har95v503cyg}, while the current observations showed much more
frequent ones (every $\sim$10~d).  \citet{kat02v503cyg} suggested that
mechanisms for suppressing normal outbursts may have worked when 
normal outbursts were very infrequent.  As discussed by various authors 
(\cite{can10v344lyr}; \cite{Pdot3}; \cite{ohs12eruma}),
the state with negative superhumps prevents the disk-instability
to occur.  The condition to produce negative superhumps (likely
a disk tilt) seems to naturally explain the association of
the presence of negative superhumps with the reduced number of
normal outbursts in V503 Cyg.

\begin{table}
\caption{Superhump maxima of V503 Cyg (2011 July).}\label{tab:v503cygoc2011}
\begin{center}
\begin{tabular}{ccccc}
\hline
$E$ & max\commenta & error & $O-C$\commentb & $N$\commentc \\
\hline
0 & 55744.5173 & 0.0005 & $-$0.0041 & 89 \\
1 & 55744.6008 & 0.0008 & $-$0.0017 & 47 \\
10 & 55745.3315 & 0.0016 & $-$0.0007 & 21 \\
12 & 55745.4959 & 0.0009 & 0.0016 & 31 \\
24 & 55746.4683 & 0.0008 & 0.0010 & 92 \\
25 & 55746.5524 & 0.0009 & 0.0039 & 82 \\
35 & 55747.3620 & 0.0023 & 0.0027 & 16 \\
36 & 55747.4410 & 0.0006 & 0.0006 & 75 \\
37 & 55747.5227 & 0.0008 & 0.0012 & 75 \\
39 & 55747.6836 & 0.0004 & $-$0.0000 & 129 \\
40 & 55747.7639 & 0.0005 & $-$0.0008 & 150 \\
41 & 55747.8465 & 0.0005 & 0.0007 & 133 \\
47 & 55748.3339 & 0.0014 & 0.0016 & 30 \\
61 & 55749.4612 & 0.0024 & $-$0.0063 & 30 \\
73 & 55750.4418 & 0.0186 & 0.0013 & 101 \\
74 & 55750.5206 & 0.0021 & $-$0.0010 & 126 \\
77 & 55750.7689 & 0.0011 & 0.0041 & 136 \\
78 & 55750.8419 & 0.0026 & $-$0.0040 & 116 \\
\hline
  \multicolumn{5}{l}{\commenta BJD$-$2400000.} \\
  \multicolumn{5}{l}{\commentb Against max $= 2455744.5214 + 0.081084 E$.} \\
  \multicolumn{5}{l}{\commentc Number of points used to determine the maximum.} \\
\end{tabular}
\end{center}
\end{table}

\begin{table}
\caption{Superhump maxima of V503 Cyg (2011 October).}\label{tab:v503cygoc2011b}
\begin{center}
\begin{tabular}{ccccc}
\hline
$E$ & max\commenta & error & $O-C$\commentb & $N$\commentc \\
\hline
0 & 55831.1911 & 0.0014 & $-$0.0055 & 33 \\
1 & 55831.2718 & 0.0004 & $-$0.0059 & 58 \\
2 & 55831.3537 & 0.0004 & $-$0.0049 & 59 \\
3 & 55831.4366 & 0.0005 & $-$0.0030 & 58 \\
4 & 55831.5186 & 0.0007 & $-$0.0020 & 36 \\
25 & 55833.2234 & 0.0012 & 0.0018 & 24 \\
26 & 55833.3128 & 0.0020 & 0.0101 & 28 \\
27 & 55833.3946 & 0.0019 & 0.0110 & 29 \\
87 & 55838.2578 & 0.0016 & 0.0142 & 45 \\
114 & 55840.4147 & 0.0042 & $-$0.0159 & 45 \\
\hline
  \multicolumn{5}{l}{\commenta BJD$-$2400000.} \\
  \multicolumn{5}{l}{\commentb Against max $= 2455831.1966 + 0.081000 E$.} \\
  \multicolumn{5}{l}{\commentc Number of points used to determine the maximum.} \\
\end{tabular}
\end{center}
\end{table}

\begin{table}
\caption{Superhump maxima of V503 Cyg (2011 July) (late superhumps).}\label{tab:v503cygoc2011late}
\begin{center}
\begin{tabular}{ccccc}
\hline
$E$ & max\commenta & error & $O-C$\commentb & $N$\commentc \\
\hline
0 & 55753.4128 & 0.0018 & $-$0.0076 & 30 \\
1 & 55753.4995 & 0.0019 & $-$0.0015 & 30 \\
10 & 55754.2291 & 0.0006 & 0.0041 & 82 \\
16 & 55754.7127 & 0.0018 & 0.0050 & 60 \\
17 & 55754.7878 & 0.0008 & $-$0.0003 & 145 \\
24 & 55755.3518 & 0.0012 & 0.0006 & 30 \\
25 & 55755.4300 & 0.0021 & $-$0.0017 & 30 \\
26 & 55755.5117 & 0.0010 & $-$0.0004 & 29 \\
38 & 55756.4835 & 0.0016 & 0.0060 & 30 \\
42 & 55756.8034 & 0.0007 & 0.0040 & 144 \\
43 & 55756.8821 & 0.0008 & 0.0023 & 75 \\
47 & 55757.2020 & 0.0006 & 0.0004 & 146 \\
48 & 55757.2768 & 0.0008 & $-$0.0052 & 148 \\
60 & 55758.2381 & 0.0010 & $-$0.0094 & 74 \\
66 & 55758.7351 & 0.0010 & 0.0050 & 136 \\
67 & 55758.8125 & 0.0008 & 0.0019 & 141 \\
70 & 55759.0506 & 0.0008 & $-$0.0013 & 65 \\
71 & 55759.1311 & 0.0007 & $-$0.0013 & 160 \\
72 & 55759.2122 & 0.0010 & $-$0.0007 & 179 \\
\hline
  \multicolumn{5}{l}{\commenta BJD$-$2400000.} \\
  \multicolumn{5}{l}{\commentb Against max $= 2455753.4205 + 0.080450 E$.} \\
  \multicolumn{5}{l}{\commentc Number of points used to determine the maximum.} \\
\end{tabular}
\end{center}
\end{table}

\subsection{V1454 Cygni}\label{obj:v1454cyg}

   This SU UMa-type dwarf nova undergoes outbursts relatively rarely
and the last outburst was in 2009 \citep{Pdot2}.  The new observation
during the 2012 superoutburst confirmed the period selection as
stated in \citet{Pdot2}.  The times of superhump maxima are
listed in table \ref{tab:v1454cygoc2012}.
A. Henden reported that there is a $V\sim20.5$-mag blue quiescent
counterpart (cf. vsnet-alert 14568), whose position is in good
agreement with the astrometry (\timeform{19h 53m 38.47s},
\timeform{+35D 21' 45.8''}) measured during the outburst 
(vsnet-alert 14566).

\begin{table}
\caption{Superhump maxima of V1454 Cyg (2012).}\label{tab:v1454cygoc2012}
\begin{center}
\begin{tabular}{ccccc}
\hline
$E$ & max\commenta & error & $O-C$\commentb & $N$\commentc \\
\hline
0 & 56059.5294 & 0.0005 & $-$0.0001 & 60 \\
1 & 56059.5872 & 0.0006 & 0.0002 & 58 \\
17 & 56060.5067 & 0.0006 & $-$0.0002 & 60 \\
18 & 56060.5647 & 0.0004 & 0.0002 & 61 \\
\hline
  \multicolumn{5}{l}{\commenta BJD$-$2400000.} \\
  \multicolumn{5}{l}{\commentb Against max $= 2456059.5296 + 0.057494 E$.} \\
  \multicolumn{5}{l}{\commentc Number of points used to determine the maximum.} \\
\end{tabular}
\end{center}
\end{table}

\subsection{AQ Eridani}\label{obj:aqeri}

   The 2011 superoutburst of AQ Eri was observed only for its
early and late stages.  Although well-developed superhumps were
observed on the first night, we could not measure the superhump
period precisely.  The late stage of the superoutburst and
post-superoutburst stage were well observed.  The superhumps
apparently persisted after the rapid decline.  The times of
superhump maxima are listed in table \ref{tab:aqerioc2011}.
By using the PDM analysis, the signal of the superhumps
was detected until BJD 2455586.  The signal, however, was not
significantly detected after this epoch.  The present case appears
to be different from long-persisting stage C superhumps
in many short-$P_{\rm orb}$ dwarf novae, such as QZ Vir \citep{ohs11qzvir}.

\begin{table}
\caption{Superhump maxima of AQ Eri (2011).}\label{tab:aqerioc2011}
\begin{center}
\begin{tabular}{ccccc}
\hline
$E$ & max\commenta & error & $O-C$\commentb & $N$\commentc \\
\hline
0 & 55875.8290 & 0.0005 & $-$0.0016 & 121 \\
1 & 55875.8930 & 0.0003 & 0.0001 & 103 \\
143 & 55884.7372 & 0.0006 & 0.0079 & 77 \\
144 & 55884.7964 & 0.0007 & 0.0049 & 98 \\
145 & 55884.8561 & 0.0009 & 0.0024 & 80 \\
159 & 55885.7152 & 0.0013 & $-$0.0098 & 24 \\
160 & 55885.7813 & 0.0029 & $-$0.0059 & 22 \\
161 & 55885.8513 & 0.0008 & 0.0019 & 15 \\
\hline
  \multicolumn{5}{l}{\commenta BJD$-$2400000.} \\
  \multicolumn{5}{l}{\commentb Against max $= 2455875.8306 + 0.062228 E$.} \\
  \multicolumn{5}{l}{\commentc Number of points used to determine the maximum.} \\
\end{tabular}
\end{center}
\end{table}

\subsection{UV Geminorum}\label{obj:uvgem}

\begin{figure}
  \begin{center}
    \FigureFile(88mm,70mm){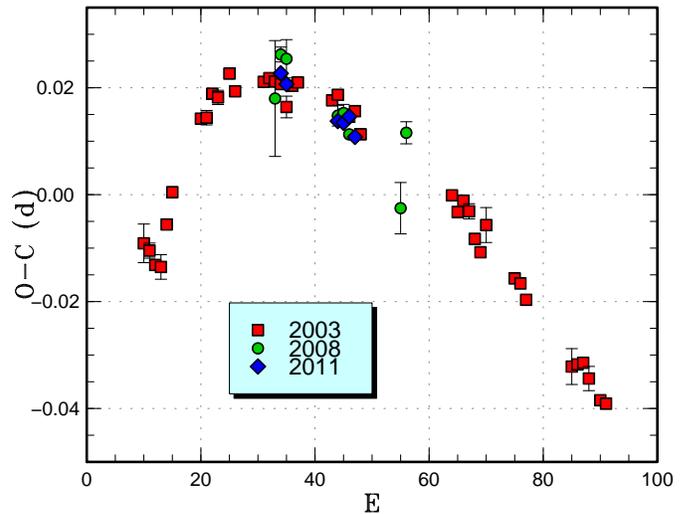}
  \end{center}
  \caption{Comparison of $O-C$ diagrams of UV Gem between different
  superoutbursts.  A period of 0.0936~d was used to draw this figure.
  Approximate cycle counts ($E$) after the start of the superoutburst
  were used.
  }
  \label{fig:uvgemcomp}
\end{figure}

   We observed the middle part of the 2011 superoutburst.
The times of superhump maxima are listed as table \ref{tab:awgemoc2011}.
A comparison of $O-C$ diagram between different superoutbursts
is shown in figure \ref{fig:uvgemcomp}.  Despite the large variation
of the superhump period, the periods during the middle stage
of superoutbursts were almost the same in different superoutbursts.

\begin{table}
\caption{Superhump maxima of AW Gem (2011).}\label{tab:awgemoc2011}
\begin{center}
\begin{tabular}{ccccc}
\hline
$E$ & max\commenta & error & $O-C$\commentb & $N$\commentc \\
\hline
0 & 55892.6080 & 0.0004 & 0.0006 & 62 \\
1 & 55892.6996 & 0.0005 & $-$0.0007 & 80 \\
10 & 55893.5351 & 0.0005 & $-$0.0005 & 84 \\
11 & 55893.6284 & 0.0004 & $-$0.0001 & 83 \\
12 & 55893.7231 & 0.0007 & 0.0019 & 51 \\
13 & 55893.8129 & 0.0004 & $-$0.0012 & 60 \\
\hline
  \multicolumn{5}{l}{\commenta BJD$-$2400000.} \\
  \multicolumn{5}{l}{\commentb Against max $= 2455892.6074 + 0.092822 E$.} \\
  \multicolumn{5}{l}{\commentc Number of points used to determine the maximum.} \\
\end{tabular}
\end{center}
\end{table}

\subsection{NY Herculis}\label{obj:nyher}

   NY Her was discovered by \citet{hof49newvar} as a Mira-type
variable with a photographic range of 15.0 to fainter than 16.5.
\citet{ges66VS8} classified this object as a Cepheid (likely
a W Vir-type variable) with a period of 6.3146~d.
\citet{pas88nyher}, however, did not confirm this classification.
\citet{pas88nyher} identified the object as an 18-mag blue object
on POSS plates and obtained a mean period of 67.7067~d.
In addition to this mean period, short outbursts were irregularly
observed.  The object varied at a rate up to 2 mag d$^{-1}$,
and \citet{pas88nyher} classified the object to be a blue irregular
variable.

\begin{figure}
  \begin{center}
    \FigureFile(88mm,110mm){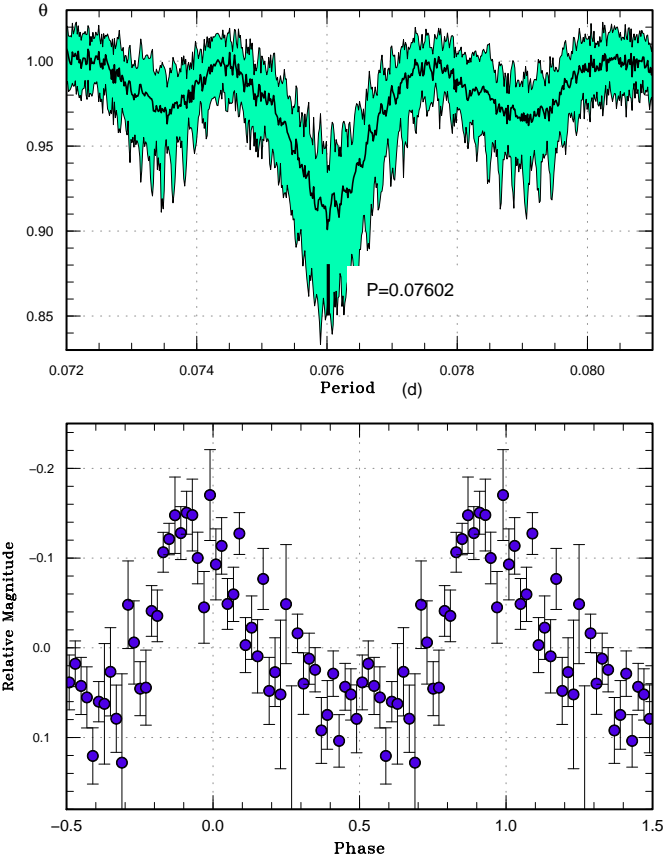}
  \end{center}
  \caption{Superhumps in NY Her (2011). (Upper): PDM analysis.
     (Lower): Phase-averaged profile.}
  \label{fig:nyhershpdm}
\end{figure}

   On 2011 June 10, CRTS detected an outburst of this object.
T. Kato suggested that the known behavior of this object resembles
that of ER UMa (cf. vsnet-alert 13410).  
Follow-up observation indicated the presence of superhumps 
(vsnet-alert 13418).  The best superhump period with the PDM
method was 0.07602(14)~d (figure \ref{fig:nyhershpdm}).

   The times of superhump maxima are listed in table \ref{tab:nyheroc2011}.
CRTS data suggest a supercycle of 80--90~d.  As judged from
the relatively long superhump period, the object may be more
similar to V503 Cyg \citep{har95v503cyg} rather than ER UMa.
Further intensive observations are particularly needed to
determine the true cycle length for such a rare variety of
SU UMa-type dwarf novae.

\begin{table}
\caption{Superhump maxima of NY Her (2011).}\label{tab:nyheroc2011}
\begin{center}
\begin{tabular}{ccccc}
\hline
$E$ & max\commenta & error & $O-C$\commentb & $N$\commentc \\
\hline
0 & 55724.6932 & 0.0013 & $-$0.0023 & 77 \\
1 & 55724.7690 & 0.0009 & $-$0.0022 & 79 \\
2 & 55724.8466 & 0.0009 & $-$0.0005 & 78 \\
3 & 55724.9220 & 0.0006 & $-$0.0009 & 78 \\
13 & 55725.6829 & 0.0014 & 0.0020 & 78 \\
14 & 55725.7585 & 0.0072 & 0.0018 & 40 \\
26 & 55726.6702 & 0.0027 & 0.0039 & 65 \\
27 & 55726.7455 & 0.0031 & 0.0034 & 34 \\
29 & 55726.8994 & 0.0040 & 0.0057 & 77 \\
37 & 55727.4892 & 0.0044 & $-$0.0109 & 28 \\
\hline
  \multicolumn{5}{l}{\commenta BJD$-$2400000.} \\
  \multicolumn{5}{l}{\commentb Against max $= 2455724.6955 + 0.075802 E$.} \\
  \multicolumn{5}{l}{\commentc Number of points used to determine the maximum.} \\
\end{tabular}
\end{center}
\end{table}

\subsection{PR Herculis}\label{obj:prher}

   PR Her was discovered as a dwarf nova (S 4247) by \citet{hof49prher}
with a photographic range of 14.0 to fainter than 17.5.
Although this star was monitored by amateur observes since 
the early 1990s, no outburst had been recorded.
In the meantime, A. Henden identified the object
as a $V=21$-mag blue star in 1999 (vsnet-chat 1800).\footnote{
See also
$<$ftp://ftp.aavso.org/upload/chartteam/MISC/seq/\\Her\%20PR.txt$>$.
}
The large outburst amplitude made the object a good candidate
for a WZ Sge-type dwarf nova.

\begin{figure}
  \begin{center}
    \FigureFile(88mm,110mm){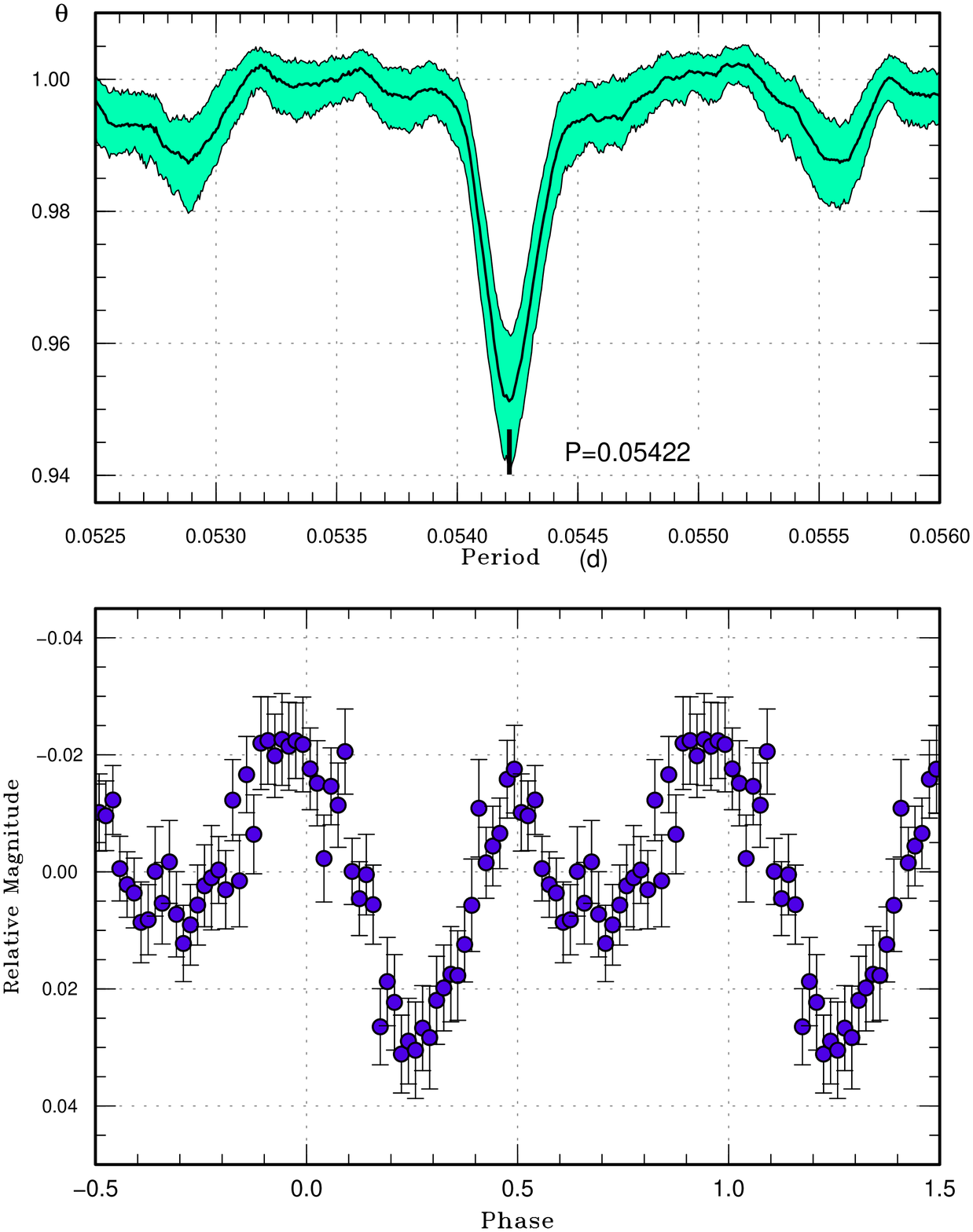}
  \end{center}
  \caption{Early superhumps in PR Her (2011). (Upper): PDM analysis.
     (Lower): Phase-averaged profile.}
  \label{fig:prhereshpdm}
\end{figure}

\begin{figure}
  \begin{center}
    \FigureFile(88mm,110mm){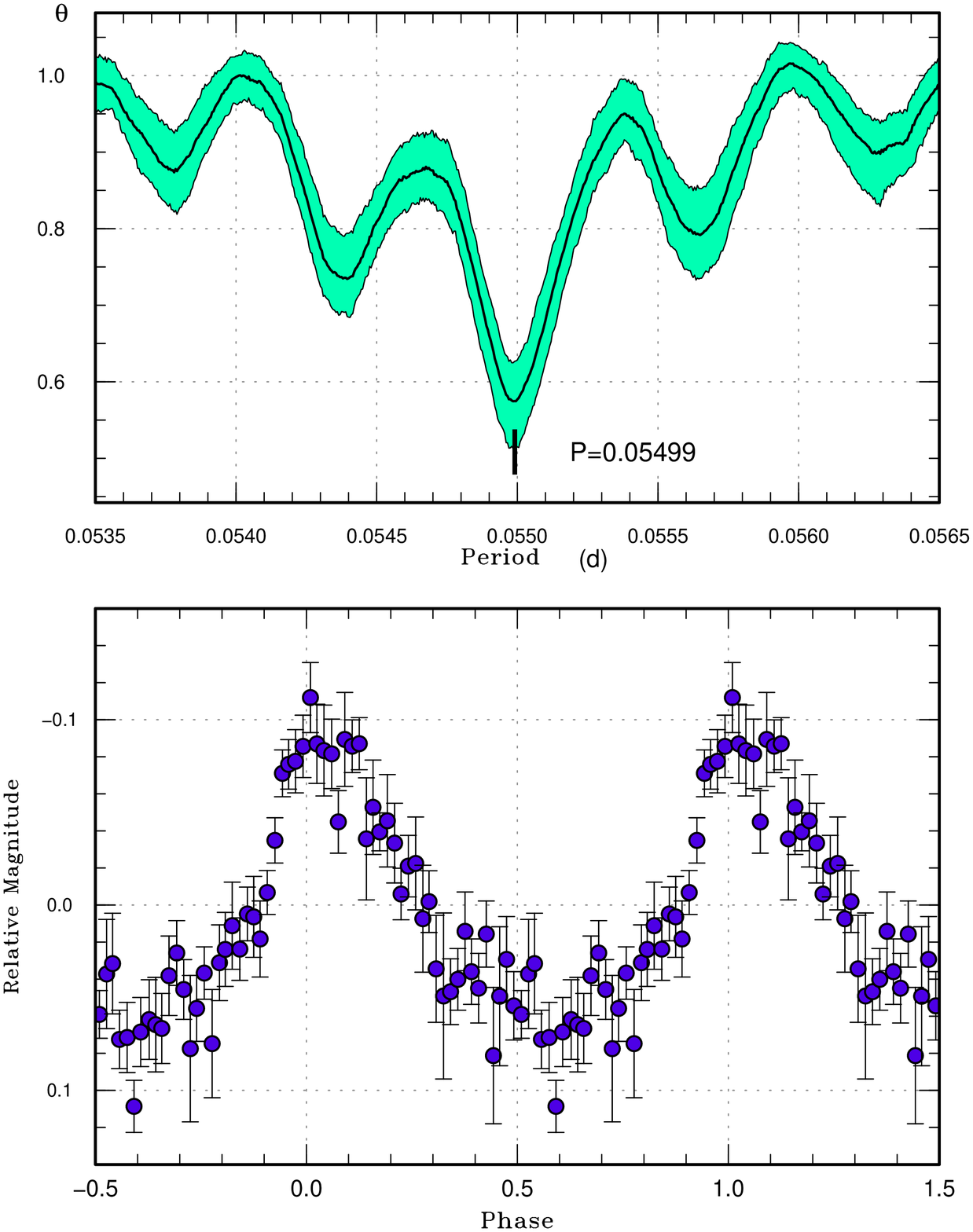}
  \end{center}
  \caption{Ordinary superhumps in PR Her (2011). (Upper): PDM analysis.
     (Lower): Phase-averaged profile.}
  \label{fig:prhershpdm}
\end{figure}

   On 2011 November 21, Walter MacDonald II reported a very bright
outburst at a magnitude of $V=12.84$ (cf. cvnet-outburst 4406).
Subsequent observations confirmed the presence of typical
double-wave early superhumps (figure \ref{fig:prhereshpdm}).
Due to the unfavorable location, the object soon became hard to
access in the low evening sky.  Ordinary superhumps were detected
despite this unfavorable condition (vsnet-alert 13932; figure
\ref{fig:prhershpdm}).  The times of superhump maxima are listed
in table \ref{tab:prheroc2011}.  The large outburst amplitude,
the low frequency of outbursts, and the existence of prominent
early superhumps qualify PR Her as a WZ Sge-type dwarf nova.

\begin{table}
\caption{Superhump maxima of PR Her (2011).}\label{tab:prheroc2011}
\begin{center}
\begin{tabular}{ccccc}
\hline
$E$ & max\commenta & error & $O-C$\commentb & $N$\commentc \\
\hline
0 & 55900.2456 & 0.0006 & 0.0008 & 74 \\
1 & 55900.3034 & 0.0006 & 0.0035 & 54 \\
11 & 55900.8507 & 0.0015 & 0.0006 & 31 \\
12 & 55900.9020 & 0.0021 & $-$0.0031 & 51 \\
19 & 55901.2905 & 0.0005 & 0.0002 & 43 \\
37 & 55902.2771 & 0.0008 & $-$0.0036 & 51 \\
91 & 55905.2512 & 0.0004 & $-$0.0008 & 114 \\
92 & 55905.3094 & 0.0011 & 0.0025 & 58 \\
\hline
  \multicolumn{5}{l}{\commenta BJD$-$2400000.} \\
  \multicolumn{5}{l}{\commentb Against max $= 2455900.2449 + 0.055022 E$.} \\
  \multicolumn{5}{l}{\commentc Number of points used to determine the maximum.} \\
\end{tabular}
\end{center}
\end{table}

\subsection{V611 Herculis}\label{obj:v611her}

\begin{figure}
  \begin{center}
    \FigureFile(88mm,60mm){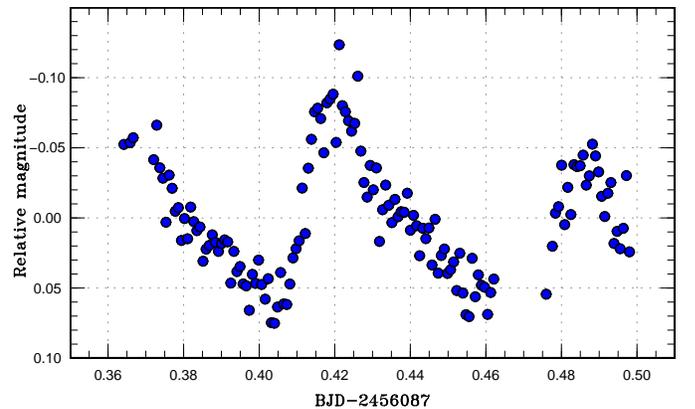}
  \end{center}
  \caption{Superhumps in V611 Her.}
  \label{fig:v611herlc}
\end{figure}

   Little had been known about this dwarf nova since its discovery
\citep{hof68an290277}.  CRTS detected four past outbursts.
An analysis of the SDSS colors of the quiescent counterpart
suggested an object below the period gap \citep{kat12DNSDSS}.
A new outburst was detected by CRTS on 2012 June 8
(cf. vsnet-alert 14647).  Subsequent observations detected
superhumps (vsnet-alert 14648; figure \ref{fig:v611herlc}).
We detected two superhump maxima at BJD 2456087.4232(5) ($N=62$) 
and 2456087.4877(6) ($N=27$).
The best period determined by the PDM method was 0.0636(4)~d.

\subsection{V844 Herculis}\label{obj:v844her}

   The well-known SU UMa-type dwarf nova V844 Her underwent
a superoutburst in 2012 May (vsnet-alert 14525).  After a period
of frequent outburst in 2009--2011, the object again entered
a relatively inactive phase in 2011--2012 and the superoutburst
occurred $\sim$370~d after the 2011 superoutburst.
The times of superhump maxima are listed in table \ref{tab:v844heroc2012}.
Although a clear pattern of stages A--C was observed, the period of
stage A was not determined due to the limited observations
in this stage.  The $P_{\rm dot}$ for stage B was clearly positive
as in other superoutbursts in this object.

\begin{figure}
  \begin{center}
    \FigureFile(88mm,70mm){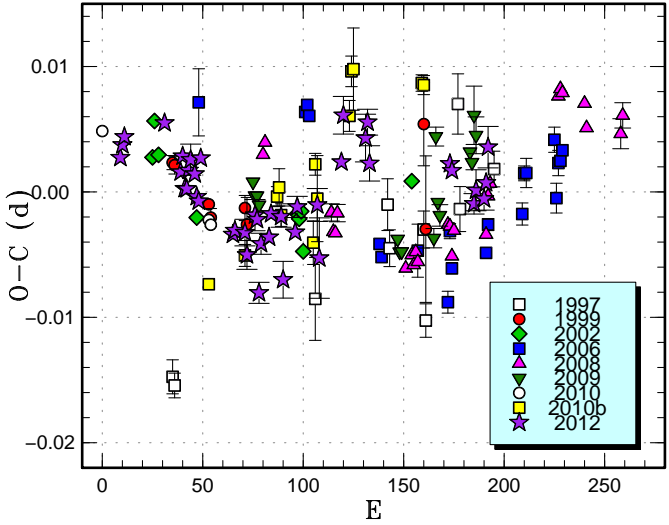}
  \end{center}
  \caption{Comparison of $O-C$ diagrams of V844 Her between different
  superoutbursts.  A period of 0.05590~d was used to draw this figure.
  Approximate cycle counts ($E$) after the start of the
  superoutburst were used.  For descriptions of the 2009, 2010 and 2010b
  superoutburst see \citet{Pdot3}.
  }
  \label{fig:v844hercomp4}
\end{figure}

   Figure \ref{fig:v844hercomp4} illustrates a comparison of $O-C$
diagrams between different superoutbursts.  As noted in \citet{Pdot3},
the epoch of stage B--C transition is different between different
superoutbursts.  The B--C transition in the 2012 superoutburst appears
to have occurred earlier than in other superoutbursts.

\begin{table}
\caption{Superhump maxima of V844 Her (2012).}\label{tab:v844heroc2012}
\begin{center}
\begin{tabular}{ccccc}
\hline
$E$ & max\commenta & error & $O-C$\commentb & $N$\commentc \\
\hline
0 & 56050.3428 & 0.0002 & 0.0028 & 108 \\
1 & 56050.3997 & 0.0001 & 0.0038 & 117 \\
2 & 56050.4563 & 0.0001 & 0.0044 & 71 \\
22 & 56051.5754 & 0.0004 & 0.0055 & 51 \\
30 & 56052.0187 & 0.0004 & 0.0016 & 91 \\
31 & 56052.0758 & 0.0003 & 0.0029 & 92 \\
32 & 56052.1291 & 0.0003 & 0.0003 & 93 \\
33 & 56052.1850 & 0.0003 & 0.0002 & 81 \\
34 & 56052.2424 & 0.0008 & 0.0017 & 23 \\
35 & 56052.2992 & 0.0012 & 0.0026 & 22 \\
37 & 56052.4098 & 0.0002 & 0.0014 & 68 \\
38 & 56052.4639 & 0.0002 & $-$0.0004 & 75 \\
39 & 56052.5194 & 0.0003 & $-$0.0007 & 61 \\
40 & 56052.5788 & 0.0004 & 0.0027 & 41 \\
56 & 56053.4671 & 0.0003 & $-$0.0034 & 76 \\
57 & 56053.5233 & 0.0002 & $-$0.0031 & 76 \\
62 & 56053.8026 & 0.0011 & $-$0.0032 & 11 \\
63 & 56053.8568 & 0.0011 & $-$0.0050 & 12 \\
67 & 56054.0835 & 0.0012 & $-$0.0018 & 67 \\
68 & 56054.1391 & 0.0005 & $-$0.0022 & 92 \\
69 & 56054.1891 & 0.0008 & $-$0.0080 & 108 \\
70 & 56054.2490 & 0.0008 & $-$0.0041 & 57 \\
74 & 56054.4731 & 0.0003 & $-$0.0036 & 74 \\
75 & 56054.5308 & 0.0003 & $-$0.0018 & 76 \\
80 & 56054.8101 & 0.0008 & $-$0.0020 & 12 \\
81 & 56054.8610 & 0.0015 & $-$0.0070 & 12 \\
87 & 56055.2002 & 0.0007 & $-$0.0032 & 61 \\
88 & 56055.2581 & 0.0008 & $-$0.0012 & 62 \\
98 & 56055.8172 & 0.0013 & $-$0.0010 & 12 \\
99 & 56055.8689 & 0.0032 & $-$0.0053 & 12 \\
110 & 56056.4915 & 0.0006 & 0.0024 & 75 \\
111 & 56056.5511 & 0.0015 & 0.0061 & 69 \\
122 & 56057.1642 & 0.0019 & 0.0043 & 35 \\
123 & 56057.2213 & 0.0010 & 0.0055 & 40 \\
124 & 56057.2740 & 0.0014 & 0.0023 & 38 \\
164 & 56059.5099 & 0.0007 & 0.0022 & 46 \\
165 & 56059.5653 & 0.0004 & 0.0017 & 42 \\
176 & 56060.1775 & 0.0008 & $-$0.0010 & 42 \\
177 & 56060.2345 & 0.0016 & 0.0001 & 40 \\
181 & 56060.4574 & 0.0009 & $-$0.0006 & 59 \\
182 & 56060.5146 & 0.0008 & 0.0007 & 62 \\
183 & 56060.5734 & 0.0017 & 0.0035 & 62 \\
\hline
  \multicolumn{5}{l}{\commenta BJD$-$2400000.} \\
  \multicolumn{5}{l}{\commentb Against max $= 2456050.3401 + 0.055900 E$.} \\
  \multicolumn{5}{l}{\commentc Number of points used to determine the maximum.} \\
\end{tabular}
\end{center}
\end{table}

\subsection{MM Hydrae}\label{obj:mmhya}

\begin{figure}
  \begin{center}
    \FigureFile(88mm,70mm){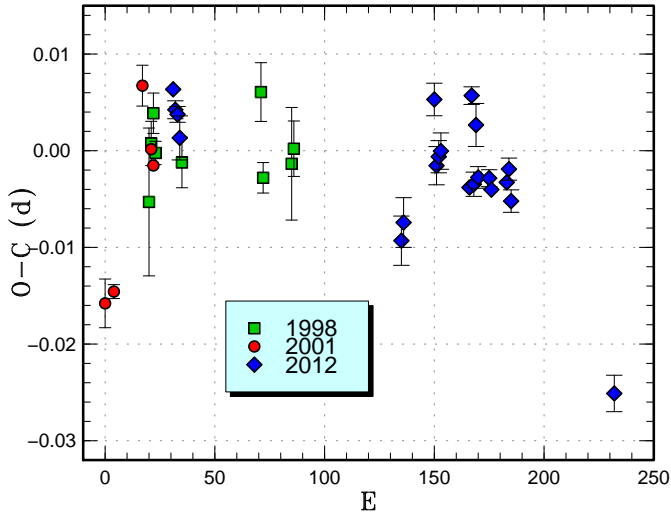}
  \end{center}
  \caption{Comparison of $O-C$ diagrams of MM Hya between different
  superoutbursts.  A period of 0.05892~d was used to draw this figure.
  Approximate cycle counts ($E$) after the start of the superoutburst
  were used.
  }
  \label{fig:mmhyacomp2}
\end{figure}

   We observed the early and late stages of the 2012 superoutburst of 
this object.  The times of superhump maxima are listed in
table \ref{tab:mmhyaoc2012}.  The $O-C$ diagram indicates that
we missed the middle-to-end part of stage B, and it was impossible
to determine $P_{\rm dot}$.  Although a comparison of $O-C$ diagrams
can be drawn (figure \ref{fig:mmhyacomp2}), middle to late part of
stage B has not yet been well recorded in this object.

\begin{table}
\caption{Superhump maxima of MM Hya (2012).}\label{tab:mmhyaoc2012}
\begin{center}
\begin{tabular}{ccccc}
\hline
$E$ & max\commenta & error & $O-C$\commentb & $N$\commentc \\
\hline
0 & 55993.5863 & 0.0008 & 0.0008 & 12 \\
1 & 55993.6431 & 0.0009 & $-$0.0012 & 17 \\
2 & 55993.7015 & 0.0008 & $-$0.0016 & 18 \\
3 & 55993.7580 & 0.0023 & $-$0.0040 & 20 \\
104 & 55999.6983 & 0.0025 & $-$0.0078 & 9 \\
105 & 55999.7591 & 0.0026 & $-$0.0058 & 14 \\
119 & 56000.5967 & 0.0017 & 0.0078 & 16 \\
120 & 56000.6488 & 0.0020 & 0.0011 & 19 \\
121 & 56000.7086 & 0.0017 & 0.0021 & 14 \\
122 & 56000.7681 & 0.0019 & 0.0027 & 10 \\
135 & 56001.5303 & 0.0005 & $-$0.0002 & 44 \\
136 & 56001.5987 & 0.0009 & 0.0094 & 23 \\
137 & 56001.6485 & 0.0013 & 0.0003 & 18 \\
138 & 56001.7135 & 0.0022 & 0.0065 & 6 \\
139 & 56001.7670 & 0.0011 & 0.0011 & 15 \\
144 & 56002.0616 & 0.0009 & 0.0014 & 122 \\
145 & 56002.1193 & 0.0008 & 0.0003 & 99 \\
152 & 56002.5325 & 0.0007 & 0.0015 & 40 \\
153 & 56002.5928 & 0.0011 & 0.0029 & 23 \\
154 & 56002.6484 & 0.0012 & $-$0.0003 & 18 \\
201 & 56005.3977 & 0.0019 & $-$0.0170 & 61 \\
\hline
  \multicolumn{5}{l}{\commenta BJD$-$2400000.} \\
  \multicolumn{5}{l}{\commentb Against max $= 2455993.5854 + 0.058852 E$.} \\
  \multicolumn{5}{l}{\commentc Number of points used to determine the maximum.} \\
\end{tabular}
\end{center}
\end{table}

\subsection{VW Hydri}\label{obj:vwhyi}

   Although this object is one of the best and oldest known prototypical
SU UMa-type dwarf novae, no high-quality photometric data
for superhumps had been publicly available.  The present observation 
\citep{ham12ROADaavso} recorded the 2011 November--December superoutburst and 
two normal outbursts in 2011 December and 2012 January.  
Although the data were not as uninterrupted as Kepler observations, 
the data provide an opportunity to analyze observations of this 
well-known object in a modern way and with modern knowledge.

   The times of superhump maxima during the superoutburst are listed 
in table \ref{tab:vwhyioc2011}.

\begin{figure*}
  \begin{center}
    \FigureFile(160mm,200mm){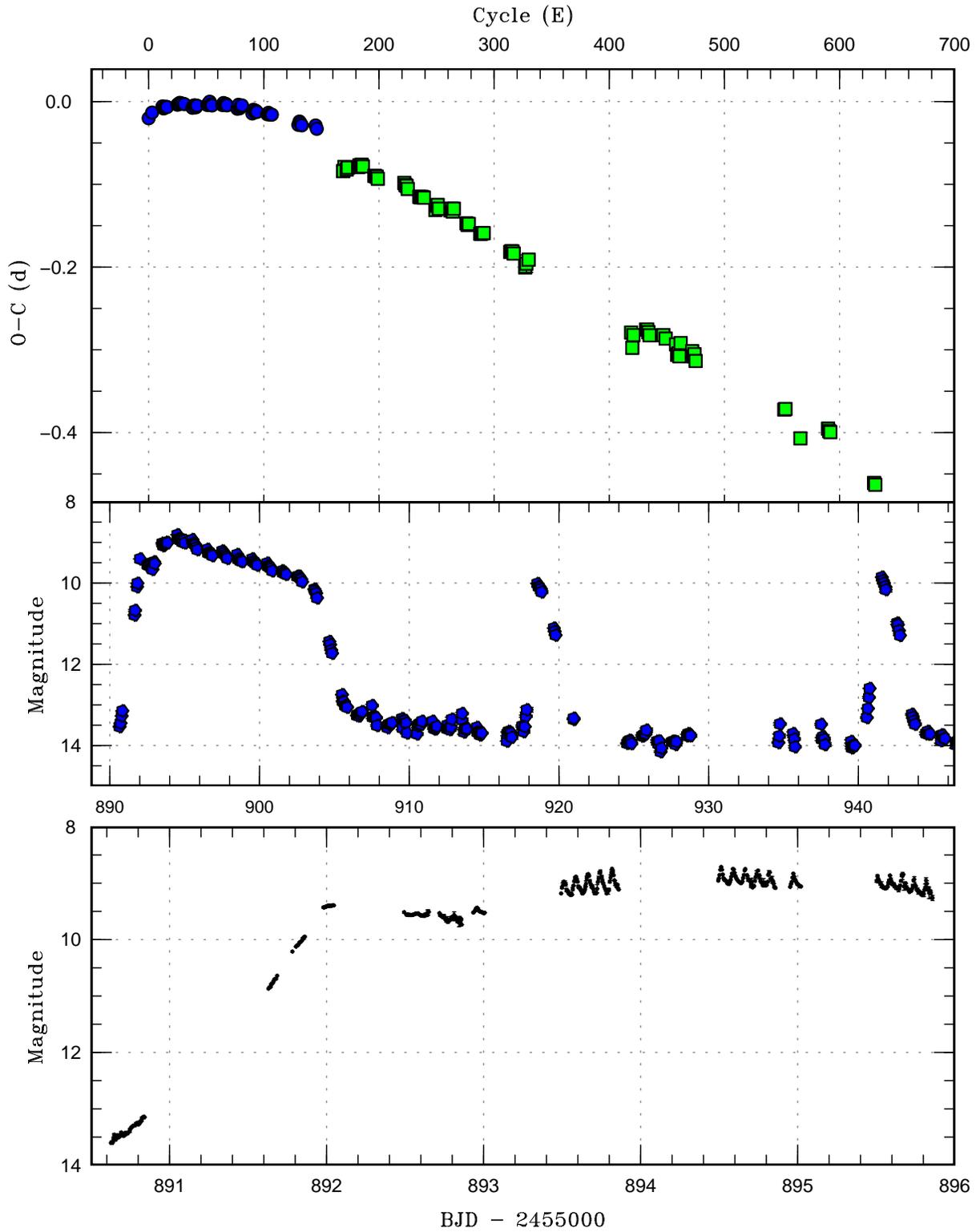}
  \end{center}
  \caption{$O-C$ diagram of superhumps in VW Hyi (2011).
     (Upper): $O-C$.
     Filled circles and filled squares represent superhumps
     and late superhumps after the rapid fading.
     We used a period of 0.076914~d for calculating the $O-C$ residuals.
     (Middle): Light curve.
     (Lower): Enlarged light curve of showing the precursor and
     evolution of superhumps.
  }
  \label{fig:vwhyihumpall}
\end{figure*}

   The outburst started with a precursor (figure \ref{fig:vwhyihumpall},
lower pnel), after a stage of short fading branch and entered 
the plateau phase.  During the plateau
phase, stage A and two segments of almost constant periods,
which we attribute to stage B and C.  The stage B--C transition
occurred between $E=68$ and $E=77$ and was apparently relatively 
smooth compared to short-$P_{\rm orb}$ systems (cf. \cite{Pdot}
figure 4).

   During the rapid fading stage of the superoutburst a phase 
reversal occurred as described as for ``traditional''
late superhumps (\cite{sch80vwhyi}; \cite{vog83lateSH}),
and this signal persisted during the quiescent period after
this superoutburst (figure \ref{fig:vwhyihumpall}).
The times of maxima of these superhumps
are listed in table \ref{tab:vwhyioc2011late}.  In contrast to
V344 Lyr (\cite{Pdot3}; \cite{woo11v344lyr}), there was no
prominent signal of ``secondary maxima'' during the late plateau 
stage of the superoutburst, and it looks like that the phase suddenly
jumped by an $\sim$0.5 $P_{\rm SH}$.  Although ``traditional''
late superhumps were usually considered to arise from
an ordinary stream-impact hot spot,\footnote{
   See also a discussion in \citet{hes92lateSH}, who reported that the
   traditional model of late superhumps by \citet{vog83lateSH} did not
   trivially explain the observed eclipse depths in OY Car.
} the apparent absence of the corresponding signal before
the rapid fading, as recorded in V344 Lyr, would make this
traditional explanation worth reconsideration.  The unavoidable
gap between BJD 2455904.9 and 2455905.4 made it difficult
to examine how this phase jump occurred.

\begin{figure}
  \begin{center}
    \FigureFile(88mm,110mm){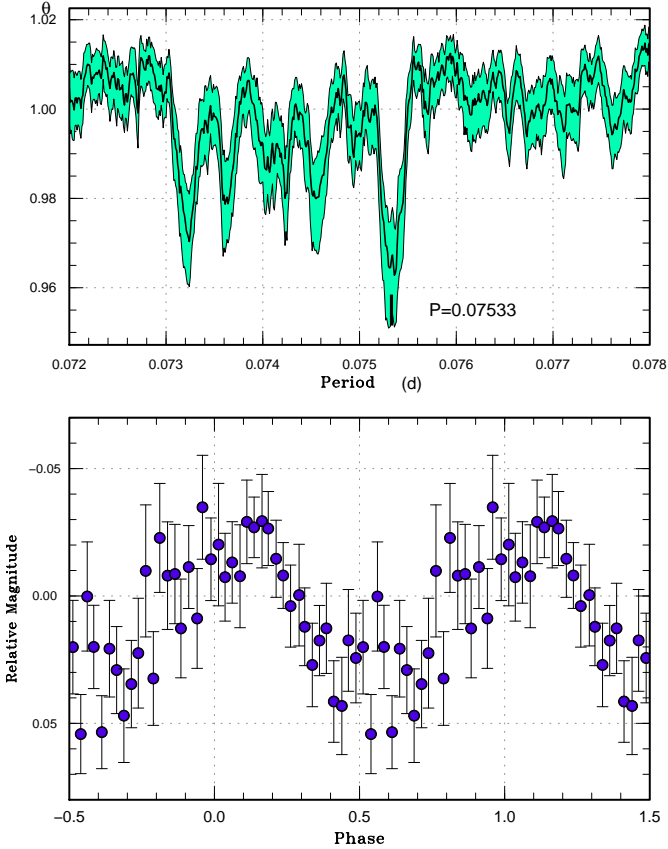}
  \end{center}
  \caption{Late superhumps in VW Hyi after a normal outburst.
     (Upper): PDM analysis after removing the mean orbital variation.
     The rejection rate for bootstrapping was reduced to 0.2 for
     better visualization.
     (Lower): Phase-averaged profile.}
  \label{fig:vwhyilateshpdm}
\end{figure}

   The times of the late superhumps, measured after subtracting
the mean orbital variation, are listed in table \ref{tab:vwhyioc2011late}.
These late superhumps persisted until the second next normal outburst,
as observed in V344 Lyr (\cite{Pdot3}; \cite{woo11v344lyr}).
After this second normal outburst, superhumps still persisted 
with a shorter period [0.075333(4)~d] and there was a well-recognizable 
signal in PDM analysis (figure \ref{fig:vwhyilateshpdm}).

\begin{table}
\caption{Superhump maxima of VW Hyi (2011).}\label{tab:vwhyioc2011}
\begin{center}
\begin{tabular}{ccccc}
\hline
$E$ & max\commenta & error & $O-C$\commentb & $N$\commentc \\
\hline
0 & 55892.5748 & 0.0007 & $-$0.0202 & 840 \\
3 & 55892.8124 & 0.0024 & $-$0.0128 & 26 \\
12 & 55893.5120 & 0.0006 & $-$0.0041 & 33 \\
13 & 55893.5865 & 0.0004 & $-$0.0064 & 42 \\
14 & 55893.6647 & 0.0005 & $-$0.0050 & 24 \\
15 & 55893.7424 & 0.0004 & $-$0.0040 & 22 \\
16 & 55893.8190 & 0.0005 & $-$0.0042 & 25 \\
25 & 55894.5140 & 0.0001 & $-$0.0001 & 721 \\
26 & 55894.5919 & 0.0001 & 0.0011 & 834 \\
27 & 55894.6701 & 0.0003 & 0.0025 & 371 \\
28 & 55894.7459 & 0.0007 & 0.0016 & 19 \\
29 & 55894.8231 & 0.0013 & 0.0020 & 21 \\
31 & 55894.9762 & 0.0002 & 0.0016 & 180 \\
38 & 55895.5102 & 0.0015 & $-$0.0018 & 28 \\
39 & 55895.5896 & 0.0008 & 0.0009 & 42 \\
40 & 55895.6670 & 0.0012 & 0.0015 & 29 \\
41 & 55895.7415 & 0.0012 & $-$0.0008 & 21 \\
42 & 55895.8203 & 0.0014 & 0.0013 & 22 \\
51 & 55896.5134 & 0.0004 & 0.0034 & 647 \\
52 & 55896.5924 & 0.0002 & 0.0057 & 839 \\
53 & 55896.6712 & 0.0004 & 0.0078 & 353 \\
54 & 55896.7447 & 0.0011 & 0.0045 & 20 \\
55 & 55896.8205 & 0.0024 & 0.0035 & 22 \\
64 & 55897.5133 & 0.0010 & 0.0054 & 32 \\
65 & 55897.5926 & 0.0013 & 0.0079 & 42 \\
66 & 55897.6672 & 0.0014 & 0.0058 & 21 \\
67 & 55897.7454 & 0.0017 & 0.0072 & 22 \\
68 & 55897.8205 & 0.0010 & 0.0056 & 22 \\
77 & 55898.5086 & 0.0013 & 0.0028 & 26 \\
78 & 55898.5903 & 0.0009 & 0.0077 & 42 \\
79 & 55898.6630 & 0.0008 & 0.0036 & 30 \\
80 & 55898.7420 & 0.0011 & 0.0058 & 31 \\
81 & 55898.8203 & 0.0017 & 0.0074 & 34 \\
90 & 55899.5028 & 0.0038 & $-$0.0010 & 25 \\
91 & 55899.5842 & 0.0007 & 0.0036 & 42 \\
92 & 55899.6586 & 0.0006 & 0.0013 & 30 \\
93 & 55899.7368 & 0.0014 & 0.0027 & 31 \\
94 & 55899.8123 & 0.0010 & 0.0015 & 33 \\
103 & 55900.5017 & 0.0022 & $-$0.0000 & 25 \\
104 & 55900.5804 & 0.0013 & 0.0019 & 21 \\
105 & 55900.6553 & 0.0015 & 0.0000 & 30 \\
106 & 55900.7322 & 0.0011 & 0.0002 & 30 \\
107 & 55900.8089 & 0.0012 & 0.0001 & 34 \\
130 & 55902.5656 & 0.0020 & $-$0.0088 & 25 \\
131 & 55902.6465 & 0.0038 & $-$0.0046 & 30 \\
132 & 55902.7214 & 0.0016 & $-$0.0066 & 30 \\
133 & 55902.7957 & 0.0015 & $-$0.0090 & 34 \\
145 & 55903.7186 & 0.0016 & $-$0.0073 & 30 \\
146 & 55903.7915 & 0.0015 & $-$0.0112 & 34 \\
\hline
  \multicolumn{5}{l}{\commenta BJD$-$2400000.} \\
  \multicolumn{5}{l}{\commentb Against max $= 2455892.5949 + 0.076765 E$.} \\
  \multicolumn{5}{l}{\commentc Number of points used to determine the maximum.} \\
\end{tabular}
\end{center}
\end{table}

\begin{table}
\caption{Late superhumps in VW Hyi (2011).}\label{tab:vwhyioc2011late}
\begin{center}
\begin{tabular}{ccccc}
\hline
$E$ & max\commenta & error & $O-C$\commentb & $N$\commentc \\
\hline
0 & 55905.5093 & 0.0016 & $-$0.0199 & 32 \\
1 & 55905.5913 & 0.0018 & $-$0.0140 & 21 \\
0 & 55905.5093 & 0.0015 & $-$0.0199 & 30 \\
3 & 55905.7420 & 0.0014 & $-$0.0155 & 32 \\
4 & 55905.8211 & 0.0019 & $-$0.0124 & 33 \\
13 & 55906.5142 & 0.0015 & $-$0.0042 & 33 \\
14 & 55906.5930 & 0.0015 & $-$0.0015 & 19 \\
15 & 55906.6703 & 0.0014 & $-$0.0004 & 28 \\
16 & 55906.7476 & 0.0012 & 0.0009 & 34 \\
17 & 55906.8225 & 0.0021 & $-$0.0003 & 33 \\
27 & 55907.5803 & 0.0011 & $-$0.0035 & 21 \\
28 & 55907.6569 & 0.0013 & $-$0.0030 & 30 \\
29 & 55907.7331 & 0.0009 & $-$0.0029 & 33 \\
30 & 55907.8075 & 0.0025 & $-$0.0046 & 34 \\
53 & 55909.5714 & 0.0009 & 0.0090 & 22 \\
54 & 55909.6458 & 0.0010 & 0.0073 & 28 \\
55 & 55909.7218 & 0.0015 & 0.0072 & 31 \\
56 & 55909.7949 & 0.0014 & 0.0042 & 32 \\
66 & 55910.5545 & 0.0008 & 0.0029 & 20 \\
67 & 55910.6310 & 0.0010 & 0.0032 & 21 \\
68 & 55910.7084 & 0.0008 & 0.0045 & 22 \\
69 & 55910.7847 & 0.0010 & 0.0048 & 27 \\
70 & 55910.8612 & 0.0023 & 0.0051 & 14 \\
80 & 55911.6155 & 0.0027 & $-$0.0016 & 15 \\
81 & 55911.6975 & 0.0007 & 0.0044 & 24 \\
82 & 55911.7755 & 0.0007 & 0.0062 & 27 \\
83 & 55911.8477 & 0.0016 & 0.0024 & 20 \\
93 & 55912.6148 & 0.0021 & 0.0085 & 16 \\
94 & 55912.6937 & 0.0012 & 0.0113 & 20 \\
95 & 55912.7673 & 0.0019 & 0.0088 & 26 \\
96 & 55912.8477 & 0.0009 & 0.0130 & 18 \\
107 & 55913.6761 & 0.0015 & 0.0043 & 20 \\
108 & 55913.7510 & 0.0010 & 0.0031 & 25 \\
109 & 55913.8291 & 0.0016 & 0.0052 & 26 \\
119 & 55914.5864 & 0.0022 & 0.0015 & 15 \\
120 & 55914.6634 & 0.0008 & 0.0024 & 20 \\
121 & 55914.7410 & 0.0019 & 0.0039 & 27 \\
122 & 55914.8180 & 0.0014 & 0.0048 & 26 \\
145 & 55916.5642 & 0.0016 & 0.0007 & 15 \\
146 & 55916.6421 & 0.0019 & 0.0025 & 21 \\
147 & 55916.7187 & 0.0015 & 0.0030 & 26 \\
148 & 55916.7930 & 0.0020 & 0.0012 & 26 \\
158 & 55917.5454 & 0.0023 & $-$0.0074 & 28 \\
159 & 55917.6268 & 0.0105 & $-$0.0021 & 13 \\
161 & 55917.7855 & 0.0047 & 0.0044 & 19 \\
250 & 55924.5427 & 0.0012 & $-$0.0112 & 46 \\
251 & 55924.6012 & 0.0035 & $-$0.0288 & 22 \\
252 & 55924.6934 & 0.0016 & $-$0.0126 & 25 \\
263 & 55925.5463 & 0.0011 & 0.0031 & 41 \\
264 & 55925.6232 & 0.0026 & 0.0039 & 23 \\
\hline
  \multicolumn{5}{l}{\commenta BJD$-$2400000.} \\
  \multicolumn{5}{l}{\commentb Against max $= 2455905.5291 + 0.076099 E$.} \\
  \multicolumn{5}{l}{\commentc Number of points used to determine the maximum.} \\
\end{tabular}
\end{center}
\end{table}

\addtocounter{table}{-1}
\begin{table}
\caption{Late superhumps in VW Hyi (2011) (continued).}
\begin{center}
\begin{tabular}{ccccc}
\hline
$E$ & max\commenta & error & $O-C$\commentb & $N$\commentc \\
\hline
265 & 55925.6969 & 0.0017 & 0.0016 & 26 \\
266 & 55925.7700 & 0.0022 & $-$0.0015 & 32 \\
278 & 55926.6932 & 0.0046 & 0.0085 & 13 \\
280 & 55926.8429 & 0.0028 & 0.0060 & 9 \\
289 & 55927.5281 & 0.0021 & 0.0063 & 23 \\
290 & 55927.5923 & 0.0015 & $-$0.0056 & 18 \\
291 & 55927.6709 & 0.0013 & $-$0.0031 & 19 \\
292 & 55927.7444 & 0.0025 & $-$0.0056 & 21 \\
293 & 55927.8374 & 0.0021 & 0.0112 & 14 \\
303 & 55928.5967 & 0.0027 & 0.0095 & 17 \\
304 & 55928.6673 & 0.0010 & 0.0040 & 18 \\
305 & 55928.7469 & 0.0010 & 0.0075 & 22 \\
306 & 55928.8156 & 0.0025 & 0.0002 & 13 \\
383 & 55934.6792 & 0.0017 & 0.0042 & 17 \\
384 & 55934.7569 & 0.0012 & 0.0057 & 14 \\
397 & 55935.7212 & 0.0034 & $-$0.0192 & 18 \\
421 & 55937.5789 & 0.0010 & 0.0121 & 16 \\
422 & 55937.6529 & 0.0007 & 0.0099 & 18 \\
423 & 55937.7285 & 0.0014 & 0.0095 & 18 \\
461 & 55940.5898 & 0.0008 & $-$0.0210 & 19 \\
462 & 55940.6646 & 0.0015 & $-$0.0223 & 22 \\
\hline
  \multicolumn{5}{l}{\commenta BJD$-$2400000.} \\
  \multicolumn{5}{l}{\commentb Against max $= 2455905.5291 + 0.076099 E$.} \\
  \multicolumn{5}{l}{\commentc Number of points used to determine the maximum.} \\
\end{tabular}
\end{center}
\end{table}

\subsection{RZ Leonis Minoris}\label{obj:rzlmi}

\begin{figure}
  \begin{center}
    \FigureFile(88mm,70mm){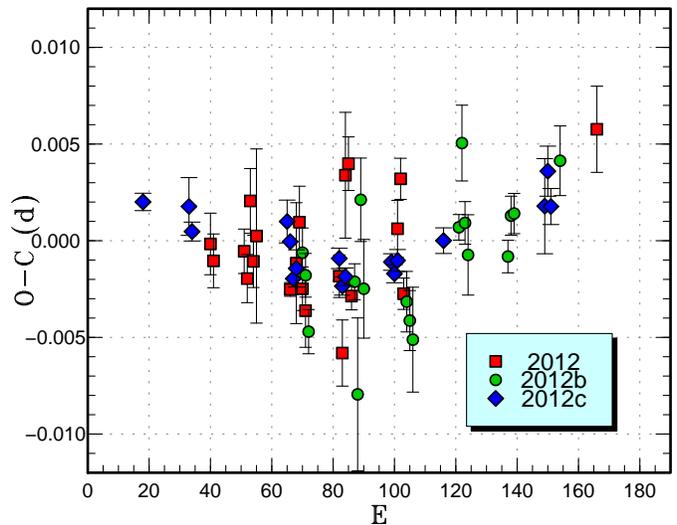}
  \end{center}
  \caption{Comparison of $O-C$ diagrams of RZ LMi between different
  superoutbursts.  A period of 0.05940~d was used to draw this figure.
  Approximate cycle counts ($E$) after the start of the superoutburst
  were used.
  }
  \label{fig:rzlmicomp2}
\end{figure}

\begin{figure}
  \begin{center}
    \FigureFile(88mm,60mm){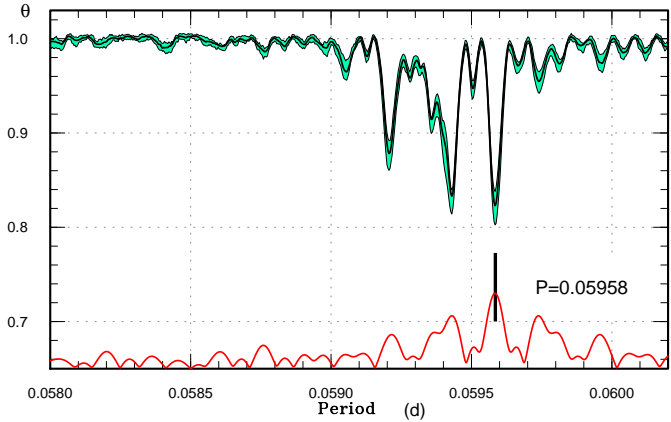}
  \end{center}
  \caption{Period analysis of plateau phases of three subsequent 
  superoutbursts of RZ LMi.  The curve at the bottom of the figure 
  represents the window function.}
  \label{fig:rzlmiallshpdm}
\end{figure}

   We analyzed three superoutbursts in 2012 from the AAVSO data
(tables \ref{tab:rzlmioc2012}, \ref{tab:rzlmioc2012b},
\ref{tab:rzlmioc2012c}).  The first two superoutbursts were
observed for their later parts and the last superoutburst was
mainly observed for the earlier part.  In measuring $P_{\rm dot}$,
we did not use $E \ge 176$ for the first outburst, which were
obtained during the fading stage and the identification of
the phases was ambiguous.
A comparison of $O-C$ diagrams is shown in figure \ref{fig:rzlmicomp2}.
Although a combined $O-C$ analysis of \citet{ole08rzlmi} in
\citet{Pdot} was suggestive of a positive $P_{\rm dot}$,
the current analysis of the new data more strongly supports
the positive $P_{\rm dot}$ in this very unusual object.
Although there was a hint of emergence of double-wave modulations
during the fading stage, we could not detect secure stage C
superhumps.  It would be worth noting that the epochs of
superhump maxima for these three superoutbursts can be reasonably
well (within 0.005~d) expressed by a single period of 0.059432(2)~d,
which might strengthen the finding by \citet{ole08rzlmi} that there
was no phase shift of superhumps between different superhumps.
A direct analysis of the photometric data (PDM method, figure
\ref{fig:rzlmiallshpdm}), however,
strongly preferred a period of 0.059585(1)~d with larger (0.010~d) and
systematically variable $O-C$ values.  Since the $O-C$ analysis 
of individual superoutbursts gives only small residuals for the period
of 0.05940~d, this preference of a different period over the
0.05940~d is an unnatural behavior.  This suggests that the apparent 
coherence of superhumps in the combined $O-C$ analysis with a period 
of 0.059432(2)~d may simply be superficial, and that the true underlying 
period may be different.  This possibility should be clarified by 
a larger set of data.

\begin{figure}
  \begin{center}
    \FigureFile(88mm,60mm){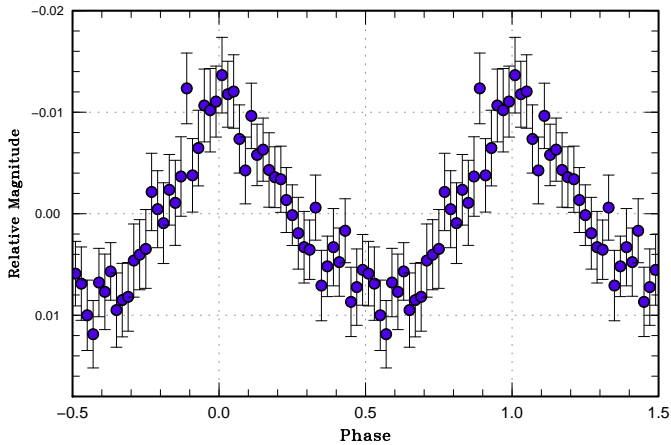}
  \end{center}
  \caption{Waveform of the candidate orbital period (0.059053~d) of RZ LMi.}
  \label{fig:rzlmiporb}
\end{figure}

   While most of the weak signal in figure \ref{fig:rzlmiallshpdm}
corresponds to aliases of the main superhump signal as is evident
from the window function, the period at 0.059053(2)~d does not
arise from an alias.  Since $\epsilon$ for objects around these
$P_{\rm SH}$ is usually 1.0\% or slightly less (cf. \cite{Pdot3}),
we regard this period to be a candidate orbital period.
The waveform of this periodicity is shown in figure \ref{fig:rzlmiporb}.
If this is the true orbital period, the $\epsilon$ for stage B
superhumps is 0.6\%.  Further testing for the stability of
this signal needs to be confirmed.

\begin{table}
\caption{Superhump maxima of RZ LMi (2012 February--March).}\label{tab:rzlmioc2012}
\begin{center}
\begin{tabular}{ccccc}
\hline
$E$ & max\commenta & error & $O-C$\commentb & $N$\commentc \\
\hline
0 & 55985.8946 & 0.0016 & 0.0046 & 10 \\
1 & 55985.9531 & 0.0014 & 0.0036 & 7 \\
11 & 55986.5476 & 0.0011 & 0.0027 & 18 \\
12 & 55986.6056 & 0.0012 & 0.0011 & 14 \\
13 & 55986.6690 & 0.0017 & 0.0050 & 14 \\
14 & 55986.7253 & 0.0014 & 0.0017 & 15 \\
15 & 55986.7866 & 0.0057 & 0.0035 & 13 \\
26 & 55987.4366 & 0.0004 & $-$0.0014 & 40 \\
28 & 55987.5568 & 0.0031 & $-$0.0003 & 12 \\
29 & 55987.6183 & 0.0019 & 0.0016 & 14 \\
30 & 55987.6743 & 0.0011 & $-$0.0019 & 14 \\
31 & 55987.7325 & 0.0019 & $-$0.0032 & 13 \\
42 & 55988.3877 & 0.0011 & $-$0.0030 & 38 \\
43 & 55988.4431 & 0.0017 & $-$0.0071 & 34 \\
44 & 55988.5117 & 0.0033 & 0.0020 & 44 \\
45 & 55988.5717 & 0.0014 & 0.0024 & 57 \\
46 & 55988.6243 & 0.0007 & $-$0.0045 & 55 \\
61 & 55989.5188 & 0.0015 & $-$0.0032 & 42 \\
62 & 55989.5808 & 0.0011 & $-$0.0007 & 42 \\
63 & 55989.6342 & 0.0008 & $-$0.0068 & 41 \\
126 & 55993.3849 & 0.0022 & $-$0.0071 & 38 \\
176 & 55996.3719 & 0.0012 & 0.0030 & 35 \\
177 & 55996.4281 & 0.0040 & $-$0.0004 & 35 \\
178 & 55996.4921 & 0.0011 & 0.0040 & 43 \\
179 & 55996.5485 & 0.0008 & 0.0009 & 43 \\
180 & 55996.6107 & 0.0011 & 0.0036 & 43 \\
\hline
  \multicolumn{5}{l}{\commenta BJD$-$2400000.} \\
  \multicolumn{5}{l}{\commentb Against max $= 2455985.8900 + 0.059540 E$.} \\
  \multicolumn{5}{l}{\commentc Number of points used to determine the maximum.} \\
\end{tabular}
\end{center}
\end{table}

\begin{table}
\caption{Superhump maxima of RZ LMi (2012 March--April).}\label{tab:rzlmioc2012b}
\begin{center}
\begin{tabular}{ccccc}
\hline
$E$ & max\commenta & error & $O-C$\commentb & $N$\commentc \\
\hline
0 & 56013.6575 & 0.0013 & 0.0032 & 48 \\
1 & 56013.7157 & 0.0011 & 0.0019 & 48 \\
2 & 56013.7722 & 0.0011 & $-$0.0011 & 47 \\
17 & 56014.6658 & 0.0009 & 0.0004 & 47 \\
18 & 56014.7193 & 0.0040 & $-$0.0055 & 61 \\
19 & 56014.7888 & 0.0022 & 0.0045 & 72 \\
20 & 56014.8436 & 0.0026 & $-$0.0002 & 25 \\
34 & 56015.6745 & 0.0016 & $-$0.0018 & 47 \\
35 & 56015.7329 & 0.0016 & $-$0.0029 & 49 \\
36 & 56015.7914 & 0.0027 & $-$0.0040 & 47 \\
50 & 56016.6273 & 0.0017 & $-$0.0006 & 64 \\
51 & 56016.6882 & 0.0006 & 0.0008 & 63 \\
52 & 56016.7519 & 0.0020 & 0.0051 & 91 \\
53 & 56016.8072 & 0.0011 & 0.0009 & 76 \\
54 & 56016.8649 & 0.0021 & $-$0.0009 & 20 \\
67 & 56017.6371 & 0.0009 & $-$0.0018 & 62 \\
68 & 56017.6986 & 0.0010 & 0.0002 & 62 \\
69 & 56017.7581 & 0.0010 & 0.0002 & 63 \\
84 & 56018.6517 & 0.0017 & 0.0017 & 60 \\
\hline
  \multicolumn{5}{l}{\commenta BJD$-$2400000.} \\
  \multicolumn{5}{l}{\commentb Against max $= 2456013.6543 + 0.059472 E$.} \\
  \multicolumn{5}{l}{\commentc Number of points used to determine the maximum.} \\
\end{tabular}
\end{center}
\end{table}

\begin{table}
\caption{Superhump maxima of RZ LMi (2012 April).}\label{tab:rzlmioc2012c}
\begin{center}
\begin{tabular}{ccccc}
\hline
$E$ & max\commenta & error & $O-C$\commentb & $N$\commentc \\
\hline
0 & 56030.8288 & 0.0004 & 0.0026 & 18 \\
15 & 56031.7195 & 0.0015 & 0.0022 & 37 \\
16 & 56031.7776 & 0.0005 & 0.0009 & 32 \\
47 & 56033.6196 & 0.0011 & 0.0012 & 34 \\
48 & 56033.6779 & 0.0004 & 0.0001 & 81 \\
49 & 56033.7354 & 0.0003 & $-$0.0018 & 88 \\
50 & 56033.7953 & 0.0010 & $-$0.0013 & 21 \\
64 & 56034.6275 & 0.0005 & $-$0.0009 & 63 \\
65 & 56034.6854 & 0.0005 & $-$0.0023 & 99 \\
66 & 56034.7453 & 0.0005 & $-$0.0018 & 96 \\
81 & 56035.6371 & 0.0004 & $-$0.0012 & 67 \\
82 & 56035.6959 & 0.0005 & $-$0.0018 & 94 \\
83 & 56035.7560 & 0.0006 & $-$0.0011 & 74 \\
98 & 56036.6480 & 0.0007 & $-$0.0002 & 62 \\
131 & 56038.6100 & 0.0025 & 0.0013 & 44 \\
132 & 56038.6712 & 0.0013 & 0.0031 & 63 \\
133 & 56038.7287 & 0.0009 & 0.0012 & 61 \\
\hline
  \multicolumn{5}{l}{\commenta BJD$-$2400000.} \\
  \multicolumn{5}{l}{\commentb Against max $= 2456030.8262 + 0.059408 E$.} \\
  \multicolumn{5}{l}{\commentc Number of points used to determine the maximum.} \\
\end{tabular}
\end{center}
\end{table}

\subsection{BK Lyncis}\label{obj:bklyn}

   BK Lyn has been a well-known permanent superhumper below
the period gap \citep{ski93bklyn}.  The object, however, has recently
been demonstrated to show dwarf nova-type outbursts, and
the pattern of outbursts is quite similar to those of
ER UMa stars (E. de Miguel, see also \cite{kem12bklynsass}).  
According to the Northern Sky Variability Survey (NSVS),
the object was still in novalike (NL)-type state in 2002.\footnote{
  $<$http://skydot.lanl.gov/nsvs/star.php?\\num=7454712\&mask=32004$>$
}  The CRTS data indicate that the object already entered
a DN state in 2005.  The outburst-like variations were also
recorded in AAVSO observations in 2005--2006.  We here analyze
observations in 2012, mostly from the AAVSO database.

\begin{figure}
  \begin{center}
    \FigureFile(88mm,110mm){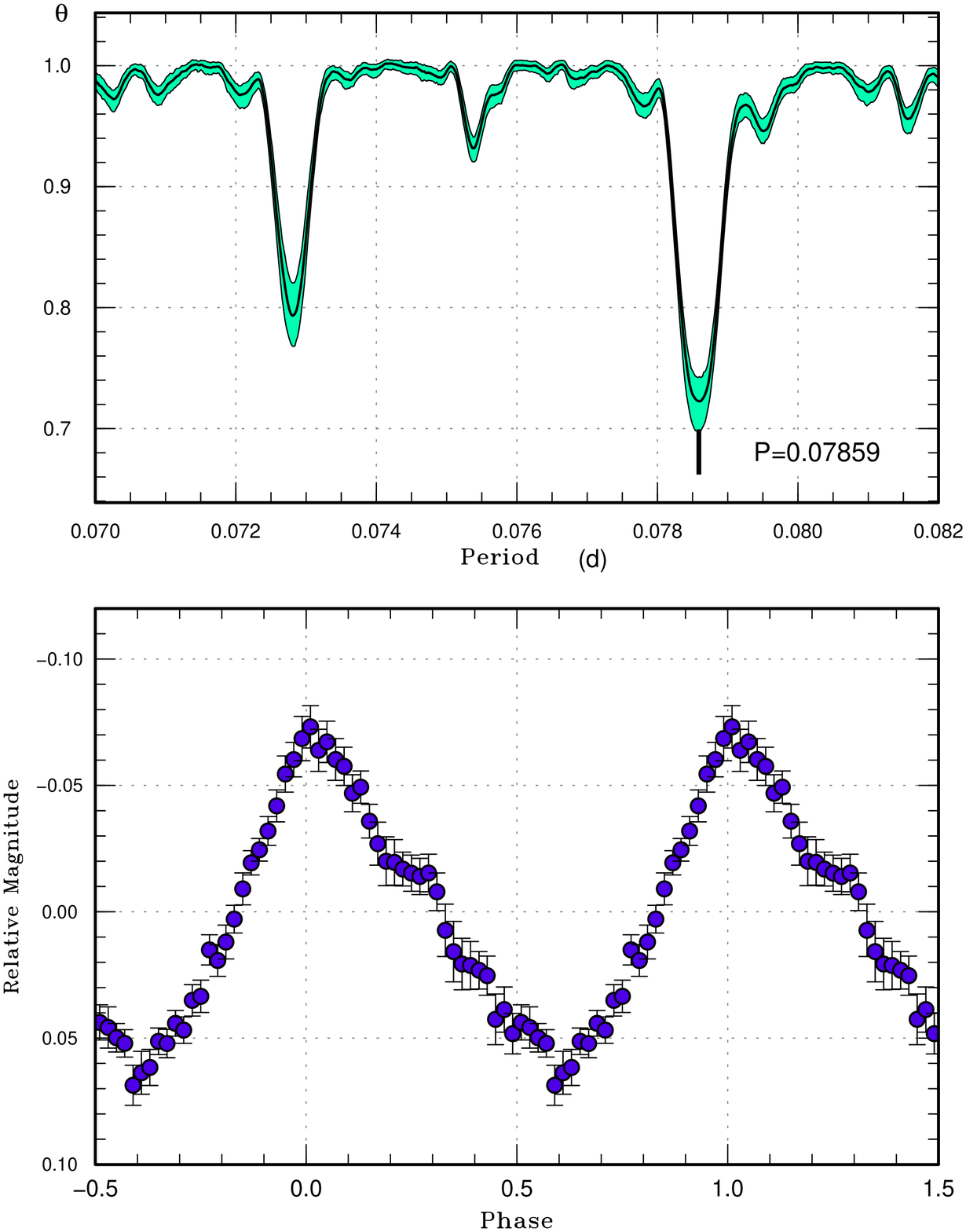}
  \end{center}
  \caption{Positive superhumps in BK Lyn (2012 April). (Upper): PDM analysis.
     A period at 0.0728~d is a one-day alias of the superhump period.
     This period coincided the period of negative superhumps by chance.
     (Lower): Phase-averaged profile.}
  \label{fig:bklynshpdm}
\end{figure}

   As in recent ER UMa \citep{ohs12eruma}, the object showed 
negative superhumps during most of its outburst cycle, and showed
positive superhumps during the $\sim$10~d initial part of
superoutbursts.  We first identified the period of positive superhumps 
using the best observed superoutburst in 2012 April.  
The period was identified to be 0.07859(1)~d (figure \ref{fig:bklynshpdm}).
With the help of this period, we could identify the times of
superhump maxima during the less-observed 2012 February--March
superoutburst (table \ref{tab:bklynoc2012}).  Although the maxima for
$E \le 3$ were those of negative superhumps [$P=$0.071(2)~d],
we included these epochs to illustrate the smooth transition
from negative superhumps to positive superhumps in phase
as recorded in ER UMa \citep{ohs12eruma}.
The times of superhump maxima during the 2012 April
superoutburst are shown in table \ref{tab:bklynoc2012b}.
Since the epochs $E=1,2$ were obtained before the maximum of 
the superoutburst, we did not use them in calculating the period 
and $P_{\rm dot}$.  During the later stage ($E \ge 114$), 
the structure of superhumps became complex
and both negative and positive superhumps appeared to coexist.
Although the superhump period during the superoutburst was not very 
different from those recorded during its NL-type state
\citep{ski93bklyn}, the amplitudes of superhumps were much larger
than those in its former NL-type state, implying that 3:1 resonance is
more strongly excited during a superoutburst.

\begin{figure}
  \begin{center}
    \FigureFile(88mm,70mm){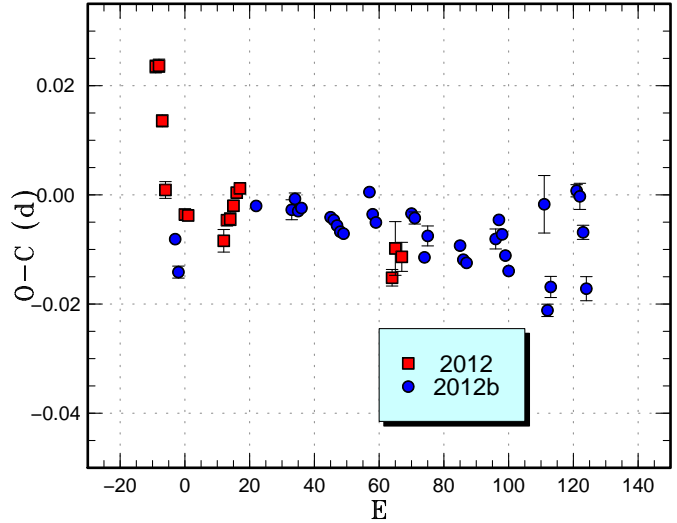}
  \end{center}
  \caption{Comparison of $O-C$ diagrams of positive superhumps of
  BK Lyn between different superoutbursts.  The abbreviation 2012
  refers to the 2012 February--March superoutburst and 2012b
  the 2012 April one, respectively.  A period of 0.07859~d was 
  used to draw this figure.  Approximate cycle counts ($E$) after
  the appearance of the positive superhumps were used.
  The maxima for $E < 0$ are negative superhumps.
  As known in ER UMa \citep{ohs12eruma}, there were relatively
  large intranight $O-C$ variations against the mean period of
  positive superhumps.  This can be interpreted as a result of
  the coexistence of negative superhumps. 
  }
  \label{fig:bklyncomp}
\end{figure}

   A comparison of the $O-C$ diagrams of positive superhumps
between different superhumps is shown in figure \ref{fig:bklyncomp}.
The disagreement between the $O-C$ diagrams was slightly larger
than in other SU UMa-type dwarf novae, which may be a result of
remnant, overlapping negative superhumps.  Particularly, the relatively
large scatter in the $O-C$ diagram in the later part of this figure
was caused by profile variations caused by evolving negative superhumps.

\begin{figure*}
  \begin{center}
    \FigureFile(160mm,190mm){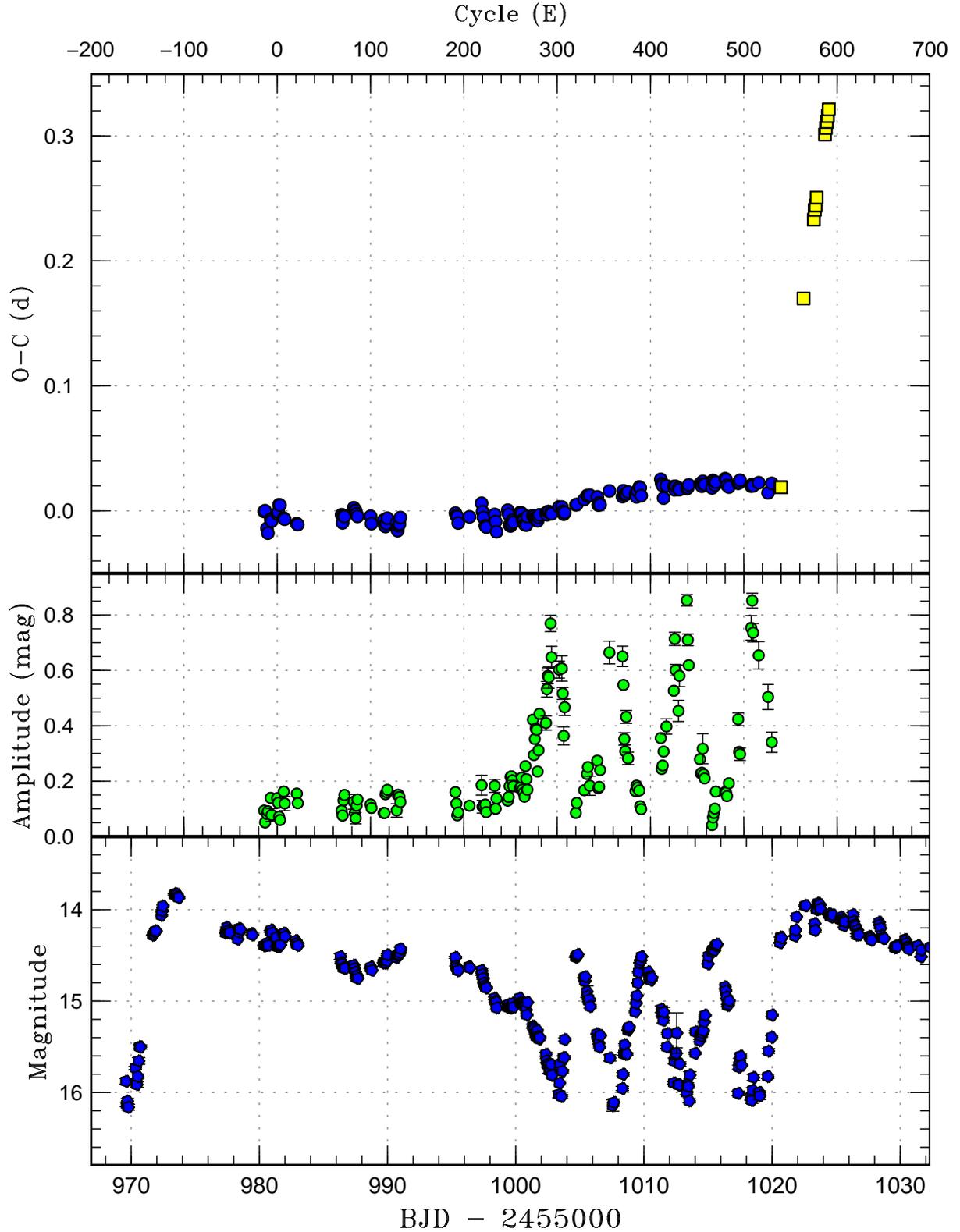}
  \end{center}
  \caption{$O-C$ diagram of negative superhumps in BK Lyn (2012).
     (Upper:) $O-C$.
     Filled circles and filled squares represent negative superhumps
     and positive superhumps, respectively.  The positive superhumps
     appeared as the next superoutburst started, and the phase of
     the hump maximum was continuous with that of the preceding
     negative superhumps.  The maxima of positive superhumps during
     the first superoutburst are not shown.
     We used a period of 0.07280~d for calculating the $O-C$ residuals.
     (Middle:) Amplitudes of negative superhumps.  The amplitudes
     become larger when the system fades.
     (Lower:) Light curve.  The supercycle is $\sim$50~d and there were
     three normal outbursts between the superoutbursts.
  }
  \label{fig:bklynnegsh}
\end{figure*}

   The behavior of negative superhumps was very similar to that of
ER UMa (figure \ref{fig:bklynnegsh}; cf. figure 2 of \cite{ohs12eruma}).
The mean period of negative superhumps during the superoutburst
was 0.072793(7)~d ($0 \le E \le 280$).  The period slightly lengthened 
later, and stabilized to a slightly longer period
during the phase showing normal outbursts [mean period 0.072922(6)~d
for $280 \le E \le 544$] (table \ref{tab:bklynocneg}).  It is noteworthy that
there was no jump in phase when superhumps switched from negative ones
to positive ones.  The same phenomenon was observed in ER UMa
\citep{ohs12eruma}.  The amplitudes of negative superhumps were well
correlated with the system magnitude, and the amplitudes became larger
when the system gets fainter.  This relation was also observed in
V344 Lyr (cf. figure 79 of \cite{Pdot3}), V503 Cyg \citep{har95v503cyg},
MN Dra \citep{pav10mndraproc} and ER UMa, although \citet{ohs12eruma}
did not present the corresponding figure.

   \citet{kem12bklynsass} proposed that a transition from a permanent 
superhumper to a dwarf nova may a result of cooling of the white
dwarf following a nova eruption.  The time-scale (several years) of 
this transition, however, appears to be too short compared to the
proposed duration ($\sim$ 1900~yr) of the post-nova state.
The change of state may also be a result of variable mass-transfer rate
as recorded in other ER UMa-type dwarf novae such as V1159 Ori
\citep{kat01v1159ori} rather than secular evolution.

\begin{table}
\caption{Superhump maxma of BK Lyn (2012 February--March).}\label{tab:bklynoc2012}
\begin{center}
\begin{tabular}{ccccc}
\hline
$E$ & max\commenta & error & $O-C$\commentb & $N$\commentc \\
\hline
0 & 55971.6983 & 0.0012 & 0.0155 & 69 \\
1 & 55971.7771 & 0.0012 & 0.0159 & 75 \\
2 & 55971.8456 & 0.0011 & 0.0061 & 76 \\
3 & 55971.9115 & 0.0015 & $-$0.0062 & 71 \\
9 & 55972.3785 & 0.0006 & $-$0.0088 & 39 \\
10 & 55972.4569 & 0.0005 & $-$0.0087 & 41 \\
21 & 55973.3168 & 0.0021 & $-$0.0099 & 21 \\
22 & 55973.3992 & 0.0008 & $-$0.0059 & 40 \\
23 & 55973.4779 & 0.0012 & $-$0.0054 & 40 \\
24 & 55973.5590 & 0.0008 & $-$0.0026 & 39 \\
25 & 55973.6400 & 0.0009 & 0.0001 & 39 \\
26 & 55973.7193 & 0.0010 & 0.0012 & 20 \\
73 & 55977.3967 & 0.0015 & $-$0.0006 & 72 \\
74 & 55977.4807 & 0.0049 & 0.0051 & 43 \\
76 & 55977.6363 & 0.0027 & 0.0041 & 71 \\
\hline
  \multicolumn{5}{l}{\commenta BJD$-$2400000.} \\
  \multicolumn{5}{l}{\commentb Against max $= 2455971.6828 + 0.078280 E$.} \\
  \multicolumn{5}{l}{\commentc Number of points used to determine the maximum.} \\
\end{tabular}
\end{center}
\end{table}

\begin{table}
\caption{Times of negative superhumps in BK Lyn.}\label{tab:bklynocneg}
\begin{center}
\begin{tabular}{ccccc}
\hline
$E$ & max\commenta & error & $O-C$\commentb & $N$\commentc \\
\hline
0 & 55980.3754 & 0.0007 & 0.0147 & 38 \\
1 & 55980.4487 & 0.0017 & 0.0153 & 38 \\
3 & 55980.5802 & 0.0018 & 0.0010 & 37 \\
4 & 55980.6492 & 0.0011 & $-$0.0028 & 38 \\
7 & 55980.8787 & 0.0011 & 0.0080 & 33 \\
8 & 55980.9500 & 0.0028 & 0.0065 & 35 \\
14 & 55981.3926 & 0.0006 & 0.0118 & 39 \\
15 & 55981.4669 & 0.0007 & 0.0133 & 37 \\
16 & 55981.5455 & 0.0014 & 0.0191 & 37 \\
17 & 55981.6180 & 0.0020 & 0.0186 & 38 \\
21 & 55981.8987 & 0.0005 & 0.0079 & 33 \\
22 & 55981.9707 & 0.0020 & 0.0071 & 35 \\
35 & 55982.9136 & 0.0006 & 0.0027 & 37 \\
36 & 55982.9856 & 0.0014 & 0.0018 & 35 \\
83 & 55986.4151 & 0.0016 & 0.0066 & 38 \\
84 & 55986.4814 & 0.0014 & 0.0001 & 33 \\
85 & 55986.5601 & 0.0013 & 0.0059 & 34 \\
86 & 55986.6320 & 0.0009 & 0.0050 & 37 \\
96 & 55987.3671 & 0.0016 & 0.0114 & 34 \\
97 & 55987.4377 & 0.0017 & 0.0091 & 37 \\
98 & 55987.5104 & 0.0029 & 0.0089 & 20 \\
99 & 55987.5807 & 0.0010 & 0.0064 & 37 \\
100 & 55987.6514 & 0.0012 & 0.0042 & 74 \\
114 & 55988.6707 & 0.0007 & 0.0034 & 70 \\
115 & 55988.7376 & 0.0010 & $-$0.0026 & 70 \\
128 & 55989.6868 & 0.0011 & $-$0.0006 & 107 \\
129 & 55989.7556 & 0.0009 & $-$0.0047 & 109 \\
130 & 55989.8273 & 0.0006 & $-$0.0059 & 76 \\
131 & 55989.9025 & 0.0008 & $-$0.0036 & 38 \\
132 & 55989.9795 & 0.0006 & 0.0006 & 27 \\
142 & 55990.7038 & 0.0018 & $-$0.0037 & 29 \\
143 & 55990.7704 & 0.0006 & $-$0.0100 & 28 \\
144 & 55990.8470 & 0.0009 & $-$0.0063 & 37 \\
145 & 55990.9213 & 0.0009 & $-$0.0049 & 36 \\
146 & 55990.9992 & 0.0013 & 0.0002 & 32 \\
205 & 55995.2980 & 0.0015 & $-$0.0001 & 37 \\
206 & 55995.3687 & 0.0010 & $-$0.0022 & 105 \\
207 & 55995.4409 & 0.0019 & $-$0.0029 & 61 \\
208 & 55995.5085 & 0.0016 & $-$0.0082 & 30 \\
220 & 55996.3870 & 0.0015 & $-$0.0040 & 37 \\
233 & 55997.3443 & 0.0015 & 0.0060 & 28 \\
234 & 55997.4104 & 0.0020 & $-$0.0008 & 37 \\
235 & 55997.4785 & 0.0015 & $-$0.0056 & 36 \\
237 & 55997.6176 & 0.0013 & $-$0.0122 & 37 \\
238 & 55997.6893 & 0.0015 & $-$0.0134 & 26 \\
247 & 55998.3545 & 0.0010 & $-$0.0039 & 33 \\
248 & 55998.4217 & 0.0017 & $-$0.0096 & 36 \\
249 & 55998.4863 & 0.0012 & $-$0.0179 & 37 \\
261 & 55999.3772 & 0.0008 & $-$0.0014 & 149 \\
262 & 55999.4466 & 0.0006 & $-$0.0049 & 155 \\
263 & 55999.5109 & 0.0006 & $-$0.0134 & 85 \\
264 & 55999.5829 & 0.0004 & $-$0.0142 & 83 \\
\hline
  \multicolumn{5}{l}{\commenta BJD$-$2400000.} \\
  \multicolumn{5}{l}{\commentb Against max $= 2455980.3606 + 0.072866 E$.} \\
  \multicolumn{5}{l}{\commentc Number of points used to determine the maximum.} \\
\end{tabular}
\end{center}
\end{table}

\addtocounter{table}{-1}
\begin{table}
\caption{Times of negative superhumps in BK Lyn.}
\begin{center}
\begin{tabular}{ccccc}
\hline
$E$ & max\commenta & error & $O-C$\commentb & $N$\commentc \\
\hline
265 & 55999.6572 & 0.0009 & $-$0.0128 & 80 \\
266 & 55999.7331 & 0.0010 & $-$0.0098 & 76 \\
267 & 55999.8042 & 0.0010 & $-$0.0115 & 69 \\
274 & 56000.3215 & 0.0007 & $-$0.0043 & 59 \\
275 & 56000.3944 & 0.0005 & $-$0.0043 & 94 \\
276 & 56000.4666 & 0.0004 & $-$0.0049 & 96 \\
277 & 56000.5342 & 0.0008 & $-$0.0102 & 74 \\
278 & 56000.6064 & 0.0006 & $-$0.0109 & 118 \\
279 & 56000.6762 & 0.0006 & $-$0.0139 & 144 \\
280 & 56000.7488 & 0.0003 & $-$0.0142 & 169 \\
281 & 56000.8213 & 0.0004 & $-$0.0146 & 159 \\
282 & 56000.9008 & 0.0010 & $-$0.0079 & 65 \\
288 & 56001.3384 & 0.0004 & $-$0.0076 & 90 \\
289 & 56001.4108 & 0.0003 & $-$0.0080 & 165 \\
290 & 56001.4818 & 0.0003 & $-$0.0099 & 164 \\
291 & 56001.5541 & 0.0004 & $-$0.0104 & 138 \\
292 & 56001.6257 & 0.0004 & $-$0.0117 & 160 \\
293 & 56001.6982 & 0.0005 & $-$0.0120 & 154 \\
294 & 56001.7739 & 0.0004 & $-$0.0092 & 118 \\
295 & 56001.8487 & 0.0003 & $-$0.0073 & 127 \\
302 & 56002.3597 & 0.0005 & $-$0.0064 & 52 \\
303 & 56002.4311 & 0.0005 & $-$0.0078 & 60 \\
304 & 56002.5064 & 0.0005 & $-$0.0054 & 61 \\
305 & 56002.5792 & 0.0006 & $-$0.0054 & 60 \\
307 & 56002.7236 & 0.0004 & $-$0.0068 & 70 \\
308 & 56002.7958 & 0.0006 & $-$0.0074 & 70 \\
316 & 56003.3838 & 0.0005 & $-$0.0024 & 67 \\
319 & 56003.6021 & 0.0007 & $-$0.0026 & 77 \\
320 & 56003.6714 & 0.0004 & $-$0.0062 & 96 \\
321 & 56003.7420 & 0.0008 & $-$0.0085 & 70 \\
322 & 56003.8161 & 0.0006 & $-$0.0073 & 61 \\
334 & 56004.6958 & 0.0008 & $-$0.0020 & 71 \\
335 & 56004.7690 & 0.0007 & $-$0.0016 & 70 \\
343 & 56005.3553 & 0.0007 & 0.0018 & 52 \\
346 & 56005.5768 & 0.0005 & 0.0047 & 31 \\
347 & 56005.6492 & 0.0006 & 0.0041 & 58 \\
349 & 56005.7955 & 0.0019 & 0.0048 & 40 \\
357 & 56006.3766 & 0.0006 & 0.0030 & 150 \\
358 & 56006.4429 & 0.0005 & $-$0.0036 & 161 \\
359 & 56006.5174 & 0.0005 & $-$0.0020 & 150 \\
360 & 56006.5887 & 0.0007 & $-$0.0036 & 119 \\
370 & 56007.3278 & 0.0005 & 0.0069 & 38 \\
384 & 56008.3424 & 0.0005 & 0.0014 & 99 \\
385 & 56008.4199 & 0.0003 & 0.0060 & 99 \\
386 & 56008.4893 & 0.0006 & 0.0025 & 90 \\
387 & 56008.5628 & 0.0007 & 0.0032 & 132 \\
388 & 56008.6355 & 0.0005 & 0.0030 & 82 \\
390 & 56008.7829 & 0.0007 & 0.0047 & 64 \\
398 & 56009.3636 & 0.0008 & 0.0025 & 67 \\
399 & 56009.4342 & 0.0006 & 0.0002 & 67 \\
400 & 56009.5119 & 0.0008 & 0.0051 & 66 \\
402 & 56009.6606 & 0.0008 & 0.0080 & 65 \\
403 & 56009.7327 & 0.0013 & 0.0072 & 70 \\
404 & 56009.7992 & 0.0014 & 0.0008 & 66 \\
\hline
  \multicolumn{5}{l}{\commenta BJD$-$2400000.} \\
  \multicolumn{5}{l}{\commentb Against max $= 2455980.3606 + 0.072866 E$.} \\
  \multicolumn{5}{l}{\commentc Number of points used to determine the maximum.} \\
\end{tabular}
\end{center}
\end{table}

\addtocounter{table}{-1}
\begin{table}
\caption{Times of negative superhumps in BK Lyn.}
\begin{center}
\begin{tabular}{ccccc}
\hline
$E$ & max\commenta & error & $O-C$\commentb & $N$\commentc \\
\hline
425 & 56011.3411 & 0.0003 & 0.0125 & 162 \\
426 & 56011.4104 & 0.0004 & 0.0090 & 233 \\
427 & 56011.4817 & 0.0003 & 0.0074 & 238 \\
428 & 56011.5444 & 0.0005 & $-$0.0027 & 136 \\
431 & 56011.7727 & 0.0005 & 0.0069 & 48 \\
439 & 56012.3546 & 0.0003 & 0.0059 & 220 \\
440 & 56012.4247 & 0.0003 & 0.0031 & 212 \\
441 & 56012.5003 & 0.0003 & 0.0059 & 124 \\
444 & 56012.7173 & 0.0008 & 0.0043 & 70 \\
445 & 56012.7888 & 0.0006 & 0.0029 & 61 \\
453 & 56013.3721 & 0.0002 & 0.0033 & 213 \\
454 & 56013.4456 & 0.0003 & 0.0040 & 215 \\
455 & 56013.5206 & 0.0003 & 0.0061 & 111 \\
467 & 56014.3951 & 0.0004 & 0.0062 & 196 \\
468 & 56014.4673 & 0.0004 & 0.0056 & 163 \\
469 & 56014.5387 & 0.0004 & 0.0041 & 131 \\
470 & 56014.6154 & 0.0011 & 0.0080 & 23 \\
471 & 56014.6861 & 0.0008 & 0.0057 & 56 \\
472 & 56014.7585 & 0.0008 & 0.0053 & 56 \\
480 & 56015.3382 & 0.0010 & 0.0020 & 299 \\
481 & 56015.4170 & 0.0006 & 0.0080 & 281 \\
482 & 56015.4887 & 0.0007 & 0.0068 & 201 \\
483 & 56015.5588 & 0.0006 & 0.0041 & 162 \\
484 & 56015.6340 & 0.0005 & 0.0064 & 50 \\
494 & 56016.3649 & 0.0006 & 0.0086 & 211 \\
495 & 56016.4368 & 0.0006 & 0.0077 & 182 \\
496 & 56016.5051 & 0.0005 & 0.0031 & 125 \\
498 & 56016.6493 & 0.0007 & 0.0016 & 46 \\
508 & 56017.3803 & 0.0005 & 0.0039 & 71 \\
509 & 56017.4536 & 0.0007 & 0.0044 & 108 \\
510 & 56017.5283 & 0.0008 & 0.0062 & 74 \\
522 & 56018.3972 & 0.0005 & 0.0007 & 60 \\
523 & 56018.4714 & 0.0003 & 0.0020 & 62 \\
524 & 56018.5430 & 0.0004 & 0.0008 & 61 \\
530 & 56018.9824 & 0.0006 & 0.0030 & 129 \\
540 & 56019.7023 & 0.0011 & $-$0.0058 & 71 \\
544 & 56020.0010 & 0.0010 & 0.0015 & 150 \\
\hline
  \multicolumn{5}{l}{\commenta BJD$-$2400000.} \\
  \multicolumn{5}{l}{\commentb Against max $= 2455980.3606 + 0.072866 E$.} \\
  \multicolumn{5}{l}{\commentc Number of points used to determine the maximum.} \\
\end{tabular}
\end{center}
\end{table}

\begin{table}
\caption{Superhump maxima of BK Lyn (2012 April).}\label{tab:bklynoc2012b}
\begin{center}
\begin{tabular}{ccccc}
\hline
$E$ & max\commenta & error & $O-C$\commentb & $N$\commentc \\
\hline
0 & 56020.6533 & 0.0009 & $-$0.0043 & 67 \\
1 & 56020.7259 & 0.0011 & $-$0.0102 & 66 \\
25 & 56022.6242 & 0.0004 & 0.0029 & 142 \\
36 & 56023.4880 & 0.0018 & 0.0027 & 62 \\
37 & 56023.5685 & 0.0011 & 0.0047 & 95 \\
38 & 56023.6449 & 0.0005 & 0.0025 & 71 \\
39 & 56023.7240 & 0.0006 & 0.0031 & 70 \\
48 & 56024.4297 & 0.0005 & 0.0018 & 66 \\
49 & 56024.5077 & 0.0006 & 0.0013 & 73 \\
50 & 56024.5853 & 0.0006 & 0.0003 & 94 \\
51 & 56024.6628 & 0.0004 & $-$0.0007 & 146 \\
52 & 56024.7410 & 0.0005 & $-$0.0011 & 139 \\
60 & 56025.3774 & 0.0005 & 0.0069 & 82 \\
61 & 56025.4519 & 0.0005 & 0.0029 & 82 \\
62 & 56025.5290 & 0.0005 & 0.0014 & 90 \\
73 & 56026.3951 & 0.0008 & 0.0035 & 75 \\
74 & 56026.4729 & 0.0011 & 0.0027 & 75 \\
77 & 56026.7014 & 0.0009 & $-$0.0043 & 76 \\
78 & 56026.7839 & 0.0018 & $-$0.0004 & 43 \\
88 & 56027.5681 & 0.0009 & $-$0.0017 & 43 \\
89 & 56027.6441 & 0.0005 & $-$0.0043 & 70 \\
90 & 56027.7221 & 0.0005 & $-$0.0048 & 66 \\
99 & 56028.4338 & 0.0018 & $-$0.0000 & 88 \\
100 & 56028.5159 & 0.0008 & 0.0035 & 50 \\
101 & 56028.5918 & 0.0007 & 0.0009 & 74 \\
102 & 56028.6665 & 0.0006 & $-$0.0030 & 69 \\
103 & 56028.7423 & 0.0008 & $-$0.0057 & 63 \\
114 & 56029.6190 & 0.0053 & 0.0069 & 51 \\
115 & 56029.6781 & 0.0011 & $-$0.0124 & 66 \\
116 & 56029.7610 & 0.0019 & $-$0.0081 & 40 \\
124 & 56030.4074 & 0.0011 & 0.0098 & 153 \\
125 & 56030.4849 & 0.0024 & 0.0089 & 104 \\
126 & 56030.5569 & 0.0013 & 0.0023 & 69 \\
127 & 56030.6252 & 0.0022 & $-$0.0080 & 91 \\
\hline
  \multicolumn{5}{l}{\commenta BJD$-$2400000.} \\
  \multicolumn{5}{l}{\commentb Against max $= 2456020.6576 + 0.078548 E$.} \\
  \multicolumn{5}{l}{\commentc Number of points used to determine the maximum.} \\
\end{tabular}
\end{center}
\end{table}

\subsection{V585 Lyrae}\label{obj:v585lyr}

   Although the object was extensively observed during the 2003
superoutburst (cf. \cite{Pdot}), no secure record of an outburst
had been recorded until 2012.  The 2012 superoutburst was detected
by P. A. Dubovsky (vsnet-alert 14494).  We obtained two nights of
observations and listed the times of maxima (table \ref{tab:v585lyroc2012}).
The period in table \ref{tab:perlist} was obtained by the PDM method.

\begin{table}
\caption{Superhump maxima of V585 Lyr (2012).}\label{tab:v585lyroc2012}
\begin{center}
\begin{tabular}{ccccc}
\hline
$E$ & max\commenta & error & $O-C$\commentb & $N$\commentc \\
\hline
0 & 56045.1580 & 0.0005 & $-$0.0005 & 123 \\
1 & 56045.2195 & 0.0006 & 0.0006 & 71 \\
18 & 56046.2441 & 0.0008 & $-$0.0025 & 124 \\
19 & 56046.3094 & 0.0023 & 0.0023 & 68 \\
\hline
  \multicolumn{5}{l}{\commenta BJD$-$2400000.} \\
  \multicolumn{5}{l}{\commentb Against max $= 2456045.1584 + 0.060454 E$.} \\
  \multicolumn{5}{l}{\commentc Number of points used to determine the maximum.} \\
\end{tabular}
\end{center}
\end{table}

\subsection{FQ Monocerotis}\label{obj:fqmon}

   Only a fragment of the 2011 superoutburst was observed.
The times of superhump maxima are listed in table \ref{tab:fqmonoc2011}.
Since the object quickly faded three days after the observation,
it is likely we only observed stage C superhumps.

\begin{table}
\caption{Superhump maxima of FQ Mon (2011).}\label{tab:fqmonoc2011}
\begin{center}
\begin{tabular}{ccccc}
\hline
$E$ & max\commenta & error & $O-C$\commentb & $N$\commentc \\
\hline
0 & 55922.0918 & 0.0011 & $-$0.0015 & 130 \\
1 & 55922.1676 & 0.0033 & 0.0015 & 67 \\
13 & 55923.0405 & 0.0020 & 0.0018 & 91 \\
14 & 55923.1096 & 0.0015 & $-$0.0018 & 130 \\
\hline
  \multicolumn{5}{l}{\commenta BJD$-$2400000.} \\
  \multicolumn{5}{l}{\commentb Against max $= 2455922.0934 + 0.072718 E$.} \\
  \multicolumn{5}{l}{\commentc Number of points used to determine the maximum.} \\
\end{tabular}
\end{center}
\end{table}

\subsection{V1032 Ophiuchi}\label{obj:v1032oph}

   This object is an eclipsing SU UMa-type dwarf nova \citep{Pdot2}.
By applying Markov-Chain Monte Carlo (MCMC) method to the phased
data using the period and epoch as trial variables
(see appendix \ref{sec:app:mcmcecl}), we obtained an updated
orbital ephemeris of
\begin{equation}
{\rm Min(BJD)} = 2455286.68256(7) + 0.081055386(10) E
\label{equ:v1032ophecl}
\end{equation}
based on 2010 and 2012 observations.
The times of superhump maxima are listed in table \ref{tab:v1032ophoc2012}.
A PDM analysis yielded a consistent result of 0.08599(5)~d.

\begin{table}
\caption{Superhump maxima of V1032 Oph (2012).}\label{tab:v1032ophoc2012}
\begin{center}
\begin{tabular}{cccccc}
\hline
$E$ & max\commenta & error & $O-C$\commentb & phase\commentc & $N$\commentd \\
\hline
0 & 56076.3539 & 0.0046 & $-$0.0101 & 0.52 & 80 \\
9 & 56077.1475 & 0.0015 & 0.0098 & 0.23 & 127 \\
12 & 56077.3984 & 0.0009 & 0.0029 & 0.15 & 148 \\
47 & 56080.4017 & 0.0012 & $-$0.0026 & 0.29 & 150 \\
\hline
  \multicolumn{6}{l}{\commenta BJD$-$2400000.} \\
  \multicolumn{6}{l}{\commentb Against max $= 2456076.3640 + 0.085965 E$.} \\
  \multicolumn{6}{l}{\commentc Orbital phase.} \\
  \multicolumn{6}{l}{\commentd Number of points used to determine the maximum.} \\
\end{tabular}
\end{center}
\end{table}

\subsection{V2051 Ophiuchi}\label{obj:v2051oph}

   Only one superhump was recorded during the 2012 February
superoutburst: BJD 2455985.3378(2) ($N=157$).

\subsection{V1159 Orionis}\label{obj:v1159ori}

   V1159 Ori is one of the member of the ER UMa stars
(\cite{rob95eruma}; \cite{nog95v1159ori}; \cite{pat95v1159ori}).
Although the object generally follows the ER UMa-type
pattern with a short supercycle \citep{kat95eruma}, the object
is known to show variations of supercycles with a range of
44.6--53.3~d \citep{kat01v1159ori}.  Since it has been demonstrated
that the prototype ER UMa has recently been in a state of
``negative superhumps'' \citep{ohs12eruma}, it would be worth
examining the current state of superhumps in V1159 Ori.

   The observations were taken during its 2012 March superoutburst
(the data were mainly from the AAVSO).  The times of superhump
maxima are listed in table \ref{tab:v1159orioc2012}.
There was a $\sim$0.5 phase shift between $E=31$ and $E=77$,
as seen in ER UMa in its ``positive superhump'' state
\citep{kat03erumaSH}.  A period analysis for $E \ge 61$ yielded
a period of 0.06430(5)~d, indicating that the period after
the phase shift was that of positive superhumps, and not that of
negative superhumps.  The behavior well reproduced ER UMa in
its ``positive superhump'' state (in the 1990s).
V1159 Ori is currently the best target to investigate
the ER UMa-type phenomena in ``positive superhump'' state
and further detailed observations are required.

\begin{table}
\caption{Superhump maxima of V1159 Ori (2012).}\label{tab:v1159orioc2012}
\begin{center}
\begin{tabular}{ccccc}
\hline
$E$ & max\commenta & error & $O-C$\commentb & $N$\commentc \\
\hline
0 & 55991.6169 & 0.0004 & 0.0065 & 123 \\
1 & 55991.6826 & 0.0004 & 0.0074 & 124 \\
16 & 55992.6410 & 0.0010 & $-$0.0046 & 57 \\
30 & 55993.5407 & 0.0012 & $-$0.0106 & 42 \\
31 & 55993.6059 & 0.0023 & $-$0.0101 & 57 \\
61 & 55995.5631 & 0.0005 & 0.0064 & 54 \\
62 & 55995.6264 & 0.0009 & 0.0049 & 58 \\
77 & 55996.5919 & 0.0011 & 0.0000 & 58 \\
\hline
  \multicolumn{5}{l}{\commenta BJD$-$2400000.} \\
  \multicolumn{5}{l}{\commentb Against max $= 2455991.6105 + 0.064694 E$.} \\
  \multicolumn{5}{l}{\commentc Number of points used to determine the maximum.} \\
\end{tabular}
\end{center}
\end{table}

\subsection{AR Pictoris}\label{obj:arpic}

\begin{figure}
  \begin{center}
    \FigureFile(88mm,70mm){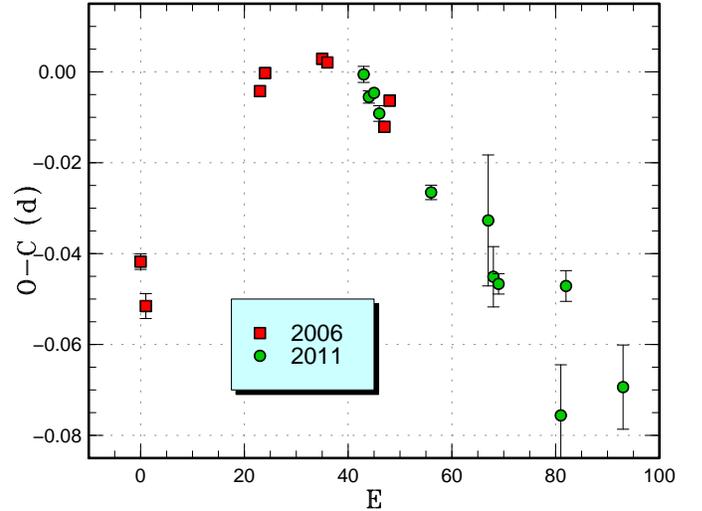}
  \end{center}
  \caption{Comparison of $O-C$ diagrams of AR Pic between different
  superoutbursts.  A period of 0.08458~d was used to draw this figure.
  Approximate cycle counts ($E$) after the start of the superoutburst
  were used.
  }
  \label{fig:arpiccomp}
\end{figure}

   We observed the 2011 superoutburst of this object
(=CTCV J0549$-$4921, \cite{ima08fltractcv0549}).  \citet{Pdot}
identified this object with a large negative $P_{\rm dot}$.
The times of superhump maxima are listed in table
\ref{tab:arpicoc2011}.  We only observed the terminal portion
of the superoutburst, and the mean superhump period
[0.08315(15)~d] was much shorter than the value obtained
during the 2006 superoutburst \citep{Pdot}.  During the
post-superoutburst phase, we could detect a superhump period of 
0.08225(9)~d (PDM method).  A lasso analysis yielded a combination
of the superhump period and a period of 0.0801(1)~d, which is
potentially the orbital period.  If this is indeed $P_{\rm orb}$,
$\epsilon$ for the 2006 and 2011 superoutburst were 5.4\%
(average) and 3.7\%, respectively.
A comparison of $O-C$ diagrams (figure \ref{fig:arpiccomp})
indicates that the $O-C$ diagram in 2011 is in smooth extension
of the 2006 one, which was obtained only during the early stage.

\begin{table}
\caption{Superhump maxima of AR Pic (2011).}\label{tab:arpicoc2011}
\begin{center}
\begin{tabular}{ccccc}
\hline
$E$ & max\commenta & error & $O-C$\commentb & $N$\commentc \\
\hline
0 & 55910.5774 & 0.0018 & 0.0038 & 20 \\
1 & 55910.6570 & 0.0014 & 0.0002 & 25 \\
2 & 55910.7425 & 0.0011 & 0.0026 & 29 \\
3 & 55910.8225 & 0.0017 & $-$0.0005 & 29 \\
13 & 55911.6510 & 0.0016 & $-$0.0036 & 25 \\
24 & 55912.5752 & 0.0144 & 0.0059 & 19 \\
25 & 55912.6474 & 0.0066 & $-$0.0051 & 17 \\
26 & 55912.7304 & 0.0022 & $-$0.0052 & 23 \\
38 & 55913.7164 & 0.0111 & $-$0.0171 & 27 \\
39 & 55913.8294 & 0.0034 & 0.0128 & 29 \\
50 & 55914.7376 & 0.0093 & 0.0063 & 28 \\
\hline
  \multicolumn{5}{l}{\commenta BJD$-$2400000.} \\
  \multicolumn{5}{l}{\commentb Against max $= 2455910.5736 + 0.083154 E$.} \\
  \multicolumn{5}{l}{\commentc Number of points used to determine the maximum.} \\
\end{tabular}
\end{center}
\end{table}

\subsection{GV Piscium}\label{obj:gvpsc}

   The 2011 superoutburst of this object was detected by CRTS
on October 17.  Subsequent observations confirmed the presence
of superhumps (vsnet-alert 13768).  The times of superhump maxima are
listed in table \ref{tab:gvpscoc2011}.  There was little hint of
period variation, and the mean period was close to that obtained
during the 2008 superoutburst \citep{Pdot}.  Since the object faded
relatively soon after the outburst detection, it looks likely that
we also observed only stage C superhumps as in 2008.

\begin{table}
\caption{Superhump maxima of GV Psc (2011).}\label{tab:gvpscoc2011}
\begin{center}
\begin{tabular}{ccccc}
\hline
$E$ & max\commenta & error & $O-C$\commentb & $N$\commentc \\
\hline
0 & 55852.3350 & 0.0003 & $-$0.0003 & 95 \\
1 & 55852.4291 & 0.0003 & $-$0.0005 & 86 \\
4 & 55852.7127 & 0.0003 & 0.0001 & 94 \\
5 & 55852.8055 & 0.0003 & $-$0.0014 & 87 \\
11 & 55853.3728 & 0.0003 & $-$0.0000 & 90 \\
12 & 55853.4688 & 0.0005 & 0.0016 & 99 \\
25 & 55854.6952 & 0.0003 & 0.0020 & 99 \\
26 & 55854.7867 & 0.0004 & $-$0.0008 & 90 \\
61 & 55858.0892 & 0.0008 & 0.0007 & 62 \\
62 & 55858.1814 & 0.0007 & $-$0.0014 & 61 \\
\hline
  \multicolumn{5}{l}{\commenta BJD$-$2400000.} \\
  \multicolumn{5}{l}{\commentb Against max $= 2455852.3354 + 0.094313 E$.} \\
  \multicolumn{5}{l}{\commentc Number of points used to determine the maximum.} \\
\end{tabular}
\end{center}
\end{table}

\subsection{BW Sculptoris}\label{obj:bwscl}

   This object (=HE 2350$-$3908, RX J2353.0$-$3852) was initially
discovered in the Hamburg/ESO quasar survey \citep{aug97bwscl}
and was also selected as a ROSAT CV \citep{abb97bwscl}.
Its remarkable similarity with WZ Sge
was already noted at its very early history \citep{aug97bwscl}.
Despite monitoring, there had been no outbursts until 2011.
\citet{uth12j1457bwscl} reported the ZZ Cet-type pulsation
of the white dwarf and the presence of quiescent superhumps
11\% longer than $P_{\rm orb}$.

\begin{figure}
  \begin{center}
    \FigureFile(88mm,110mm){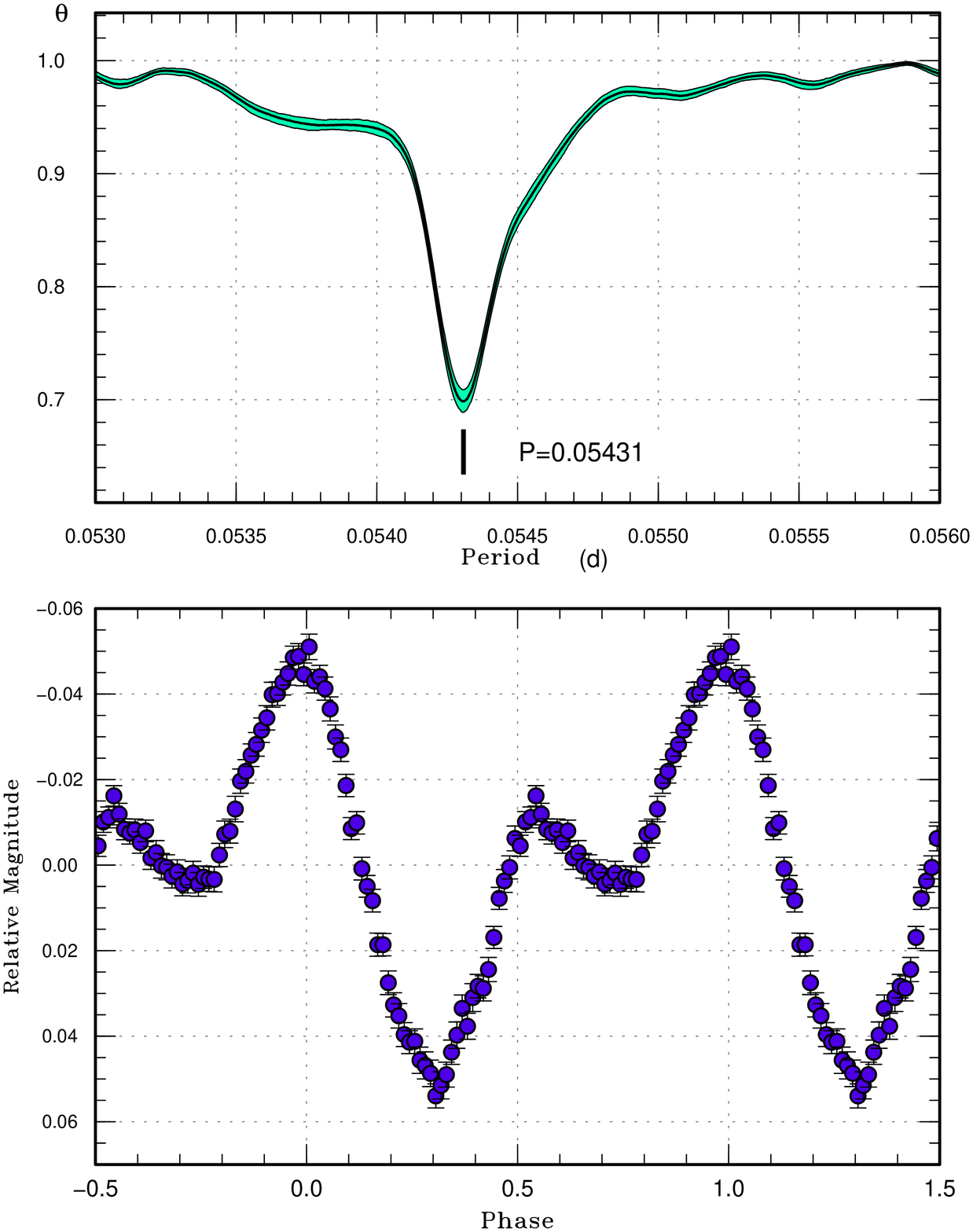}
  \end{center}
  \caption{Early superhumps in BW Scl (2011). (Upper): PDM analysis.
     (Lower): Phase-averaged profile.}
  \label{fig:bwscleshpdm}
\end{figure}

\begin{figure}
  \begin{center}
    \FigureFile(88mm,110mm){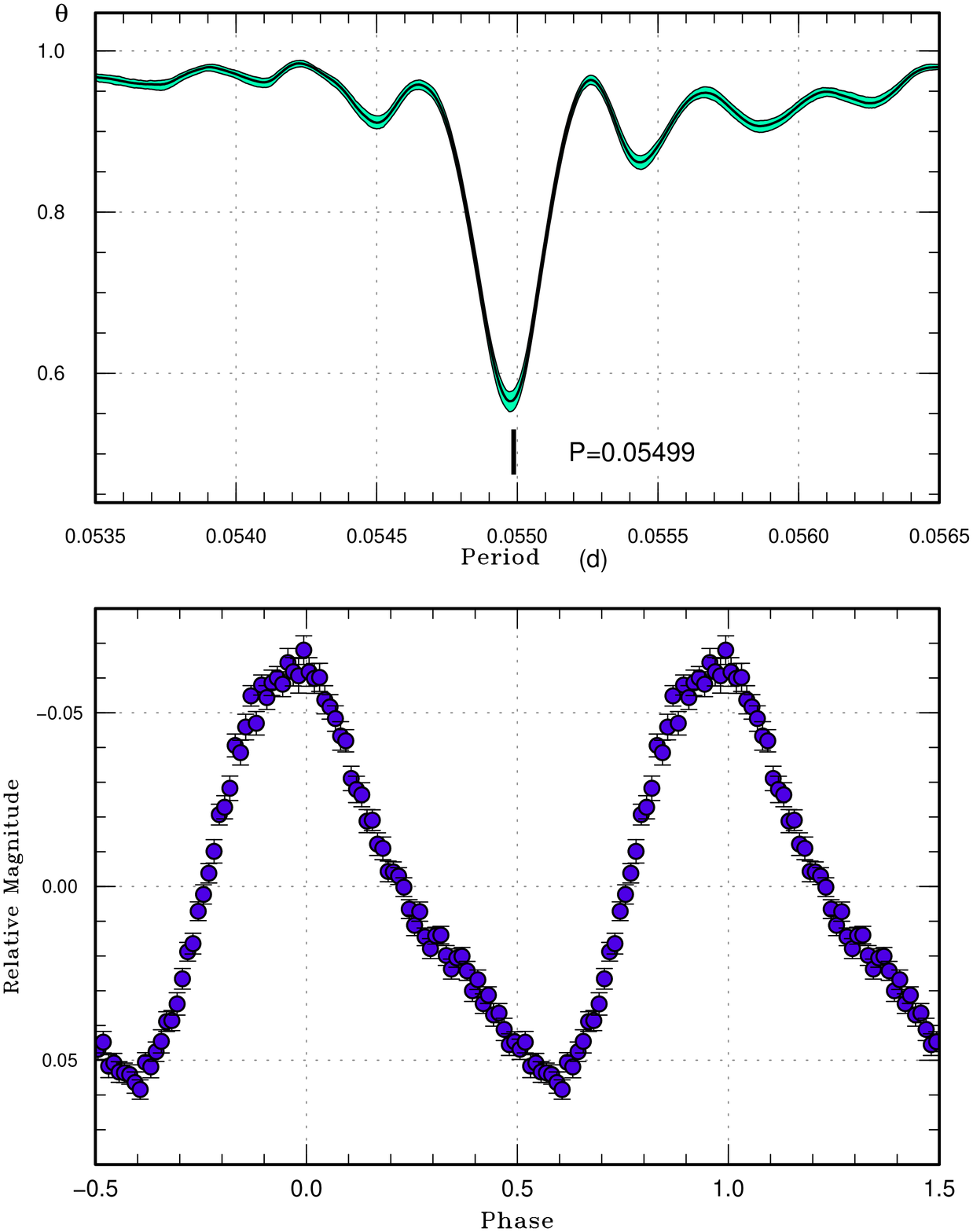}
  \end{center}
  \caption{Ordinary superhumps in BW Scl (2011). (Upper): PDM analysis.
     (Lower): Phase-averaged profile.}
  \label{fig:bwsclshpdm}
\end{figure}

\begin{figure}
  \begin{center}
    \FigureFile(88mm,110mm){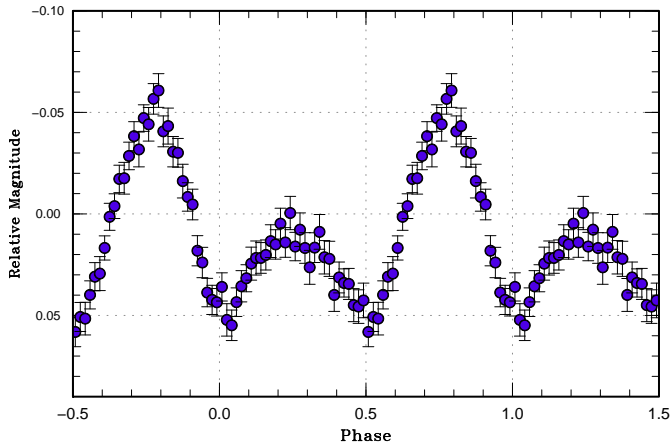}
  \end{center}
  \caption{Averaged orbital profile of BW Scl (2011) after the rapid decline.}
  \label{fig:bwsclorbph}
\end{figure}

   The 2011 outburst was detected by M. Linnolt on October 21
at a visual magnitude of 9.6 (posting to AAVSO discussion),
and subsequent observation soon confirmed early superhumps
(vsnet-alert 13786; figure \ref{fig:bwscleshpdm}).
The last observation before this outburst was on October 15
(by J. Hambsch; see also \cite{ham12ROADaavso}) when the object was 
still in quiescence.
On October 31, ordinary superhumps developed (vsnet-alert
13815, 13819; figure \ref{fig:bwsclshpdm}).  The object entered
the rapid fading stage on November 12 (vsnet-alert 13847, 13850).
The times of maxima of ordinary superhumps are listed in
table \ref{tab:bwscloc2011}.  Following a period of stage A
($E \le 25$), there was stage B with $P_{\rm dot} = +4.3(0.3) \times 10^{-5}$.
Although there was a suggestion of sudden shortening
of the superhump period after $E=210$, as seen in other WZ Sge-type
dwarf novae [e.g. GW Lib and V455 And \citep{Pdot}; OT J012059.6$+$325545 and 
SDSS J080434.20$+$510349.2 = EZ Lyn, \citep{Pdot3}], 
a discontinuity in the observation made 
the identification of hump phasing uncertain.  We list times
of hump maxima after this rapid fading in table 
\ref{tab:bwscloc2011late}, which were measured after subtracting
the mean orbital variation (figure \ref{fig:bwsclorbph}).  
The overall behavior of the outburst and $O-C$ diagram were
very similar to those of GW Lib and V455 And.
The period of early superhumps was 0.054308(2)~d, 0.03\%
shorter than $P_{\rm orb}$.  The $\epsilon$ for stage B superhumps
was 1.3\%.

A full analysis will be presented by Ohshima et al., in preparation.

\begin{figure*}
  \begin{center}
    \FigureFile(160mm,190mm){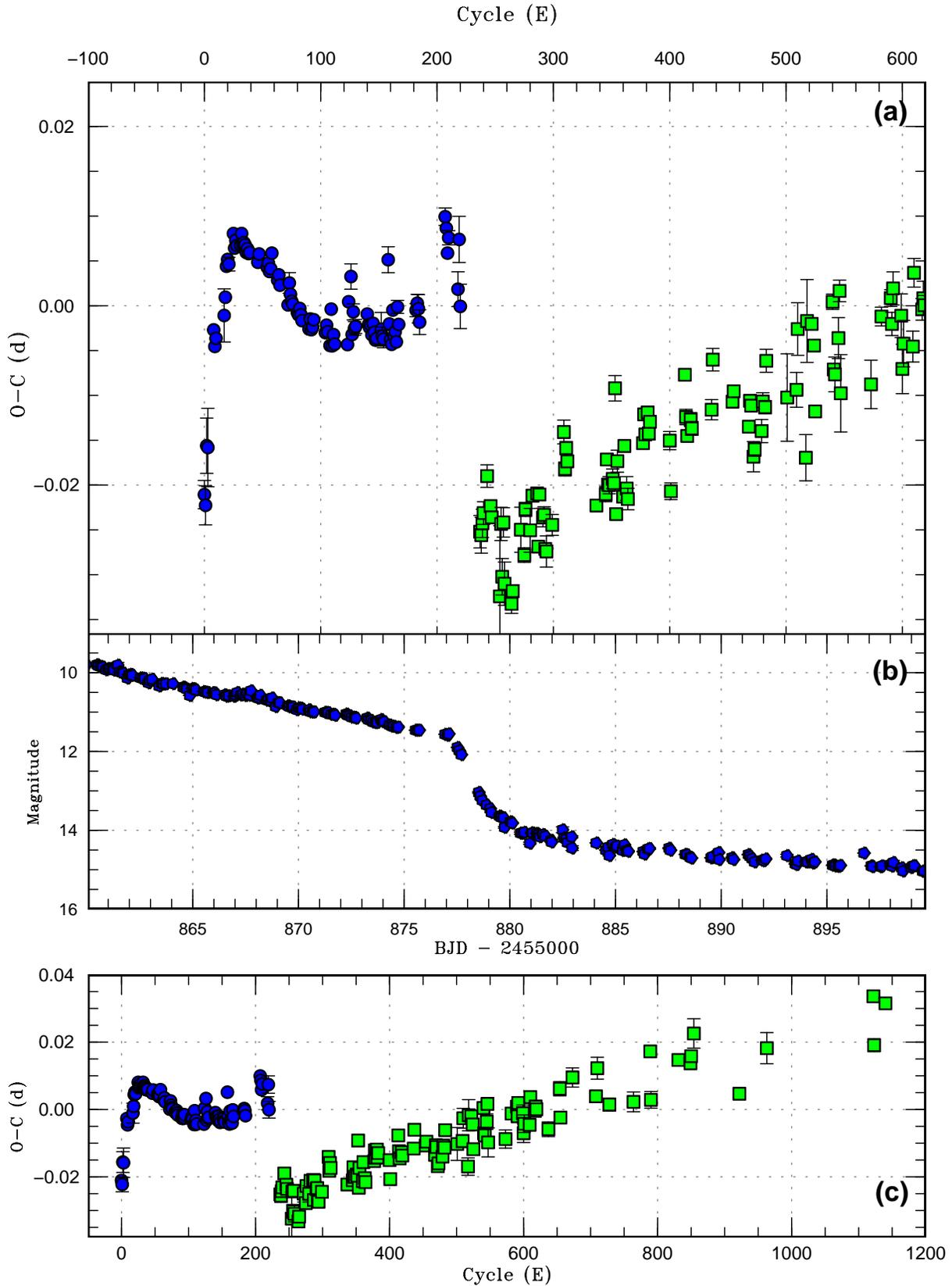}
  \end{center}
  \caption{$O-C$ diagram of superhumps in BW Scl (2011).
     (a) $O-C$.
     Filled circles and filled squares represent ordinary superhumps
     and late-stage superhumps after the rapid fading.
     We used a period of 0.055036~d for calculating the $O-C$ residuals.
     (b) Light curve.
     (c) $O-C$ diagram of the entire observation.
     The global evolution of the $O-C$ diagram is remarkably similar
     to those of GW Lib and V455 And \citep{Pdot}.
  }
  \label{fig:bwsclhumpall}
\end{figure*}

\begin{table}
\caption{Superhump maxima of BW Scl (2011).}\label{tab:bwscloc2011}
\begin{center}
\begin{tabular}{ccccc}
\hline
$E$ & max\commenta & error & $O-C$\commentb & $N$\commentc \\
\hline
0 & 55865.5288 & 0.0016 & $-$0.0209 & 89 \\
1 & 55865.5827 & 0.0022 & $-$0.0221 & 61 \\
2 & 55865.6444 & 0.0031 & $-$0.0155 & 89 \\
3 & 55865.6992 & 0.0044 & $-$0.0157 & 89 \\
8 & 55865.9875 & 0.0003 & $-$0.0026 & 426 \\
9 & 55866.0407 & 0.0004 & $-$0.0044 & 187 \\
10 & 55866.0966 & 0.0006 & $-$0.0035 & 154 \\
17 & 55866.4844 & 0.0030 & $-$0.0010 & 50 \\
18 & 55866.5415 & 0.0005 & 0.0011 & 89 \\
19 & 55866.6000 & 0.0004 & 0.0045 & 88 \\
20 & 55866.6558 & 0.0004 & 0.0053 & 89 \\
21 & 55866.7103 & 0.0008 & 0.0048 & 89 \\
25 & 55866.9339 & 0.0003 & 0.0082 & 279 \\
26 & 55866.9873 & 0.0002 & 0.0065 & 305 \\
27 & 55867.0432 & 0.0003 & 0.0074 & 274 \\
28 & 55867.0976 & 0.0004 & 0.0068 & 223 \\
31 & 55867.2628 & 0.0001 & 0.0069 & 238 \\
32 & 55867.3191 & 0.0001 & 0.0081 & 238 \\
33 & 55867.3729 & 0.0001 & 0.0069 & 236 \\
34 & 55867.4282 & 0.0003 & 0.0071 & 125 \\
35 & 55867.4830 & 0.0001 & 0.0069 & 281 \\
36 & 55867.5371 & 0.0001 & 0.0060 & 302 \\
37 & 55867.5926 & 0.0002 & 0.0065 & 75 \\
38 & 55867.6471 & 0.0002 & 0.0059 & 89 \\
39 & 55867.7022 & 0.0002 & 0.0060 & 89 \\
46 & 55868.0864 & 0.0002 & 0.0049 & 89 \\
47 & 55868.1424 & 0.0002 & 0.0058 & 68 \\
54 & 55868.5261 & 0.0004 & 0.0044 & 89 \\
55 & 55868.5816 & 0.0003 & 0.0048 & 88 \\
56 & 55868.6357 & 0.0003 & 0.0039 & 89 \\
57 & 55868.6911 & 0.0003 & 0.0042 & 88 \\
58 & 55868.7479 & 0.0005 & 0.0059 & 55 \\
63 & 55869.0200 & 0.0002 & 0.0029 & 144 \\
64 & 55869.0756 & 0.0003 & 0.0035 & 182 \\
65 & 55869.1295 & 0.0003 & 0.0023 & 102 \\
72 & 55869.5126 & 0.0006 & 0.0001 & 91 \\
73 & 55869.5701 & 0.0012 & 0.0026 & 65 \\
74 & 55869.6238 & 0.0004 & 0.0013 & 89 \\
75 & 55869.6781 & 0.0005 & 0.0005 & 89 \\
76 & 55869.7328 & 0.0005 & 0.0002 & 84 \\
80 & 55869.9520 & 0.0005 & $-$0.0008 & 141 \\
81 & 55870.0068 & 0.0004 & $-$0.0010 & 165 \\
82 & 55870.0626 & 0.0004 & $-$0.0003 & 237 \\
83 & 55870.1169 & 0.0002 & $-$0.0010 & 93 \\
84 & 55870.1713 & 0.0003 & $-$0.0017 & 84 \\
90 & 55870.5005 & 0.0004 & $-$0.0026 & 33 \\
91 & 55870.5567 & 0.0004 & $-$0.0015 & 37 \\
92 & 55870.6106 & 0.0003 & $-$0.0027 & 44 \\
93 & 55870.6659 & 0.0003 & $-$0.0024 & 43 \\
94 & 55870.7217 & 0.0003 & $-$0.0016 & 44 \\
104 & 55871.2706 & 0.0001 & $-$0.0030 & 238 \\
105 & 55871.3265 & 0.0002 & $-$0.0022 & 237 \\
106 & 55871.3808 & 0.0003 & $-$0.0029 & 197 \\
108 & 55871.4894 & 0.0006 & $-$0.0045 & 24 \\
109 & 55871.5484 & 0.0006 & $-$0.0004 & 33 \\
110 & 55871.5994 & 0.0003 & $-$0.0045 & 40 \\
\hline
  \multicolumn{5}{l}{\commenta BJD$-$2400000.} \\
  \multicolumn{5}{l}{\commentb Against max $= 2455865.5497 + 0.055038 E$.} \\
  \multicolumn{5}{l}{\commentc Number of points used to determine the maximum.} \\
\end{tabular}
\end{center}
\end{table}

\addtocounter{table}{-1}
\begin{table}
\caption{Superhump maxima of BW Scl (2011). (continued)}
\begin{center}
\begin{tabular}{ccccc}
\hline
$E$ & max\commenta & error & $O-C$\commentb & $N$\commentc \\
\hline
111 & 55871.6557 & 0.0004 & $-$0.0033 & 40 \\
112 & 55871.7096 & 0.0006 & $-$0.0043 & 40 \\
123 & 55872.3150 & 0.0003 & $-$0.0044 & 219 \\
124 & 55872.3748 & 0.0004 & 0.0004 & 237 \\
126 & 55872.4877 & 0.0014 & 0.0032 & 222 \\
127 & 55872.5363 & 0.0005 & $-$0.0033 & 152 \\
128 & 55872.5938 & 0.0009 & $-$0.0007 & 40 \\
129 & 55872.6470 & 0.0008 & $-$0.0026 & 40 \\
130 & 55872.7023 & 0.0008 & $-$0.0024 & 41 \\
140 & 55873.2540 & 0.0005 & $-$0.0010 & 155 \\
141 & 55873.3079 & 0.0002 & $-$0.0022 & 238 \\
142 & 55873.3626 & 0.0002 & $-$0.0025 & 235 \\
143 & 55873.4179 & 0.0002 & $-$0.0022 & 240 \\
144 & 55873.4718 & 0.0004 & $-$0.0033 & 161 \\
145 & 55873.5282 & 0.0006 & $-$0.0021 & 32 \\
146 & 55873.5822 & 0.0003 & $-$0.0031 & 40 \\
147 & 55873.6364 & 0.0004 & $-$0.0039 & 41 \\
148 & 55873.6916 & 0.0004 & $-$0.0037 & 40 \\
152 & 55873.9127 & 0.0019 & $-$0.0027 & 76 \\
153 & 55873.9670 & 0.0010 & $-$0.0035 & 115 \\
154 & 55874.0217 & 0.0010 & $-$0.0038 & 114 \\
158 & 55874.2507 & 0.0014 & 0.0050 & 133 \\
159 & 55874.2986 & 0.0002 & $-$0.0021 & 238 \\
160 & 55874.3519 & 0.0004 & $-$0.0039 & 235 \\
161 & 55874.4064 & 0.0003 & $-$0.0044 & 237 \\
162 & 55874.4653 & 0.0004 & $-$0.0006 & 237 \\
163 & 55874.5174 & 0.0004 & $-$0.0035 & 232 \\
164 & 55874.5729 & 0.0004 & $-$0.0030 & 40 \\
165 & 55874.6268 & 0.0007 & $-$0.0042 & 40 \\
166 & 55874.6858 & 0.0007 & $-$0.0003 & 40 \\
167 & 55874.7388 & 0.0005 & $-$0.0022 & 32 \\
182 & 55875.5660 & 0.0007 & $-$0.0006 & 40 \\
183 & 55875.6218 & 0.0010 & 0.0001 & 40 \\
184 & 55875.6762 & 0.0016 & $-$0.0005 & 40 \\
185 & 55875.7297 & 0.0014 & $-$0.0020 & 40 \\
207 & 55876.9523 & 0.0010 & 0.0097 & 60 \\
208 & 55877.0060 & 0.0005 & 0.0084 & 50 \\
209 & 55877.0583 & 0.0005 & 0.0057 & 100 \\
210 & 55877.1150 & 0.0008 & 0.0074 & 58 \\
218 & 55877.5496 & 0.0019 & 0.0016 & 24 \\
219 & 55877.6102 & 0.0026 & 0.0072 & 25 \\
220 & 55877.6578 & 0.0025 & $-$0.0003 & 25 \\
\hline
  \multicolumn{5}{l}{\commenta BJD$-$2400000.} \\
  \multicolumn{5}{l}{\commentb Against max $= 2455865.5497 + 0.055038 E$.} \\
  \multicolumn{5}{l}{\commentc Number of points used to determine the maximum.} \\
\end{tabular}
\end{center}
\end{table}

\begin{table}
\caption{Late-stage superhumps in BW Scl (2011).}\label{tab:bwscloc2011late}
\begin{center}
\begin{tabular}{ccccc}
\hline
$E$ & max\commenta & error & $O-C$\commentb & $N$\commentc \\
\hline
0 & 55878.5682 & 0.0018 & 0.0005 & 25 \\
1 & 55878.6229 & 0.0020 & 0.0000 & 25 \\
2 & 55878.6792 & 0.0016 & 0.0013 & 25 \\
3 & 55878.7354 & 0.0013 & 0.0024 & 22 \\
6 & 55878.9046 & 0.0013 & 0.0063 & 106 \\
9 & 55879.0664 & 0.0003 & 0.0028 & 83 \\
10 & 55879.1202 & 0.0006 & 0.0014 & 84 \\
17 & 55879.4966 & 0.0079 & $-$0.0078 & 12 \\
18 & 55879.5597 & 0.0019 & 0.0002 & 19 \\
19 & 55879.6089 & 0.0020 & $-$0.0058 & 20 \\
20 & 55879.6700 & 0.0016 & 0.0002 & 19 \\
21 & 55879.7182 & 0.0024 & $-$0.0067 & 19 \\
27 & 55880.0462 & 0.0011 & $-$0.0093 & 110 \\
28 & 55880.1026 & 0.0007 & $-$0.0079 & 86 \\
35 & 55880.4947 & 0.0025 & $-$0.0015 & 10 \\
38 & 55880.6570 & 0.0008 & $-$0.0045 & 19 \\
39 & 55880.7171 & 0.0008 & 0.0005 & 16 \\
43 & 55880.9349 & 0.0024 & $-$0.0021 & 61 \\
45 & 55881.0489 & 0.0004 & 0.0016 & 97 \\
49 & 55881.2693 & 0.0001 & 0.0016 & 390 \\
50 & 55881.3184 & 0.0002 & $-$0.0044 & 476 \\
51 & 55881.3792 & 0.0002 & 0.0014 & 474 \\
54 & 55881.5417 & 0.0010 & $-$0.0014 & 18 \\
55 & 55881.5971 & 0.0009 & $-$0.0011 & 19 \\
56 & 55881.6483 & 0.0005 & $-$0.0050 & 20 \\
57 & 55881.7031 & 0.0017 & $-$0.0054 & 19 \\
62 & 55881.9812 & 0.0012 & $-$0.0027 & 76 \\
72 & 55882.5419 & 0.0013 & 0.0070 & 20 \\
73 & 55882.5929 & 0.0008 & 0.0028 & 20 \\
74 & 55882.6502 & 0.0007 & 0.0051 & 20 \\
75 & 55882.7038 & 0.0004 & 0.0035 & 19 \\
100 & 55884.0747 & 0.0005 & $-$0.0030 & 55 \\
108 & 55884.5163 & 0.0008 & $-$0.0022 & 57 \\
109 & 55884.5752 & 0.0006 & 0.0016 & 59 \\
110 & 55884.6275 & 0.0005 & $-$0.0013 & 58 \\
111 & 55884.6824 & 0.0010 & $-$0.0014 & 58 \\
114 & 55884.8482 & 0.0011 & $-$0.0009 & 41 \\
115 & 55884.9028 & 0.0005 & $-$0.0014 & 38 \\
116 & 55884.9684 & 0.0014 & 0.0091 & 31 \\
117 & 55885.0094 & 0.0007 & $-$0.0050 & 59 \\
118 & 55885.0703 & 0.0012 & 0.0008 & 37 \\
123 & 55885.3416 & 0.0004 & $-$0.0034 & 127 \\
124 & 55885.4023 & 0.0006 & 0.0021 & 99 \\
126 & 55885.5076 & 0.0013 & $-$0.0027 & 51 \\
127 & 55885.5614 & 0.0012 & $-$0.0040 & 59 \\
140 & 55886.2831 & 0.0004 & 0.0014 & 110 \\
141 & 55886.3414 & 0.0004 & 0.0046 & 127 \\
142 & 55886.3942 & 0.0007 & 0.0023 & 39 \\
144 & 55886.5067 & 0.0004 & 0.0046 & 48 \\
145 & 55886.5594 & 0.0007 & 0.0022 & 59 \\
146 & 55886.6157 & 0.0006 & 0.0034 & 44 \\
163 & 55887.5493 & 0.0010 & 0.0003 & 59 \\
164 & 55887.5986 & 0.0009 & $-$0.0055 & 59 \\
176 & 55888.2721 & 0.0003 & 0.0068 & 94 \\
177 & 55888.3224 & 0.0009 & 0.0020 & 127 \\
\hline
  \multicolumn{5}{l}{\commenta BJD$-$2400000.} \\
  \multicolumn{5}{l}{\commentb Against max $= 2455878.5677 + 0.055100 E$.} \\
  \multicolumn{5}{l}{\commentc Number of points used to determine the maximum.} \\
\end{tabular}
\end{center}
\end{table}

\addtocounter{table}{-1}
\begin{table}
\caption{Late-stage superhumps in BW Scl (2011). (continued)}
\begin{center}
\begin{tabular}{ccccc}
\hline
$E$ & max\commenta & error & $O-C$\commentb & $N$\commentc \\
\hline
178 & 55888.3753 & 0.0004 & $-$0.0002 & 108 \\
181 & 55888.5423 & 0.0008 & 0.0015 & 59 \\
182 & 55888.5963 & 0.0008 & 0.0004 & 59 \\
199 & 55889.5340 & 0.0011 & 0.0014 & 58 \\
200 & 55889.5946 & 0.0013 & 0.0069 & 59 \\
217 & 55890.5255 & 0.0005 & 0.0012 & 58 \\
218 & 55890.5817 & 0.0006 & 0.0023 & 59 \\
231 & 55891.2933 & 0.0005 & $-$0.0025 & 127 \\
232 & 55891.3512 & 0.0004 & 0.0003 & 127 \\
233 & 55891.4057 & 0.0004 & $-$0.0003 & 127 \\
235 & 55891.5101 & 0.0017 & $-$0.0061 & 56 \\
236 & 55891.5659 & 0.0008 & $-$0.0054 & 59 \\
242 & 55891.8982 & 0.0013 & $-$0.0037 & 55 \\
243 & 55891.9565 & 0.0010 & $-$0.0005 & 68 \\
245 & 55892.0659 & 0.0007 & $-$0.0012 & 22 \\
246 & 55892.1261 & 0.0013 & 0.0039 & 19 \\
264 & 55893.1127 & 0.0049 & $-$0.0014 & 29 \\
272 & 55893.5538 & 0.0019 & $-$0.0010 & 31 \\
273 & 55893.6157 & 0.0029 & 0.0057 & 24 \\
280 & 55893.9866 & 0.0026 & $-$0.0091 & 27 \\
281 & 55894.0568 & 0.0046 & 0.0061 & 19 \\
285 & 55894.2767 & 0.0004 & 0.0055 & 104 \\
287 & 55894.3843 & 0.0004 & 0.0030 & 126 \\
288 & 55894.4320 & 0.0004 & $-$0.0044 & 127 \\
303 & 55895.2698 & 0.0009 & 0.0068 & 90 \\
304 & 55895.3173 & 0.0005 & $-$0.0008 & 127 \\
305 & 55895.3718 & 0.0019 & $-$0.0014 & 127 \\
308 & 55895.5409 & 0.0023 & 0.0025 & 30 \\
309 & 55895.6012 & 0.0012 & 0.0077 & 30 \\
310 & 55895.6448 & 0.0043 & $-$0.0038 & 30 \\
336 & 55897.0768 & 0.0027 & $-$0.0045 & 31 \\
345 & 55897.5796 & 0.0010 & 0.0025 & 30 \\
353 & 55898.0220 & 0.0009 & 0.0041 & 22 \\
354 & 55898.0742 & 0.0013 & 0.0011 & 30 \\
355 & 55898.1331 & 0.0018 & 0.0050 & 20 \\
362 & 55898.5154 & 0.0024 & 0.0015 & 26 \\
363 & 55898.5645 & 0.0028 & $-$0.0045 & 30 \\
364 & 55898.6223 & 0.0026 & $-$0.0017 & 17 \\
372 & 55899.0623 & 0.0017 & $-$0.0026 & 29 \\
373 & 55899.1255 & 0.0016 & 0.0056 & 28 \\
380 & 55899.5067 & 0.0012 & 0.0011 & 22 \\
381 & 55899.5630 & 0.0019 & 0.0022 & 30 \\
382 & 55899.6172 & 0.0011 & 0.0014 & 19 \\
400 & 55900.6020 & 0.0022 & $-$0.0057 & 15 \\
417 & 55901.5496 & 0.0021 & 0.0053 & 20 \\
418 & 55901.5961 & 0.0017 & $-$0.0033 & 16 \\
436 & 55902.5986 & 0.0028 & 0.0074 & 18 \\
471 & 55904.5193 & 0.0017 & $-$0.0004 & 30 \\
473 & 55904.6377 & 0.0032 & 0.0078 & 19 \\
491 & 55905.6176 & 0.0019 & $-$0.0041 & 15 \\
527 & 55907.5997 & 0.0029 & $-$0.0056 & 16 \\
552 & 55908.9906 & 0.0015 & 0.0078 & 30 \\
553 & 55909.0312 & 0.0025 & $-$0.0067 & 30 \\
594 & 55911.2995 & 0.0009 & 0.0025 & 126 \\
612 & 55912.2892 & 0.0012 & 0.0004 & 102 \\
\hline
  \multicolumn{5}{l}{\commenta BJD$-$2400000.} \\
  \multicolumn{5}{l}{\commentb Against max $= 2455878.5677 + 0.055100 E$.} \\
  \multicolumn{5}{l}{\commentc Number of points used to determine the maximum.} \\
\end{tabular}
\end{center}
\end{table}

\addtocounter{table}{-1}
\begin{table}
\caption{Late-stage superhumps in BW Scl (2011). (continued)}
\begin{center}
\begin{tabular}{ccccc}
\hline
$E$ & max\commenta & error & $O-C$\commentb & $N$\commentc \\
\hline
613 & 55912.3463 & 0.0008 & 0.0024 & 126 \\
617 & 55912.5732 & 0.0044 & 0.0090 & 12 \\
685 & 55916.2978 & 0.0005 & $-$0.0133 & 116 \\
726 & 55918.5678 & 0.0046 & $-$0.0024 & 14 \\
885 & 55927.3339 & 0.0011 & 0.0029 & 126 \\
886 & 55927.3744 & 0.0019 & $-$0.0117 & 105 \\
903 & 55928.3225 & 0.0003 & $-$0.0003 & 85 \\
\hline
  \multicolumn{5}{l}{\commenta BJD$-$2400000.} \\
  \multicolumn{5}{l}{\commentb Against max $= 2455878.5677 + 0.055100 E$.} \\
  \multicolumn{5}{l}{\commentc Number of points used to determine the maximum.} \\
\end{tabular}
\end{center}
\end{table}

\subsection{CC Sculptoris}\label{obj:ccscl}

   CC Scl was discovered as a ROSAT-selected CV \citep{sch00RASSID}.
During its 2000 October outburst (the second known),
\citet{ish01j2315} detected likely superhumps with a period
of 0.078~d and amplitudes of $\sim$0.3 mag.  However, 
Augusteijn et al. (2000, vsnet-campaign 544) reported the detection
of a photometric period of 0.058~d, which was considered as being
the orbital period.  Based on the discrepancy between the apparent
period of superhumps and the orbital period, \citet{ish01j2315}
suggested that the object may be an intermediate polar.
The unusual short duration of the superoutburst was also noted.
The 0.058~d period was later confirmed to be the orbital period
(\cite{che01ECCV}; \cite{tap04CTCV}).
Although there have been several outbursts since then, no confirmatory
observations of superhumps have been reported.

   The 2011 superoutburst was detected by CRTS Siding Spring Survey (SSS)
and subsequent observations indicated the presence of low-amplitude 
(up to 0.1 mag) variations similar to superhumps with a period
of 0.0603~d (vsnet-alert 13832).  Although the observed variations
had a definite underlying periodicity (vsnet-alert 13841, 13846),
individual waveforms were rather irregular (vsnet-alert 13840;
see also actual observations in figure \ref{fig:ccsclperdub})
unlike most of SU UMa-type dwarf novae.  Even after the object
faded, the superhump signal persisted for at least eight days.

\begin{figure}
  \begin{center}
    \FigureFile(88mm,110mm){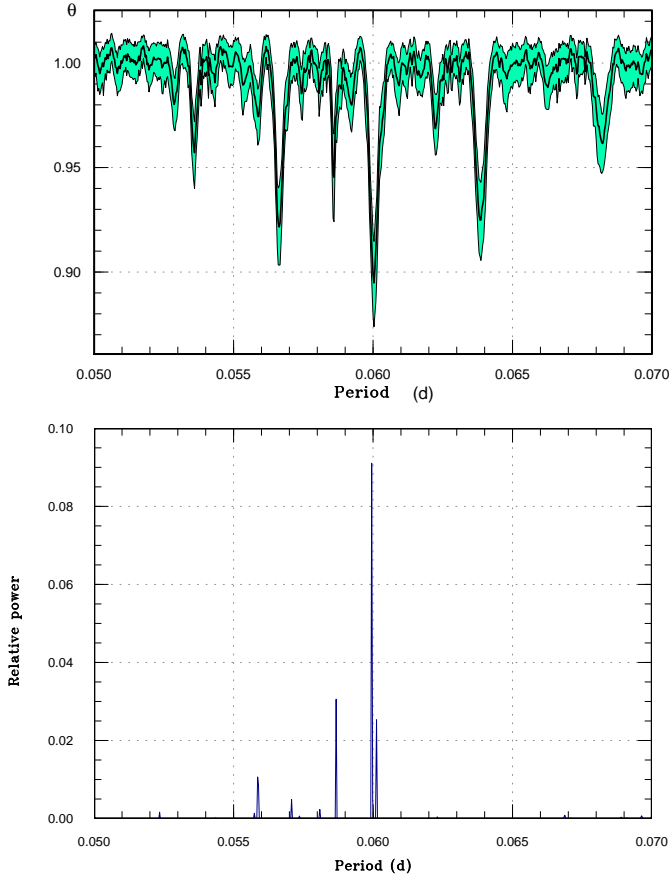}
  \end{center}
  \caption{Period analysis in CC Scl (2011). (Upper): PDM analysis.
     (Lower): lasso analysis ($\log \lambda=-5.05$).}
  \label{fig:ccsclshpdm}
\end{figure}

\begin{figure}
  \begin{center}
    \FigureFile(88mm,110mm){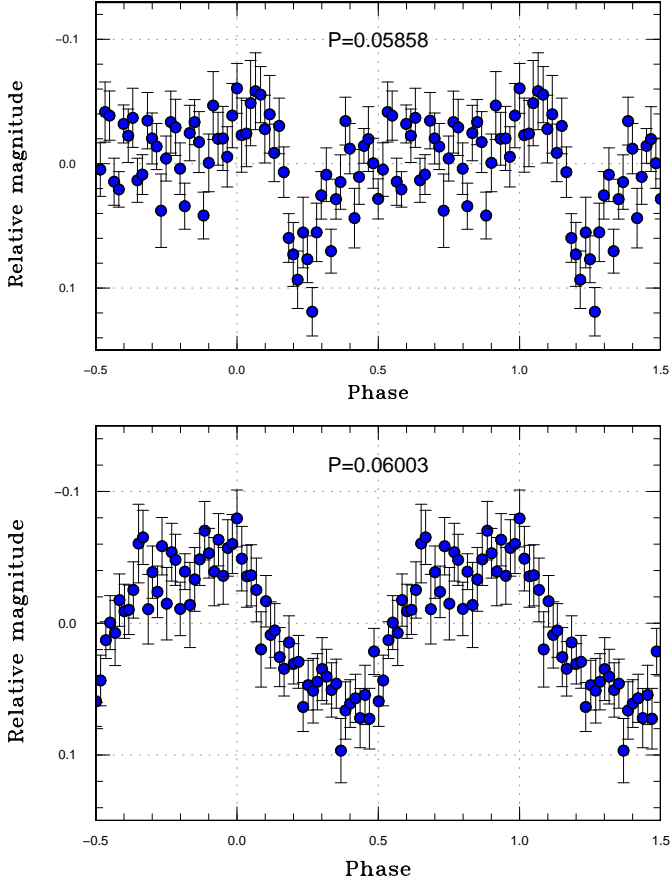}
  \end{center}
  \caption{Profiles of two periodicities in CC Scl (2011).
     (Upper) orbital variation.
     (Lower) superhump.}
  \label{fig:ccsclmeanc}
\end{figure}

\begin{figure}
  \begin{center}
    \FigureFile(88mm,110mm){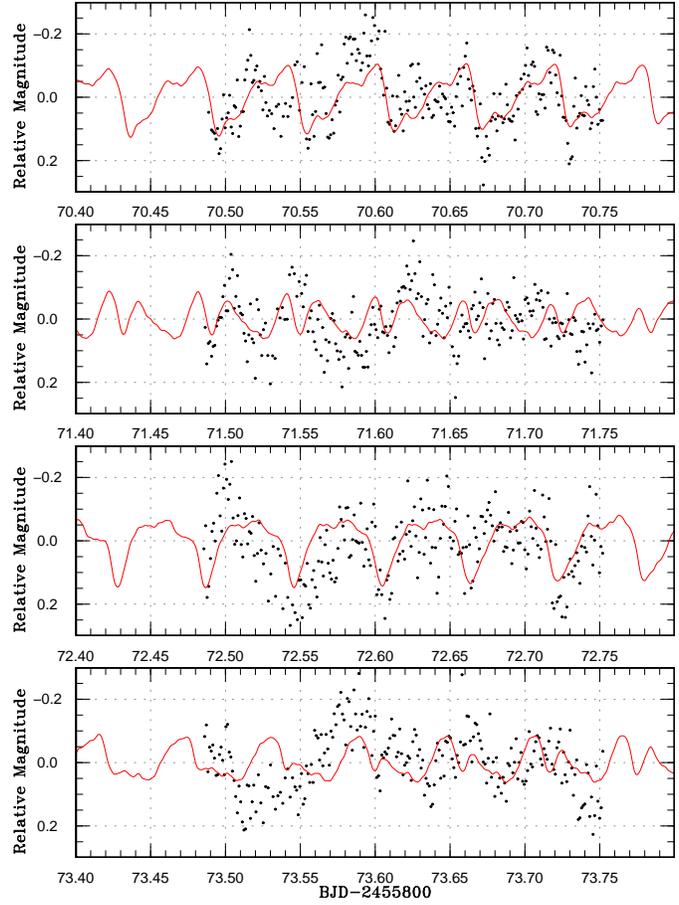}
  \end{center}
  \caption{Synthesized light curve of CC Scl (2011).
    The points represent observations.  The curves represent
    the expected light curve by adding two waves in figure
    \ref{fig:ccsclmeanc}.
  }
  \label{fig:ccsclperdub}
\end{figure}

   A PDM analysis yielded the stronger superhump signal and weaker
orbital signal (figure \ref{fig:ccsclshpdm} upper).  A lasso analysis,
which is less affected by aliasing, also indicated the presence of
both signals (figure \ref{fig:ccsclshpdm} lower).  We adopted
a refined orbital period of 0.0585845(10)~d.  We decomposed
the observations into these two periodicities (figure \ref{fig:ccsclmeanc}),
and tried to reproduce the observed light curve by combining
these waves (figure \ref{fig:ccsclperdub}).  Although the result
was not as remarkable as in OT J173516.9$+$154708, as we will see
later (subsection \ref{obj:j1735}), a part of the complex structure 
in the light curve appears to be understood as an effect of 
the orbital signal.  We therefore subtracted the orbital variation 
and determined the time of superhump maxima (table \ref{tab:ccscloc2011}).  
The relatively large scatter in the $O-C$ residuals suggests the presence of
irregularities not attributable to the orbital variation.
Despite these irregularities, the $O-C$ residuals itself did not
show a strong trend of variation.
Considering that the initial part of the outburst was likely
missed, we probably observed only the stage C superhumps.
We listed the value in table \ref{tab:perlist} based on this
interpretation.

   The unusualbehavior of superhumps in this system, as well as 
the strong presence of the orbital signal, might have
led to a detection of a different period in \citep{ish01j2315}.
Such unusual behavior may be related to a likely high
orbital inclination \citep{tap04CTCV}.  The $\epsilon$, however,
was 2.4\%, a normal value for this $P_{\rm orb}$.
Future dense monitoring to detect the early stage of a superoutburst 
is desired.

   \citet{wou12ccscl} recently established that this object is
an intermediate polar similar to HT Cam.  CC Scl is an intriguing
case since HT Cam has not yet shown superoutbursts \citep{ish02htcam}.
The unusual behavior of the superhumps in CC Scl may be related
to the magnetism of the white dwarf.

\begin{table}
\caption{Superhump maxima of CC Scl (2011).}\label{tab:ccscloc2011}
\begin{center}
\begin{tabular}{ccccc}
\hline
$E$ & max\commenta & error & $O-C$\commentb & $N$\commentc \\
\hline
0 & 55870.5914 & 0.0009 & $-$0.0043 & 47 \\
1 & 55870.6551 & 0.0040 & $-$0.0006 & 48 \\
2 & 55870.7078 & 0.0017 & $-$0.0079 & 48 \\
15 & 55871.5040 & 0.0011 & 0.0082 & 38 \\
16 & 55871.5482 & 0.0011 & $-$0.0077 & 37 \\
17 & 55871.6251 & 0.0014 & 0.0092 & 44 \\
18 & 55871.6732 & 0.0023 & $-$0.0027 & 44 \\
32 & 55872.5019 & 0.0011 & $-$0.0142 & 36 \\
33 & 55872.5826 & 0.0019 & 0.0065 & 45 \\
51 & 55873.6608 & 0.0014 & 0.0045 & 44 \\
52 & 55873.7157 & 0.0031 & $-$0.0006 & 44 \\
66 & 55874.5598 & 0.0009 & 0.0033 & 41 \\
67 & 55874.6204 & 0.0020 & 0.0039 & 44 \\
69 & 55874.7376 & 0.0014 & 0.0011 & 35 \\
82 & 55875.5213 & 0.0034 & 0.0046 & 22 \\
83 & 55875.5765 & 0.0014 & $-$0.0002 & 43 \\
84 & 55875.6400 & 0.0014 & 0.0033 & 43 \\
85 & 55875.7007 & 0.0037 & 0.0040 & 43 \\
99 & 55876.5243 & 0.0047 & $-$0.0126 & 24 \\
100 & 55876.5904 & 0.0086 & $-$0.0065 & 27 \\
117 & 55877.6246 & 0.0030 & 0.0076 & 28 \\
118 & 55877.6874 & 0.0060 & 0.0103 & 26 \\
119 & 55877.7551 & 0.0023 & 0.0181 & 14 \\
133 & 55878.5726 & 0.0013 & $-$0.0046 & 27 \\
134 & 55878.6307 & 0.0027 & $-$0.0065 & 27 \\
149 & 55879.5368 & 0.0016 & $-$0.0006 & 20 \\
150 & 55879.5937 & 0.0048 & $-$0.0037 & 21 \\
151 & 55879.6535 & 0.0029 & $-$0.0040 & 21 \\
152 & 55879.7095 & 0.0011 & $-$0.0080 & 20 \\
\hline
  \multicolumn{5}{l}{\commenta BJD$-$2400000.} \\
  \multicolumn{5}{l}{\commentb Against max $= 2455870.5957 + 0.060012 E$.} \\
  \multicolumn{5}{l}{\commentc Number of points used to determine the maximum.} \\
\end{tabular}
\end{center}
\end{table}

\subsection{V1208 Tauri}\label{obj:v1208tau}

   We observed the 2011 December superoutburst of this object.
The initial part of the outburst was likely missed.
The times of superhump maxima are listed in table \ref{tab:v1208tauoc2011}.
The $O-C$ values were close to zero, which strengthens the identification
of these superhump to be stage C superhumps.  It is likely both
2000 and 2002 observations \citep{Pdot} also recorded stage C
superhumps, which was not labelled as such in table 2 of \citet{Pdot}.

\begin{table}
\caption{Superhump maxima of V1208 Tau (2011).}\label{tab:v1208tauoc2011}
\begin{center}
\begin{tabular}{ccccc}
\hline
$E$ & max\commenta & error & $O-C$\commentb & $N$\commentc \\
\hline
0 & 55919.4074 & 0.0007 & 0.0021 & 75 \\
1 & 55919.4775 & 0.0008 & 0.0016 & 54 \\
9 & 55920.0347 & 0.0008 & $-$0.0050 & 75 \\
10 & 55920.1086 & 0.0006 & $-$0.0015 & 58 \\
34 & 55921.8064 & 0.0009 & 0.0047 & 73 \\
35 & 55921.8728 & 0.0007 & 0.0006 & 73 \\
48 & 55922.7901 & 0.0007 & 0.0017 & 73 \\
49 & 55922.8548 & 0.0012 & $-$0.0041 & 73 \\
\hline
  \multicolumn{5}{l}{\commenta BJD$-$2400000.} \\
  \multicolumn{5}{l}{\commentb Against max $= 2455919.4054 + 0.070481 E$.} \\
  \multicolumn{5}{l}{\commentc Number of points used to determine the maximum.} \\
\end{tabular}
\end{center}
\end{table}

\subsection{V1212 Tauri}\label{obj:v1212tau}

   Although the object underwent a superoutburst in 2011 January--February,
it again underwent another superoutburst in September--October.
The times of superhump maxima are listed in table \ref{tab:v1212tauoc2011b}.
The period given in table \ref{tab:perlist} was determined with
the PDM method.  The value of the period suggests that we observed either
the very start of stage B or stage C.  The supercycle length
of this object is about 240~d.

\begin{table}
\caption{Superhump maxima of V1212 Tau (2011b).}\label{tab:v1212tauoc2011b}
\begin{center}
\begin{tabular}{ccccc}
\hline
$E$ & max\commenta & error & $O-C$\commentb & $N$\commentc \\
\hline
0 & 55834.4345 & 0.0034 & $-$0.0001 & 21 \\
1 & 55834.5065 & 0.0016 & 0.0021 & 63 \\
2 & 55834.5735 & 0.0008 & $-$0.0005 & 107 \\
3 & 55834.6423 & 0.0010 & $-$0.0016 & 90 \\
15 & 55835.4805 & 0.0017 & $-$0.0001 & 29 \\
16 & 55835.5497 & 0.0019 & $-$0.0006 & 66 \\
17 & 55835.6201 & 0.0010 & 0.0000 & 35 \\
18 & 55835.6906 & 0.0019 & 0.0008 & 19 \\
\hline
  \multicolumn{5}{l}{\commenta BJD$-$2400000.} \\
  \multicolumn{5}{l}{\commentb Against max $= 2455834.4346 + 0.069731 E$.} \\
  \multicolumn{5}{l}{\commentc Number of points used to determine the maximum.} \\
\end{tabular}
\end{center}
\end{table}

\subsection{DI Ursae Majoris}\label{obj:diuma}

   Since online data for \citet{rut09diuma} are available, we extracted
times of superhump maxima for two superoutbursts in 2007 using 
our method (tables \ref{tab:diumaoc2007} and \ref{tab:diumaoc2007b}).
The resultant values of $P_{\rm dot}$ were not very different from
the analysis by \citet{rut09diuma}.  Although we listed times of
maxima before the superoutburst ($E \le 2$) and after the superoutburst
($E \ge 216$) for the first superoutburst, these maxima may not be 
equivalent to stage A and C superhumps in other SU UMa-type 
dwarf novae.  Although there may be either a discontinuous period change
or a phase shift between $E=182$ and $E=216$, we could not make
a distinction from the available data.  The second superoutburst was
less observed and the resultant $P_{\rm dot}$ was less reliable.
We have also analyzed the entire 2007 light curve to determine
the orbital period.  We detected a strong signal with a period
of 0.0545665(8)~d.  This period is closer to the spectroscopic
period of 0.054564(2)~d obtained by \citet{tho02gwlibv844herdiuma}
than the one obtained by \citet{rut09diuma} and it likely represents
the correct orbital period.  The $\epsilon$ for the better determined
first superoutburst was 1.4\%, and there was no need to modify
the $\epsilon$ by \citet{rut09diuma}.  This $\epsilon$ is fairly
common for such short-$P_{\rm orb}$ objects and we cannot discriminate
DI UMa from other SU UMa-type dwarf novae by $\epsilon$ only.

\begin{table}
\caption{Superhump maxima of DI UMa (2007).}\label{tab:diumaoc2007}
\begin{center}
\begin{tabular}{ccccc}
\hline
$E$ & max\commenta & error & $O-C$\commentb & $N$\commentc \\
\hline
0 & 54204.3594 & 0.0028 & $-$0.0185 & 20 \\
1 & 54204.4108 & 0.0015 & $-$0.0224 & 24 \\
2 & 54204.4641 & 0.0011 & $-$0.0244 & 44 \\
18 & 54205.3793 & 0.0007 & 0.0070 & 20 \\
19 & 54205.4336 & 0.0005 & 0.0061 & 25 \\
20 & 54205.4894 & 0.0009 & 0.0066 & 35 \\
35 & 54206.3165 & 0.0003 & 0.0050 & 39 \\
36 & 54206.3715 & 0.0002 & 0.0048 & 37 \\
37 & 54206.4246 & 0.0003 & 0.0027 & 39 \\
53 & 54207.3080 & 0.0006 & 0.0022 & 79 \\
54 & 54207.3630 & 0.0003 & 0.0020 & 102 \\
55 & 54207.4200 & 0.0003 & 0.0037 & 69 \\
56 & 54207.4745 & 0.0003 & 0.0030 & 67 \\
57 & 54207.5293 & 0.0003 & 0.0025 & 66 \\
71 & 54208.3073 & 0.0005 & 0.0072 & 40 \\
72 & 54208.3535 & 0.0014 & $-$0.0020 & 32 \\
91 & 54209.4102 & 0.0013 & 0.0051 & 20 \\
127 & 54211.4019 & 0.0010 & 0.0081 & 28 \\
128 & 54211.4583 & 0.0016 & 0.0093 & 16 \\
143 & 54212.2886 & 0.0020 & 0.0110 & 18 \\
144 & 54212.3452 & 0.0008 & 0.0123 & 28 \\
182 & 54214.4491 & 0.0010 & 0.0170 & 31 \\
216 & 54216.3046 & 0.0006 & $-$0.0058 & 39 \\
217 & 54216.3615 & 0.0005 & $-$0.0042 & 46 \\
218 & 54216.4143 & 0.0009 & $-$0.0065 & 28 \\
219 & 54216.4696 & 0.0005 & $-$0.0065 & 24 \\
235 & 54217.3548 & 0.0018 & $-$0.0052 & 11 \\
236 & 54217.3952 & 0.0035 & $-$0.0200 & 11 \\
\hline
  \multicolumn{5}{l}{\commenta BJD$-$2400000.} \\
  \multicolumn{5}{l}{\commentb Against max $= 2454204.3779 + 0.055243 E$.} \\
  \multicolumn{5}{l}{\commentc Number of points used to determine the maximum.} \\
\end{tabular}
\end{center}
\end{table}

\begin{table}
\caption{Superhump maxima of DI UMa (2007b).}\label{tab:diumaoc2007b}
\begin{center}
\begin{tabular}{ccccc}
\hline
$E$ & max\commenta & error & $O-C$\commentb & $N$\commentc \\
\hline
0 & 54237.4668 & 0.0002 & 0.0044 & 29 \\
16 & 54238.3514 & 0.0005 & 0.0035 & 36 \\
17 & 54238.4042 & 0.0003 & 0.0009 & 39 \\
18 & 54238.4571 & 0.0006 & $-$0.0015 & 41 \\
34 & 54239.3435 & 0.0003 & $-$0.0005 & 33 \\
35 & 54239.3951 & 0.0005 & $-$0.0043 & 33 \\
36 & 54239.4538 & 0.0004 & $-$0.0009 & 30 \\
71 & 54241.3914 & 0.0008 & $-$0.0002 & 22 \\
89 & 54242.3861 & 0.0008 & $-$0.0016 & 29 \\
90 & 54242.4357 & 0.0016 & $-$0.0074 & 28 \\
107 & 54243.3918 & 0.0007 & 0.0079 & 19 \\
108 & 54243.4289 & 0.0018 & $-$0.0103 & 26 \\
126 & 54244.4451 & 0.0016 & 0.0098 & 27 \\
\hline
  \multicolumn{5}{l}{\commenta BJD$-$2400000.} \\
  \multicolumn{5}{l}{\commentb Against max $= 2454237.4624 + 0.055340 E$.} \\
  \multicolumn{5}{l}{\commentc Number of points used to determine the maximum.} \\
\end{tabular}
\end{center}
\end{table}

\subsection{IY Ursae Majoris}\label{obj:iyuma}

   We observed a superoutburst in 2011 June.
Only two superhump maxima were recorded: BJD 2455717.0312(4) ($N=60$)
and BJD 2455718.0930(17) ($N=68$).

\subsection{KS Ursae Majoris}\label{obj:ksuma}

   We observed a superoutburst in 2012 May.
Only two superhump maxima were recorded: BJD 2456052.0468(13) ($N=52$)
and BJD 2456052.1159(18) ($N=50$).

\subsection{MR Ursae Majoris}\label{obj:mruma}

\begin{figure}
  \begin{center}
    \FigureFile(88mm,70mm){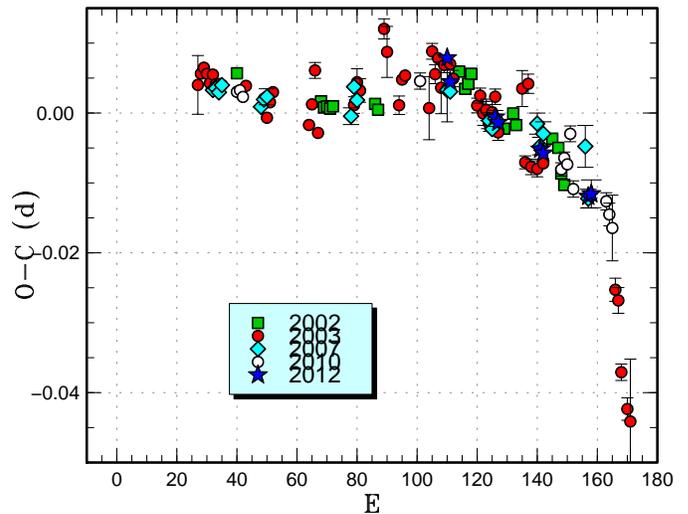}
  \end{center}
  \caption{Comparison of $O-C$ diagrams of MR UMa between different
  superoutbursts.  A period of 0.06512~d was used to draw this figure.
  Approximate cycle counts ($E$) after the start of the
  2007 superoutburst were used.  Since the starts of the other
  superoutbursts were not well constrained, we shifted the $O-C$ diagrams
  to best fit the 2007 one.
  }
  \label{fig:mrumacomp3}
\end{figure}

   The times of superhump maxima during the 2012 superoutburst are
listed in table \ref{tab:mrumaoc2012}.  Only the late stage of the
outburst was observed and we recorded typical stage C superhumps.
A comparison of $O-C$ diagrams between different superoutbursts
is shown in figure \ref{fig:mrumacomp3}.

\begin{table}
\caption{Superhump maxima of MR UMa (2012).}\label{tab:mrumaoc2012}
\begin{center}
\begin{tabular}{ccccc}
\hline
$E$ & max\commenta & error & $O-C$\commentb & $N$\commentc \\
\hline
0 & 56094.3735 & 0.0005 & 0.0019 & 129 \\
1 & 56094.4353 & 0.0006 & $-$0.0011 & 135 \\
16 & 56095.4070 & 0.0008 & $-$0.0006 & 124 \\
17 & 56095.4713 & 0.0010 & $-$0.0010 & 131 \\
31 & 56096.3792 & 0.0009 & 0.0004 & 133 \\
32 & 56096.4438 & 0.0010 & 0.0003 & 134 \\
47 & 56097.4144 & 0.0012 & $-$0.0003 & 133 \\
48 & 56097.4799 & 0.0020 & 0.0004 & 104 \\
\hline
  \multicolumn{5}{l}{\commenta BJD$-$2400000.} \\
  \multicolumn{5}{l}{\commentb Against max $= 2456094.3716 + 0.064746 E$.} \\
  \multicolumn{5}{l}{\commentc Number of points used to determine the maximum.} \\
\end{tabular}
\end{center}
\end{table}

\subsection{PU Ursae Majoris}\label{obj:puuma}

   PU UMa (=SDSS J090103.93$+$480911.1) is a deeply eclipsing
CV below the period gap \citep{dil08SDSSCV}, which was originally
discovered by \citet{szk03SDSSCV2}.  Three past outbursts had been
recorded before 2012: 2007 October (likely normal outburst),
2009 May (superoutburst; although superhumps were detected,
the duration of the observation was insufficient to determine the
period) and 2009 December (likely normal outburst).

   The 2012 outburst was detected by J. Shears (BAAVSS alert 2830).
Subsequent observations detected developing superhumps and eclipses
(vsnet-alert 14201, 14214, 14215).
The times of recorded eclipses were determined with
the Kwee and van Woerden (KW) method (\cite{KWmethod};
modified by the author, see appendix \ref{sec:app:mcmcecl}),
after removing linearly approximated trends around eclipses in order to
minimize the effect of superhumps, and are summarized in table 
\ref{tab:puumaecl}.
We obtained an updated ephemeris of
\begin{equation}
{\rm Min(BJD)} = 2453773.4875(3) + 0.07788054(1) E
\label{equ:puumaecl}.
\end{equation}

   The times of superhump maxima outside the eclipses are listed
in table \ref{tab:puumaoc2012}.  Except $E=0$ (stage A), there is
a hint of a stage B--C transition around $E=48$.  The overall
pattern is similar to relatively long $P_{\rm orb}$-systems
such as EG Aqr \citep{ima08egaqr} and NSV 4838 \citep{ima09nsv4838}.

\begin{figure}
  \begin{center}
    \FigureFile(88mm,110mm){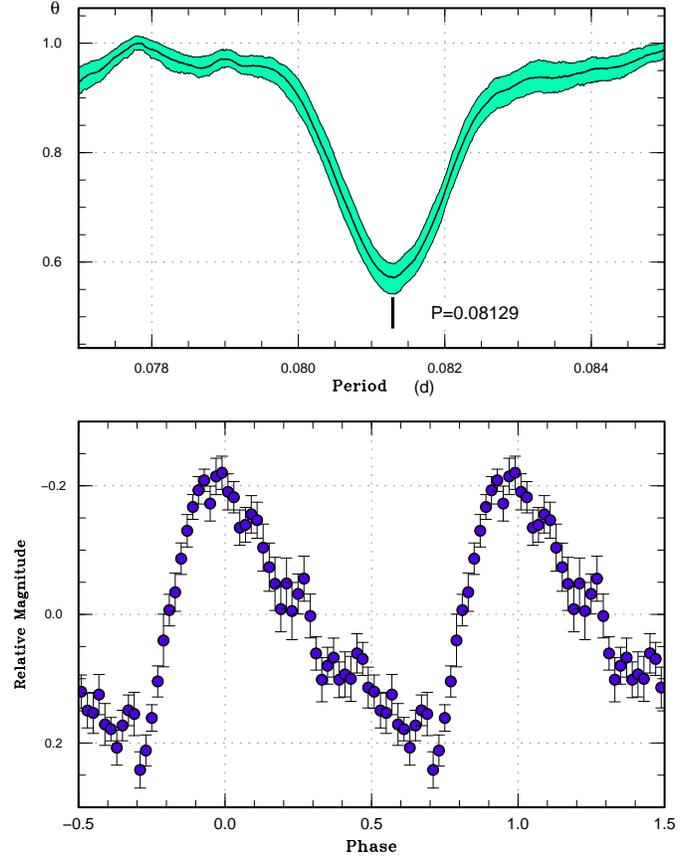}
  \end{center}
  \caption{Stage B superhumps in PU UMa (2012). (Upper): PDM analysis.
     (Lower): Phase-averaged profile.}
  \label{fig:puumashpdm}
\end{figure}

   The $\epsilon$ for stage B and C superhumps were 4.1\% and 3.7\%,
respectively, and is slightly larger than typical values of
SU UMa-type dwarf novae with these $P_{\rm orb}$.
A mean profile of stage B superhumps is shown in figure
\ref{fig:puumashpdm}.  \citet{she12puuma} also reported observations
of the same superoutburst, although they did not distinguish
stage B and C superhumps.

\begin{table}
\caption{Eclipse minima of PU UMa (2012).}\label{tab:puumaecl}
\begin{center}
\begin{tabular}{cccc}
\hline
$E$ & Minimum\commenta & error & $O-C$\commentb \\
\hline
28068 & 55959.43896 & 0.00006 & 0.00019 \\
28069 & 55959.51692 & 0.00003 & 0.00027 \\
28080 & 55960.37350 & 0.00005 & 0.00017 \\
28081 & 55960.45131 & 0.00006 & 0.00010 \\
28092 & 55961.30786 & 0.00002 & -0.00004 \\
28093 & 55961.38604 & 0.00004 & 0.00026 \\
28110 & 55962.70976 & 0.00004 & 0.00001 \\
28111 & 55962.78746 & 0.00004 & -0.00017 \\
28112 & 55962.86561 & 0.00003 & 0.00009 \\
28118 & 55963.33222 & 0.00004 & -0.00058 \\
28119 & 55963.41032 & 0.00005 & -0.00036 \\
28173 & 55967.61633 & 0.00003 & 0.00010 \\
\hline
  \multicolumn{4}{l}{\commenta BJD$-$2400000.} \\
  \multicolumn{4}{l}{\commentb Against equation \ref{equ:puumaecl}.} \\
\end{tabular}
\end{center}
\end{table}

\begin{table}
\caption{Superhump maxima of PU UMa (2012).}\label{tab:puumaoc2012}
\begin{center}
\begin{tabular}{cccccc}
\hline
$E$ & max\commenta & error & $O-C$\commentb & phase\commentc & $N$\commentd \\
\hline
0 & 55959.4659 & 0.0007 & $-$0.0331 & 0.13 & 183 \\
11 & 55960.3880 & 0.0009 & $-$0.0029 & 0.16 & 123 \\
12 & 55960.4712 & 0.0008 & $-$0.0007 & 0.18 & 101 \\
22 & 55961.2845 & 0.0012 & 0.0019 & 0.54 & 63 \\
23 & 55961.3685 & 0.0006 & 0.0049 & 0.17 & 239 \\
24 & 55961.4507 & 0.0015 & 0.0060 & 0.30 & 77 \\
35 & 55962.3444 & 0.0011 & 0.0079 & 0.46 & 34 \\
36 & 55962.4242 & 0.0011 & 0.0067 & 0.55 & 35 \\
40 & 55962.7460 & 0.0005 & 0.0042 & 0.18 & 132 \\
41 & 55962.8265 & 0.0008 & 0.0036 & 0.30 & 115 \\
47 & 55963.3178 & 0.0014 & 0.0085 & 0.41 & 67 \\
48 & 55963.3998 & 0.0011 & 0.0094 & 0.30 & 69 \\
83 & 55966.2293 & 0.0045 & 0.0015 & 0.41 & 28 \\
84 & 55966.3101 & 0.0015 & 0.0013 & 0.33 & 40 \\
100 & 55967.5984 & 0.0006 & $-$0.0076 & 0.19 & 96 \\
121 & 55969.2966 & 0.0039 & $-$0.0118 & 0.68 & 132 \\
\hline
  \multicolumn{6}{l}{\commenta BJD$-$2400000.} \\
  \multicolumn{6}{l}{\commentb Against max $= 2455959.4991 + 0.081068 E$.} \\
  \multicolumn{6}{l}{\commentc Orbital phase.} \\
  \multicolumn{6}{l}{\commentd Number of points used to determine the maximum.} \\
\end{tabular}
\end{center}
\end{table}

\subsection{SS Ursae Minoris}\label{obj:ssumi}

   \citet{ole06ssumi} reported observations of the 2004 superoutburst
and obtained a mean period of 0.070149(16)~d with a negative (global)
$P_{\rm dot}$.
We observed the late stage of a superoutburst in 2012 March.
As in \citet{ole06ssumi}, the the main peak of the superhump
was already diminishing and the secondary maximum was developing.
We list times of maxima of these humps in table \ref{tab:ssumioc2012}.
At this and following stages, the secondary humps were the dominant
signal, which are listed in table \ref{tab:ssumioc2012sec}.
These secondary humps persisted during the quiescent state
following the superoutburst, and the behavior is similar to
the late stage of VW Hyi (subsection \ref{obj:vwhyi}) and
V344 Lyr (\cite{Pdot3}; \cite{woo11v344lyr}).  Since SS UMi
is considered to have a high mass-transfer rate comparable to
ER UMa stars (\cite{kat00ssumi}; \cite{ole06ssumi}), these
secondary humps are possibly ``traditional'' late superhumps
arising from the stream impact point.  The signal became
less convincing after the next normal outburst.
A period analysis of the entire observation (BJD 2456007--2456038)
yielded a photometric orbital period of 0.067855(7)~d,
which is slightly longer than the spectroscopic period
of 0.06778(4)~d in \citet{tho96Porb}.

\begin{table}
\caption{Superhump maxima of SS UMi (2012).}\label{tab:ssumioc2012}
\begin{center}
\begin{tabular}{ccccc}
\hline
$E$ & max\commenta & error & $O-C$\commentb & $N$\commentc \\
\hline
0 & 56009.5129 & 0.0007 & 0.0029 & 53 \\
1 & 56009.5786 & 0.0009 & $-$0.0017 & 53 \\
2 & 56009.6468 & 0.0014 & $-$0.0039 & 50 \\
5 & 56009.8662 & 0.0043 & 0.0044 & 80 \\
6 & 56009.9295 & 0.0014 & $-$0.0026 & 80 \\
7 & 56010.0062 & 0.0020 & 0.0037 & 54 \\
15 & 56010.5616 & 0.0015 & $-$0.0038 & 52 \\
33 & 56011.8328 & 0.0053 & 0.0010 & 80 \\
\hline
  \multicolumn{5}{l}{\commenta BJD$-$2400000.} \\
  \multicolumn{5}{l}{\commentb Against max $= 2456009.5100 + 0.070358 E$.} \\
  \multicolumn{5}{l}{\commentc Number of points used to determine the maximum.} \\
\end{tabular}
\end{center}
\end{table}

\begin{table}
\caption{Superhump maxima of SS UMi (2012) (secondary humps).}\label{tab:ssumioc2012sec}
\begin{center}
\begin{tabular}{ccccc}
\hline
$E$ & max\commenta & error & $O-C$\commentb & $N$\commentc \\
\hline
0 & 56009.5410 & 0.0012 & $-$0.0003 & 52 \\
1 & 56009.6083 & 0.0007 & $-$0.0029 & 52 \\
4 & 56009.8206 & 0.0009 & $-$0.0004 & 80 \\
5 & 56009.8899 & 0.0010 & $-$0.0010 & 79 \\
6 & 56009.9611 & 0.0006 & 0.0002 & 80 \\
14 & 56010.5134 & 0.0007 & $-$0.0070 & 53 \\
15 & 56010.5876 & 0.0009 & $-$0.0028 & 52 \\
16 & 56010.6528 & 0.0010 & $-$0.0075 & 50 \\
18 & 56010.7969 & 0.0006 & $-$0.0033 & 71 \\
19 & 56010.8669 & 0.0006 & $-$0.0032 & 79 \\
20 & 56010.9348 & 0.0009 & $-$0.0053 & 79 \\
28 & 56011.4990 & 0.0009 & $-$0.0006 & 52 \\
29 & 56011.5696 & 0.0016 & 0.0000 & 86 \\
30 & 56011.6398 & 0.0005 & 0.0003 & 74 \\
33 & 56011.8523 & 0.0006 & 0.0030 & 79 \\
34 & 56011.9197 & 0.0008 & 0.0004 & 79 \\
35 & 56011.9931 & 0.0006 & 0.0039 & 69 \\
42 & 56012.4833 & 0.0011 & 0.0044 & 52 \\
43 & 56012.5527 & 0.0009 & 0.0039 & 64 \\
44 & 56012.6247 & 0.0004 & 0.0059 & 104 \\
45 & 56012.6965 & 0.0045 & 0.0078 & 48 \\
56 & 56013.4677 & 0.0006 & 0.0097 & 53 \\
57 & 56013.5267 & 0.0027 & $-$0.0013 & 27 \\
58 & 56013.6075 & 0.0015 & 0.0096 & 53 \\
59 & 56013.6717 & 0.0011 & 0.0038 & 36 \\
71 & 56014.5123 & 0.0012 & 0.0051 & 21 \\
72 & 56014.5843 & 0.0008 & 0.0071 & 27 \\
73 & 56014.6504 & 0.0011 & 0.0033 & 34 \\
78 & 56014.9951 & 0.0015 & $-$0.0017 & 46 \\
85 & 56015.4873 & 0.0017 & 0.0009 & 27 \\
86 & 56015.5525 & 0.0007 & $-$0.0038 & 53 \\
87 & 56015.6182 & 0.0014 & $-$0.0081 & 53 \\
88 & 56015.6914 & 0.0027 & $-$0.0048 & 18 \\
92 & 56015.9750 & 0.0008 & $-$0.0010 & 52 \\
100 & 56016.5337 & 0.0012 & $-$0.0019 & 20 \\
101 & 56016.5996 & 0.0009 & $-$0.0059 & 28 \\
102 & 56016.6762 & 0.0012 & 0.0008 & 27 \\
106 & 56016.9480 & 0.0006 & $-$0.0072 & 55 \\
\hline
  \multicolumn{5}{l}{\commenta BJD$-$2400000.} \\
  \multicolumn{5}{l}{\commentb Against max $= 2456009.5412 + 0.069943 E$.} \\
  \multicolumn{5}{l}{\commentc Number of points used to determine the maximum.} \\
\end{tabular}
\end{center}
\end{table}

\subsection{1RXS J231935.0$+$364705}\label{obj:j2319}

\begin{figure}
  \begin{center}
    \FigureFile(88mm,110mm){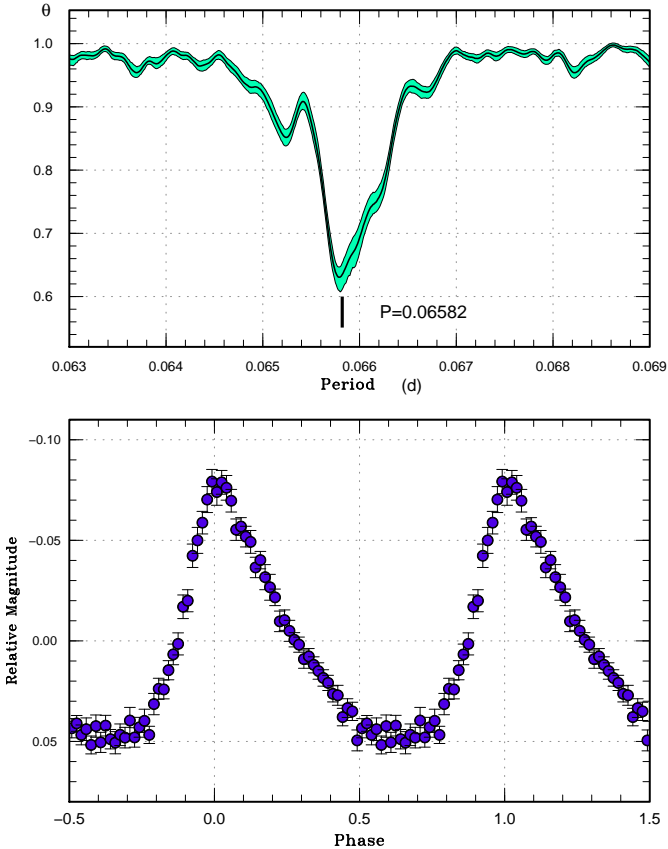}
  \end{center}
  \caption{Superhumps in 1RXS J231935 (2011). (Upper): PDM analysis.
     (Lower): Phase-averaged profile.}
  \label{fig:j2319shpdm}
\end{figure}

   This object (hereafter 1RXS J231935) was selected as a variable
star, likely a dwarf nova, during the course of identification
of the ROSAT sources \citep{den11ROSATCVs}.  The two previously
known outbursts occurred in 2009 November and 2010 December, and
both appear to be normal outbursts (H. Maehara detected no
superhumps during the 2009 outburst).
The 2011 September outburst was detected by E. Muyllaert (BAAVSS alert 2710).
Subsequent observations confirmed the presence of superhumps
(vsnet-alert 13711, 13712, 13719; figure \ref{fig:j2319shpdm}).
The times of superhump maxima are listed in table \ref{tab:j2319oc2011}.
The $O-C$ diagram clearly shows the familiar pattern of stages B and C.
The $P_{\rm dot}$ for stage B superhumps was large 
[$+11.6(1.7) \times 10^{-5}$], typical for an object with this
$P_{\rm SH}$.

\begin{table}
\caption{Superhump maxima of 1RXS J231935.}\label{tab:j2319oc2011}
\begin{center}
\begin{tabular}{ccccc}
\hline
$E$ & max\commenta & error & $O-C$\commentb & $N$\commentc \\
\hline
0 & 55835.4280 & 0.0003 & $-$0.0040 & 65 \\
1 & 55835.4930 & 0.0003 & $-$0.0048 & 68 \\
2 & 55835.5596 & 0.0002 & $-$0.0039 & 102 \\
3 & 55835.6240 & 0.0002 & $-$0.0052 & 203 \\
4 & 55835.6920 & 0.0004 & $-$0.0030 & 84 \\
14 & 55836.3491 & 0.0004 & $-$0.0035 & 49 \\
15 & 55836.4142 & 0.0003 & $-$0.0042 & 213 \\
16 & 55836.4793 & 0.0002 & $-$0.0049 & 362 \\
17 & 55836.5457 & 0.0002 & $-$0.0043 & 389 \\
18 & 55836.6113 & 0.0002 & $-$0.0044 & 217 \\
19 & 55836.6764 & 0.0003 & $-$0.0051 & 132 \\
29 & 55837.3357 & 0.0003 & $-$0.0035 & 157 \\
30 & 55837.4024 & 0.0002 & $-$0.0025 & 169 \\
31 & 55837.4658 & 0.0002 & $-$0.0049 & 250 \\
32 & 55837.5312 & 0.0003 & $-$0.0052 & 267 \\
33 & 55837.5978 & 0.0002 & $-$0.0044 & 267 \\
34 & 55837.6649 & 0.0005 & $-$0.0031 & 93 \\
44 & 55838.3270 & 0.0007 & 0.0014 & 86 \\
45 & 55838.3920 & 0.0006 & 0.0006 & 126 \\
46 & 55838.4587 & 0.0006 & 0.0015 & 125 \\
47 & 55838.5220 & 0.0038 & $-$0.0009 & 41 \\
56 & 55839.1186 & 0.0011 & 0.0038 & 94 \\
57 & 55839.1859 & 0.0006 & 0.0054 & 123 \\
59 & 55839.3162 & 0.0006 & 0.0041 & 98 \\
60 & 55839.3808 & 0.0007 & 0.0030 & 116 \\
61 & 55839.4481 & 0.0006 & 0.0045 & 121 \\
62 & 55839.5133 & 0.0006 & 0.0039 & 85 \\
63 & 55839.5871 & 0.0025 & 0.0120 & 77 \\
64 & 55839.6491 & 0.0015 & 0.0082 & 41 \\
75 & 55840.3753 & 0.0005 & 0.0110 & 51 \\
76 & 55840.4449 & 0.0014 & 0.0149 & 50 \\
77 & 55840.5081 & 0.0007 & 0.0123 & 57 \\
78 & 55840.5728 & 0.0006 & 0.0112 & 57 \\
79 & 55840.6379 & 0.0006 & 0.0106 & 44 \\
102 & 55842.1444 & 0.0006 & 0.0045 & 112 \\
103 & 55842.2125 & 0.0005 & 0.0068 & 119 \\
104 & 55842.2761 & 0.0008 & 0.0047 & 66 \\
118 & 55843.1927 & 0.0005 & 0.0005 & 95 \\
140 & 55844.6357 & 0.0010 & $-$0.0033 & 31 \\
141 & 55844.7006 & 0.0011 & $-$0.0041 & 31 \\
142 & 55844.7650 & 0.0008 & $-$0.0054 & 36 \\
143 & 55844.8324 & 0.0009 & $-$0.0038 & 33 \\
144 & 55844.8970 & 0.0009 & $-$0.0050 & 35 \\
153 & 55845.4853 & 0.0036 & $-$0.0086 & 54 \\
154 & 55845.5512 & 0.0010 & $-$0.0084 & 59 \\
155 & 55845.6205 & 0.0010 & $-$0.0049 & 83 \\
157 & 55845.7555 & 0.0039 & $-$0.0014 & 34 \\
159 & 55845.8803 & 0.0010 & $-$0.0081 & 28 \\
\hline
  \multicolumn{5}{l}{\commenta BJD$-$2400000.} \\
  \multicolumn{5}{l}{\commentb Against max $= 2455835.4320 + 0.065764 E$.} \\
  \multicolumn{5}{l}{\commentc Number of points used to determine the maximum.} \\
\end{tabular}
\end{center}
\end{table}

\subsection{ASAS J224349$+$0809.5}\label{obj:asas2243}

   The 2011 June outburst of this object (hereafter ASAS J224349)
was detected by Y. Maeda at a visual magnitude of 13.2
(vsnet-alert 13458).  Due to the unfavorable
seasonal condition, we obtained only two superhump maxima:
BJD 2455740.6014(5) ($N$=40) and 2455741.5730(5) ($N$=49).

\subsection{DDE 19}\label{obj:dde19}

\begin{figure}
  \begin{center}
    \FigureFile(88mm,110mm){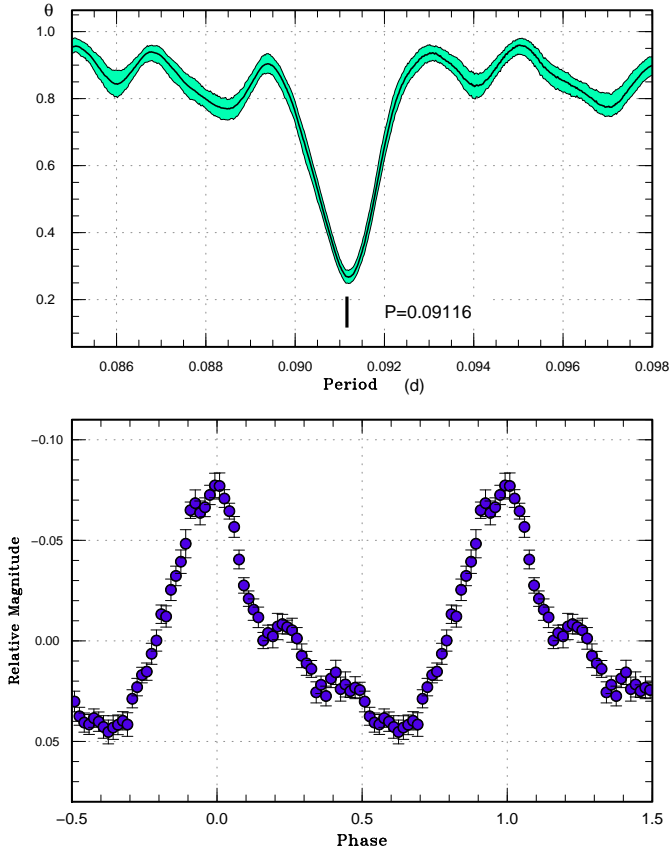}
  \end{center}
  \caption{Superhumps in DDE 19 (2011). (Upper): PDM analysis.
     (Lower): Phase-averaged profile.}
  \label{fig:dde19shpdm}
\end{figure}

   DDE 19 is a CV discovered by D. Denisenko.\footnote{
$<$http://hea.iki.rssi.ru/$\sim$denis/VarDDE.html$>$
}  The object is located at \timeform{00h 38m 37.40s},
\timeform{+79D 21' 37.5''} (J2000.0).  During its outburst
in 2011 November, superhumps were detected (vsnet-alert 13886,
13890; figure \ref{fig:dde19shpdm}).
The times of superhump maxima are listed in table \ref{tab:dde19oc2011}.
The object faded quickly after these observations, and
we only observed the late stage of this superoutburst.
We attributed the superhumps to stage C superhumps in table
\ref{tab:perlist}.

\begin{table}
\caption{Superhump maxima of DDE 19 (2011).}\label{tab:dde19oc2011}
\begin{center}
\begin{tabular}{ccccc}
\hline
$E$ & max\commenta & error & $O-C$\commentb & $N$\commentc \\
\hline
0 & 55888.6444 & 0.0004 & 0.0018 & 97 \\
2 & 55888.8256 & 0.0004 & 0.0006 & 96 \\
11 & 55889.6458 & 0.0009 & $-$0.0001 & 87 \\
12 & 55889.7356 & 0.0008 & $-$0.0014 & 96 \\
13 & 55889.8258 & 0.0006 & $-$0.0025 & 96 \\
33 & 55891.6541 & 0.0007 & 0.0017 & 96 \\
34 & 55891.7445 & 0.0013 & 0.0008 & 96 \\
35 & 55891.8339 & 0.0010 & $-$0.0010 & 96 \\
\hline
  \multicolumn{5}{l}{\commenta BJD$-$2400000.} \\
  \multicolumn{5}{l}{\commentb Against max $= 2455888.6425 + 0.091210 E$.} \\
  \multicolumn{5}{l}{\commentc Number of points used to determine the maximum.} \\
\end{tabular}
\end{center}
\end{table}

\subsection{MASTER OT J072948.66$+$593824.4}\label{obj:j0729}

   This object (hereafter MASTER J072948) is a transient detected
at an unfiltered CCD magnitude of 13.3 on 2012 February 17
(\cite{bal12j0729atl3935}; see also vsnet-alert 14249).
Although subsequent observations detected superhump-like
modulations (vsnet-alert 14252), their waveform was rather
irregular and the variation did not appear to be expressed
by a single period (vsnet-alert 14253, 14258, 14263).
The observed maxima could not be expressed by any single
period, and there was likely a superposition of two close
periods (vsnet-alert 14265).

\begin{figure}
  \begin{center}
    \FigureFile(88mm,110mm){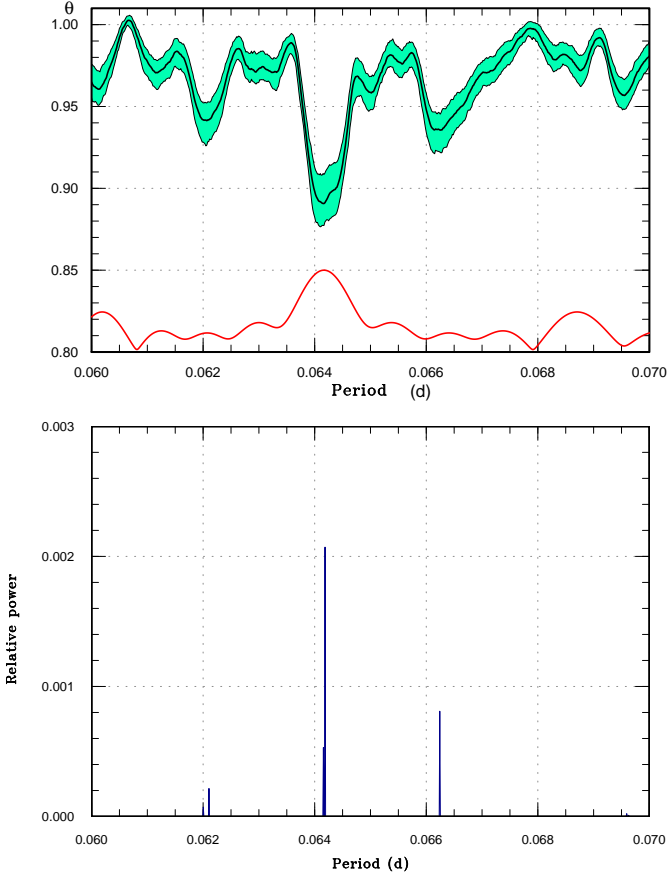}
  \end{center}
  \caption{Period analysis in MASTER J072948 (2012). (Upper): PDM analysis.
     The lower curve at the bottom indicates the window function.
     (Lower): lasso analysis ($\log \lambda=-4.34$).}
  \label{fig:j0729shpdm}
\end{figure}

\begin{figure}
  \begin{center}
    \FigureFile(88mm,110mm){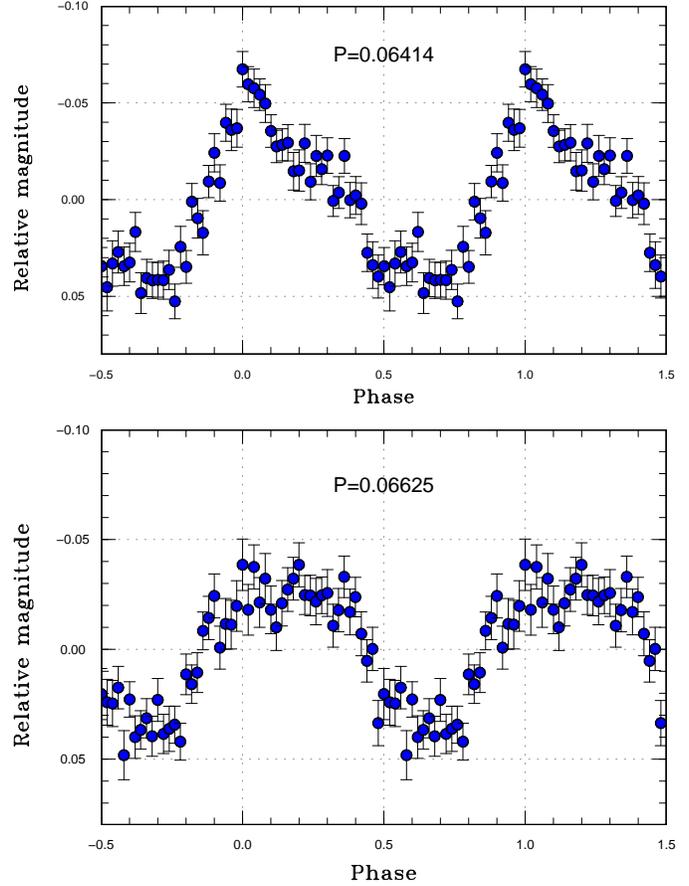}
  \end{center}
  \caption{Profiles of two periodicities in MASTER J072948 (2012).}
  \label{fig:j0729meanc}
\end{figure}

\begin{figure}
  \begin{center}
    \FigureFile(88mm,95mm){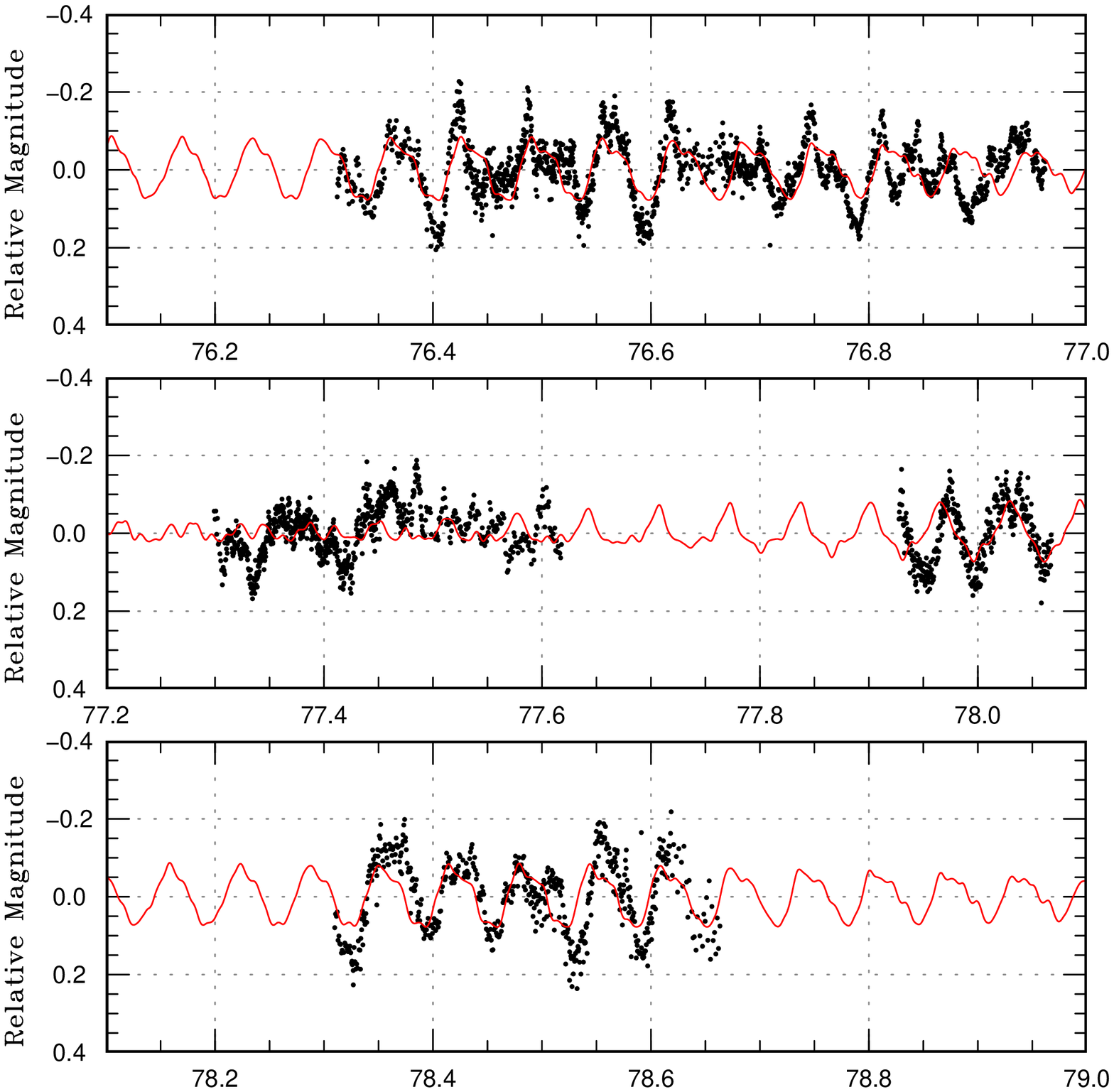}
  \end{center}
  \caption{Synthesized light curve of MASTER J072948 (2011).
    The points represent observations.  The curves represent
    the expected light curve by adding two waves in figure
    \ref{fig:j0729meanc}}.
  \label{fig:j0729perdub}
\end{figure}

   PDM and lasso analysis (see subsection \ref{obj:j1735} for
the application of lasso and separation of two signals)
is shown in figure \ref{fig:j0729shpdm}.
The lasso analysis favored the co-existence of 
two periods 0.06416(4)~d and 0.06625(4)~d.  A period of
0.06208(3)~d, a one-day alias of the 0.06625-d period, cannot be
excluded instead of the 0.06625-d period.
The mean profiles of these signals are shown in figure
\ref{fig:j0729meanc}.  While the 0.06416-d signal resembles a profile
of superhumps (faster rise, sharper maximum), the other signal
has a sharper minimum.  A combination of these signals partly
reproduced the actual light curve during the plateau phase
(figure \ref{fig:j0729perdub}).  Based on the profile, we may 
identify the 0.06416-d signal as superhumps.  We might then
interpret that the 0.06625-d period is the orbital period, 
and the 0.06416-d signal is 3.2\% shorter than $P_{\rm orb}$.
Although negative superhumps are unexpected in ordinary
SU UMa-type dwarf novae, this interpretation might explain 
why the profile of superhumps was so unstable 
(as in ER UMa, cf. \cite{ohs12eruma}), and
why the orbital signal is so strongly visible in a non-eclipsing
system.  Since the object started rapidly fading only three days
after the start of our observation, the baseline for period
analysis was insufficient to distinguish other possible periods
or interpretations.  Future observations in quiescence and
in superoutbursts are absolutely needed.

\subsection{MASTER OT J174305.70$+$231107.8}\label{obj:j1743}

\begin{figure}
  \begin{center}
    \FigureFile(88mm,110mm){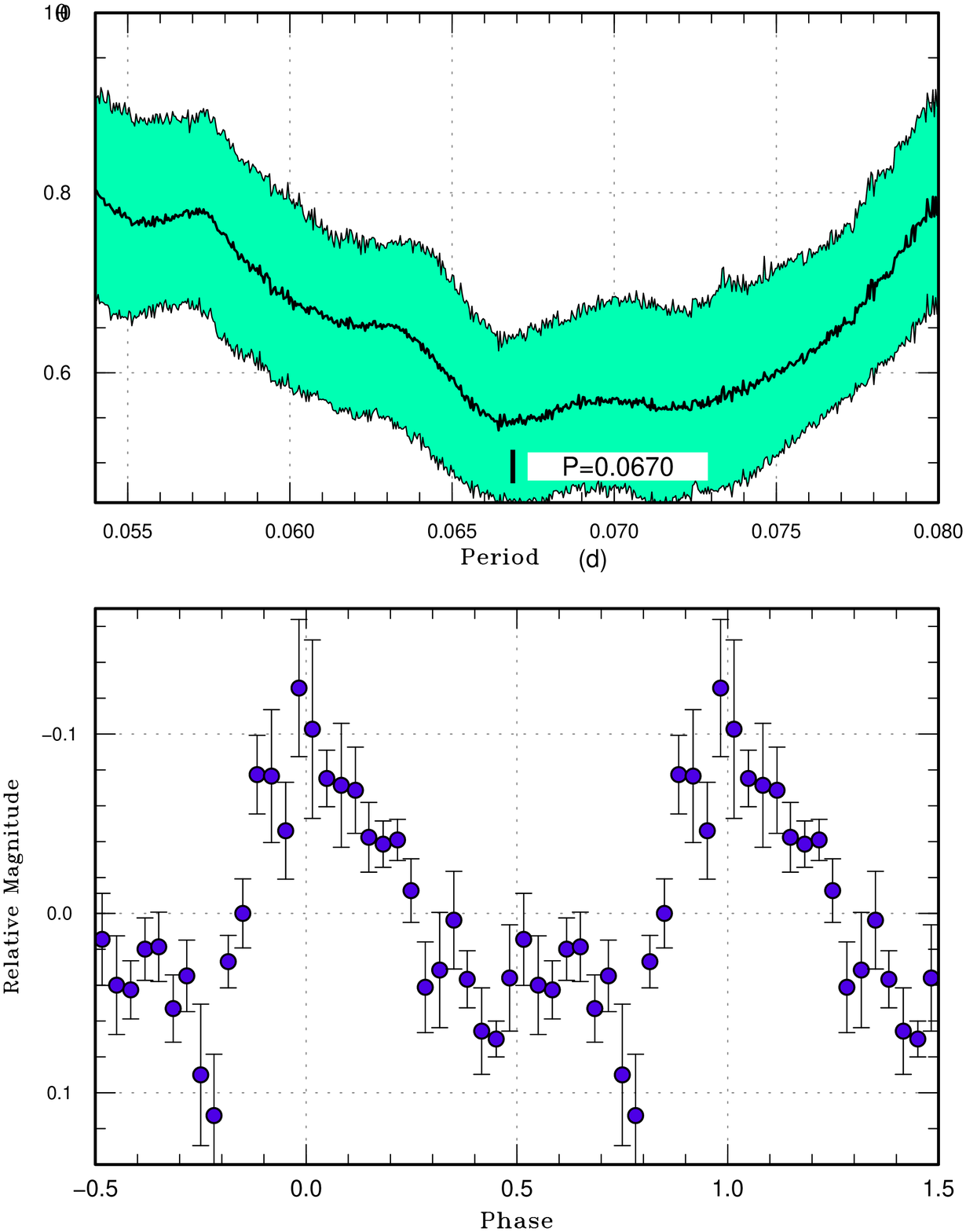}
  \end{center}
  \caption{Superhumps in MASTER J174305 (2012). (Upper): PDM analysis.
     (Lower): Phase-averaged profile.}
  \label{fig:j1743shpdm}
\end{figure}

   This object (hereafter MASTER J174305) is a transient detected
at an unfiltered CCD magnitude of 15.6 on 2012 April 5
\citep{bal12j1743atel4022}.  Subsequent observations detected
superhumps (vsnet-alert 14428; figure \ref{fig:j1743shpdm}).
Two superhump maxima were recorded: BJD 2456027.8604(12) ($N=79$)
and BJD 2456027.9281(9) ($N=82$).  The superhump period
by the PDM method was 0.0670(5)~d.

\subsection{MASTER OT J182201.93$+$324906.7}\label{obj:j1822}

\begin{figure}
  \begin{center}
    \FigureFile(88mm,110mm){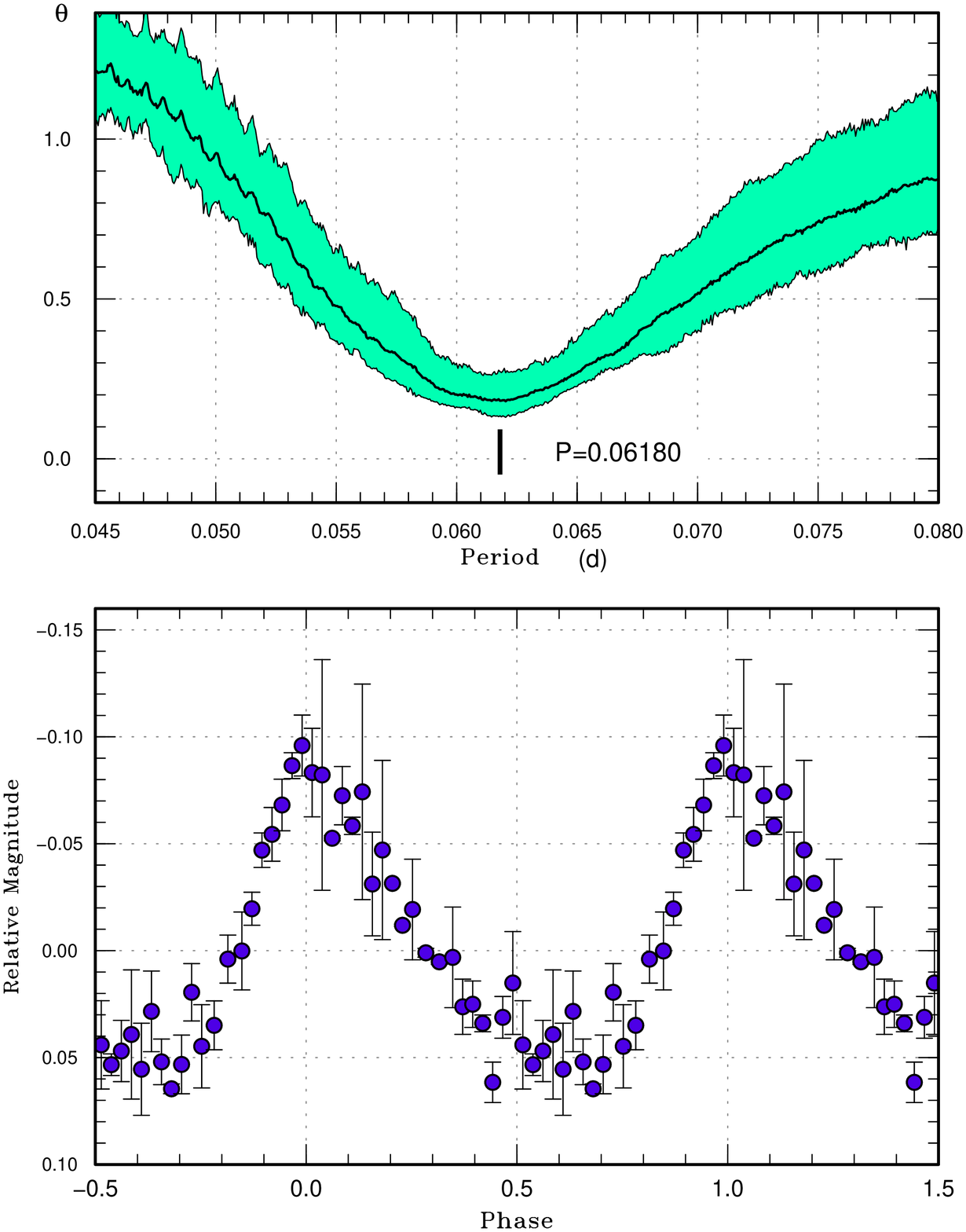}
  \end{center}
  \caption{Superhumps in MASTER J182201 (2012). (Upper): PDM analysis.
     (Lower): Phase-averaged profile.}
  \label{fig:j1822shpdm}
\end{figure}

   This object (hereafter MASTER J182201) is a transient detected
at an unfiltered CCD magnitude of 15.4 on 2012 April 29
\citep{bal12j1822atel4084}.  Subsequent observations detected
superhumps (vsnet-alert 14529; figure \ref{fig:j1822shpdm}).
Two superhump maxima were recorded: BJD 2456050.4464(6) ($N=33$)
and 2456050.5081(4) ($N=33$).  The superhump period
by the PDM method was 0.0618(2)~d (figure \ref{fig:j1822shpdm}).

\subsection{MisV 1446}\label{obj:misv1446}

\begin{figure}
  \begin{center}
    \FigureFile(88mm,110mm){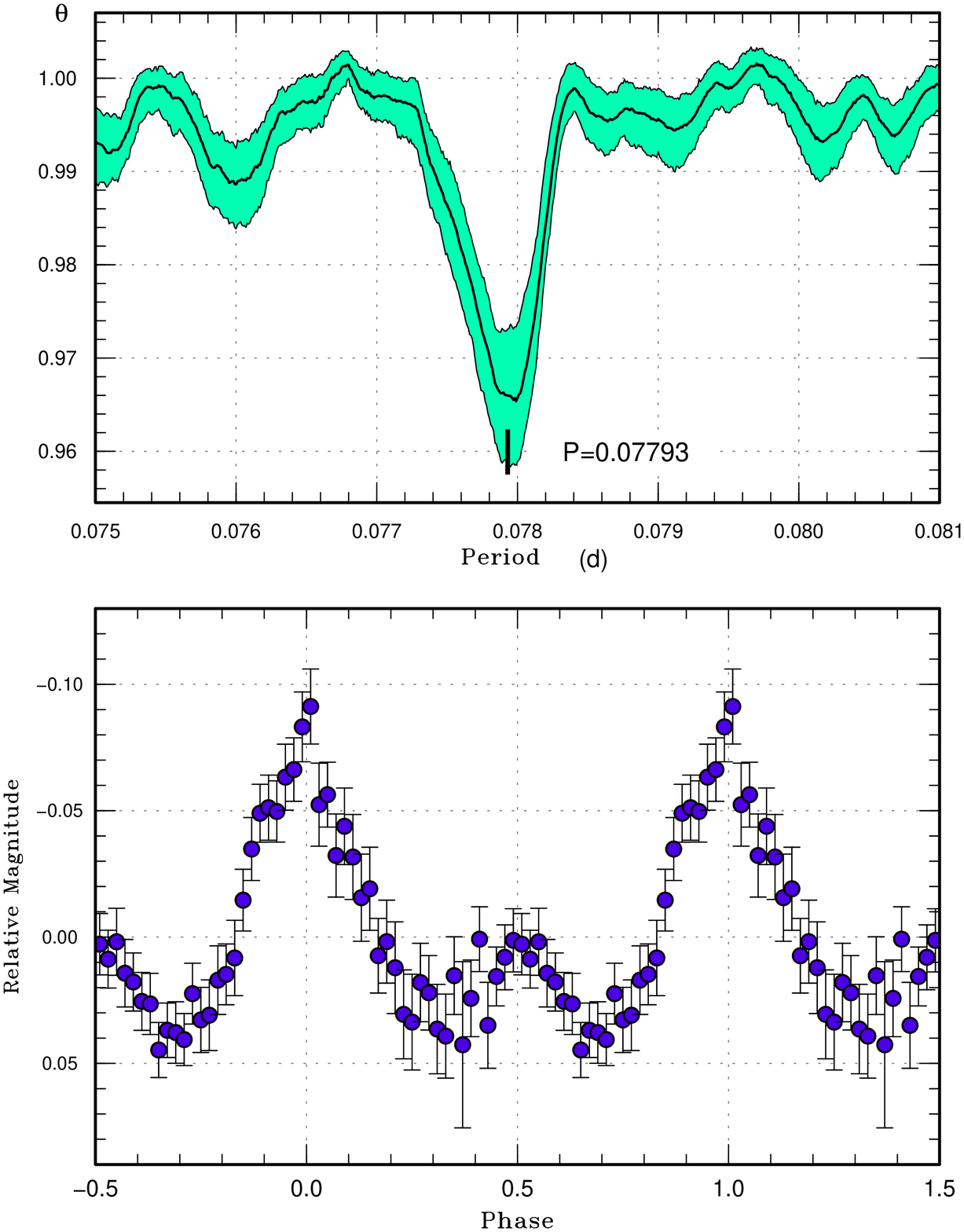}
  \end{center}
  \caption{Superhumps in MisV 1446 (2012). (Upper): PDM analysis.
     (Lower): Phase-averaged profile.}
  \label{fig:misv1446shpdm}
\end{figure}

   MisV 1446 was detected as a transient by the MISAO project, and
it could be probably identified with the X-ray source 
1RXS J074112.2$-$094529 (vsnet-alert 14080).  The coordinates
of the object are \timeform{07h 41m 12.70s}, \timeform{-09D 45' 55.9''}.
Multicolor photometry by H. Sato was consistent with that of
a color of a dwarf nova in outburst (vsnet-alert 14085).
Subsequent observations recorded superhumps (vsnet-alert 14096,
14102, 14104; figure \ref{fig:misv1446shpdm}).
The times of superhump maxima are listed in table
\ref{tab:misv1446oc2012}.  It appears that the observations recorded
the late stage of a superoutburst, and that late part of stage B and
stage C were recorded.  It was impossible to measure $P_{\rm dot}$
for stage B.

\begin{table}
\caption{Superhump maxima of MisV 1446 (2012).}\label{tab:misv1446oc2012}
\begin{center}
\begin{tabular}{ccccc}
\hline
$E$ & max\commenta & error & $O-C$\commentb & $N$\commentc \\
\hline
0 & 55938.0155 & 0.0031 & $-$0.0006 & 134 \\
1 & 55938.0949 & 0.0006 & 0.0010 & 207 \\
2 & 55938.1682 & 0.0005 & $-$0.0034 & 312 \\
3 & 55938.2450 & 0.0006 & $-$0.0044 & 281 \\
4 & 55938.3233 & 0.0006 & $-$0.0040 & 279 \\
5 & 55938.4019 & 0.0015 & $-$0.0031 & 113 \\
13 & 55939.0278 & 0.0012 & 0.0003 & 178 \\
14 & 55939.1040 & 0.0007 & $-$0.0013 & 287 \\
15 & 55939.1822 & 0.0007 & $-$0.0009 & 158 \\
16 & 55939.2639 & 0.0012 & 0.0030 & 155 \\
17 & 55939.3371 & 0.0008 & $-$0.0016 & 124 \\
26 & 55940.0351 & 0.0031 & $-$0.0038 & 138 \\
27 & 55940.1209 & 0.0010 & 0.0042 & 108 \\
35 & 55940.7515 & 0.0016 & 0.0123 & 81 \\
36 & 55940.8206 & 0.0008 & 0.0035 & 67 \\
37 & 55940.9001 & 0.0009 & 0.0053 & 82 \\
40 & 55941.1329 & 0.0011 & 0.0046 & 225 \\
41 & 55941.2098 & 0.0012 & 0.0038 & 157 \\
42 & 55941.2873 & 0.0010 & 0.0034 & 155 \\
44 & 55941.4427 & 0.0006 & 0.0032 & 301 \\
45 & 55941.5184 & 0.0007 & 0.0011 & 407 \\
48 & 55941.7484 & 0.0104 & $-$0.0023 & 18 \\
49 & 55941.8340 & 0.0123 & 0.0055 & 10 \\
56 & 55942.3688 & 0.0037 & $-$0.0043 & 191 \\
57 & 55942.4462 & 0.0025 & $-$0.0047 & 408 \\
58 & 55942.5173 & 0.0010 & $-$0.0114 & 336 \\
69 & 55943.3793 & 0.0014 & $-$0.0053 & 31 \\
\hline
  \multicolumn{5}{l}{\commenta BJD$-$2400000.} \\
  \multicolumn{5}{l}{\commentb Against max $= 2455938.0160 + 0.077806 E$.} \\
  \multicolumn{5}{l}{\commentc Number of points used to determine the maximum.} \\
\end{tabular}
\end{center}
\end{table}

\subsection{SBS 1108$+$574}\label{obj:sbs1108}

\begin{figure}
  \begin{center}
    \FigureFile(88mm,110mm){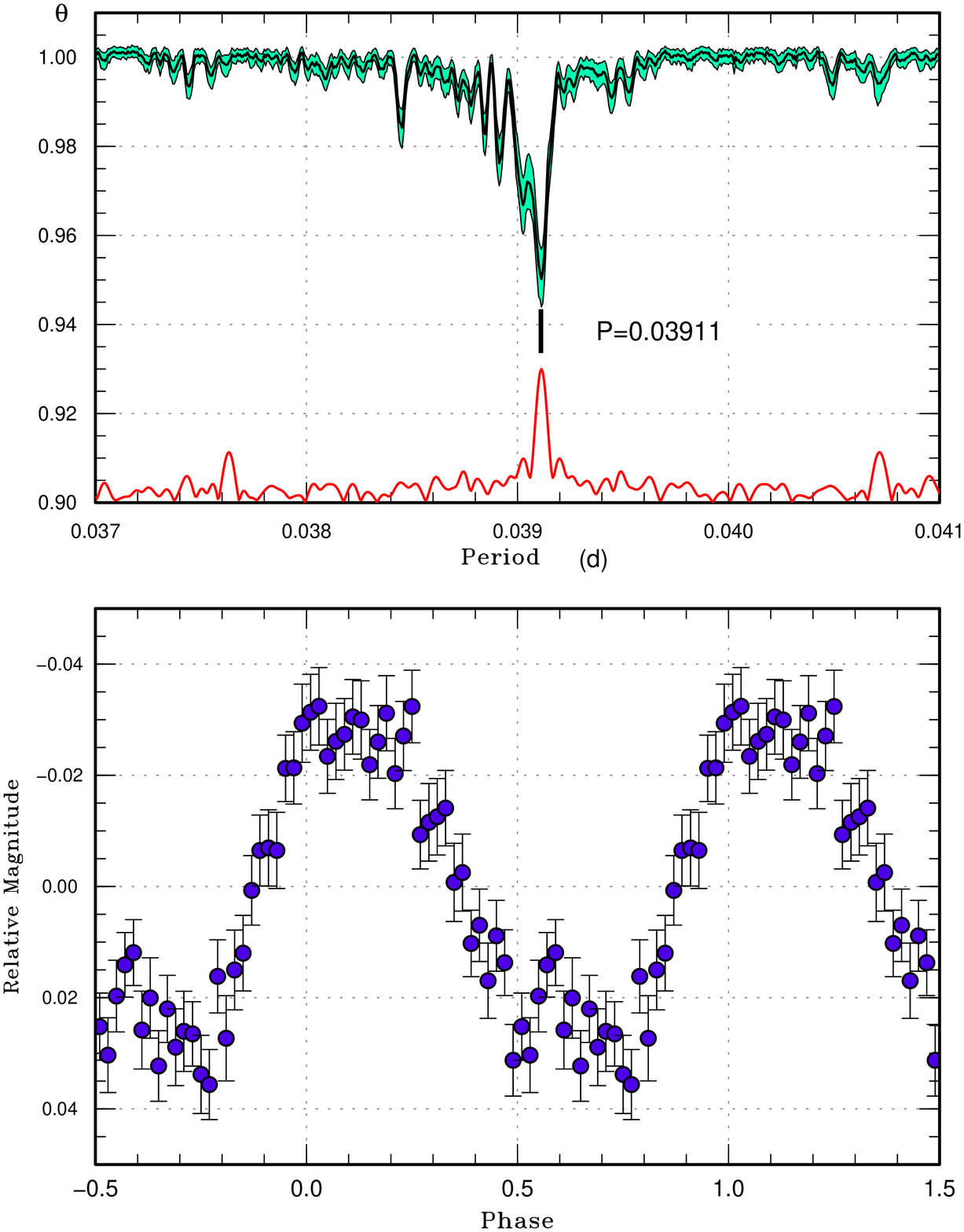}
  \end{center}
  \caption{Superhumps in SBS 1108 (2012). (Upper): PDM analysis.
    The curve at the bottom of the figure represents 
    the window function.  The signal at $P=0.038449$~d is the candidate
    orbital period.
    (Lower): Phase-averaged profile.}
  \label{fig:sbs1108shpdm}
\end{figure}

   This object (hereafter SBS 1108) was originally selected as an
ultraviolet-excess object during the course of the Second Byurakan Survey
(SBS, \cite{mar83SBS1}).  An outburst of this object was detected
by CRTS on 2012 April 22 (=CSS120422:111127$+$571239).
The very blue color ($u-g = -0.3$) in quiescence was very notable
(vsnet-alert 14475, 14483).  Subsequent observations clarified
that this object is an ultra-short period SU UMa-type dwarf nova
showing superhumps (vsnet-alert 14480, 14484, 14493; figure
\ref{fig:sbs1108shpdm}).
Although it was not initially clear whether this object belongs to 
AM CVn-type objects or hydrogen-rich objects, spectroscopic observation
\citep{gar12sbs1108atel4112} confirmed that the object is hydrogen-rich.

\begin{figure}
  \begin{center}
    \FigureFile(88mm,100mm){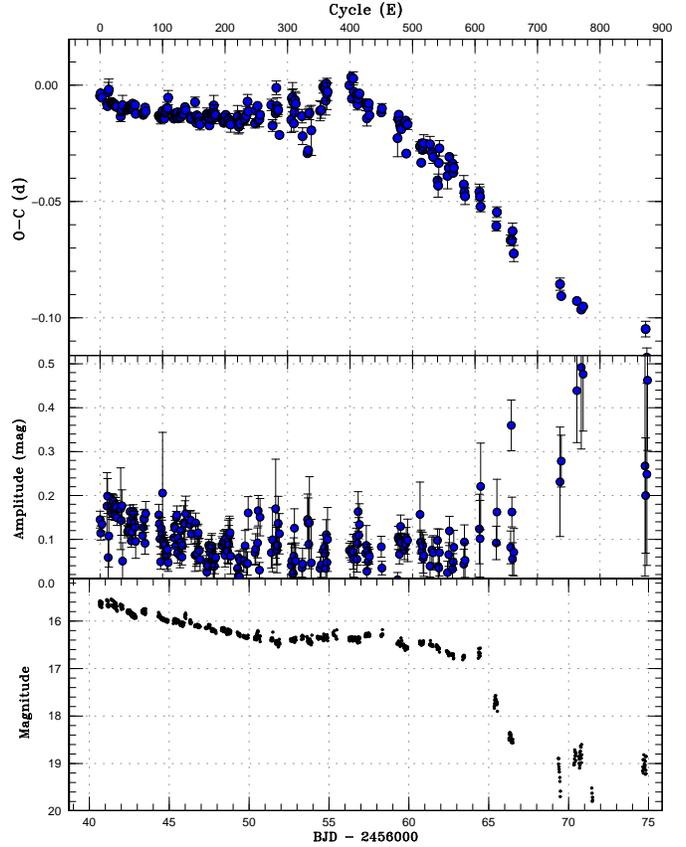}
  \end{center}
  \caption{$O-C$ diagram of superhumps in SBS 1108.
     (Upper:) $O-C$ diagram.
     We used a period of 0.03912~d for calculating the $O-C$ residuals.
     (Middle:) Amplitudes of superhumps.  There was a slight tendency
     of regrowth of superhumps around the stage B--C transition.
     (Lower:) Light curve.  The object slightly brightened after
     the stage B--C transition.
  }
  \label{fig:sbs1108humpall}
\end{figure}

\begin{figure}
  \begin{center}
    \FigureFile(88mm,70mm){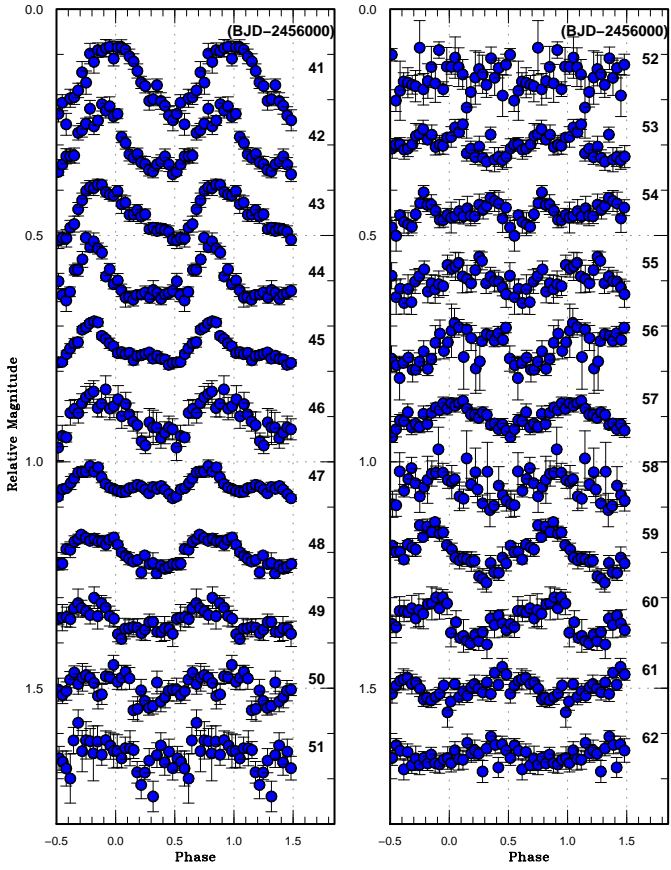}
  \end{center}
  \caption{Variation of superhump prfiles in SBS 1108 (2012).
    A period of 0.039111~d was assumed in phase-averaging.
    Although the amplitude of superhumps decreased for a time
    (BJD 2456050--2456056), it increased again
    (BJD 2456057--2456060).}
  \label{fig:sbs1108prof}
\end{figure}

   The times of superhump maxima are listed in table \ref{tab:sbs1108oc2012}.
Although the epoch of the start of the outburst is unknown,
the $O-C$ variation was very similar to that of ordinary 
short-$P_{\rm orb}$ SU UMa-type dwarf novae: consisting of
stage B with a longer $P_{\rm SH}$ and a positive $P_{\rm dot}$ and
stage C with a shorter $P_{\rm SH}$ with a relatively constant period
(figure \ref{fig:sbs1108humpall}).
The amplitudes of superhumps became smaller near the end of stage B
as in ordinary short-$P_{\rm orb}$ SU UMa-type dwarf novae
\citep{Pdot3} and became larger at the start of stage C
(figure \ref{fig:sbs1108prof}).  The object also slightly brightened 
after the stage B--C transition (figure \ref{fig:sbs1108humpall}).
This feature is commonly seen in objects with distinct stage B--C
transitions (cf. \cite{kat03hodel}; \cite{Pdot3}).
The transition from stage B to C was abrupt, as in ordinary 
short-$P_{\rm orb}$ SU UMa-type dwarf novae.
The stage C superhumps persisted after the rapid decline
without a phase shift.

\begin{figure}
  \begin{center}
    \FigureFile(88mm,60mm){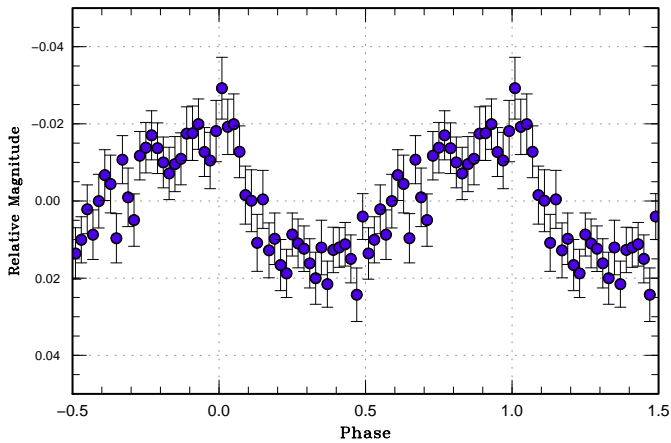}
  \end{center}
  \caption{Waveform of the candidate orbital period (0.038449~d) of SBS 1108.}
  \label{fig:sbs1108orb}
\end{figure}

\begin{figure}
  \begin{center}
    \FigureFile(88mm,60mm){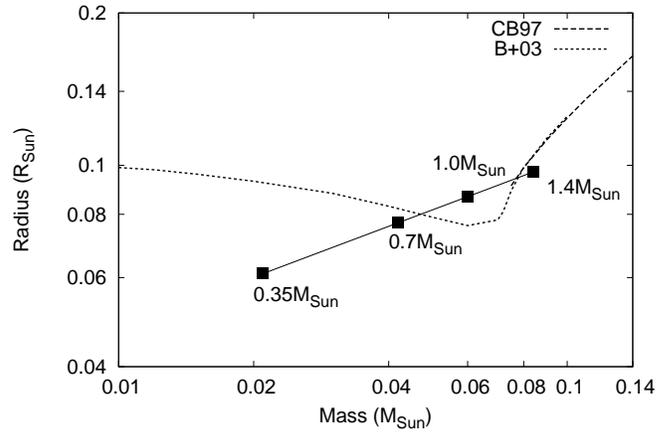}
  \end{center}
  \caption{Mass-radius relation of the secondary of SBS 1108.
  The volume radii of the Roche lobe of the secondary for various
  masses of the primary are plotted against the mass-radius
  relationship of 10 Gyr brown dwarfs and low-mass main-sequence stars
  by \citet{bar03BDmodel} (B03) and \citet{cha97lowmassstar} (CB97).}
  \label{fig:sbs1108mrr}
\end{figure}

   In addition to superhumps, we detected a stable period of
0.038449(6)~d, which we identified to be the orbital period
(figure \ref{fig:sbs1108orb}).
The $\epsilon$ for stage B and C superhumps were 1.74(2)\% and
1.11(2)\%, respectively.  By applying the $\epsilon$--$q$ relation
in \citet{Pdot} to the $\epsilon$ of stage C superhumps,
we obtained $q = 0.06$.  The $\epsilon$ or $q$ is larger than
those of many extreme WZ Sge-type dwarf novae, and this implies
that the secondary is denser, or more massive, than in ordinary 
dwarf novae.  The estimated volume radii of the Roche lobe of 
the secondary is located below the theoretical radius of a brown
dwarf when we assume a typical mass (0.7~M$_{\odot}$) for a white
dwarf in a dwarf nova (figure \ref{fig:sbs1108mrr}).  Only with
a massive ($\ge$1.0~M$_{\odot}$) white dwarf, the secondary can be
a normal lower main-sequence star.
Since sustained nuclear burning is not expected for a brown dwarf,
this finding suggests that this secondary is a somewhat evolved
star whose hydrogen envelope was mostly stripped during 
the mass-exchange.  The spectrum taken in quiescence
(Pavlenko et al. in preparation) showing the enhanced abundance of
helium is consistent with this interpretation.
The object may be analogous to OT J112253.3$-$111037
(= CSS100603:112253$-$111037) (\cite{Pdot2}; \cite{bre12j1122};
note also the unusual $u-g$ color mentioned in \cite{kat12DNSDSS}).
Further detailed radial-velocity study would enable to clarify
the nature of this binary.

\begin{table}
\caption{Superhump maxima of SBS 1108 (2012).}\label{tab:sbs1108oc2012}
\begin{center}
\begin{tabular}{ccccc}
\hline
$E$ & max\commenta & error & $O-C$\commentb & $N$\commentc \\
\hline
0 & 56040.6654 & 0.0007 & $-$0.0064 & 67 \\
1 & 56040.7056 & 0.0007 & $-$0.0053 & 56 \\
3 & 56040.7820 & 0.0006 & $-$0.0070 & 63 \\
11 & 56041.0923 & 0.0021 & $-$0.0090 & 59 \\
12 & 56041.1304 & 0.0010 & $-$0.0099 & 80 \\
13 & 56041.1755 & 0.0042 & $-$0.0039 & 80 \\
14 & 56041.2158 & 0.0045 & $-$0.0027 & 53 \\
18 & 56041.3662 & 0.0007 & $-$0.0085 & 32 \\
19 & 56041.4061 & 0.0005 & $-$0.0076 & 40 \\
20 & 56041.4448 & 0.0004 & $-$0.0079 & 43 \\
21 & 56041.4834 & 0.0005 & $-$0.0084 & 42 \\
22 & 56041.5227 & 0.0004 & $-$0.0082 & 43 \\
23 & 56041.5617 & 0.0005 & $-$0.0082 & 42 \\
24 & 56041.6002 & 0.0004 & $-$0.0087 & 108 \\
25 & 56041.6390 & 0.0005 & $-$0.0090 & 71 \\
26 & 56041.6778 & 0.0004 & $-$0.0093 & 69 \\
27 & 56041.7171 & 0.0006 & $-$0.0089 & 69 \\
33 & 56041.9475 & 0.0022 & $-$0.0128 & 62 \\
34 & 56041.9887 & 0.0012 & $-$0.0107 & 121 \\
35 & 56042.0287 & 0.0010 & $-$0.0097 & 122 \\
36 & 56042.0697 & 0.0042 & $-$0.0078 & 35 \\
44 & 56042.3807 & 0.0006 & $-$0.0091 & 42 \\
45 & 56042.4206 & 0.0005 & $-$0.0083 & 42 \\
46 & 56042.4600 & 0.0005 & $-$0.0079 & 42 \\
47 & 56042.4995 & 0.0005 & $-$0.0075 & 43 \\
48 & 56042.5390 & 0.0004 & $-$0.0070 & 43 \\
49 & 56042.5768 & 0.0005 & $-$0.0083 & 42 \\
50 & 56042.6166 & 0.0007 & $-$0.0075 & 37 \\
51 & 56042.6560 & 0.0007 & $-$0.0072 & 67 \\
52 & 56042.6963 & 0.0005 & $-$0.0059 & 69 \\
53 & 56042.7332 & 0.0005 & $-$0.0081 & 61 \\
54 & 56042.7724 & 0.0004 & $-$0.0079 & 14 \\
55 & 56042.8130 & 0.0011 & $-$0.0064 & 10 \\
56 & 56042.8522 & 0.0011 & $-$0.0062 & 14 \\
57 & 56042.8875 & 0.0019 & $-$0.0099 & 14 \\
68 & 56043.3183 & 0.0006 & $-$0.0087 & 26 \\
69 & 56043.3565 & 0.0005 & $-$0.0095 & 26 \\
70 & 56043.3964 & 0.0007 & $-$0.0086 & 24 \\
71 & 56043.4355 & 0.0008 & $-$0.0086 & 26 \\
72 & 56043.4770 & 0.0014 & $-$0.0061 & 26 \\
73 & 56043.5145 & 0.0009 & $-$0.0077 & 26 \\
94 & 56044.3342 & 0.0007 & $-$0.0080 & 75 \\
95 & 56044.3729 & 0.0006 & $-$0.0083 & 98 \\
96 & 56044.4121 & 0.0008 & $-$0.0081 & 96 \\
97 & 56044.4503 & 0.0024 & $-$0.0089 & 21 \\
98 & 56044.4894 & 0.0009 & $-$0.0089 & 12 \\
99 & 56044.5289 & 0.0012 & $-$0.0085 & 12 \\
100 & 56044.5709 & 0.0016 & $-$0.0055 & 8 \\
101 & 56044.6077 & 0.0009 & $-$0.0077 & 38 \\
102 & 56044.6465 & 0.0008 & $-$0.0080 & 53 \\
103 & 56044.6850 & 0.0010 & $-$0.0085 & 50 \\
104 & 56044.7266 & 0.0009 & $-$0.0060 & 53 \\
\hline
  \multicolumn{5}{l}{\commenta BJD$-$2400000.} \\
  \multicolumn{5}{l}{\commentb Against max $= 2456040.6718 + 0.039046 E$.} \\
  \multicolumn{5}{l}{\commentc Number of points used to determine the maximum.} \\
\end{tabular}
\end{center}
\end{table}

\addtocounter{table}{-1}
\begin{table}
\caption{Superhump maxima of SBS 1108 (2012) (continued).}
\begin{center}
\begin{tabular}{ccccc}
\hline
$E$ & max\commenta & error & $O-C$\commentb & $N$\commentc \\
\hline
105 & 56044.7671 & 0.0014 & $-$0.0045 & 36 \\
106 & 56044.8053 & 0.0014 & $-$0.0054 & 14 \\
107 & 56044.8460 & 0.0025 & $-$0.0038 & 14 \\
108 & 56044.8851 & 0.0014 & $-$0.0037 & 14 \\
109 & 56044.9287 & 0.0031 & 0.0009 & 10 \\
118 & 56045.2718 & 0.0008 & $-$0.0074 & 20 \\
119 & 56045.3110 & 0.0009 & $-$0.0073 & 21 \\
120 & 56045.3523 & 0.0007 & $-$0.0051 & 19 \\
121 & 56045.3914 & 0.0012 & $-$0.0050 & 19 \\
122 & 56045.4288 & 0.0004 & $-$0.0066 & 20 \\
123 & 56045.4678 & 0.0007 & $-$0.0067 & 22 \\
124 & 56045.5092 & 0.0016 & $-$0.0043 & 38 \\
125 & 56045.5457 & 0.0009 & $-$0.0068 & 35 \\
128 & 56045.6634 & 0.0020 & $-$0.0063 & 27 \\
129 & 56045.7028 & 0.0009 & $-$0.0060 & 31 \\
130 & 56045.7438 & 0.0011 & $-$0.0040 & 35 \\
131 & 56045.7821 & 0.0015 & $-$0.0048 & 36 \\
132 & 56045.8224 & 0.0008 & $-$0.0035 & 35 \\
134 & 56045.8998 & 0.0013 & $-$0.0042 & 12 \\
136 & 56045.9809 & 0.0017 & $-$0.0012 & 72 \\
137 & 56046.0187 & 0.0012 & $-$0.0024 & 81 \\
144 & 56046.2898 & 0.0011 & $-$0.0046 & 14 \\
145 & 56046.3279 & 0.0007 & $-$0.0056 & 20 \\
146 & 56046.3673 & 0.0010 & $-$0.0053 & 20 \\
152 & 56046.6089 & 0.0022 & 0.0021 & 12 \\
153 & 56046.6420 & 0.0008 & $-$0.0039 & 13 \\
154 & 56046.6783 & 0.0019 & $-$0.0066 & 14 \\
155 & 56046.7206 & 0.0017 & $-$0.0033 & 13 \\
156 & 56046.7576 & 0.0010 & $-$0.0054 & 13 \\
157 & 56046.7970 & 0.0008 & $-$0.0050 & 14 \\
158 & 56046.8373 & 0.0016 & $-$0.0038 & 14 \\
159 & 56046.8747 & 0.0019 & $-$0.0055 & 14 \\
160 & 56046.9123 & 0.0024 & $-$0.0068 & 14 \\
170 & 56047.3054 & 0.0016 & $-$0.0042 & 20 \\
171 & 56047.3472 & 0.0025 & $-$0.0015 & 20 \\
172 & 56047.3861 & 0.0013 & $-$0.0017 & 35 \\
173 & 56047.4238 & 0.0015 & $-$0.0029 & 30 \\
174 & 56047.4636 & 0.0008 & $-$0.0022 & 13 \\
175 & 56047.4987 & 0.0007 & $-$0.0062 & 13 \\
176 & 56047.5407 & 0.0007 & $-$0.0032 & 9 \\
178 & 56047.6210 & 0.0016 & $-$0.0010 & 12 \\
179 & 56047.6593 & 0.0017 & $-$0.0017 & 14 \\
180 & 56047.6971 & 0.0020 & $-$0.0030 & 14 \\
181 & 56047.7396 & 0.0029 & 0.0005 & 9 \\
182 & 56047.7811 & 0.0043 & 0.0029 & 13 \\
183 & 56047.8170 & 0.0036 & $-$0.0002 & 14 \\
184 & 56047.8543 & 0.0029 & $-$0.0020 & 14 \\
185 & 56047.8944 & 0.0018 & $-$0.0009 & 14 \\
195 & 56048.2844 & 0.0011 & $-$0.0014 & 20 \\
196 & 56048.3227 & 0.0010 & $-$0.0021 & 20 \\
197 & 56048.3627 & 0.0010 & $-$0.0012 & 20 \\
\hline
  \multicolumn{5}{l}{\commenta BJD$-$2400000.} \\
  \multicolumn{5}{l}{\commentb Against max $= 2456040.6718 + 0.039046 E$.} \\
  \multicolumn{5}{l}{\commentc Number of points used to determine the maximum.} \\
\end{tabular}
\end{center}
\end{table}

\addtocounter{table}{-1}
\begin{table}
\caption{Superhump maxima of SBS 1108 (2012) (continued).}
\begin{center}
\begin{tabular}{ccccc}
\hline
$E$ & max\commenta & error & $O-C$\commentb & $N$\commentc \\
\hline
198 & 56048.3996 & 0.0013 & $-$0.0034 & 15 \\
199 & 56048.4412 & 0.0012 & $-$0.0007 & 19 \\
200 & 56048.4785 & 0.0022 & $-$0.0025 & 18 \\
201 & 56048.5203 & 0.0010 & 0.0003 & 20 \\
204 & 56048.6349 & 0.0018 & $-$0.0023 & 14 \\
205 & 56048.6749 & 0.0025 & $-$0.0014 & 14 \\
206 & 56048.7141 & 0.0025 & $-$0.0012 & 14 \\
207 & 56048.7513 & 0.0019 & $-$0.0031 & 14 \\
208 & 56048.7908 & 0.0014 & $-$0.0026 & 13 \\
209 & 56048.8291 & 0.0033 & $-$0.0033 & 14 \\
220 & 56049.2630 & 0.0047 & 0.0010 & 14 \\
221 & 56049.2994 & 0.0049 & $-$0.0016 & 17 \\
222 & 56049.3368 & 0.0014 & $-$0.0033 & 20 \\
223 & 56049.3781 & 0.0048 & $-$0.0010 & 20 \\
224 & 56049.4163 & 0.0021 & $-$0.0018 & 20 \\
230 & 56049.6538 & 0.0026 & 0.0014 & 14 \\
232 & 56049.7313 & 0.0024 & 0.0008 & 11 \\
234 & 56049.8109 & 0.0040 & 0.0023 & 14 \\
236 & 56049.8953 & 0.0031 & 0.0086 & 14 \\
237 & 56049.9301 & 0.0010 & 0.0044 & 7 \\
248 & 56050.3554 & 0.0020 & 0.0001 & 82 \\
249 & 56050.4003 & 0.0022 & 0.0060 & 85 \\
252 & 56050.5193 & 0.0017 & 0.0079 & 14 \\
253 & 56050.5521 & 0.0010 & 0.0016 & 14 \\
255 & 56050.6310 & 0.0033 & 0.0024 & 13 \\
256 & 56050.6719 & 0.0013 & 0.0043 & 13 \\
274 & 56051.3804 & 0.0024 & 0.0100 & 44 \\
276 & 56051.4498 & 0.0025 & 0.0013 & 33 \\
281 & 56051.6506 & 0.0042 & 0.0069 & 9 \\
282 & 56051.7007 & 0.0029 & 0.0179 & 14 \\
283 & 56051.7320 & 0.0035 & 0.0101 & 13 \\
284 & 56051.7690 & 0.0024 & 0.0081 & 13 \\
285 & 56051.8086 & 0.0020 & 0.0087 & 13 \\
287 & 56051.8760 & 0.0019 & $-$0.0020 & 12 \\
306 & 56052.6258 & 0.0054 & 0.0059 & 13 \\
307 & 56052.6741 & 0.0038 & 0.0151 & 14 \\
308 & 56052.7138 & 0.0063 & 0.0158 & 32 \\
309 & 56052.7508 & 0.0057 & 0.0137 & 35 \\
310 & 56052.7808 & 0.0034 & 0.0047 & 31 \\
311 & 56052.8246 & 0.0017 & 0.0094 & 34 \\
312 & 56052.8690 & 0.0038 & 0.0148 & 29 \\
313 & 56052.9066 & 0.0045 & 0.0134 & 25 \\
323 & 56053.2924 & 0.0029 & 0.0088 & 16 \\
324 & 56053.3230 & 0.0036 & 0.0002 & 16 \\
332 & 56053.6286 & 0.0019 & $-$0.0065 & 9 \\
333 & 56053.6690 & 0.0019 & $-$0.0052 & 14 \\
334 & 56053.7237 & 0.0024 & 0.0105 & 13 \\
335 & 56053.7632 & 0.0029 & 0.0110 & 10 \\
338 & 56053.8731 & 0.0108 & 0.0038 & 11 \\
352 & 56054.4292 & 0.0043 & 0.0132 & 46 \\
354 & 56054.5078 & 0.0035 & 0.0137 & 29 \\
\hline
  \multicolumn{5}{l}{\commenta BJD$-$2400000.} \\
  \multicolumn{5}{l}{\commentb Against max $= 2456040.6718 + 0.039046 E$.} \\
  \multicolumn{5}{l}{\commentc Number of points used to determine the maximum.} \\
\end{tabular}
\end{center}
\end{table}

\addtocounter{table}{-1}
\begin{table}
\caption{Superhump maxima of SBS 1108 (2012) (continued).}
\begin{center}
\begin{tabular}{ccccc}
\hline
$E$ & max\commenta & error & $O-C$\commentb & $N$\commentc \\
\hline
357 & 56054.6351 & 0.0027 & 0.0239 & 10 \\
359 & 56054.7126 & 0.0010 & 0.0233 & 30 \\
360 & 56054.7527 & 0.0020 & 0.0243 & 28 \\
361 & 56054.7857 & 0.0062 & 0.0182 & 30 \\
362 & 56054.8322 & 0.0016 & 0.0257 & 27 \\
363 & 56054.8668 & 0.0067 & 0.0213 & 21 \\
364 & 56054.9069 & 0.0038 & 0.0223 & 17 \\
399 & 56056.2788 & 0.0015 & 0.0277 & 12 \\
402 & 56056.3998 & 0.0016 & 0.0315 & 24 \\
403 & 56056.4295 & 0.0015 & 0.0222 & 33 \\
405 & 56056.5165 & 0.0028 & 0.0311 & 23 \\
406 & 56056.5499 & 0.0038 & 0.0254 & 26 \\
410 & 56056.7049 & 0.0009 & 0.0242 & 20 \\
411 & 56056.7437 & 0.0026 & 0.0240 & 18 \\
412 & 56056.7791 & 0.0024 & 0.0204 & 20 \\
413 & 56056.8207 & 0.0013 & 0.0229 & 20 \\
414 & 56056.8615 & 0.0018 & 0.0247 & 20 \\
415 & 56056.9012 & 0.0017 & 0.0253 & 21 \\
425 & 56057.2882 & 0.0014 & 0.0218 & 16 \\
426 & 56057.3266 & 0.0025 & 0.0212 & 31 \\
427 & 56057.3600 & 0.0033 & 0.0155 & 80 \\
428 & 56057.4052 & 0.0009 & 0.0216 & 104 \\
429 & 56057.4426 & 0.0015 & 0.0201 & 88 \\
430 & 56057.4837 & 0.0013 & 0.0221 & 61 \\
431 & 56057.5178 & 0.0014 & 0.0171 & 60 \\
450 & 56058.2623 & 0.0018 & 0.0198 & 10 \\
451 & 56058.3031 & 0.0021 & 0.0216 & 16 \\
476 & 56059.2684 & 0.0080 & 0.0106 & 14 \\
477 & 56059.3158 & 0.0012 & 0.0190 & 17 \\
478 & 56059.3567 & 0.0007 & 0.0209 & 31 \\
479 & 56059.3930 & 0.0016 & 0.0182 & 58 \\
480 & 56059.4306 & 0.0010 & 0.0167 & 61 \\
481 & 56059.4684 & 0.0009 & 0.0154 & 41 \\
482 & 56059.5069 & 0.0012 & 0.0149 & 15 \\
487 & 56059.7054 & 0.0014 & 0.0182 & 20 \\
488 & 56059.7442 & 0.0009 & 0.0179 & 18 \\
489 & 56059.7845 & 0.0013 & 0.0191 & 20 \\
490 & 56059.8094 & 0.0012 & 0.0050 & 20 \\
492 & 56059.9002 & 0.0022 & 0.0178 & 20 \\
512 & 56060.6731 & 0.0020 & 0.0098 & 8 \\
513 & 56060.7125 & 0.0012 & 0.0100 & 20 \\
514 & 56060.7444 & 0.0013 & 0.0029 & 18 \\
515 & 56060.7918 & 0.0017 & 0.0113 & 19 \\
516 & 56060.8281 & 0.0018 & 0.0085 & 18 \\
517 & 56060.8687 & 0.0029 & 0.0101 & 20 \\
518 & 56060.9091 & 0.0038 & 0.0114 & 14 \\
528 & 56061.3000 & 0.0034 & 0.0118 & 12 \\
531 & 56061.4138 & 0.0019 & 0.0086 & 43 \\
533 & 56061.4901 & 0.0028 & 0.0068 & 32 \\
540 & 56061.7538 & 0.0013 & $-$0.0028 & 20 \\
541 & 56061.7907 & 0.0050 & $-$0.0050 & 18 \\
\hline
  \multicolumn{5}{l}{\commenta BJD$-$2400000.} \\
  \multicolumn{5}{l}{\commentb Against max $= 2456040.6718 + 0.039046 E$.} \\
  \multicolumn{5}{l}{\commentc Number of points used to determine the maximum.} \\
\end{tabular}
\end{center}
\end{table}

\addtocounter{table}{-1}
\begin{table}
\caption{Superhump maxima of SBS 1108 (2012) (continued).}
\begin{center}
\begin{tabular}{ccccc}
\hline
$E$ & max\commenta & error & $O-C$\commentb & $N$\commentc \\
\hline
542 & 56061.8396 & 0.0068 & 0.0048 & 19 \\
543 & 56061.8850 & 0.0033 & 0.0112 & 17 \\
556 & 56062.3817 & 0.0056 & 0.0003 & 72 \\
558 & 56062.4634 & 0.0020 & 0.0039 & 72 \\
559 & 56062.5072 & 0.0013 & 0.0087 & 29 \\
564 & 56062.7000 & 0.0036 & 0.0062 & 20 \\
565 & 56062.7351 & 0.0019 & 0.0023 & 19 \\
566 & 56062.7764 & 0.0023 & 0.0046 & 20 \\
582 & 56063.3952 & 0.0038 & $-$0.0014 & 29 \\
583 & 56063.4310 & 0.0020 & $-$0.0046 & 28 \\
584 & 56063.4683 & 0.0036 & $-$0.0064 & 26 \\
607 & 56064.3700 & 0.0033 & $-$0.0028 & 42 \\
608 & 56064.4070 & 0.0042 & $-$0.0048 & 43 \\
609 & 56064.4419 & 0.0022 & $-$0.0090 & 42 \\
634 & 56065.4115 & 0.0021 & $-$0.0155 & 25 \\
635 & 56065.4566 & 0.0023 & $-$0.0094 & 26 \\
657 & 56066.3051 & 0.0023 & $-$0.0199 & 17 \\
658 & 56066.3447 & 0.0008 & $-$0.0194 & 29 \\
659 & 56066.3833 & 0.0011 & $-$0.0198 & 39 \\
660 & 56066.4265 & 0.0034 & $-$0.0157 & 38 \\
662 & 56066.4951 & 0.0035 & $-$0.0252 & 12 \\
736 & 56069.3768 & 0.0026 & $-$0.0329 & 9 \\
738 & 56069.4499 & 0.0015 & $-$0.0379 & 6 \\
763 & 56070.4258 & 0.0012 & $-$0.0382 & 11 \\
770 & 56070.6959 & 0.0017 & $-$0.0413 & 9 \\
773 & 56070.8146 & 0.0013 & $-$0.0398 & 8 \\
872 & 56074.6617 & 0.0049 & $-$0.0583 & 7 \\
873 & 56074.7169 & 0.0034 & $-$0.0421 & 7 \\
875 & 56074.7828 & 0.0042 & $-$0.0542 & 9 \\
876 & 56074.8171 & 0.0018 & $-$0.0591 & 9 \\
\hline
  \multicolumn{5}{l}{\commenta BJD$-$2400000.} \\
  \multicolumn{5}{l}{\commentb Against max $= 2456040.6718 + 0.039046 E$.} \\
  \multicolumn{5}{l}{\commentc Number of points used to determine the maximum.} \\
\end{tabular}
\end{center}
\end{table}

\subsection{SDSS J073208.11$+$413008.7}\label{obj:j0732}

   We observed the 2012 superoutburst of this object (hereafter 
SDSS J073208; the object was selected by \cite{wil10newCVs}, 
see a comment in \cite{Pdot2}).  The times of superhump maxima
are listed in table \ref{tab:j0732oc2012}.
The period listed in table \ref{tab:perlist} was determined
by the PDM method.  This period appears to be a global average
of stage B and C superhumps.

\begin{table}
\caption{Superhump maxima of SDSS J073208 (2012).}\label{tab:j0732oc2012}
\begin{center}
\begin{tabular}{ccccc}
\hline
$E$ & max\commenta & error & $O-C$\commentb & $N$\commentc \\
\hline
0 & 55977.6385 & 0.0015 & $-$0.0009 & 41 \\
1 & 55977.7190 & 0.0008 & 0.0000 & 51 \\
2 & 55977.7994 & 0.0012 & 0.0009 & 53 \\
72 & 55983.3671 & 0.0025 & $-$0.0000 & 43 \\
\hline
  \multicolumn{5}{l}{\commenta BJD$-$2400000.} \\
  \multicolumn{5}{l}{\commentb Against max $= 2455977.6394 + 0.079552 E$.} \\
  \multicolumn{5}{l}{\commentc Number of points used to determine the maximum.} \\
\end{tabular}
\end{center}
\end{table}

\subsection{SDSS J080303.90$+$251627.0}\label{obj:j0803}

\begin{figure}
  \begin{center}
    \FigureFile(88mm,110mm){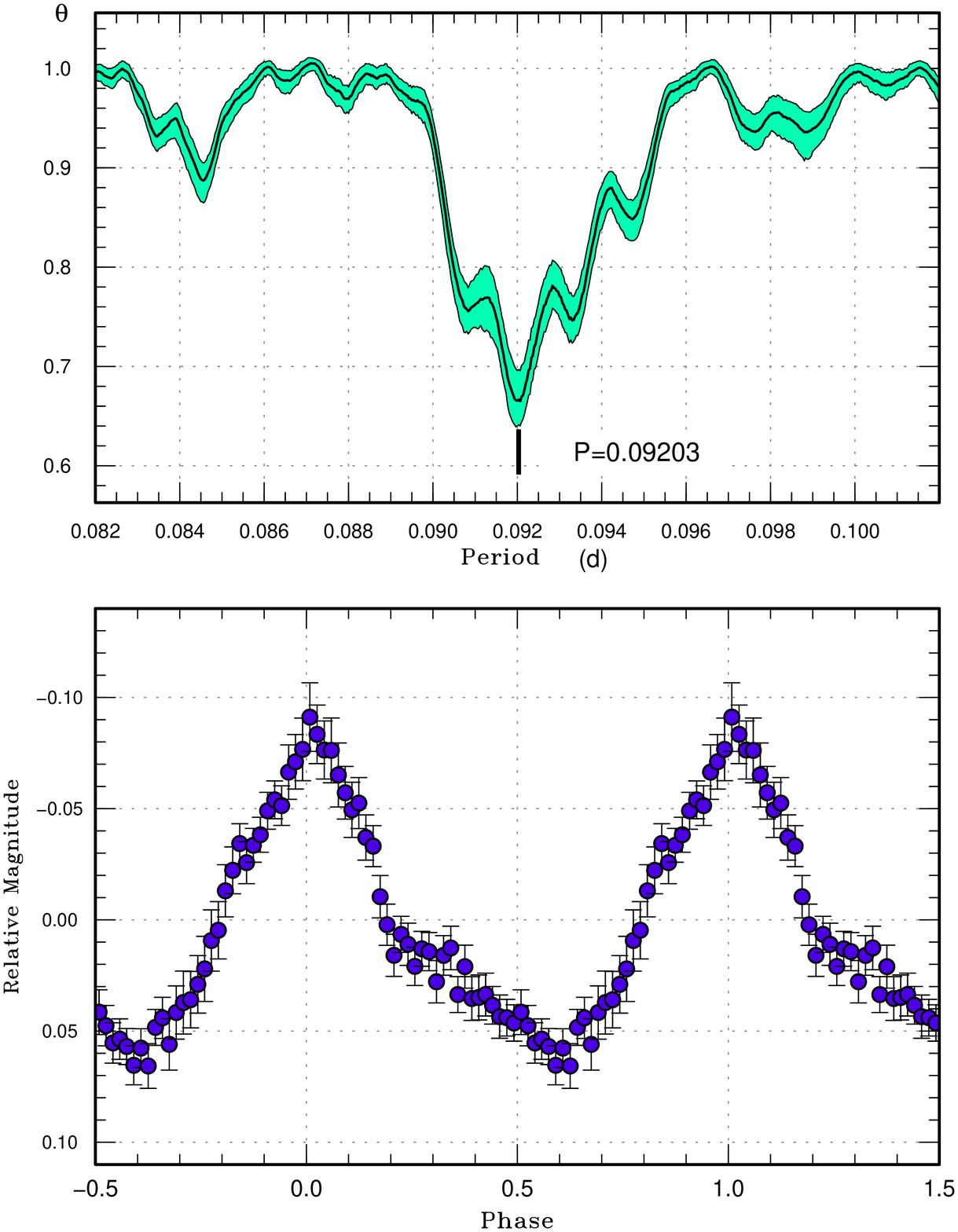}
  \end{center}
  \caption{Superhumps in SDSS J080303 (2011). (Upper): PDM analysis.
     (Lower): Phase-averaged profile.}
  \label{fig:j080303shpdm}
\end{figure}

   This object (hereafter SDSS J080303) was discovered as a CV
during the course of the SDSS \citep{szk05SDSSCV4}.
\citet{szk05SDSSCV4} identified a spectroscopic period of 0.071~d.
The object showed multiple outbursts in the CRTS data.
The 2011 outburst was detected by J. Shears (BAAVSS alert 2806).
The detection was sufficiently early to observe the early
evolution of superhumps (vsnet-alert 14006, 14015, 14033).
The object showed large variation of the superhump period
(vsnet-alert 14063).  The mean superhump profile is shown
in figure \ref{fig:j080303shpdm}.

   The times of superhump maxima are listed in table
\ref{tab:j0803oc2011}.  The initial stage A with growing superhumps
is immediately recognizable.  We considered $E=$27--31 to be
stage B--C transition and listed the period according to these
identifications in table \ref{tab:perlist}.

   The large negative $P_{\rm dot}$ [global value of
$-79(8) \times 10^{-5}$] and the long $P_{\rm SH}$ resemble
those of MN Dra \citep{pav10mndra}.

\begin{table}
\caption{Superhump maxima of SDSS J080303 (2011).}\label{tab:j0803oc2011}
\begin{center}
\begin{tabular}{ccccc}
\hline
$E$ & max\commenta & error & $O-C$\commentb & $N$\commentc \\
\hline
0 & 55921.6801 & 0.0015 & $-$0.0524 & 96 \\
1 & 55921.7738 & 0.0026 & $-$0.0499 & 56 \\
10 & 55922.6375 & 0.0004 & $-$0.0071 & 187 \\
11 & 55922.7343 & 0.0005 & $-$0.0016 & 137 \\
17 & 55923.2981 & 0.0008 & 0.0150 & 104 \\
21 & 55923.6649 & 0.0003 & 0.0169 & 99 \\
22 & 55923.7573 & 0.0005 & 0.0181 & 75 \\
27 & 55924.2195 & 0.0005 & 0.0243 & 101 \\
28 & 55924.3112 & 0.0009 & 0.0247 & 100 \\
30 & 55924.4919 & 0.0006 & 0.0230 & 59 \\
31 & 55924.5842 & 0.0006 & 0.0240 & 59 \\
43 & 55925.6680 & 0.0006 & 0.0133 & 99 \\
87 & 55929.6462 & 0.0020 & $-$0.0220 & 97 \\
88 & 55929.7332 & 0.0022 & $-$0.0262 & 98 \\
\hline
  \multicolumn{5}{l}{\commenta BJD$-$2400000.} \\
  \multicolumn{5}{l}{\commentb Against max $= 2455921.7325 + 0.091215 E$.} \\
  \multicolumn{5}{l}{\commentc Number of points used to determine the maximum.} \\
\end{tabular}
\end{center}
\end{table}

\subsection{SDSS J165359.06$+$201010.4}\label{obj:j1653}

\begin{figure}
  \begin{center}
    \FigureFile(88mm,70mm){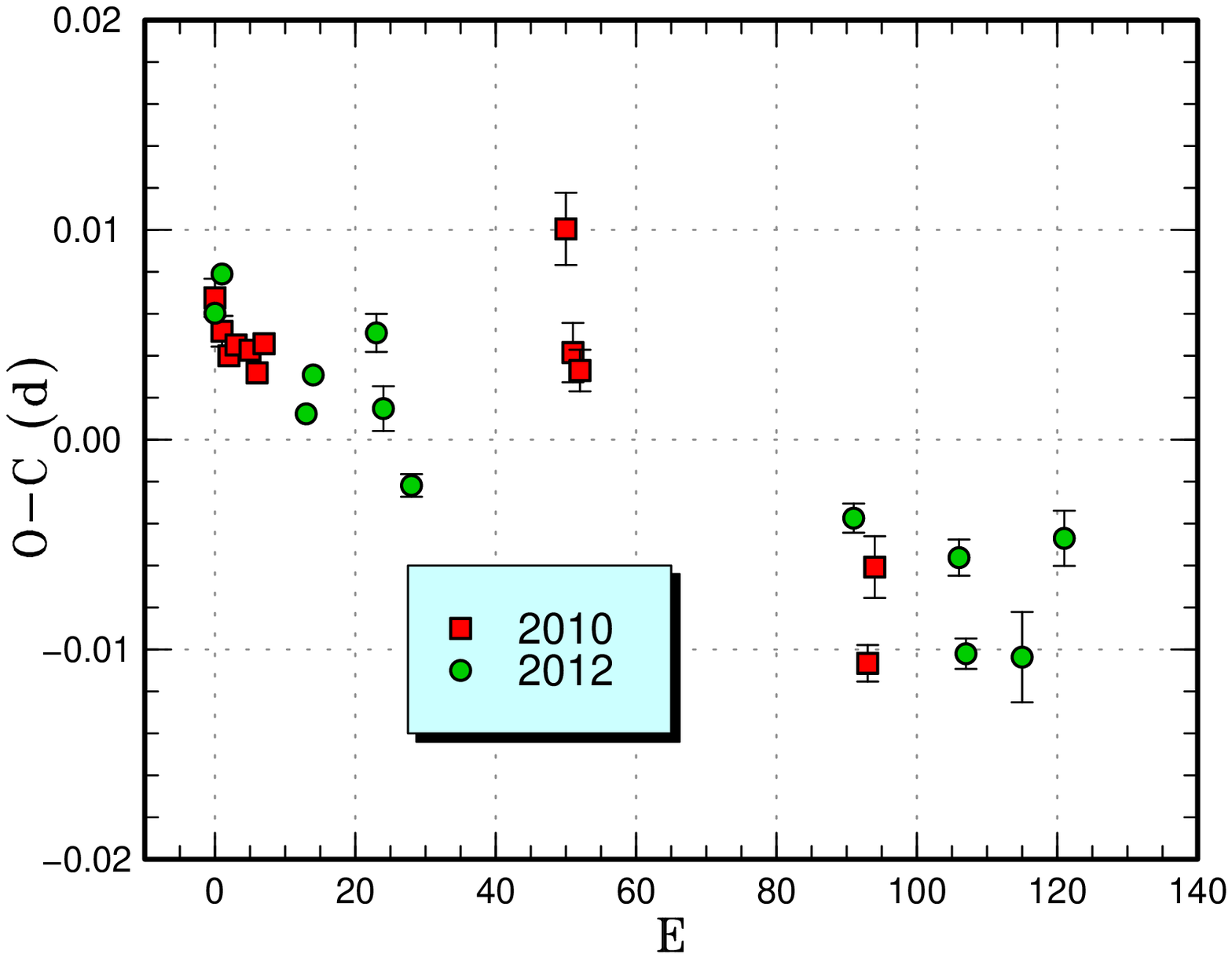}
  \end{center}
  \caption{Comparison of $O-C$ diagrams of SDSS J165359 between different
  superoutbursts.  A period of 0.06520~d was used to draw this figure.
  Approximate cycle counts ($E$) after the start of the observation
  were used.  We could not obtain a better match if we shifted the
  cycle number between these superoutbursts.
  }
  \label{fig:j1653comp}
\end{figure}

   We observed a superoutburst in 2012 May of this SU UMa-type
dwarf nova (hereafter SDSS J165359).
The times of superhumps are listed in table \ref{tab:j1653oc2012}.
Since the object faded rapidly after our final observation,
the superhumps recorded on the last three nights were most likely
stage C superhumps.  Although the identification of the stage of earlier
observations was unclear due to the long gap in the observation,
a comparison of the $O-C$ diagram with the 2010 superoutburst
suggests that we observed the earlier stage of stage B (figure
\ref{fig:j1653comp}).

\begin{table}
\caption{Superhump maxima of SDSS J165359 (2012).}\label{tab:j1653oc2012}
\begin{center}
\begin{tabular}{ccccc}
\hline
$E$ & max\commenta & error & $O-C$\commentb & $N$\commentc \\
\hline
0 & 56062.5903 & 0.0004 & 0.0011 & 67 \\
1 & 56062.6574 & 0.0004 & 0.0031 & 52 \\
13 & 56063.4328 & 0.0004 & $-$0.0025 & 105 \\
14 & 56063.4999 & 0.0004 & $-$0.0005 & 116 \\
23 & 56064.0889 & 0.0009 & 0.0027 & 70 \\
24 & 56064.1505 & 0.0011 & $-$0.0008 & 52 \\
28 & 56064.4077 & 0.0008 & $-$0.0040 & 47 \\
91 & 56068.5137 & 0.0007 & 0.0015 & 64 \\
106 & 56069.4898 & 0.0009 & 0.0013 & 57 \\
107 & 56069.5505 & 0.0007 & $-$0.0032 & 55 \\
115 & 56070.0719 & 0.0022 & $-$0.0025 & 132 \\
121 & 56070.4688 & 0.0013 & 0.0039 & 48 \\
\hline
  \multicolumn{5}{l}{\commenta BJD$-$2400000.} \\
  \multicolumn{5}{l}{\commentb Against max $= 2456062.5892 + 0.065089 E$.} \\
  \multicolumn{5}{l}{\commentc Number of points used to determine the maximum.} \\
\end{tabular}
\end{center}
\end{table}

\subsection{SDSS J170213.26$+$322954.1}\label{obj:j1702}

   This object (hereafter SDSS J170213) is an eclipsing SU UMa-type
dwarf nova in the period gap (\cite{boy06j1702}; \cite{lit06j1702}).
\citet{boy06j1702} reported an analysis of the 2005 superoutburst.
\citet{Pdot} analyzed their data and concluded that this object
showed increase in the $P_{\rm SH}$ during the middle-to-late stage of
a superoutburst, contrary to most SU UMa-type dwarf novae
with similar $P_{\rm SH}$.

   The 2011 superoutburst by detected by G. Poyner with an unfiltered
CCD magnitude of 13.93 (vsnet-outburst 13058).  Subsequent observations
confirmed the presence of superhumps and eclipses (vsnet-alert
13524, 13526, 13528, 13532).
The times of recorded eclipses, determined with the KW method,
after removing linearly approximated trends around eclipses in order to
minimize the effect of superhumps, are summarized in table \ref{tab:j1702ecl}.
We obtained an updated ephemeris of
\begin{equation}
{\rm Min(BJD)} = 2453648.23651(31) + 0.100082207(15) E
\label{equ:j1702ecl}.
\end{equation}

\begin{figure}
  \begin{center}
    \FigureFile(88mm,70mm){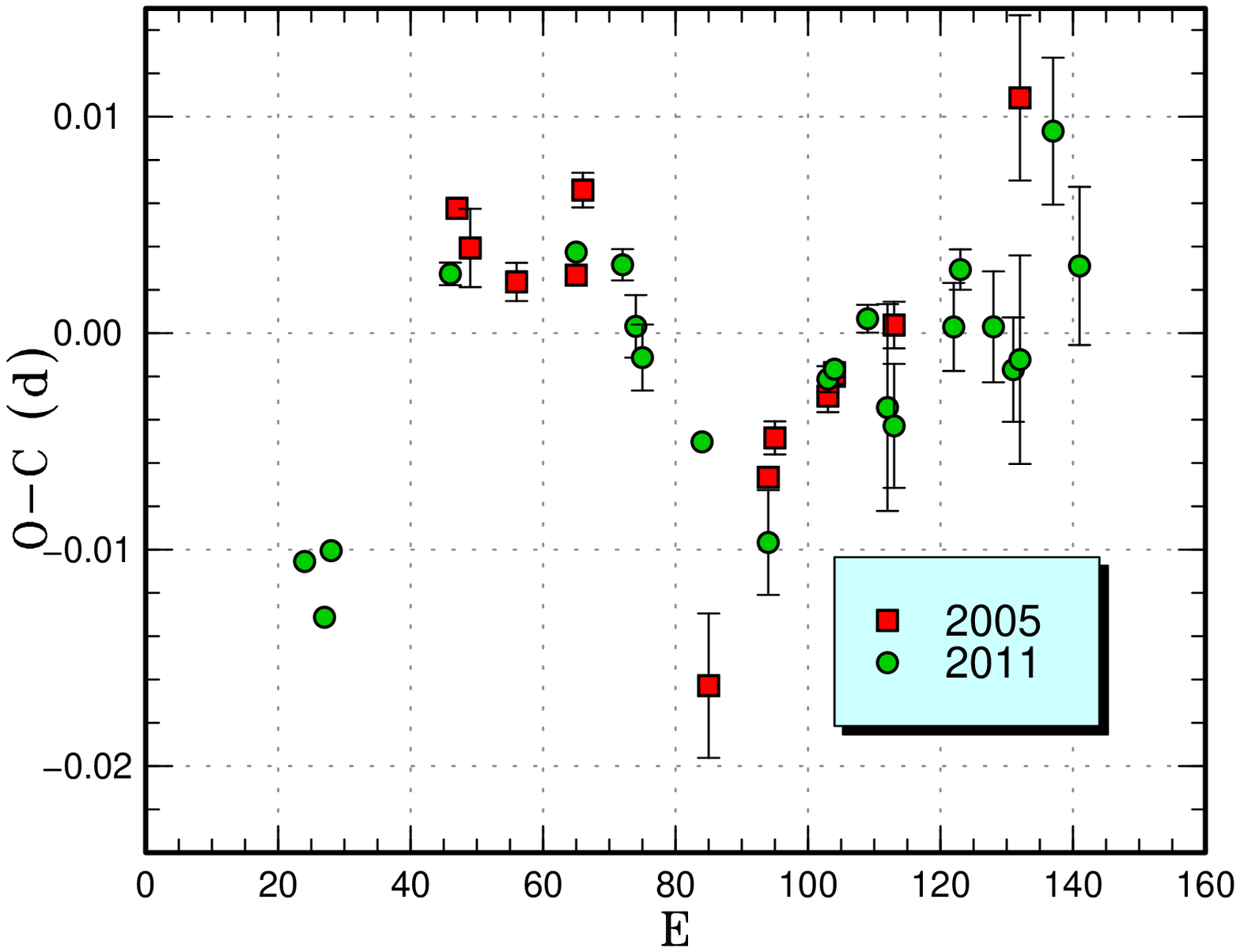}
  \end{center}
  \caption{Comparison of $O-C$ diagrams of SDSS J170213 between different
  superoutbursts.  A period of 0.10510~d was used to draw this figure.
  Approximate cycle counts ($E$) after the start of the superoutburst
  were used.
  }
  \label{fig:j1702comp}
\end{figure}

   The times of superhump maxima determined outside the eclipses
are listed in table \ref{tab:j1702oc2011}.  There were clear
stage A ($E \le 32$) and stage B with a positive $P_{\rm dot}$.
There was no indication of a transition to stage C despite that
the observation covered the early stage of the rapid decline.
The large positive $P_{\rm dot}$ confirmed the 2005 results,
and as suggested in \citet{Pdot}, this object mimics
a short-$P_{\rm orb}$ system both in $O-C$ variation and stage transitions.
The $\epsilon =$ 6.0\% is, however, much larger than those
in systems with short-$P_{\rm orb}$.  This object appears to
have relatively infrequent outbursts [the only known outbursts
have been in 2005 September--October (superoutburst), 2006 July
(normal outburst), 2007 September (superoutburst), 2009 February
(normal outburst), 2009 October (superoutburst), and
2011 July (superoutburst)] and probably indeed resembles
EF Peg (cf. \cite{how93efpeg}; \cite{kat02efpeg})
as proposed in \citet{Pdot}, rather than an unusual
system with a large $P_{\rm dot}$, GX Cas \citep{Pdot3}.
The behavior of $O-C$ variation was similar between 2005 and 2011
outbursts (figure \ref{fig:j1702comp}).  It is noteworthy that
stage A lasted much longer than in other systems.

\begin{table}
\caption{Eclipse minima of SDSS J170213 (2011).}\label{tab:j1702ecl}
\begin{center}
\begin{tabular}{cccc}
\hline
$E$ & Minimum\commenta & error & $O-C$\commentb \\
\hline
21165 & 55766.47660 & 0.00003 & 0.00018 \\
21195 & 55769.47798 & 0.00004 & -0.00091 \\
21196 & 55769.57867 & 0.00004 & -0.00029 \\
21205 & 55770.48023 & 0.00004 & 0.00052 \\
21214 & 55771.38036 & 0.00012 & -0.00009 \\
21215 & 55771.48107 & 0.00009 & 0.00054 \\
21224 & 55772.38128 & 0.00003 & 0.00001 \\
21225 & 55772.48149 & 0.00005 & 0.00013 \\
21234 & 55773.38194 & 0.00003 & -0.00015 \\
21235 & 55773.48223 & 0.00003 & 0.00005 \\
21245 & 55774.48296 & 0.00004 & -0.00004 \\
21248 & 55774.78356 & 0.00013 & 0.00032 \\
21254 & 55775.38378 & 0.00003 & 0.00004 \\
21255 & 55775.48378 & 0.00003 & -0.00004 \\
21264 & 55776.38464 & 0.00003 & 0.00008 \\
21265 & 55776.48473 & 0.00005 & 0.00009 \\
21270 & 55776.98509 & 0.00005 & 0.00004 \\
21271 & 55777.08517 & 0.00006 & 0.00003 \\
21274 & 55777.38503 & 0.00003 & -0.00035 \\
21280 & 55777.98583 & 0.00011 & -0.00005 \\
21281 & 55778.08560 & 0.00005 & -0.00036 \\
21284 & 55778.38640 & 0.00014 & 0.00020 \\
\hline
  \multicolumn{4}{l}{\commenta BJD$-$2400000.} \\
  \multicolumn{4}{l}{\commentb Against equation \ref{equ:j1702ecl}.} \\
\end{tabular}
\end{center}
\end{table}

\begin{table}
\caption{Superhump maxima of SDSS J170213 (2011).}\label{tab:j1702oc2011}
\begin{center}
\begin{tabular}{cccccc}
\hline
$E$ & max\commenta & error & $O-C$\commentb & phase\commentc & $N$\commentd \\
\hline
0 & 55766.1025 & 0.0004 & $-$0.0084 & 0.27 & 125 \\
3 & 55766.4152 & 0.0003 & $-$0.0111 & 0.39 & 187 \\
4 & 55766.5234 & 0.0004 & $-$0.0081 & 0.47 & 190 \\
22 & 55768.4280 & 0.0005 & 0.0041 & 0.50 & 89 \\
32 & 55769.4973 & 0.0010 & 0.0221 & 0.19 & 181 \\
33 & 55769.5977 & 0.0011 & 0.0174 & 0.19 & 122 \\
41 & 55770.4259 & 0.0004 & 0.0045 & 0.46 & 82 \\
48 & 55771.1610 & 0.0007 & 0.0037 & 0.81 & 141 \\
50 & 55771.3684 & 0.0014 & 0.0008 & 0.88 & 35 \\
51 & 55771.4720 & 0.0015 & $-$0.0007 & 0.92 & 77 \\
60 & 55772.4140 & 0.0005 & $-$0.0048 & 0.33 & 87 \\
70 & 55773.4604 & 0.0024 & $-$0.0098 & 0.78 & 96 \\
79 & 55774.4138 & 0.0006 & $-$0.0025 & 0.31 & 120 \\
80 & 55774.5194 & 0.0005 & $-$0.0021 & 0.37 & 110 \\
85 & 55775.0472 & 0.0006 & 0.0001 & 0.64 & 169 \\
88 & 55775.3584 & 0.0048 & $-$0.0041 & 0.75 & 53 \\
89 & 55775.4627 & 0.0029 & $-$0.0050 & 0.79 & 100 \\
98 & 55776.4132 & 0.0020 & $-$0.0007 & 0.29 & 157 \\
99 & 55776.5209 & 0.0009 & 0.0019 & 0.36 & 106 \\
104 & 55777.0438 & 0.0026 & $-$0.0009 & 0.59 & 231 \\
107 & 55777.3571 & 0.0024 & $-$0.0030 & 0.72 & 82 \\
108 & 55777.4626 & 0.0048 & $-$0.0026 & 0.77 & 79 \\
113 & 55777.9987 & 0.0034 & 0.0078 & 0.13 & 148 \\
117 & 55778.4129 & 0.0037 & 0.0015 & 0.27 & 138 \\
\hline
  \multicolumn{5}{l}{\commenta BJD$-$2400000.} \\
  \multicolumn{5}{l}{\commentb Against max $= 2455766.1110 + 0.105132 E$.} \\
  \multicolumn{6}{l}{\commentc Orbital phase.} \\
  \multicolumn{6}{l}{\commentd Number of points used to determine the maximum.} \\
\end{tabular}
\end{center}
\end{table}

\subsection{SDSS J172102.48$+$273301.2}\label{obj:j1721}

\begin{figure}
  \begin{center}
    \FigureFile(88mm,110mm){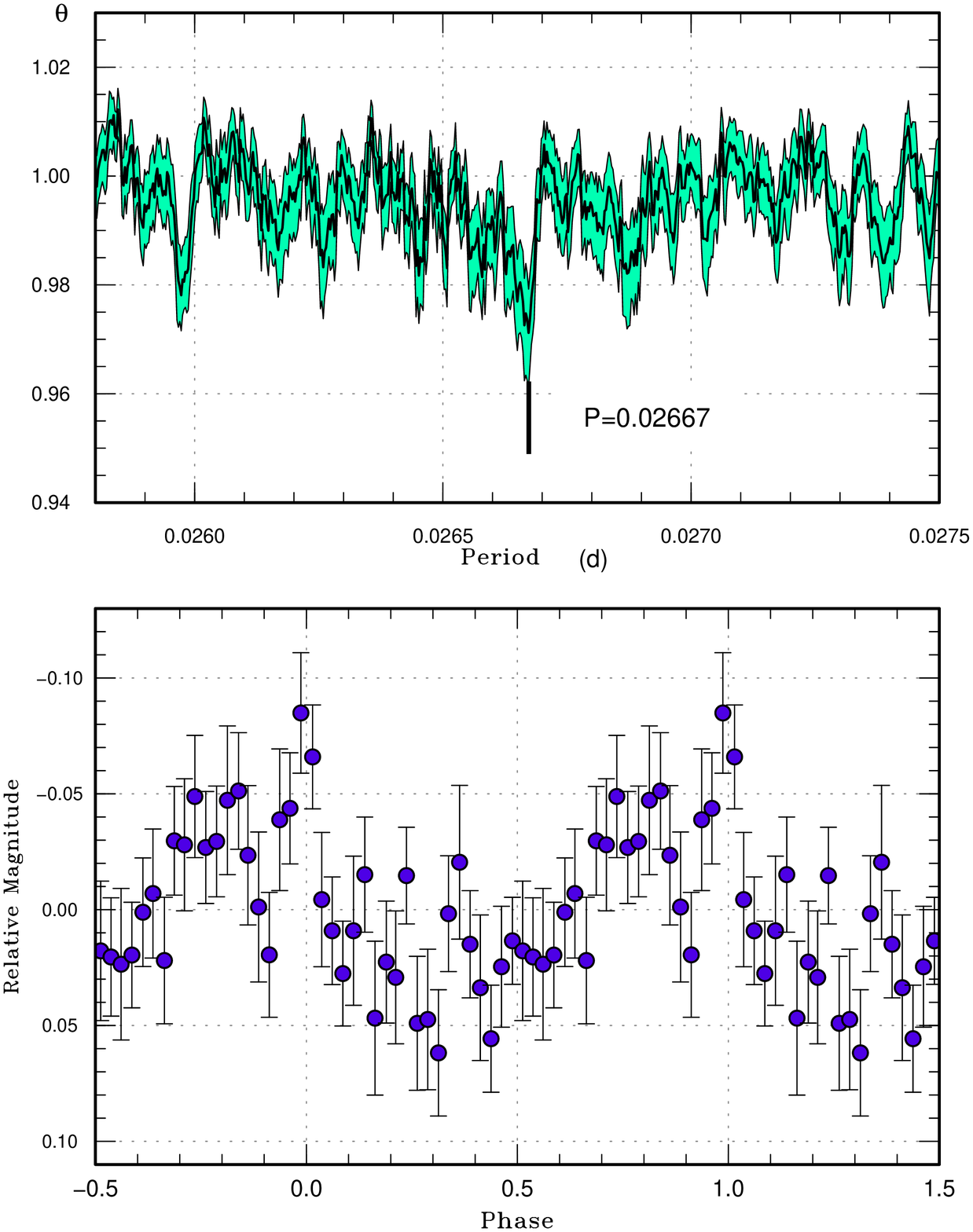}
  \end{center}
  \caption{Superhumps in SDSS J172102 (2012). (Upper): PDM analysis.
     The rejection rate for bootstrapping was reduced to 0.2 for
     better visualization.
     (Lower): Phase-averaged profile.}
  \label{fig:j1721shpdm}
\end{figure}

   This object (hereafter SDSS J172102) was originally selected
as a helium CV using the SDSS colors and confirmed by spectroscopy
\citep{rau10HeDN}.  Although there had been no record of outbursts,
CRTS detected this object in outburst on 2012 June 8 at an unfiltered
CCD magnitude of 16.4 (CSS120608:172102$+$273301).  A quick follow-up
observation confirmed the presence of superhumps on June 9
(vsnet-alert 14653).  A retrospective study indicated that the object
was recorded at unfiltered CCD magnitudes of 16.0--16.2 on June 5
at MASTER-Kislovodsk (vsnet-alert 14657).  The object rapidly faded to 
18.9 on June 11 (Goff) and 19.1 on June 15 (CRTS).
The object was thus likely a superoutburst of an AM CVn-type object
detected during its final stage.  The object underwent a short
post-superoutburst rebrightening on June 20 whose peak brightness
must have been brighter than 17.5 (Goff).
The times of superhump maxima are listed in table \ref{tab:j1721oc2012}.
Note that these observations were mostly made during the final
stage of the superoutburst and subsequent post-superoutburst phase,
and the maxima were rather difficult to identify due to the faintness.
Our best estimate of the superhump period is 0.026673(8) d
(figure \ref{fig:j1721shpdm}).  It is remarkable that both spectra
in \citet{rau10HeDN} and the SDSS public archive were continuum-dominated
(vsnet-alert 14650) in contrast to the quiescent state of
many AM CVn-type objects showing dwarf-nova type outbursts.

\begin{table}
\caption{Superhump maxima of SDSS J172102 (2012).}\label{tab:j1721oc2012}
\begin{center}
\begin{tabular}{ccccc}
\hline
$E$ & max\commenta & error & $O-C$\commentb & $N$\commentc \\
\hline
0 & 56087.5334 & 0.0007 & 0.0035 & 18 \\
1 & 56087.5601 & 0.0008 & 0.0036 & 19 \\
82 & 56089.7123 & 0.0007 & $-$0.0043 & 13 \\
83 & 56089.7401 & 0.0007 & $-$0.0032 & 14 \\
84 & 56089.7687 & 0.0010 & $-$0.0013 & 12 \\
85 & 56089.7919 & 0.0017 & $-$0.0047 & 14 \\
86 & 56089.8217 & 0.0014 & $-$0.0016 & 12 \\
87 & 56089.8449 & 0.0008 & $-$0.0050 & 10 \\
88 & 56089.8753 & 0.0011 & $-$0.0013 & 14 \\
89 & 56089.9039 & 0.0016 & 0.0006 & 13 \\
90 & 56089.9306 & 0.0015 & 0.0007 & 9 \\
120 & 56090.7290 & 0.0011 & $-$0.0010 & 13 \\
122 & 56090.7869 & 0.0019 & 0.0036 & 14 \\
123 & 56090.8112 & 0.0032 & 0.0012 & 14 \\
124 & 56090.8354 & 0.0056 & $-$0.0013 & 13 \\
127 & 56090.9128 & 0.0041 & $-$0.0039 & 13 \\
233 & 56093.7449 & 0.0021 & 0.0015 & 13 \\
234 & 56093.7751 & 0.0037 & 0.0050 & 14 \\
236 & 56093.8200 & 0.0012 & $-$0.0034 & 13 \\
237 & 56093.8512 & 0.0013 & 0.0011 & 8 \\
238 & 56093.8733 & 0.0019 & $-$0.0035 & 14 \\
239 & 56093.9027 & 0.0032 & $-$0.0007 & 13 \\
271 & 56094.7616 & 0.0007 & 0.0048 & 11 \\
272 & 56094.7900 & 0.0030 & 0.0065 & 14 \\
273 & 56094.8203 & 0.0050 & 0.0101 & 14 \\
274 & 56094.8427 & 0.0012 & 0.0059 & 8 \\
275 & 56094.8690 & 0.0033 & 0.0055 & 13 \\
345 & 56096.7276 & 0.0019 & $-$0.0026 & 11 \\
346 & 56096.7560 & 0.0020 & $-$0.0009 & 10 \\
347 & 56096.7784 & 0.0019 & $-$0.0051 & 14 \\
350 & 56096.8612 & 0.0021 & $-$0.0023 & 13 \\
384 & 56097.7699 & 0.0025 & $-$0.0004 & 11 \\
387 & 56097.8535 & 0.0022 & 0.0033 & 11 \\
388 & 56097.8736 & 0.0057 & $-$0.0033 & 13 \\
389 & 56097.9032 & 0.0033 & $-$0.0004 & 14 \\
390 & 56097.9289 & 0.0016 & $-$0.0013 & 12 \\
457 & 56099.7196 & 0.0050 & 0.0026 & 13 \\
460 & 56099.8005 & 0.0044 & 0.0035 & 13 \\
461 & 56099.8212 & 0.0034 & $-$0.0025 & 9 \\
462 & 56099.8486 & 0.0024 & $-$0.0017 & 12 \\
463 & 56099.8699 & 0.0050 & $-$0.0071 & 14 \\
\hline
  \multicolumn{5}{l}{\commenta BJD$-$2400000.} \\
  \multicolumn{5}{l}{\commentb Against max $= 2456087.5299 + 0.026668 E$.} \\
  \multicolumn{5}{l}{\commentc Number of points used to determine the maximum.} \\
\end{tabular}
\end{center}
\end{table}

\subsection{SDSS J210449.94$+$010545.8}\label{obj:j2104}

\begin{figure}
  \begin{center}
    \FigureFile(88mm,110mm){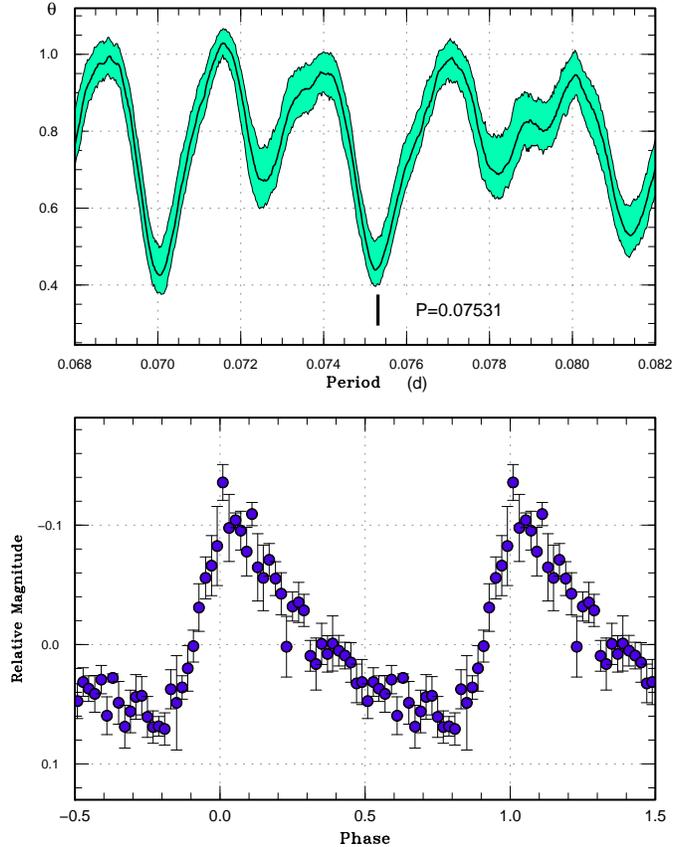}
  \end{center}
  \caption{Superhumps in SDSS J210449 (2011). (Upper): PDM analysis.
     The alias selection was based on superhump timing analysis.
     (Lower): Phase-averaged profile.}
  \label{fig:j2104shpdm}
\end{figure}

   This object (hereafter SDSS J210449) was discovered as a CV
during the course of the SDSS \citep{szk06SDSSCV5}.
\citet{szk06SDSSCV5} recorded high and low states ranging
17.1--20.6 mag.  \citet{sou07SDSSCV2} detected a photometric
period of 0.07196(8)~d.  CRTS recorded multiple outbursts, and
at least one of them (2006 November) lasted more than 18 d
and was likely a superoutburst.  During the superoutburst in
2011 September, I. Miller detected superhumps (vsnet-alert 13704).
A PDM analysis yielded two equally acceptable one-day aliases
(figure \ref{fig:j2104shpdm}).

   The times of superhump maxima are listed in table \ref{tab:j2104oc2011}
The timing analysis prefers an alias of 0.0753~d, and we obtained
a period of 0.07531(4)~d with the PDM method, which is used in
table \ref{tab:perlist}.  The $\epsilon$ of 4.7\% inferred from
this period is likely to be too large for this $P_{\rm orb}$,
and there may have been negative superhumps at the time of
observations by \citet{sou07SDSSCV2}.  The exact orbital period
needs to be determined by radial-velocity studies.

\begin{table}
\caption{Superhump maxima of SDSS J210449 (2011).}\label{tab:j2104oc2011}
\begin{center}
\begin{tabular}{ccccc}
\hline
$E$ & max\commenta & error & $O-C$\commentb & $N$\commentc \\
\hline
0 & 55834.4629 & 0.0008 & $-$0.0000 & 79 \\
26 & 55836.4217 & 0.0015 & 0.0000 & 69 \\
27 & 55836.4970 & 0.0011 & $-$0.0000 & 55 \\
\hline
  \multicolumn{5}{l}{\commenta BJD$-$2400000.} \\
  \multicolumn{5}{l}{\commentb Against max $= 2455834.4629 + 0.075338 E$.} \\
  \multicolumn{5}{l}{\commentc Number of points used to determine the maximum.} \\
\end{tabular}
\end{center}
\end{table}

\subsection{SDSS J220553.98$+$115553.7}\label{obj:j2205}

   This object (hereafter SDSS J220553) was detected as
a CV during the course of SDSS \citep{szk03SDSSCV2}.
\citet{szk03SDSSCV2} showed the presence
of the underlying white dwarf in the spectrum, suggesting
the low mass-transfer rate.  \citet{war04CVnewZZproc} indicated
that this system contains a ZZ Cet-type pulsating white dwarf
(see also \cite{szk07CVWDpuls}),
also suggesting the low surface-temperature of the white dwarf
(consistent with the low mass-transfer rate).
\citet{sou08CVperiod} obtained the spectroscopic orbital period
of 0.0575175(62)~d, and they found that the pulsation of white dwarf
ceased in 2007.  Although the spectrum and the orbital period
suggested an SU UMa-type or an even WZ Sge-type dwarf nova, 
no outburst had been recorded until 2011.

\begin{figure}
  \begin{center}
    \FigureFile(88mm,110mm){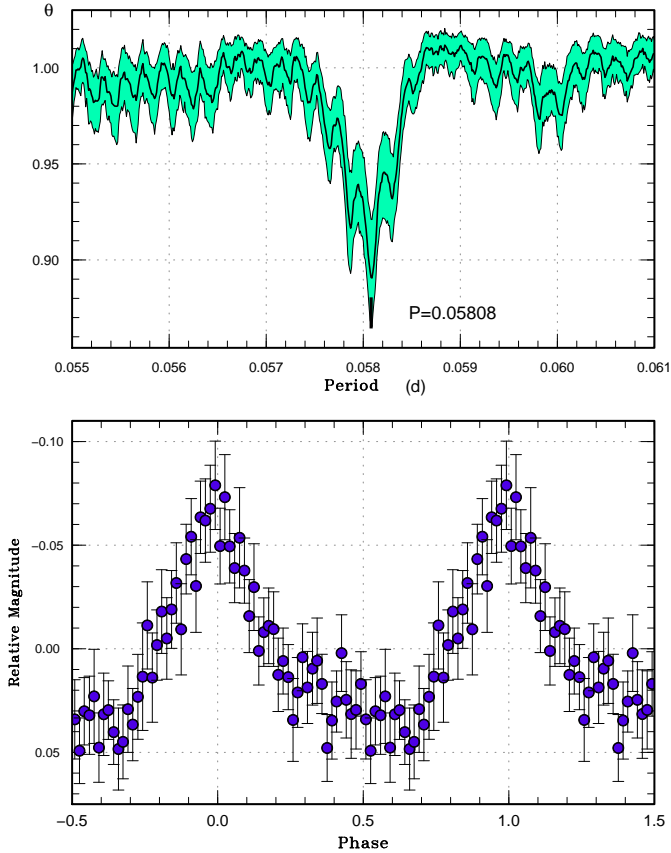}
  \end{center}
  \caption{Superhumps in SDSS J220553 (2011). (Upper): PDM analysis.
     (Lower): Phase-averaged profile.}
  \label{fig:j2205shpdm}
\end{figure}

   CRTS detected an outburst on 2011 May 20
(cf. vsnet-alert 13325) and the announcement of this detection
was immediately followed by observations.
Fully grown superhumps were soon detected (vsnet-alert 13329,
figure \ref{fig:j2205shpdm}),
and the later development suggested a low $\epsilon$,
characteristic to a WZ Sge-type dwarf nova (vsnet-alert 13332, 13336).
The $P_{\rm dot}$ was, however, unexpectedly large (vsnet-alert 13348).

   The times of superhump maxima are listed in table \ref{tab:j2205oc2011}.
The resultant $P_{\rm dot}$ was $+7.7(0.9) \times 10^{-5}$.
The $\epsilon$ for the mean period of stage B superhumps was
1.1\%, which is much smaller than what would be expected for
this large $P_{\rm dot}$.

   According to the CRTS data, the object brighter than in usual
quiescence four months after the outburst.  This, combined with
the low $\epsilon$ and the lack of previous outbursts in the CRTS
data, suggest that the object is a WZ Sge-type dwarf nova.
It may have been that the period of early superhumps was missed,
and that the true maximum was much brighter.

\begin{table}
\caption{Superhump maxima of SDSS J220553 (2011).}\label{tab:j2205oc2011}
\begin{center}
\begin{tabular}{ccccc}
\hline
$E$ & max\commenta & error & $O-C$\commentb & $N$\commentc \\
\hline
0 & 55702.2039 & 0.0015 & 0.0042 & 109 \\
1 & 55702.2618 & 0.0008 & 0.0039 & 119 \\
12 & 55702.8979 & 0.0004 & 0.0003 & 60 \\
13 & 55702.9551 & 0.0004 & $-$0.0006 & 53 \\
29 & 55703.8834 & 0.0004 & $-$0.0028 & 87 \\
30 & 55703.9436 & 0.0003 & $-$0.0007 & 86 \\
40 & 55704.5235 & 0.0010 & $-$0.0024 & 31 \\
46 & 55704.8715 & 0.0009 & $-$0.0032 & 87 \\
47 & 55704.9297 & 0.0005 & $-$0.0032 & 86 \\
63 & 55705.8626 & 0.0008 & $-$0.0007 & 60 \\
64 & 55705.9204 & 0.0005 & $-$0.0010 & 61 \\
97 & 55707.8427 & 0.0028 & 0.0023 & 36 \\
98 & 55707.9006 & 0.0007 & 0.0021 & 60 \\
99 & 55707.9586 & 0.0007 & 0.0018 & 43 \\
\hline
  \multicolumn{5}{l}{\commenta BJD$-$2400000.} \\
  \multicolumn{5}{l}{\commentb Against max $= 2455702.1998 + 0.058151 E$.} \\
  \multicolumn{5}{l}{\commentc Number of points used to determine the maximum.} \\
\end{tabular}
\end{center}
\end{table}

\subsection{OT J001952.2$+$433901}\label{obj:j0019}

\begin{figure}
  \begin{center}
    \FigureFile(88mm,110mm){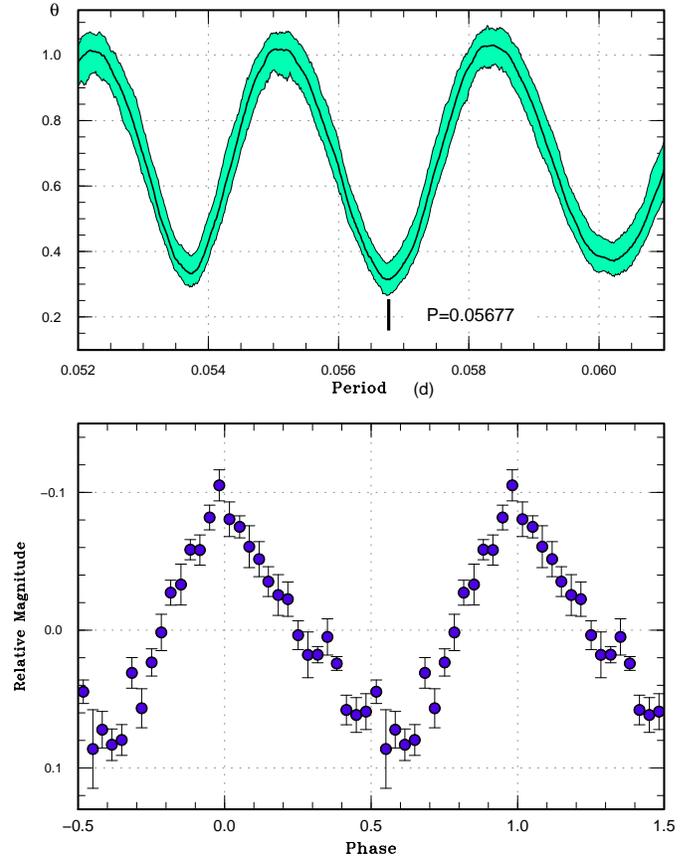}
  \end{center}
  \caption{Superhumps in OT J001952 (2011). (Upper): PDM analysis.
     The alias selection was based on continuous single-night
     observation.
     (Lower): Phase-averaged profile.}
  \label{fig:j0019shpdm}
\end{figure}

   This transient (=CSS120131:001952$+$433901; hereafter OT J001952)
was detected by CRTS on 2012 January 31.  The large outburst amplitude 
($\sim$ 6.5 mag) and the lack of previous outbursts attracted
observers' attention (vsnet-alert 14182).  Subsequent observations detected
short-period superhumps (vsnet-alert 14189, 14202; figure
\ref{fig:j0019shpdm}).  The times of superhump maxima are listed in table
\ref{tab:j0019oc2012}.  The large outburst amplitude and short
superhump period suggest a possibility of a WZ Sge-type dwarf nova.

\begin{table}
\caption{Superhump maxima of OT J001952.}\label{tab:j0019oc2012}
\begin{center}
\begin{tabular}{ccccc}
\hline
$E$ & max\commenta & error & $O-C$\commentb & $N$\commentc \\
\hline
0 & 55958.3297 & 0.0005 & 0.0005 & 61 \\
1 & 55958.3856 & 0.0007 & $-$0.0005 & 42 \\
17 & 55959.2959 & 0.0006 & 0.0005 & 50 \\
18 & 55959.3517 & 0.0008 & $-$0.0005 & 60 \\
\hline
  \multicolumn{5}{l}{\commenta BJD$-$2400000.} \\
  \multicolumn{5}{l}{\commentb Against max $= 2455958.3292 + 0.056827 E$.} \\
  \multicolumn{5}{l}{\commentc Number of points used to determine the maximum.} \\
\end{tabular}
\end{center}
\end{table}

\subsection{OT J011516.5$+$245530}\label{obj:j0115}

\begin{figure}
  \begin{center}
    \FigureFile(88mm,110mm){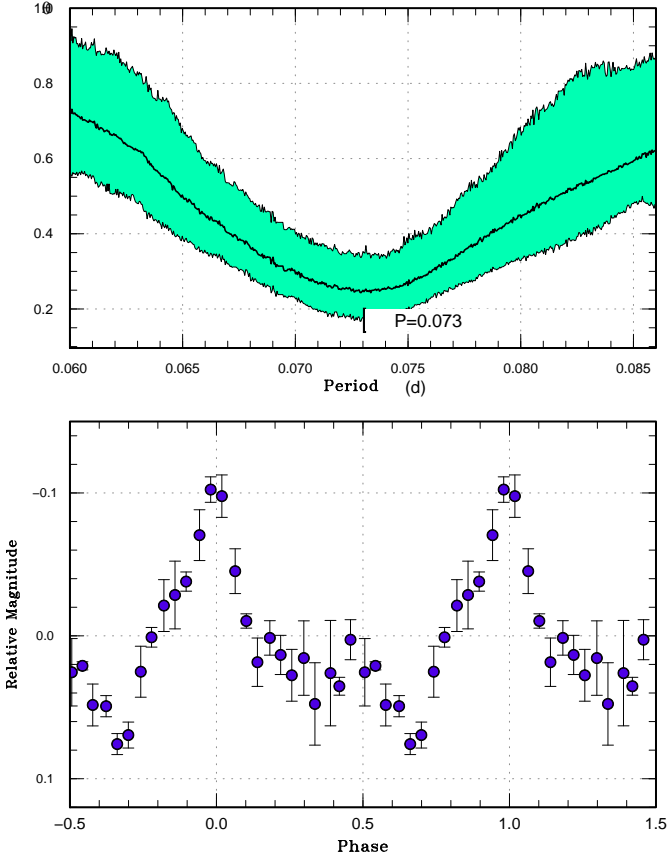}
  \end{center}
  \caption{Superhumps in OT J011516 (2012). (Upper): PDM analysis.
     (Lower): Phase-averaged profile.}
  \label{fig:j0115shpdm}
\end{figure}

   This transient (=CSS101008:011517$+$245530; hereafter OT J011516)
was detected by CRTS on 2010 October 8.  Although several outbursts
were known, the 2012 outburst was the brightest one (vsnet-alert 14142).
Subsequent observations recorded superhumps (vsnet-alert 14147, 14149).
Only single-night observation was available with superhump maxima
of BJD 2455952.2518(7) ($N$=51) and 2455952.3253(10) ($N$=44).
The best superhump period by the PDM was 0.0731(6)~d (figure
\ref{fig:j0115shpdm}).

\subsection{OT J050716.2$+$125314}\label{obj:j0507}

\begin{figure}
  \begin{center}
    \FigureFile(88mm,110mm){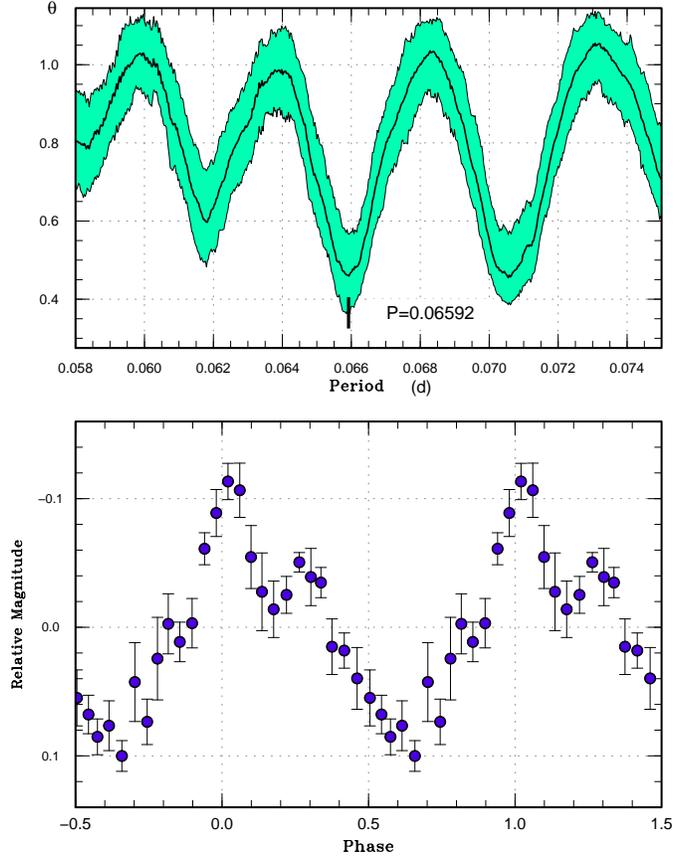}
  \end{center}
  \caption{Superhumps in OT OT J050716 (2012). (Upper): PDM analysis.
     (Lower): Phase-averaged profile.}
  \label{fig:j0507shpdm}
\end{figure}

   This transient (=CSS081221:050716$+$125314; hereafter OT J050716)
was detected by CRTS on 2008 December 21.  The 2012 January outburst
led to the detection of superhumps (vsnet-alert 14150, 14151).
The times of superhump maxima are listed in table
\ref{tab:j0507oc2012}.  Although the $O-C$ analysis favored
an alias of 0.06592(8)~d (adopted in table \ref{tab:perlist}),
the alias of 0.07055(9)~d is not excluded (figure \ref{fig:j0507shpdm}).

\begin{table}
\caption{Superhump maxima of OT J050716.}\label{tab:j0507oc2012}
\begin{center}
\begin{tabular}{ccccc}
\hline
$E$ & max\commenta & error & $O-C$\commentb & $N$\commentc \\
\hline
0 & 55952.4395 & 0.0009 & $-$0.0012 & 31 \\
1 & 55952.5075 & 0.0024 & 0.0015 & 19 \\
14 & 55953.3518 & 0.0055 & $-$0.0033 & 28 \\
15 & 55953.4234 & 0.0018 & 0.0030 & 36 \\
\hline
  \multicolumn{5}{l}{\commenta BJD$-$2400000.} \\
  \multicolumn{5}{l}{\commentb Against max $= 2455952.4407 + 0.065317 E$.} \\
  \multicolumn{5}{l}{\commentc Number of points used to determine the maximum.} \\
\end{tabular}
\end{center}
\end{table}

\subsection{OT J055721.8$-$363055}\label{obj:j0557}

\begin{figure}
  \begin{center}
    \FigureFile(88mm,110mm){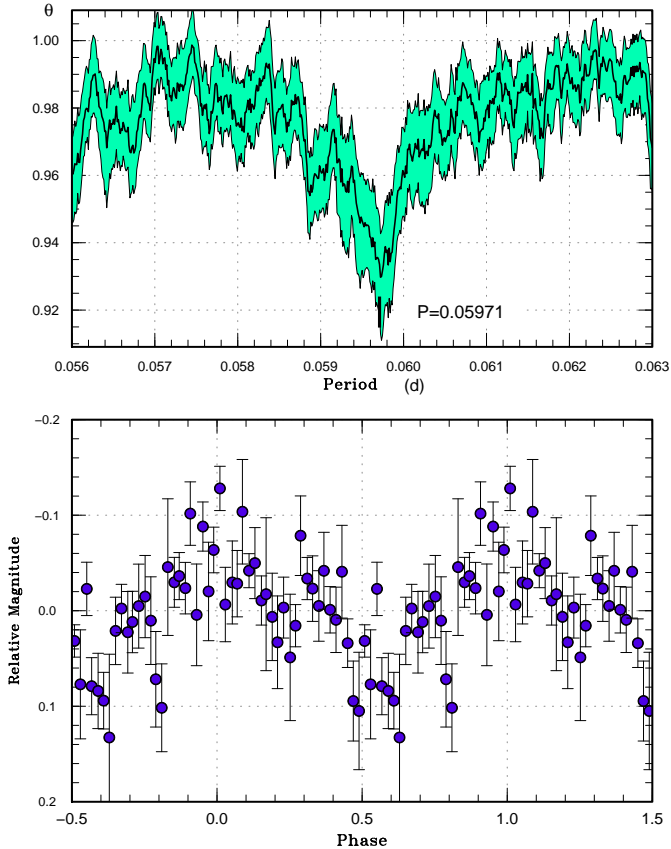}
  \end{center}
  \caption{Superhumps in OT J055721 (2011). (Upper): PDM analysis.
     The rejection rate for bootstrapping was reduced to 0.2 for
     better visualization.
     (Lower): Phase-averaged profile.}
  \label{fig:j055721shpdm}
\end{figure}

   This transient (=SSS111229:055722$-$363055; hereafter OT J055721)
was detected by CRTS SSS on 2011 December 29.  The large outburst
amplitude suggested an SU UMa-type, or even a WZ Sge-type object 
(vsnet-alert 14041).  Subsequent observations detected superhumps 
(vsnet-alert 14052).
The times of superhump maxima are listed in table
\ref{tab:j0557oc2011}.  Although the observations in the middle
of the outburst were rather sparse, we likely observed stage B
superhumps with a positive $P_{\rm dot}$.  The amplitude of
superhumps (cf. figure \ref{fig:j055721shpdm}) resembles those of 
ordinary SU UMa-type dwarf novae rather than those of extreme 
WZ Sge-type dwarf novae.  The object, however, underwent 
a post-outburst rebrightening similar to those of WZ Sge-type 
dwarf novae (rebrightening followed by a short ``dip'';
vsnet-alert 14097).

\begin{table}
\caption{Superhump maxima of OT J055721 (2011).}\label{tab:j0557oc2011}
\begin{center}
\begin{tabular}{ccccc}
\hline
$E$ & max\commenta & error & $O-C$\commentb & $N$\commentc \\
\hline
0 & 55926.5487 & 0.0010 & 0.0008 & 22 \\
1 & 55926.6101 & 0.0006 & 0.0024 & 14 \\
2 & 55926.6685 & 0.0005 & 0.0010 & 10 \\
3 & 55926.7283 & 0.0015 & 0.0010 & 15 \\
4 & 55926.7898 & 0.0011 & 0.0028 & 20 \\
17 & 55927.5657 & 0.0006 & 0.0019 & 17 \\
18 & 55927.6250 & 0.0012 & 0.0014 & 14 \\
19 & 55927.6836 & 0.0006 & 0.0002 & 13 \\
20 & 55927.7438 & 0.0005 & 0.0008 & 17 \\
21 & 55927.8028 & 0.0007 & 0.0000 & 20 \\
33 & 55928.5184 & 0.0024 & $-$0.0015 & 17 \\
34 & 55928.5806 & 0.0010 & 0.0010 & 15 \\
35 & 55928.6376 & 0.0009 & $-$0.0018 & 14 \\
36 & 55928.6967 & 0.0008 & $-$0.0024 & 16 \\
37 & 55928.7549 & 0.0005 & $-$0.0040 & 17 \\
38 & 55928.8170 & 0.0006 & $-$0.0016 & 19 \\
94 & 55932.1550 & 0.0068 & $-$0.0100 & 100 \\
152 & 55935.6360 & 0.0036 & 0.0051 & 14 \\
153 & 55935.6934 & 0.0029 & 0.0029 & 17 \\
\hline
  \multicolumn{5}{l}{\commenta BJD$-$2400000.} \\
  \multicolumn{5}{l}{\commentb Against max $= 2455926.5480 + 0.059756 E$.} \\
  \multicolumn{5}{l}{\commentc Number of points used to determine the maximum.} \\
\end{tabular}
\end{center}
\end{table}

\subsection{OT J064608.2$+$403305}\label{obj:j0646}

\begin{figure}
  \begin{center}
    \FigureFile(88mm,110mm){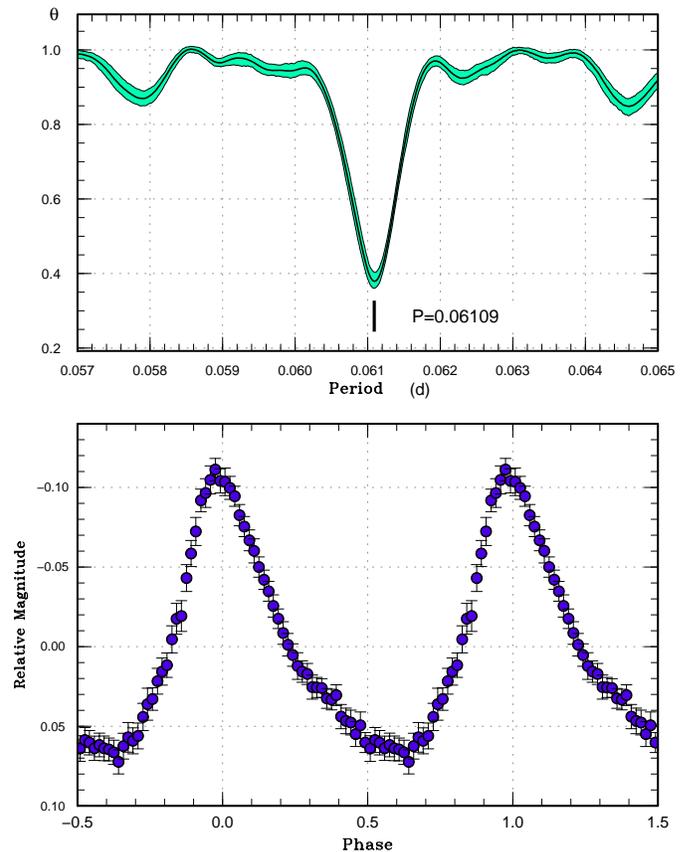}
  \end{center}
  \caption{Superhumps in OT J064608 (2011). (Upper): PDM analysis.
     (Lower): Phase-averaged profile.}
  \label{fig:j0646shpdm}
\end{figure}

   This transient (=CSS 080512:064608$+$403305; hereafter OT J064608)
was detected by CRTS on 2008 May 12.  A bright outburst in 2011
December was detected by E. Muyllaert (BAAVSS alert 2808).
Subsequent observations confirmed the presence of superhumps
(vsnet-alert 14023, 14036, 14039; figure \ref{fig:j0646shpdm}).
The time of superhump maxima are listed in table \ref{tab:j0646oc2011}.
The $P_{\rm dot}$ was $+11.1(2.6) \times 10^{-5}$, a typical value
for stage B superhumps with this $P_{\rm SH}$.

\begin{table}
\caption{Superhump maxima of OT J064608 (2011).}\label{tab:j0646oc2011}
\begin{center}
\begin{tabular}{ccccc}
\hline
$E$ & max\commenta & error & $O-C$\commentb & $N$\commentc \\
\hline
0 & 55923.3629 & 0.0006 & $-$0.0018 & 36 \\
1 & 55923.4269 & 0.0002 & 0.0010 & 64 \\
2 & 55923.4884 & 0.0002 & 0.0014 & 79 \\
3 & 55923.5492 & 0.0002 & 0.0011 & 54 \\
4 & 55923.6111 & 0.0002 & 0.0019 & 54 \\
7 & 55923.7948 & 0.0002 & 0.0023 & 63 \\
8 & 55923.8553 & 0.0002 & 0.0017 & 63 \\
9 & 55923.9152 & 0.0006 & 0.0005 & 63 \\
10 & 55923.9751 & 0.0003 & $-$0.0007 & 64 \\
17 & 55924.4049 & 0.0003 & 0.0013 & 64 \\
18 & 55924.4661 & 0.0003 & 0.0015 & 64 \\
23 & 55924.7705 & 0.0002 & 0.0004 & 64 \\
24 & 55924.8319 & 0.0003 & 0.0006 & 64 \\
25 & 55924.8923 & 0.0002 & $-$0.0001 & 65 \\
26 & 55924.9527 & 0.0003 & $-$0.0007 & 64 \\
27 & 55925.0124 & 0.0005 & $-$0.0022 & 55 \\
39 & 55925.7448 & 0.0008 & $-$0.0031 & 36 \\
40 & 55925.8078 & 0.0003 & $-$0.0012 & 64 \\
41 & 55925.8668 & 0.0003 & $-$0.0032 & 64 \\
42 & 55925.9280 & 0.0004 & $-$0.0031 & 64 \\
43 & 55925.9880 & 0.0005 & $-$0.0043 & 63 \\
56 & 55926.7855 & 0.0004 & $-$0.0011 & 64 \\
57 & 55926.8473 & 0.0003 & $-$0.0004 & 64 \\
58 & 55926.9074 & 0.0003 & $-$0.0014 & 64 \\
59 & 55926.9667 & 0.0008 & $-$0.0032 & 63 \\
60 & 55927.0366 & 0.0016 & 0.0055 & 34 \\
82 & 55928.3828 & 0.0008 & 0.0075 & 65 \\
\hline
  \multicolumn{5}{l}{\commenta BJD$-$2400000.} \\
  \multicolumn{5}{l}{\commentb Against max $= 2455923.3648 + 0.061105 E$.} \\
  \multicolumn{5}{l}{\commentc Number of points used to determine the maximum.} \\
\end{tabular}
\end{center}
\end{table}

\subsection{OT J081117.1$+$152003}\label{obj:j0811}

\begin{figure}
  \begin{center}
    \FigureFile(88mm,110mm){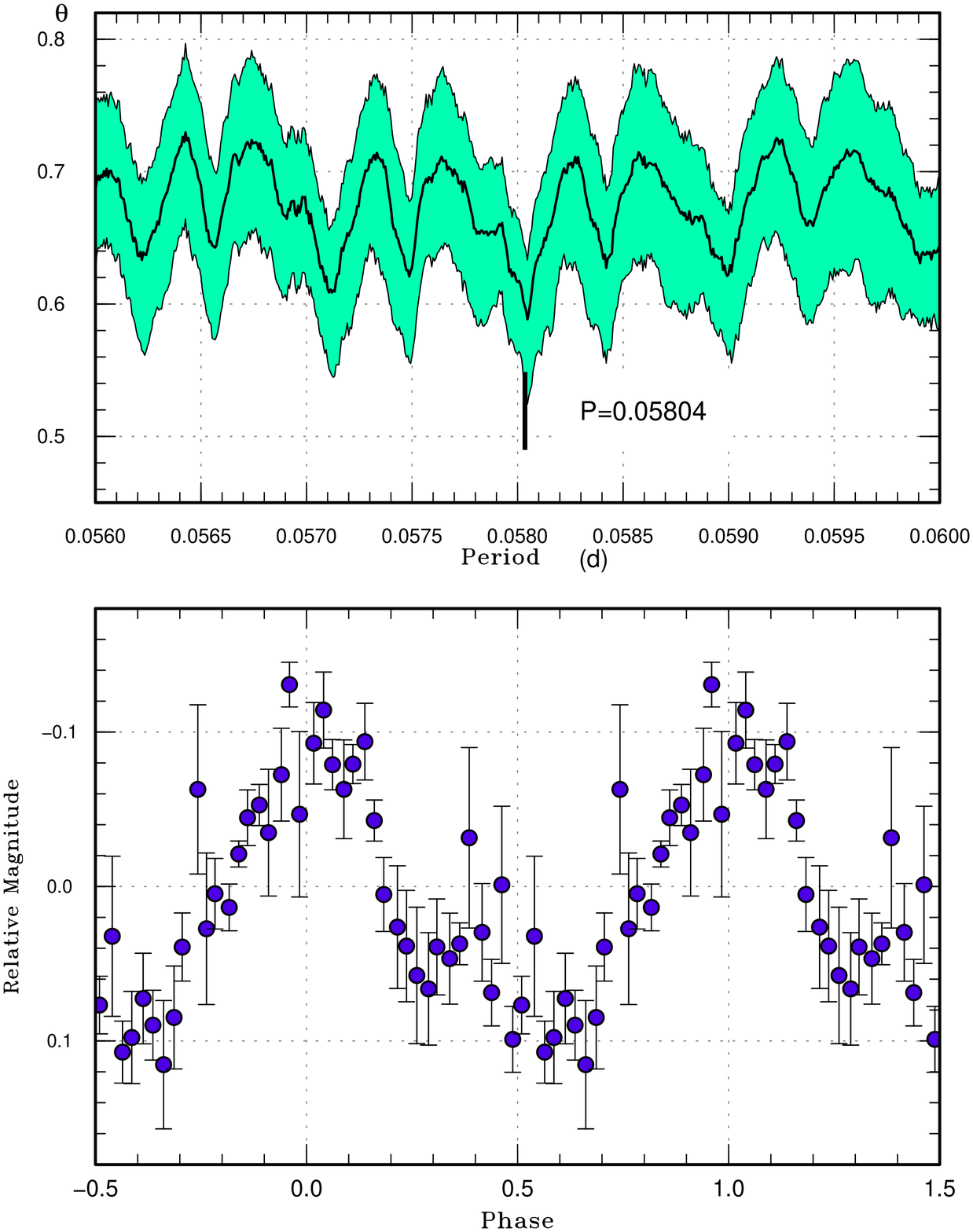}
  \end{center}
  \caption{Superhumps in OT J081117 (2011). (Upper): PDM analysis.
     The rejection rate for bootstrapping was reduced to 0.2 for
     better visualization.
     (Lower): Phase-averaged profile.}
  \label{fig:j0811shpdm}
\end{figure}

   This transient (=CSS111030:081117$+$152003; hereafter OT J081117)
was detected by CRTS on 2011 October 30.  The object had a large
($\sim$ 6 mag) outburst amplitude and was considered as a good candidate 
for an SU UMa-type dwarf nova.  As expected, short-period superhumps
were detected (vsnet-alert 13816; figure \ref{fig:j0811shpdm}). 
Due to the insufficient observation, we could not measure $P_{\rm SH}$ 
precisely.  The times of superhump maxima are listed in table 
\ref{tab:j0811oc2011}.  We listed the most likely alias determined 
with the PDM method in table \ref{tab:perlist}.

\begin{table}
\caption{Superhump maxima of OT J081117 (2011).}\label{tab:j0811oc2011}
\begin{center}
\begin{tabular}{ccccc}
\hline
$E$ & max\commenta & error & $O-C$\commentb & $N$\commentc \\
\hline
0 & 55865.6525 & 0.0003 & $-$0.0003 & 55 \\
1 & 55865.7111 & 0.0004 & 0.0003 & 36 \\
63 & 55869.3040 & 0.0031 & $-$0.0000 & 20 \\
\hline
  \multicolumn{5}{l}{\commenta BJD$-$2400000.} \\
  \multicolumn{5}{l}{\commentb Against max $= 2455865.6529 + 0.057955 E$.} \\
  \multicolumn{5}{l}{\commentc Number of points used to determine the maximum.} \\
\end{tabular}
\end{center}
\end{table}

\subsection{OT J084127.4$+$210053}\label{obj:j0841}

\begin{figure}
  \begin{center}
    \FigureFile(88mm,110mm){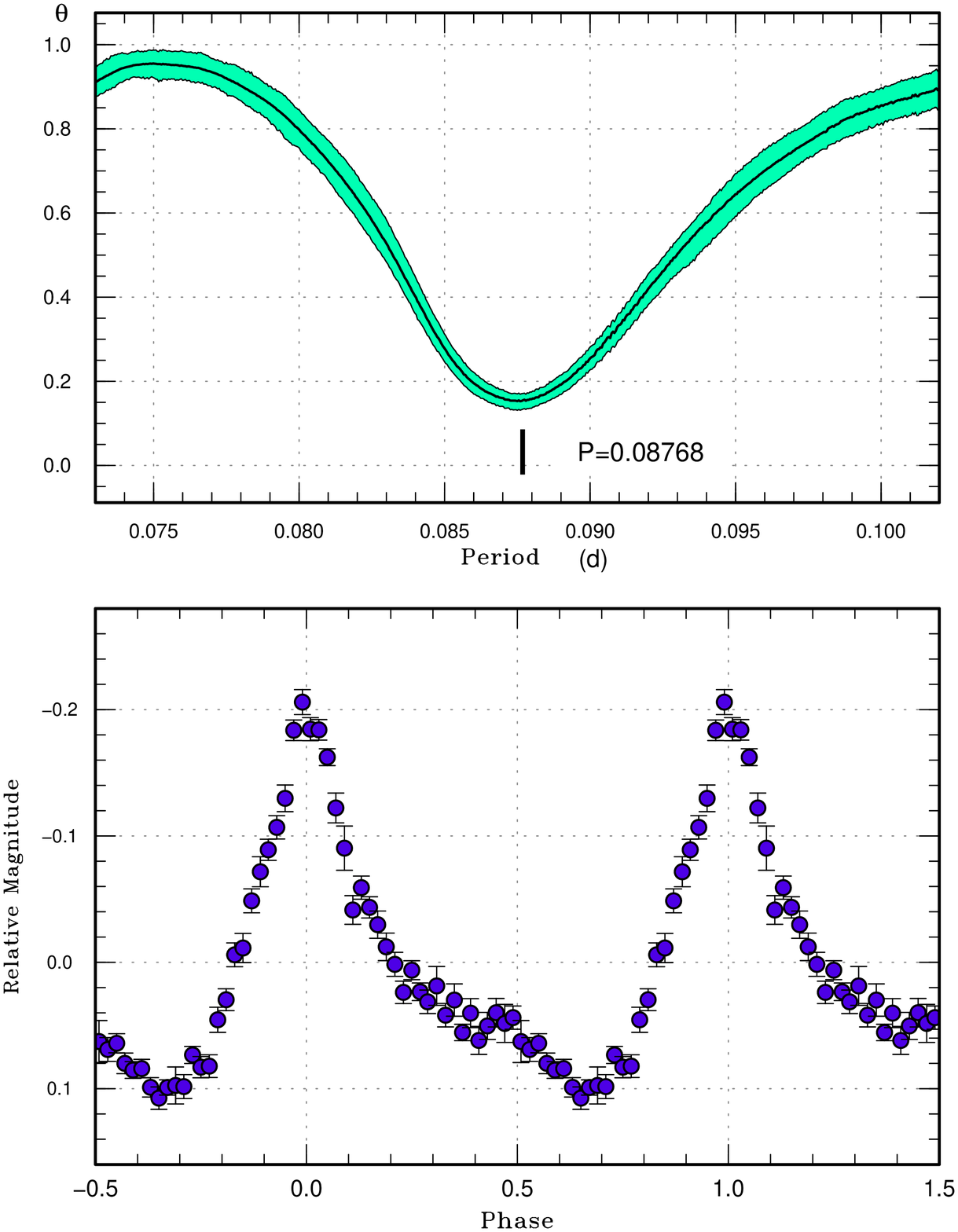}
  \end{center}
  \caption{Superhumps in OT J084127 (2012). (Upper): PDM analysis.
     (Lower): Phase-averaged profile.}
  \label{fig:j0841shpdm}
\end{figure}

   This transient (=CSS090525:084127$+$210054; hereafter OT J084127)
was detected by CRTS on 2009 May 25.  There were other known
outbursts (CRTS detections) in 2007 September, 2010 January
and 2010 May.  \citet{kat12DNSDSS} suggested $P_{\rm orb}$=0.10 d
from the SDSS colors.  The 2012 March outburst was observed and
superhumps were immediately detected (vsnet-alert 14391, 14392,
14396; figure \ref{fig:j0841shpdm}).
The times of superhump maxima are listed in table \ref{tab:j0841oc2012}.

\begin{table}
\caption{Superhump maxima of OT J084127 (2012).}\label{tab:j0841oc2012}
\begin{center}
\begin{tabular}{ccccc}
\hline
$E$ & max\commenta & error & $O-C$\commentb & $N$\commentc \\
\hline
0 & 56014.1263 & 0.0005 & $-$0.0004 & 131 \\
2 & 56014.3029 & 0.0004 & 0.0007 & 84 \\
3 & 56014.3901 & 0.0004 & 0.0003 & 177 \\
4 & 56014.4769 & 0.0007 & $-$0.0006 & 91 \\
\hline
  \multicolumn{5}{l}{\commenta BJD$-$2400000.} \\
  \multicolumn{5}{l}{\commentb Against max $= 2456014.1268 + 0.087686 E$.} \\
  \multicolumn{5}{l}{\commentc Number of points used to determine the maximum.} \\
\end{tabular}
\end{center}
\end{table}

\subsection{OT J094854.0$+$014911}\label{obj:j0948}

\begin{figure}
  \begin{center}
    \FigureFile(88mm,110mm){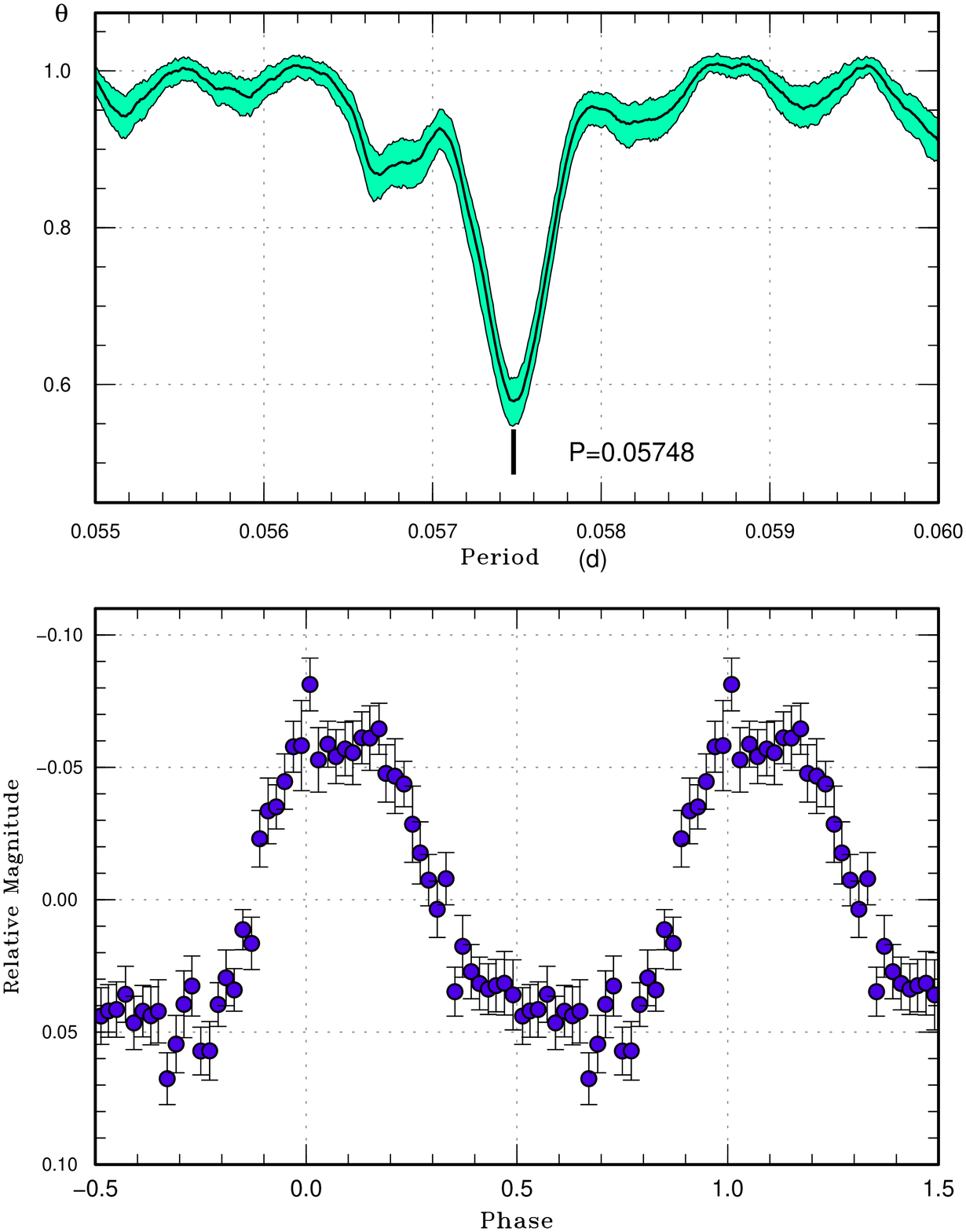}
  \end{center}
  \caption{Superhumps in OT J094854 (2012). (Upper): PDM analysis.
     (Lower): Phase-averaged profile.}
  \label{fig:j0948shpdm}
\end{figure}

   This transient (=CSS120315:094854$+$014911; hereafter OT J094854)
was detected by CRTS on 2012 March 15.  There was no previous outbursts
detected by CRTS.  Immediately following this discovery superhumps were
detected (vsnet-alert 14326, 14327; figure \ref{fig:j0948shpdm}).
The SDSS color of the quiescent counterpart resembles those of 
ordinary SU UMa-type dwarf novae rather than those of extreme 
WZ Sge-type dwarf novae (vsnet-alert 14328; see also \cite{kat12DNSDSS}).
S. Yoshida pointed out that the object was already in outburst on 
March 11 (vsnet-alert 14330).
The times of superhump maxima are listed in table \ref{tab:j0948oc2012}.
There was likely a stage B--C transition around $E=77$.  The early part
of stage B was missed and the observations were not sufficient to
determine the period of stage C superhumps.

\begin{table}
\caption{Superhump maxima of OT J094854 (2012).}\label{tab:j0948oc2012}
\begin{center}
\begin{tabular}{ccccc}
\hline
$E$ & max\commenta & error & $O-C$\commentb & $N$\commentc \\
\hline
0 & 56002.2812 & 0.0013 & 0.0024 & 24 \\
1 & 56002.3383 & 0.0005 & 0.0019 & 40 \\
2 & 56002.3946 & 0.0005 & 0.0007 & 31 \\
18 & 56003.3141 & 0.0010 & 0.0002 & 40 \\
19 & 56003.3720 & 0.0008 & 0.0007 & 40 \\
22 & 56003.5445 & 0.0013 & 0.0006 & 20 \\
23 & 56003.6031 & 0.0015 & 0.0018 & 16 \\
24 & 56003.6583 & 0.0012 & $-$0.0005 & 13 \\
25 & 56003.7166 & 0.0005 & 0.0003 & 74 \\
26 & 56003.7748 & 0.0016 & 0.0010 & 21 \\
28 & 56003.8899 & 0.0009 & 0.0011 & 30 \\
35 & 56004.2905 & 0.0007 & $-$0.0008 & 31 \\
36 & 56004.3477 & 0.0007 & $-$0.0011 & 31 \\
39 & 56004.5197 & 0.0008 & $-$0.0017 & 26 \\
40 & 56004.5798 & 0.0010 & 0.0010 & 19 \\
41 & 56004.6341 & 0.0018 & $-$0.0022 & 14 \\
42 & 56004.6889 & 0.0010 & $-$0.0049 & 13 \\
43 & 56004.7467 & 0.0018 & $-$0.0046 & 15 \\
57 & 56005.5512 & 0.0016 & $-$0.0051 & 14 \\
58 & 56005.6125 & 0.0018 & $-$0.0013 & 10 \\
59 & 56005.6712 & 0.0009 & $-$0.0001 & 9 \\
60 & 56005.7287 & 0.0033 & $-$0.0001 & 9 \\
74 & 56006.5347 & 0.0040 & 0.0009 & 14 \\
75 & 56006.5936 & 0.0016 & 0.0023 & 13 \\
76 & 56006.6552 & 0.0017 & 0.0064 & 9 \\
77 & 56006.7073 & 0.0025 & 0.0010 & 9 \\
78 & 56006.7620 & 0.0016 & $-$0.0017 & 9 \\
91 & 56007.5118 & 0.0024 & 0.0006 & 17 \\
92 & 56007.5717 & 0.0026 & 0.0030 & 14 \\
93 & 56007.6220 & 0.0022 & $-$0.0042 & 9 \\
94 & 56007.6851 & 0.0027 & 0.0014 & 58 \\
95 & 56007.7428 & 0.0013 & 0.0016 & 70 \\
96 & 56007.7971 & 0.0013 & $-$0.0016 & 60 \\
97 & 56007.8572 & 0.0019 & 0.0010 & 60 \\
\hline
  \multicolumn{5}{l}{\commenta BJD$-$2400000.} \\
  \multicolumn{5}{l}{\commentb Against max $= 2456002.2789 + 0.057498 E$.} \\
  \multicolumn{5}{l}{\commentc Number of points used to determine the maximum.} \\
\end{tabular}
\end{center}
\end{table}

\subsection{OT J102842.9$-$081927}\label{obj:j1028}

\begin{figure}
  \begin{center}
    \FigureFile(88mm,70mm){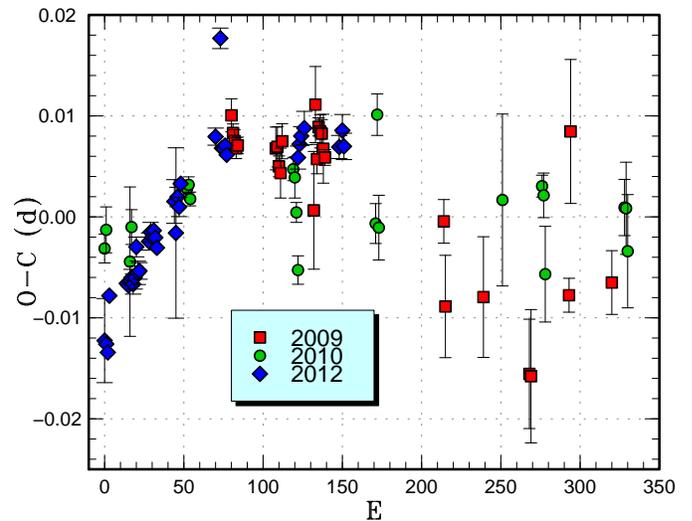}
  \end{center}
  \caption{Comparison of $O-C$ diagrams of J102842 between different
  superoutbursts.  A period of 0.03816~d was used to draw this figure.
  Approximate cycle counts ($E$) after the start of the observations
  were used for 2009 and 2012, assuming that the observations started 
  at the initial stage of the outbursts.  The $O-C$ diagram for 2010
  was shifted by 80 cycles.
  }
  \label{fig:j1028comp}
\end{figure}

   This object (=CSS090331:102843$-$081927, hereafter OT J102842)
was originally discovered by CRTS.  \citet{Pdot} indicated that
this object has a very short [0.038147(14)~d] superhump period,
suggesting an unusual evolutionary status similar to EI Psc 
(\cite{tho02j2329}; \cite{uem02j2329letter}) or V485 Cen
(\cite{aug93v485cen}; \cite{aug96v485cen}; \cite{ole97v485cen}).
The 2012 superoutburst, detected by CRTS, was also
observed.  The times of superhump maxima are listed in
table \ref{tab:j1028oc2012}.  Contrary to \citet{Pdot}, the present
observation gave a longer superhump period, particularly during
the first half of the observation.  It is difficult to reconcile
with this disagreement of the periods unless we assume that
the first half of the 2012 observation recorded stage A superhumps
(sinec CRTS observations were typically made with 10-d intervals,
it is difficult to determine the starting dates of outbursts).
In table \ref{tab:perlist} and figure \ref{fig:j1028comp}
we gave values and a comparison of $O-C$ diagrams based on this
identification.  This interpretation, however, has a problem
in that it cannot explain the large amplitudes of superhumps at 
the initial stage of the 2012 observation.  It may be either that
the evolution of superhumps in this system is unusual, or that
the period of superhumps greatly vary between superoutbursts.
Further observations, particularly regular monitoring to record
the epoch of the start of the outbursts, and to record superhumps
during the full course of the outbursts.

\begin{table}
\caption{Superhump maxima of OT J102842 (2012).}\label{tab:j1028oc2012}
\begin{center}
\begin{tabular}{ccccc}
\hline
$E$ & max\commenta & error & $O-C$\commentb & $N$\commentc \\
\hline
0 & 55958.0498 & 0.0042 & $-$0.0053 & 41 \\
1 & 55958.0876 & 0.0007 & $-$0.0058 & 64 \\
2 & 55958.1250 & 0.0005 & $-$0.0068 & 65 \\
3 & 55958.1687 & 0.0007 & $-$0.0013 & 60 \\
14 & 55958.5897 & 0.0004 & $-$0.0015 & 37 \\
15 & 55958.6277 & 0.0004 & $-$0.0018 & 36 \\
16 & 55958.6663 & 0.0004 & $-$0.0015 & 37 \\
17 & 55958.7048 & 0.0008 & $-$0.0013 & 31 \\
18 & 55958.7423 & 0.0010 & $-$0.0021 & 19 \\
19 & 55958.7811 & 0.0011 & $-$0.0016 & 20 \\
20 & 55958.8223 & 0.0010 & 0.0013 & 12 \\
21 & 55958.8579 & 0.0012 & $-$0.0014 & 20 \\
22 & 55958.8962 & 0.0008 & $-$0.0013 & 11 \\
28 & 55959.1281 & 0.0008 & 0.0008 & 68 \\
29 & 55959.1672 & 0.0010 & 0.0016 & 68 \\
30 & 55959.2048 & 0.0004 & 0.0010 & 123 \\
31 & 55959.2437 & 0.0004 & 0.0015 & 113 \\
32 & 55959.2811 & 0.0003 & 0.0007 & 113 \\
33 & 55959.3183 & 0.0005 & $-$0.0004 & 78 \\
44 & 55959.7426 & 0.0022 & 0.0027 & 8 \\
45 & 55959.7777 & 0.0084 & $-$0.0005 & 5 \\
46 & 55959.8194 & 0.0010 & 0.0029 & 7 \\
47 & 55959.8566 & 0.0009 & 0.0018 & 11 \\
48 & 55959.8970 & 0.0007 & 0.0040 & 5 \\
70 & 55960.7412 & 0.0008 & 0.0058 & 9 \\
73 & 55960.8654 & 0.0010 & 0.0151 & 24 \\
74 & 55960.8927 & 0.0003 & 0.0041 & 44 \\
75 & 55960.9308 & 0.0004 & 0.0039 & 39 \\
76 & 55960.9693 & 0.0004 & 0.0041 & 39 \\
77 & 55961.0065 & 0.0004 & 0.0031 & 40 \\
122 & 55962.7235 & 0.0011 & $-$0.0031 & 18 \\
123 & 55962.7629 & 0.0008 & $-$0.0019 & 15 \\
124 & 55962.8019 & 0.0009 & $-$0.0012 & 20 \\
126 & 55962.8790 & 0.0017 & $-$0.0007 & 20 \\
148 & 55963.7167 & 0.0011 & $-$0.0054 & 16 \\
150 & 55963.7946 & 0.0016 & $-$0.0040 & 20 \\
151 & 55963.8312 & 0.0013 & $-$0.0057 & 15 \\
\hline
  \multicolumn{5}{l}{\commenta BJD$-$2400000.} \\
  \multicolumn{5}{l}{\commentb Against max $= 2455958.0552 + 0.038290 E$.} \\
  \multicolumn{5}{l}{\commentc Number of points used to determine the maximum.} \\
\end{tabular}
\end{center}
\end{table}

\subsection{OT J105122.8$+$672528}\label{obj:j1051}

   This transient (=CSS120101:105123$+$672528; hereafter OT J105122)
was detected by CRTS on 2012 January 1.  There is an X-ray counterpart
1RXS J105120.5$+$672550.  D. Denisenko reported that this object
was recorded bright on a Palomer Sky Survey infrared plate 
taken on 1999 December 12 (vsnet-alert 14060).
Subsequent observations detected superhumps (vsnet-alert 14067,
14072).  
MASTER team also independently detected this transient
\citep{tiu12j1051atel3845}.  Although it was classified as a CV
\citep{sol12j1051atel3849}, they couldn't detect variability.
\citet{pav12j1051atel3889} further observed this object
in quiescence and recorded high-amplitude variations with
a period of 0.0596(9)~d, which was considered to be the orbital
period.

\begin{figure}
  \begin{center}
    \FigureFile(88mm,110mm){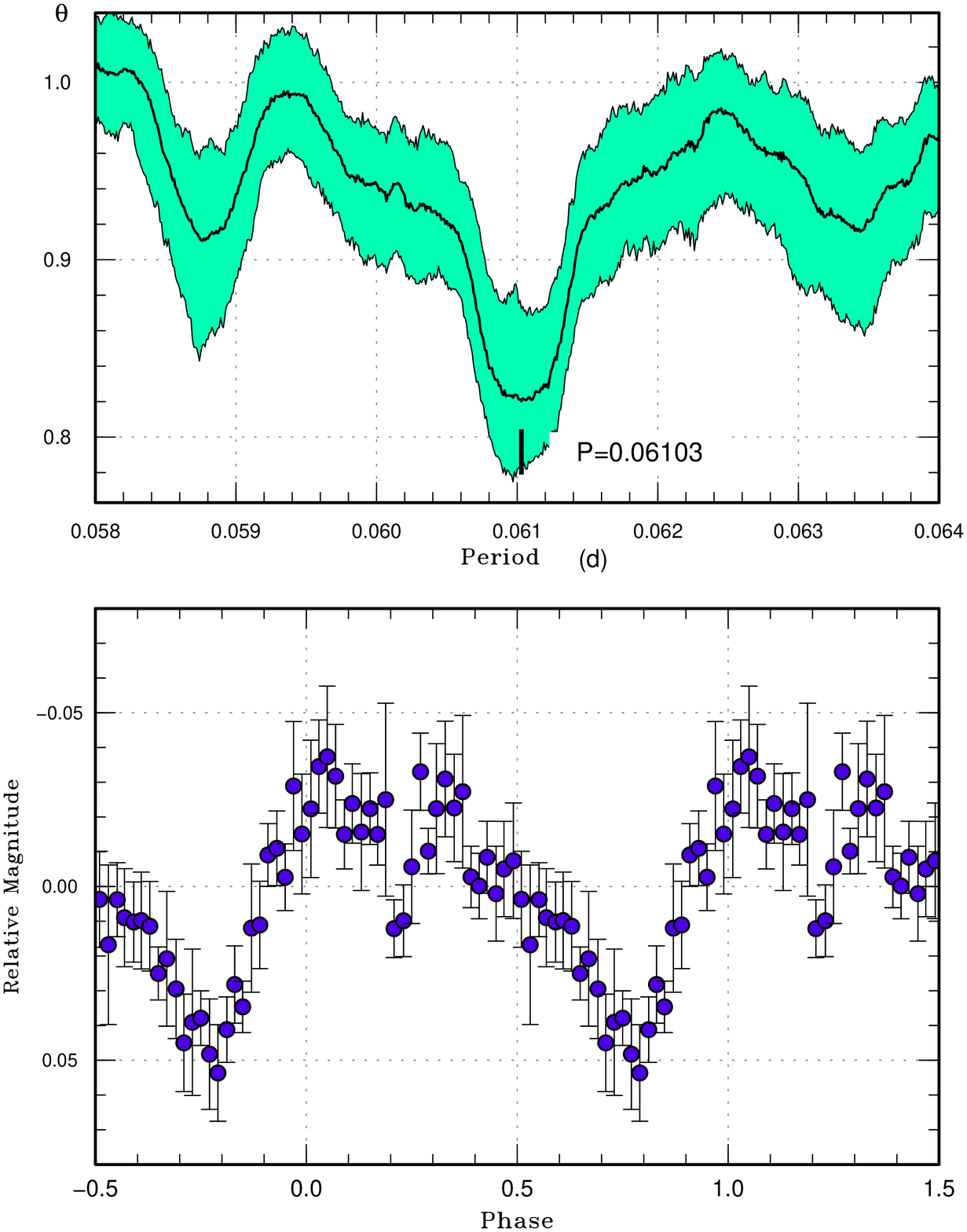}
  \end{center}
  \caption{Superhumps in OT J105122 (2012). (Upper): PDM analysis.
     (Lower): Phase-averaged profile.}
  \label{fig:j1051shpdm}
\end{figure}

   The times of superhump maxima are listed in table
\ref{tab:j1051oc2012}.  During these observations, the amplitudes
of superhumps were small (figure \ref{fig:j1051shpdm}), and 
not triangular as typically seen in early stage superhumps.
Although these superhumps were likely recorded when 
the amplitudes get smaller (particularly before the stage 
B--C transition, cf. subsection 4.7 of \cite{Pdot3}), it was
impossible to identify the stage in which they were observed.

\begin{table}
\caption{Superhump maxima of OT J105122 (2012).}\label{tab:j1051oc2012}
\begin{center}
\begin{tabular}{ccccc}
\hline
$E$ & max\commenta & error & $O-C$\commentb & $N$\commentc \\
\hline
0 & 55929.7466 & 0.0019 & 0.0007 & 64 \\
1 & 55929.8069 & 0.0015 & $-$0.0001 & 63 \\
2 & 55929.8626 & 0.0009 & $-$0.0055 & 63 \\
3 & 55929.9291 & 0.0014 & $-$0.0001 & 63 \\
4 & 55929.9894 & 0.0009 & $-$0.0008 & 57 \\
12 & 55930.4884 & 0.0023 & 0.0098 & 16 \\
26 & 55931.3292 & 0.0033 & $-$0.0041 & 16 \\
27 & 55931.3949 & 0.0024 & 0.0005 & 20 \\
29 & 55931.5163 & 0.0072 & $-$0.0003 & 23 \\
30 & 55931.5776 & 0.0038 & $-$0.0000 & 23 \\
\hline
  \multicolumn{5}{l}{\commenta BJD$-$2400000.} \\
  \multicolumn{5}{l}{\commentb Against max $= 2455929.7460 + 0.061054 E$.} \\
  \multicolumn{5}{l}{\commentc Number of points used to determine the maximum.} \\
\end{tabular}
\end{center}
\end{table}

\subsection{OT J125905.8$+$242634}\label{obj:j1259}

\begin{figure}
  \begin{center}
    \FigureFile(88mm,70mm){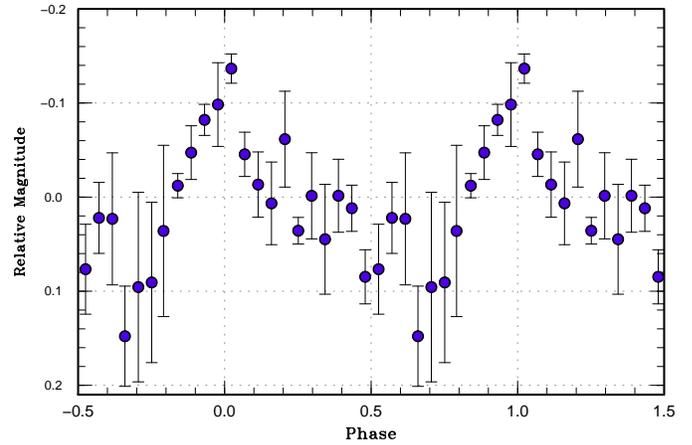}
  \end{center}
  \caption{Superhumps in OT J125905 (2012).  A period of 0.0660 d
    was assumed in phase-averaging.}
  \label{fig:j1259shprof}
\end{figure}

   This transient (=CSS120424:125906$+$242634; hereafter OT J125905)
was detected by CRTS on 2012 April 24.  There was a previous outburst
in 2009 February.  Subsequent observations detected superhumps
(vsnet-alert 14500, 14501).  The two superhumps maxima were
BJD 2456045.3135(17) ($N=28$) and 2456045.3795(14) ($N=35$).
We adopted a period of 0.0660(2)~d from this timing analysis.
The profile of the superhumps is shown in figure \ref{fig:j1259shprof}.

\subsection{OT J131625.7$-$151313}\label{obj:j1316}

   This transient (=CSS080427:131626$-$151313; hereafter OT J131625)
was detected by CRTS on 2008 April 27.  Two further outbursts were
detected by CRTS in 2010 February and 2011 April.  The 2012 March
outburst was also detected by CRTS.  Subsequent observation
clarified the presence of superhumps (vsnet-alert 14376, 14377).
Since the observation was done only on one night, we obtained
a single superhump maximum of BJD 2456012.5086(9) ($N=71$).
The best superhump period is 0.0955(8)~d (PDM method).

\subsection{OT J142548.1$+$151502}\label{obj:j1425}

   This transient (=CSS110628:142548$+$151502; hereafter OT J142548)
was detected by CRTS on 2011 June 28.  Only single-night observation
was available, which clearly showed superhumps (vsnet-alert 13474).
The best superhump period (PDM method) was 0.0984(10)~d and
we obtained only two superhump maxima:
BJD 2455742.3565(9) ($N$=34) and 2455742.4548(14) ($N$=27).

\subsection{OT J144252.0$-$225040}\label{obj:j1442}

\begin{figure}
  \begin{center}
    \FigureFile(88mm,110mm){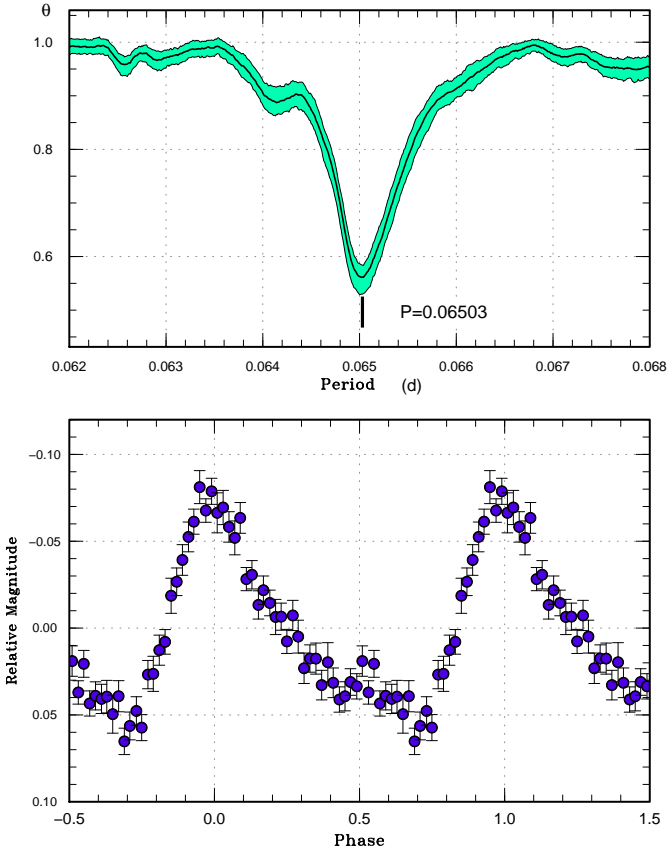}
  \end{center}
  \caption{Superhumps in OT J144252 (2012). (Upper): PDM analysis.
     (Lower): Phase-averaged profile.}
  \label{fig:j1442shpdm}
\end{figure}

   This transient (=CSS120417:144252$-$225040; hereafter OT J144252)
was detected by CRTS on 2012 April 17.  The large outburst amplitude
received attention (vsnet-alert 14448).  Subsequent observations
detected superhumps (vsnet-alert 14455, 14457; figure \ref{fig:j1442shpdm}).
The times of superhump maxima are listed in table \ref{tab:j1442oc2012}.
A clear pattern of stages B and C can be recognized.
Despite the large outburst amplitude and the lack of past outbursts
in the CRTS data, the $O-C$ diagram resembles those of ordinary
SU UMa-type dwarf novae rather than those of extreme WZ Sge-type
dwarf novae.

\begin{table}
\caption{Superhump maxima of OT J144252 (2012).}\label{tab:j1442oc2012}
\begin{center}
\begin{tabular}{ccccc}
\hline
$E$ & max\commenta & error & $O-C$\commentb & $N$\commentc \\
\hline
0 & 56035.6984 & 0.0015 & $-$0.0041 & 78 \\
1 & 56035.7648 & 0.0010 & $-$0.0027 & 89 \\
2 & 56035.8322 & 0.0007 & $-$0.0003 & 14 \\
3 & 56035.8949 & 0.0006 & $-$0.0025 & 18 \\
12 & 56036.4809 & 0.0004 & $-$0.0013 & 140 \\
13 & 56036.5455 & 0.0006 & $-$0.0017 & 152 \\
14 & 56036.6126 & 0.0010 & 0.0004 & 75 \\
15 & 56036.6742 & 0.0011 & $-$0.0030 & 12 \\
18 & 56036.8697 & 0.0009 & $-$0.0024 & 18 \\
27 & 56037.4552 & 0.0007 & $-$0.0017 & 150 \\
28 & 56037.5212 & 0.0008 & $-$0.0006 & 133 \\
29 & 56037.5855 & 0.0011 & $-$0.0014 & 13 \\
30 & 56037.6543 & 0.0016 & 0.0024 & 14 \\
33 & 56037.8448 & 0.0014 & $-$0.0020 & 17 \\
34 & 56037.9103 & 0.0021 & $-$0.0014 & 13 \\
44 & 56038.5666 & 0.0031 & 0.0051 & 13 \\
45 & 56038.6260 & 0.0017 & $-$0.0004 & 13 \\
48 & 56038.8248 & 0.0013 & 0.0034 & 14 \\
49 & 56038.8921 & 0.0032 & 0.0057 & 18 \\
59 & 56039.5453 & 0.0018 & 0.0092 & 12 \\
60 & 56039.6056 & 0.0015 & 0.0045 & 13 \\
61 & 56039.6725 & 0.0012 & 0.0064 & 11 \\
64 & 56039.8669 & 0.0013 & 0.0059 & 19 \\
75 & 56040.5781 & 0.0013 & 0.0023 & 13 \\
76 & 56040.6422 & 0.0017 & 0.0014 & 13 \\
79 & 56040.8370 & 0.0012 & 0.0013 & 18 \\
80 & 56040.9001 & 0.0021 & $-$0.0006 & 17 \\
90 & 56041.5552 & 0.0069 & 0.0048 & 9 \\
91 & 56041.6086 & 0.0013 & $-$0.0068 & 13 \\
106 & 56042.5791 & 0.0034 & $-$0.0109 & 13 \\
107 & 56042.6463 & 0.0026 & $-$0.0087 & 13 \\
\hline
  \multicolumn{5}{l}{\commenta BJD$-$2400000.} \\
  \multicolumn{5}{l}{\commentb Against max $= 2456035.7025 + 0.064977 E$.} \\
  \multicolumn{5}{l}{\commentc Number of points used to determine the maximum.} \\
\end{tabular}
\end{center}
\end{table}

\subsection{OT J144453.1$-$131118}\label{obj:j1444}

\begin{figure}
  \begin{center}
    \FigureFile(88mm,110mm){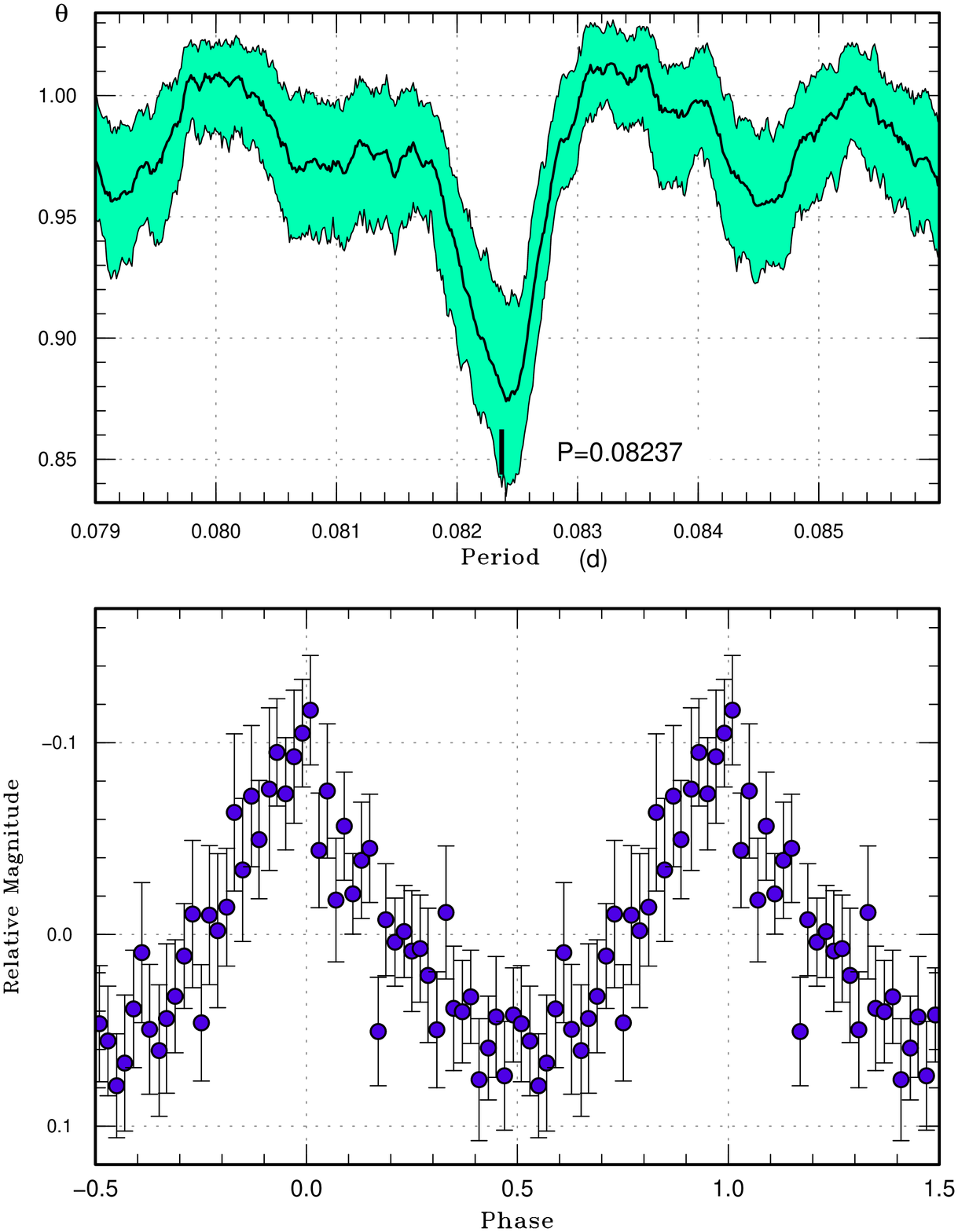}
  \end{center}
  \caption{Superhumps in OT J144453 (2011). (Upper): PDM analysis.
     (Lower): Phase-averaged profile.}
  \label{fig:j1444shpdm}
\end{figure}

   This transient (=CSS120424:144453$-$131118; hereafter OT J144453)
was detected by CRTS on 2012 April 24.  There was a past outburst
in 2005 December.  Although there was a hint of superhumps in the
first observation (vsnet-alert 14491), confirmatory observations
became available after 4~d (vsnet-alert 14507).  Later observations
well characterized superhumps (vsnet-alert 14516, 14522, 14530;
figure \ref{fig:j1444shpdm}).
The times of superhump maxima are listed in table \ref{tab:j1444oc2012}.
There was no hint of period variation.  By taking the long initial
gap of observation into account, we likely observed stage C superhumps.
It is not, however, excluded that this object has a virtually zero
$P_{\rm dot}$ as in some long-$P_{\rm orb}$ systems.

\begin{table}
\caption{Superhump maxima of OT J144453 (2012).}\label{tab:j1444oc2012}
\begin{center}
\begin{tabular}{ccccc}
\hline
$E$ & max\commenta & error & $O-C$\commentb & $N$\commentc \\
\hline
0 & 56046.1078 & 0.0009 & 0.0001 & 170 \\
17 & 56047.5119 & 0.0020 & 0.0053 & 13 \\
18 & 56047.5888 & 0.0015 & $-$0.0001 & 14 \\
20 & 56047.7561 & 0.0009 & 0.0026 & 10 \\
21 & 56047.8346 & 0.0010 & $-$0.0011 & 25 \\
32 & 56048.7402 & 0.0010 & $-$0.0007 & 12 \\
33 & 56048.8242 & 0.0012 & 0.0010 & 24 \\
34 & 56048.9017 & 0.0016 & $-$0.0039 & 19 \\
42 & 56049.5541 & 0.0030 & $-$0.0098 & 15 \\
43 & 56049.6493 & 0.0045 & 0.0031 & 15 \\
44 & 56049.7259 & 0.0205 & $-$0.0025 & 12 \\
45 & 56049.8057 & 0.0010 & $-$0.0050 & 21 \\
46 & 56049.8926 & 0.0020 & $-$0.0004 & 22 \\
55 & 56050.6386 & 0.0041 & 0.0050 & 14 \\
56 & 56050.7191 & 0.0029 & 0.0032 & 12 \\
57 & 56050.7967 & 0.0050 & $-$0.0015 & 18 \\
58 & 56050.8849 & 0.0028 & 0.0044 & 26 \\
\hline
  \multicolumn{5}{l}{\commenta BJD$-$2400000.} \\
  \multicolumn{5}{l}{\commentb Against max $= 2456046.1077 + 0.082289 E$.} \\
  \multicolumn{5}{l}{\commentc Number of points used to determine the maximum.} \\
\end{tabular}
\end{center}
\end{table}

\subsection{OT J145921.8$+$354806}\label{obj:j1459}

\begin{figure}
  \begin{center}
    \FigureFile(88mm,110mm){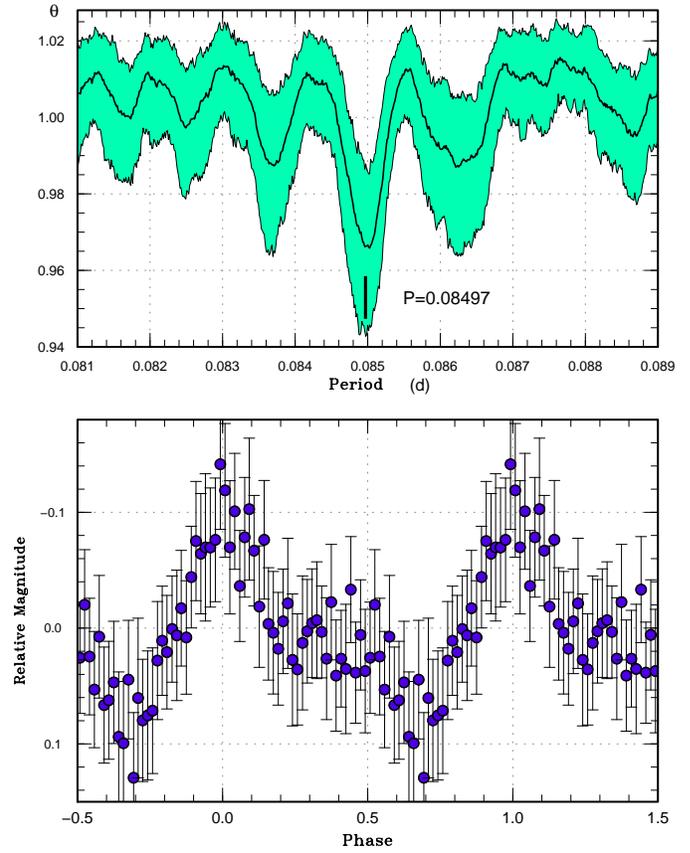}
  \end{center}
  \caption{Superhumps in OT J145921 (2011). (Upper): PDM analysis.
     (Lower): Phase-averaged profile.}
  \label{fig:j1459shpdm}
\end{figure}

   This transient (=CSS110613:145922$+$354806; hereafter OT J145921)
was detected by CRTS on 2011 June 13.  There was an earlier
outburst in 2008 April.  Subsequent observations detected
superhumps (vsnet-alert 13427; figure \ref{fig:j1459shpdm}).
The times of superhump maxima are listed in table
\ref{tab:j1459oc2011}.
The large amplitudes of the superhumps suggest that the outburst
was detected in a relatively early stage.  The resultant
$P_{\rm dot}$ was less likely negative, as expected for an object
with this $P_{\rm SH}$, and may be even positive.  The object
may be analogous to GX Cas (cf. \cite{Pdot3}) which showed
a large positive $P_{\rm dot}$ despite its long $P_{\rm SH}$.

\begin{table}
\caption{Superhump maxima of OT J145921 (2011).}\label{tab:j1459oc2011}
\begin{center}
\begin{tabular}{ccccc}
\hline
$E$ & max\commenta & error & $O-C$\commentb & $N$\commentc \\
\hline
0 & 55728.4160 & 0.0011 & 0.0033 & 69 \\
11 & 55729.3516 & 0.0003 & 0.0027 & 50 \\
12 & 55729.4362 & 0.0005 & 0.0022 & 59 \\
24 & 55730.4490 & 0.0023 & $-$0.0064 & 30 \\
25 & 55730.5413 & 0.0015 & 0.0008 & 33 \\
35 & 55731.3891 & 0.0006 & $-$0.0026 & 43 \\
36 & 55731.4725 & 0.0007 & $-$0.0043 & 36 \\
58 & 55733.3492 & 0.0016 & $-$0.0001 & 36 \\
59 & 55733.4357 & 0.0011 & 0.0013 & 46 \\
62 & 55733.6885 & 0.0018 & $-$0.0013 & 125 \\
63 & 55733.7754 & 0.0016 & 0.0005 & 126 \\
71 & 55734.4462 & 0.0018 & $-$0.0096 & 37 \\
73 & 55734.6391 & 0.0082 & 0.0131 & 78 \\
74 & 55734.7113 & 0.0032 & 0.0002 & 128 \\
\hline
  \multicolumn{5}{l}{\commenta BJD$-$2400000.} \\
  \multicolumn{5}{l}{\commentb Against max $= 2455728.4126 + 0.085114 E$.} \\
  \multicolumn{5}{l}{\commentc Number of points used to determine the maximum.} \\
\end{tabular}
\end{center}
\end{table}

\subsection{OT J155631.0$-$080440}\label{obj:j1556}

\begin{figure}
  \begin{center}
    \FigureFile(88mm,110mm){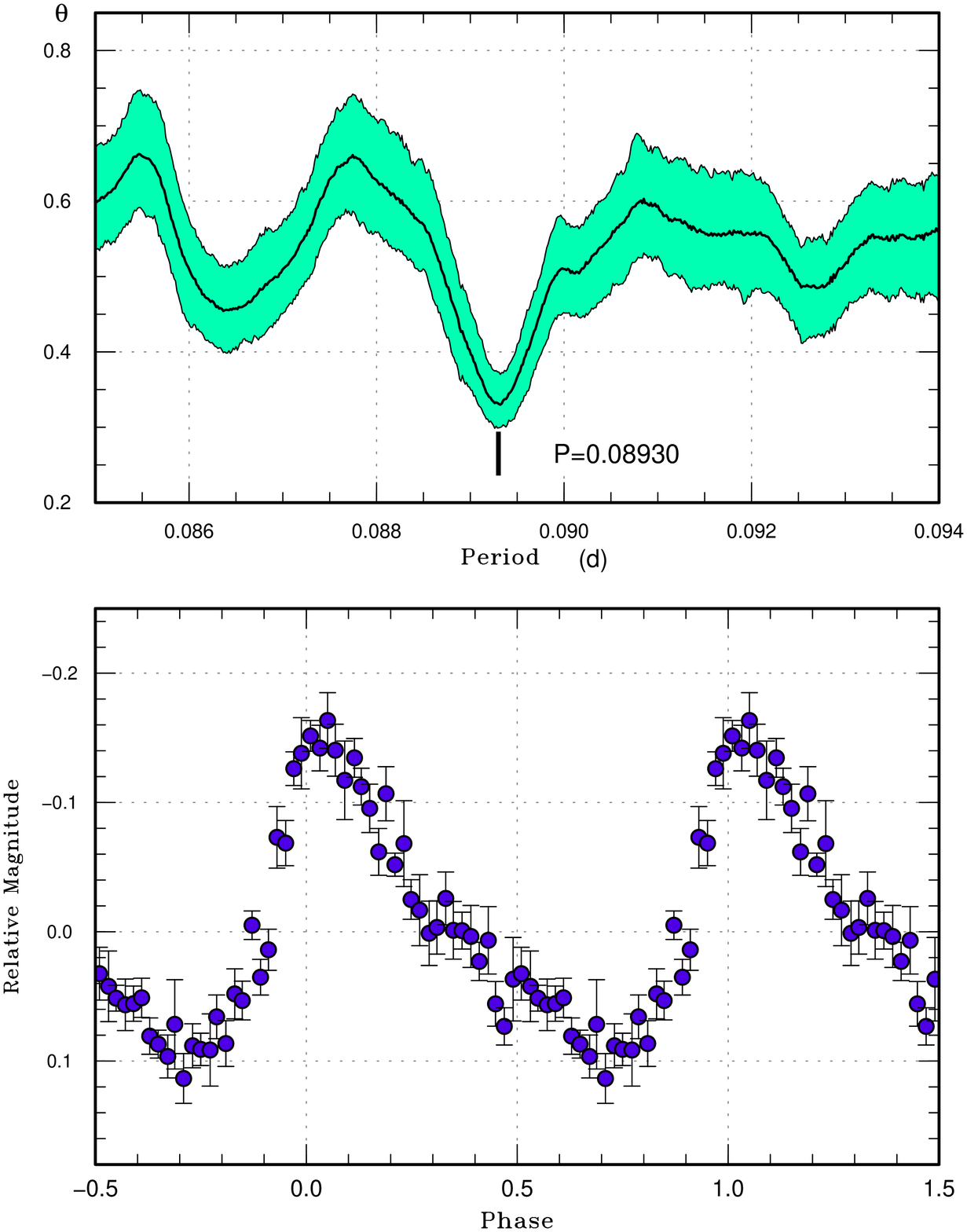}
  \end{center}
  \caption{Superhumps in OT J155631 (2012). (Upper): PDM analysis.
     (Lower): Phase-averaged profile.}
  \label{fig:j1556shpdm}
\end{figure}

   The object was detected as a transient (=CSS090321:155631$-$080440;
hereafter OT J155631) by CRTS on 2009 March 21.  Although several
outbursts were recorded since then, the 2012 March outburst was
the brightest (15.3 mag) in its history.  Superhumps were
soon detected (vsnet-alert 14406, 14416; figure \ref{fig:j1556shpdm}).
The times of superhump maxima are listed in table \ref{tab:j1556oc2012}.

\begin{table}
\caption{Superhump maxima of OT J155631 (2012).}\label{tab:j1556oc2012}
\begin{center}
\begin{tabular}{ccccc}
\hline
$E$ & max\commenta & error & $O-C$\commentb & $N$\commentc \\
\hline
0 & 56017.2322 & 0.0007 & $-$0.0021 & 92 \\
1 & 56017.3236 & 0.0005 & $-$0.0000 & 95 \\
18 & 56018.8443 & 0.0013 & 0.0024 & 13 \\
29 & 56019.8257 & 0.0020 & 0.0015 & 10 \\
30 & 56019.9149 & 0.0019 & 0.0013 & 14 \\
40 & 56020.8067 & 0.0025 & 0.0001 & 9 \\
41 & 56020.8928 & 0.0029 & $-$0.0032 & 21 \\
\hline
  \multicolumn{5}{l}{\commenta BJD$-$2400000.} \\
  \multicolumn{5}{l}{\commentb Against max $= 2456017.2343 + 0.089309 E$.} \\
  \multicolumn{5}{l}{\commentc Number of points used to determine the maximum.} \\
\end{tabular}
\end{center}
\end{table}

\subsection{OT J160410.6$+$145618}\label{obj:j1604}

   The object was detected as a transient (=CSS120326:160411$+$145618;
hereafter OT J160410) by CRTS on 2012 March 26.  There was another
outburst in 2010 July (CRTS data).  The large ($>$5 mag)
outburst amplitude was noted (vsnet-alert 14384).  Although there
was only a single-night observation, two superhump maxima were
recorded: BJD 2456014.5194(8) ($N=35$) and BJD 2456014.5841(11)
($N=34$).  The superhump period by the PDM method is 0.0656(5)~d.

\subsection{OT J162806.2$+$065316}\label{obj:j1628}

\begin{figure}
  \begin{center}
    \FigureFile(88mm,110mm){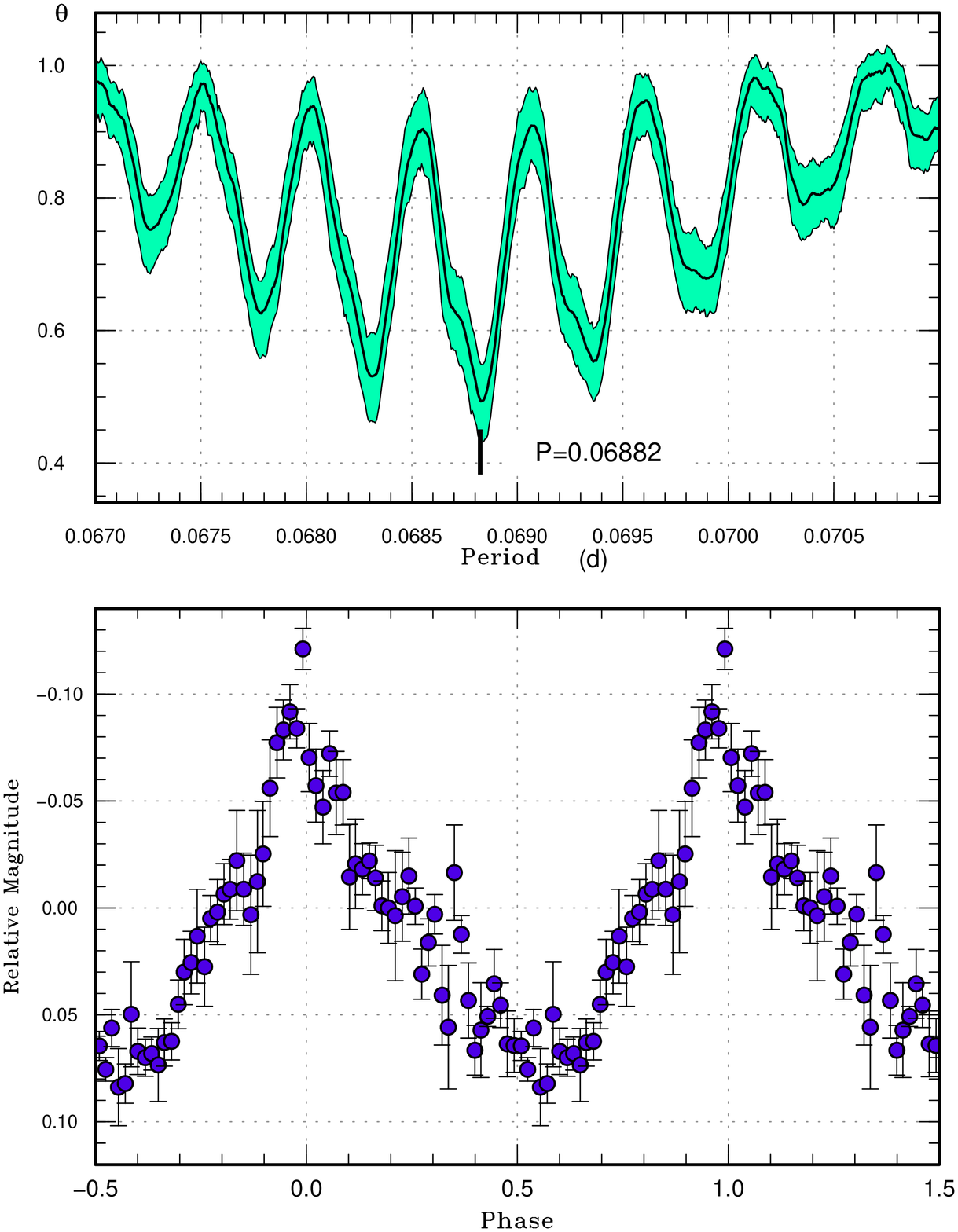}
  \end{center}
  \caption{Superhumps in OT J162806 (2011). (Upper): PDM analysis.
     (Lower): Phase-averaged profile.}
  \label{fig:j1628shpdm}
\end{figure}

   The object was detected as a transient (=CSS110611:162806$+$065316;
hereafter OT J162806)
by CRTS on 2011 June 11.  The object had been selected as a candidate
for QSO based on SDSS colors \citep{ric09SDSSQSOcand}.
The existence of two previous outbursts in the CRTS data
confirmed the DN-type nature (vsnet-alert 13413).
The object was soon confirmed to show superhumps (vsnet-alert 13416;
figure \ref{fig:j1628shpdm}).
The times of superhump maxima are listed in table \ref{tab:j1628oc2011}.

\begin{table}
\caption{Superhump maxima of OT J162806 (2011).}\label{tab:j1628oc2011}
\begin{center}
\begin{tabular}{ccccc}
\hline
$E$ & max\commenta & error & $O-C$\commentb & $N$\commentc \\
\hline
0 & 55724.4545 & 0.0003 & 0.0009 & 67 \\
13 & 55725.3473 & 0.0005 & $-$0.0013 & 38 \\
14 & 55725.4175 & 0.0003 & 0.0000 & 64 \\
15 & 55725.4866 & 0.0003 & 0.0003 & 55 \\
139 & 55734.0220 & 0.0012 & $-$0.0013 & 105 \\
140 & 55734.0935 & 0.0022 & 0.0014 & 99 \\
\hline
  \multicolumn{5}{l}{\commenta BJD$-$2400000.} \\
  \multicolumn{5}{l}{\commentb Against max $= 2455724.4536 + 0.068847 E$.} \\
  \multicolumn{5}{l}{\commentc Number of points used to determine the maximum.} \\
\end{tabular}
\end{center}
\end{table}

\subsection{OT J163942.7$+$122414}\label{obj:j1639}

\begin{figure}
  \begin{center}
    \FigureFile(88mm,110mm){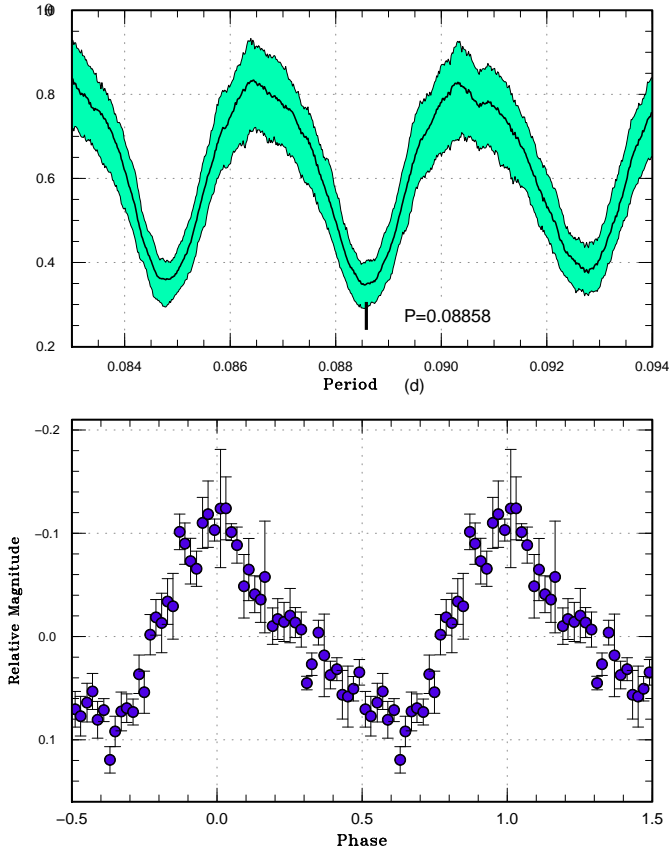}
  \end{center}
  \caption{Superhumps in OT J163942 (2012). (Upper): PDM analysis.
     The alias was selected by $O-C$ analysis of a continuous run
     on a single night.
     (Lower): Phase-averaged profile.}
  \label{fig:j1639shpdm}
\end{figure}

   The object was originally detected as a transient
(=CSS080131:163943$+$122414; hereafter OT J163942) by CRTS on 2008
January 31.  We observed the 2012 April outburst detected by the CRTS.
The observations confirmed the presence of superhumps (vsnet-alert 14474; 
figure \ref{fig:j1639shpdm}).
The times of superhump maxima are listed in table \ref{tab:j1639oc2012}.
The period in table \ref{tab:perlist} refers to the result of
the PDM analysis.  The object appears to be a long-$P_{\rm orb}$ system
with frequent outbursts based on numerous outburst detections by the CRTS.

\begin{table}
\caption{Superhump maxima of OT J163942 (2012).}\label{tab:j1639oc2012}
\begin{center}
\begin{tabular}{ccccc}
\hline
$E$ & max\commenta & error & $O-C$\commentb & $N$\commentc \\
\hline
0 & 56039.5603 & 0.0007 & $-$0.0000 & 88 \\
22 & 56041.5094 & 0.0024 & 0.0009 & 33 \\
23 & 56041.5962 & 0.0011 & $-$0.0008 & 39 \\
\hline
  \multicolumn{5}{l}{\commenta BJD$-$2400000.} \\
  \multicolumn{5}{l}{\commentb Against max $= 2456039.5603 + 0.088554 E$.} \\
  \multicolumn{5}{l}{\commentc Number of points used to determine the maximum.} \\
\end{tabular}
\end{center}
\end{table}

\subsection{OT J170609.7$+$143452}\label{obj:j1706}

   The object was originally detected as a transient 
(=CSS090205:170610$+$143452; hereafter OT J170609) by CRTS on 
2009 February 5.  Although superhumps were detected during the
2009 outburst (vsnet-alert 11061), the period was not well
determined because the object faded quickly after this observation.

\begin{figure}
  \begin{center}
    \FigureFile(88mm,110mm){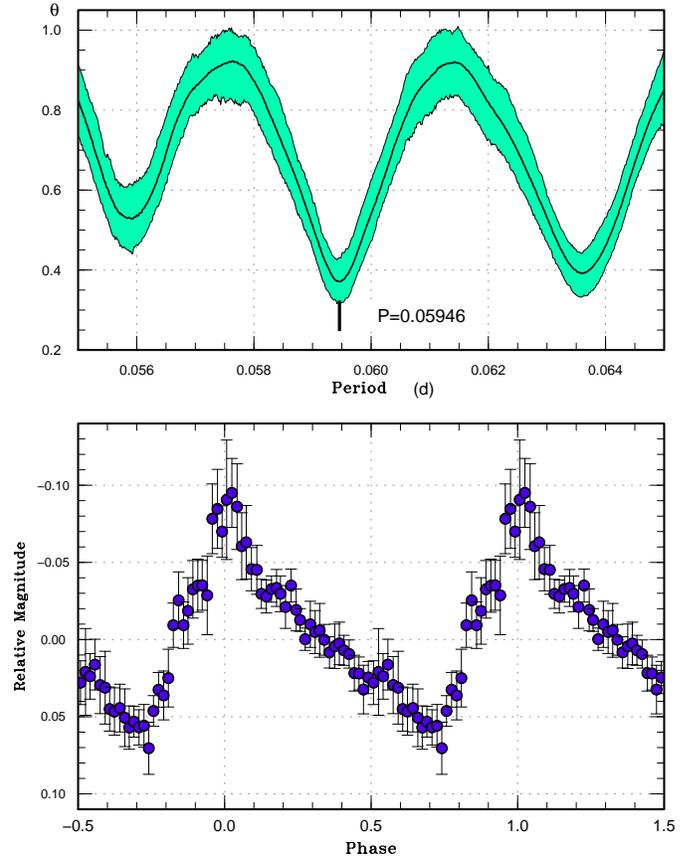}
  \end{center}
  \caption{Superhumps in OT J170609 (2011). (Upper): PDM analysis.
     (Lower): Phase-averaged profile.}
  \label{fig:j1706shpdm}
\end{figure}

   The object underwent another outburst in 2011 June (CRTS
detection, see also vsnet-alert 13456).  Although the object once
faded (vsnet-alert 13464), it showed a rebrightening in July
(vsnet-alert 13481).  The outburst turned out to be a superoutburst
preceded by a precursor.  Observations on two nights
yielded a likely superhump period of 0.05946(8)~d (PDM analysis;
figure \ref{fig:j1706shpdm}), although one-day aliases cannot be 
perfectly excluded.  The selection of the alias appears to be
justified by independently determined spectroscopic period
of 0.0582~d \citep{tho12CRTSCVs}, yielding an $\epsilon$ of 2.2\%. 
The times of superhump maxima are listed in table \ref{tab:j1706oc2011}.

\begin{table}
\caption{Superhump maxima of OT J170609 (2011).}\label{tab:j1706oc2011}
\begin{center}
\begin{tabular}{ccccc}
\hline
$E$ & max\commenta & error & $O-C$\commentb & $N$\commentc \\
\hline
0 & 55745.5333 & 0.0006 & $-$0.0017 & 59 \\
1 & 55745.5963 & 0.0015 & 0.0018 & 31 \\
15 & 55746.4283 & 0.0004 & $-$0.0003 & 57 \\
16 & 55746.4884 & 0.0004 & 0.0002 & 60 \\
\hline
  \multicolumn{5}{l}{\commenta BJD$-$2400000.} \\
  \multicolumn{5}{l}{\commentb Against max $= 2455745.5350 + 0.059578 E$.} \\
  \multicolumn{5}{l}{\commentc Number of points used to determine the maximum.} \\
\end{tabular}
\end{center}
\end{table}

\subsection{OT J173516.9$+$154708}\label{obj:j1735}

   The object was detected as a transient (=CSS110623:173517$+$154708;
hereafter OT J173516) by CRTS on 2011 June 23.  The object was
also in outburst in GSC 1.2.  Although early observations already
recorded superhump-like modulations (vsnet-alert 13465, 13468, 13470),
the times of maxima could not be well expressed by any trial
period (vsnet-alert 13473, 13482).  Although the main power of
periodicities was recorded in a range of 0.05--0.06~d, we could
not sort out a single superhump period at the time of the observation.
On July 9, the object entered a rapid decline phase.

\begin{figure}
  \begin{center}
    \FigureFile(88mm,110mm){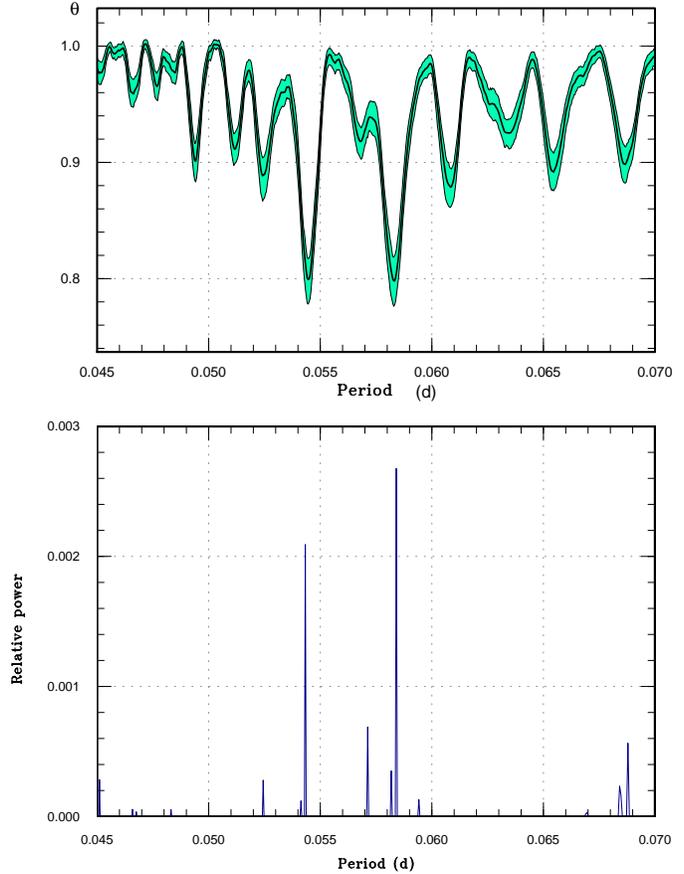}
  \end{center}
  \caption{Period analysis in OT J173516 (2011). (Upper): PDM analysis.
     (Lower): lasso analysis ($\log \lambda=-6.42$).}
  \label{fig:j1735shpdm}
\end{figure}

\begin{figure}
  \begin{center}
    \FigureFile(88mm,110mm){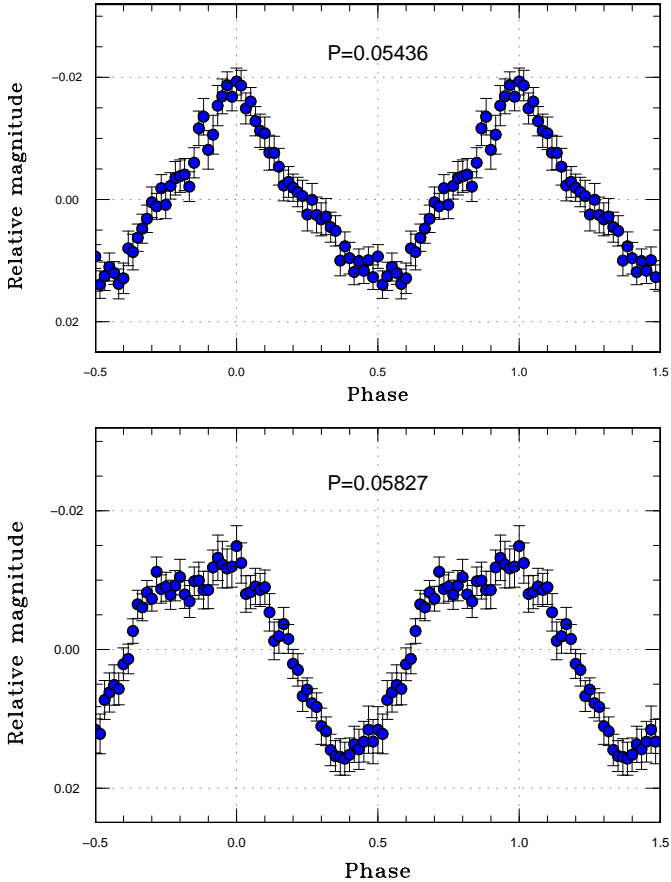}
  \end{center}
  \caption{Profiles of two periodicities in OT J173516 (2011).}
  \label{fig:j1735meanc}
\end{figure}

\begin{figure}
  \begin{center}
    \FigureFile(88mm,110mm){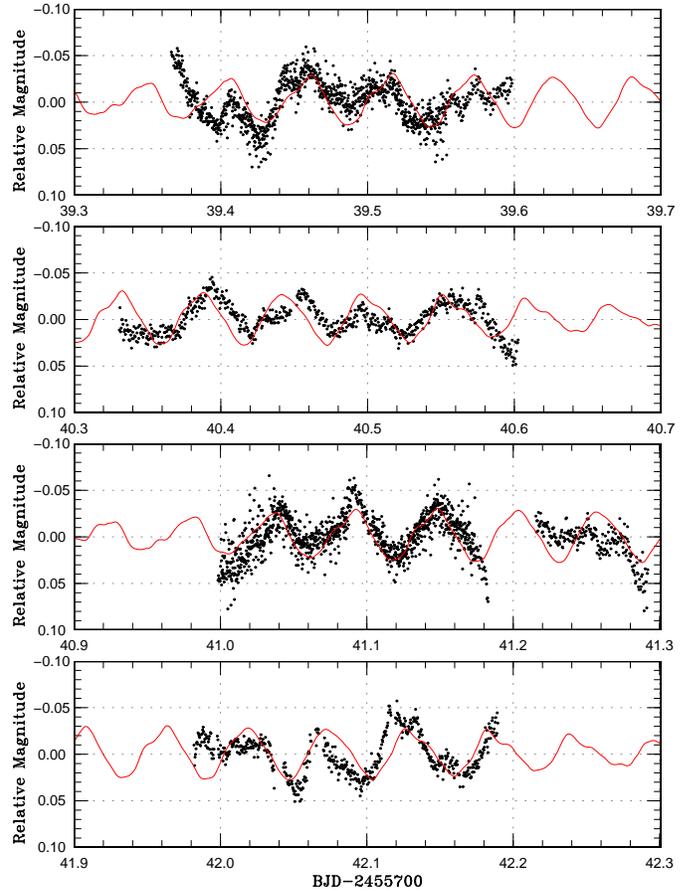}
  \end{center}
  \caption{Synthesized light curve of OT J173516 (2011).}
    The points represent observations.  The curves represent
    the expected light curve by adding two waves in figure
    \ref{fig:j1735meanc}.
  \label{fig:j1735perdub}
\end{figure}

   Using the best part (June 26--29) of our observation before
the rapid decline, there appeared to be two strong signals around
0.05436~d and 0.05827~d with the PDM analysis.  The lasso analysis,
which is less affected by the window function, yielded the same
two signals (figure \ref{fig:j1735shpdm}).  By partially subtracting
mean profiles from the observations folded by each period, we have
been able to decompose the light curve into these periods
(figure \ref{fig:j1735meanc}).
A simple superposition of these two waves has been shown to
express the observation fairly well (figure \ref{fig:j1735perdub}).

   The exact identifications of these periodicities are yet
unclear.  It might be that the shorter period is $P_{\rm orb}$
and the longer period is $P_{\rm SH}$.  In this case, however,
the $\epsilon$ is 7.2\%, extremely too large for this $P_{\rm orb}$,
and none of non-eclipsing SU UMa-type dwarf novae have yet
shown similar amplitudes of orbital humps and superhumps at the
same time.  We suggest an alternative interpretation that
the longer period is $P_{\rm orb}$ and the shorter period
represents negative superhumps.  The small amplitude of superhumps,
compared to that of orbital humps, may be reconciled if negative
superhumps were indeed excited.  The exact identification of 
the periods should await future observations.

\subsection{OT J184228.1$+$483742}\label{obj:j1842}

   This object (hereafter OT J184228) was discovered H. Nishimura
on 2011 September 5.5293 UT at an unfiltered CCD magnitude of
11.8 (=PNV J18422792$+$4837425; \cite{nak11j1842cbet2818}).
S. Kiyota's early multicolor photometry already suggested the dwarf
nova-type nature (vsnet-alert 13645).  M. Fujii, A. Ayani, C. Buil
(vsnet-alert 13650, 13651, 13655) and A. Arai\footnote{
$<$http://www.cbat.eps.harvard.edu/unconf/followups/\\J18422792$+$4837425.html$>$.
}
reported spectra, all of which indicated Balmer lines in absorption
with emission cores for H$\alpha$ and H$\beta$, indicating
that the object is indeed a dwarf nova in outburst.  The relatively
narrow absorption suggested a low-inclination of this object.
U. Munari also reported a spectrum \citep{nak11j1842cbet2818}.
Although there were small-amplitude variations, the object
started rapid fading on September 25 before the development
of superhumps (we call this outburst ``first plateau phase'';
vsnet-alert 13703).  The object brightened again
on October 3 (vsnet-alert 13713) and object further brightened
to the second plateau phase.  During this plateau phase, ordinary
superhumps finally developed (vsnet-alert 13726, 13728, 13729).
Such development of the outburst was unprecedented in dwarf novae.
On October 18, the object enter the rapid fading stage
(vsnet-alert 13775).  The object remained above quiescence
even following the rapid decline, and there was another rebrightening
following the second plateau phase (Katysheva et al. in preparation).

   The times of superhump maxima during the second plateau phase
and post-superoutburst stage are listed in table \ref{tab:j1842oc2011}.
The epochs $E=0, 1$ were recorded during the rise to the second
plateau phase.  The times for $E > 206$ were recorded during
the post-superoutburst stage.  Although the humps were clearly
detected, the cycle counts for the latter maxima were slightly
uncertain.  As seen from $E=413$ and $E=428$, there appeared
to have been double maxima during one superhump cycle.
The short visibility in the evening during this stage hindered
unambiguous identification of the nature of these maxima.
We identified $E \le 64$ as the stage A superhumps because
the superhumps evolved during this stage, and subsequent
superhumps as stage B superhumps.  These stages, however,
may be inadequate considering the peculiar evolution of
the entire outburst.  In determining the period of stage A
superhumps, we disregarded $E=0, 1$.  The period of stage B
superhumps was very stable, and $P_{\rm dot}$ was almost zero.
Although these superhumps bore more characteristics of stage C
superhumps in ordinary SU UMa-type dwarf novae, the identification
of the nature should await future research.  The amplitudes of
superhumps were small (0.045 mag in full amplitude in average),
suggesting a weak manifestation of the tidal instability.

   During the first plateau phase, there was a possible signal
of early superhumps with a period of 0.07168(1) d
(figure \ref{fig:j1842eshpdm}).  Since the signal had only
a small amplitude, exact identification
of the orbital period should await future observations.
Assuming that this period is close to the orbital period,
we obtained an $\epsilon$ of 0.9\%, comparable to those
of short-$P_{\rm orb}$ WZ Sge-type dwarf novae, but is
unusually small for a $P_{\rm orb}=0.07168$~d object.
This might suggest the presence of an anomalously undermassive
secondary and this object could be a good candidate for
a period bouncer.

   The unique feature of the outburst evolution
might be also understood if the mass-ratio is anomalously low
as explained in the following scenario.
(1) The disk initially expanded enough to trigger the 2:1 resonance.
(2) The disk started to cool down before the 3:1 resonance governs
as in ordinary WZ Sge-type dwarf novae, and the object underwent
a temporary excursion to a quiescent state, then (3) the 3:1 resonance
started to grow slowly, and triggered a second thermal instability
and entered the second plateau phase.
The second plateau phase apparently started from an inside-out type
outburst, as suggested by the slow rise.  The small amplitude
probably reflect the weak tidal torque resulting from
a low mass-ratio.

   The object resembles in its long $P_{\rm orb}$ and apparently
in its low mass-ratio, suggesting a brown-dwarf secondary,
the famous CV GD 552 (\cite{hes90gd552}; \cite{und08gd552}),
which had never been observed to undergo an outburst 
(cf. \cite{ric90gd552}).  If GD 552 were to undergo an outburst,
we might expect a phenomenon similar to OT J184228.

   A more detailed analysis will be reported in Ohshima et al.,
in preparation.

\begin{figure}
  \begin{center}
    \FigureFile(88mm,110mm){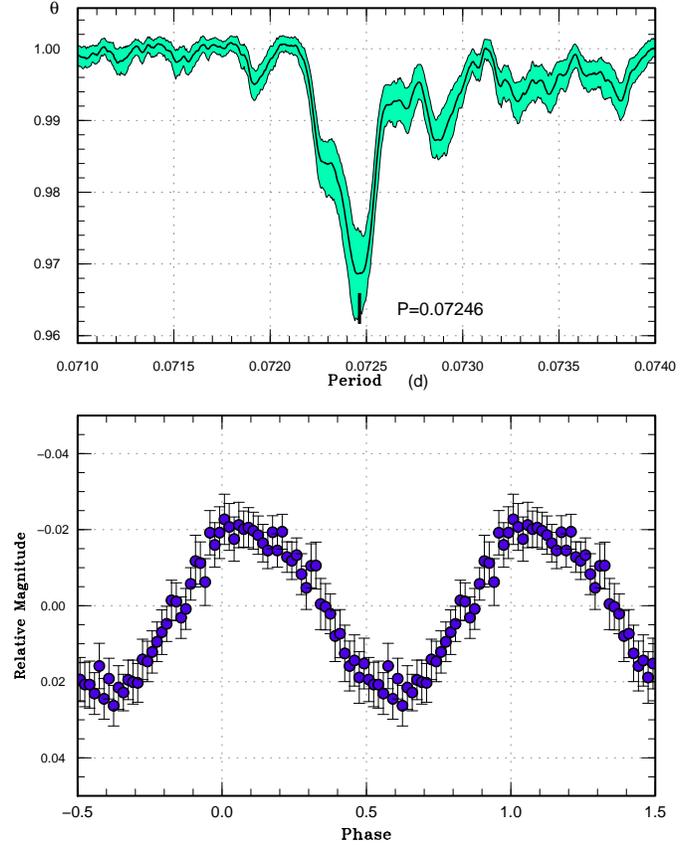}
  \end{center}
  \caption{Superhumps in OT J184228 (2011).
     (Upper): PDM analysis.
     (Lower): Phase-averaged profile.}
  \label{fig:j1842shpdm}
\end{figure}

\begin{figure}
  \begin{center}
    \FigureFile(88mm,110mm){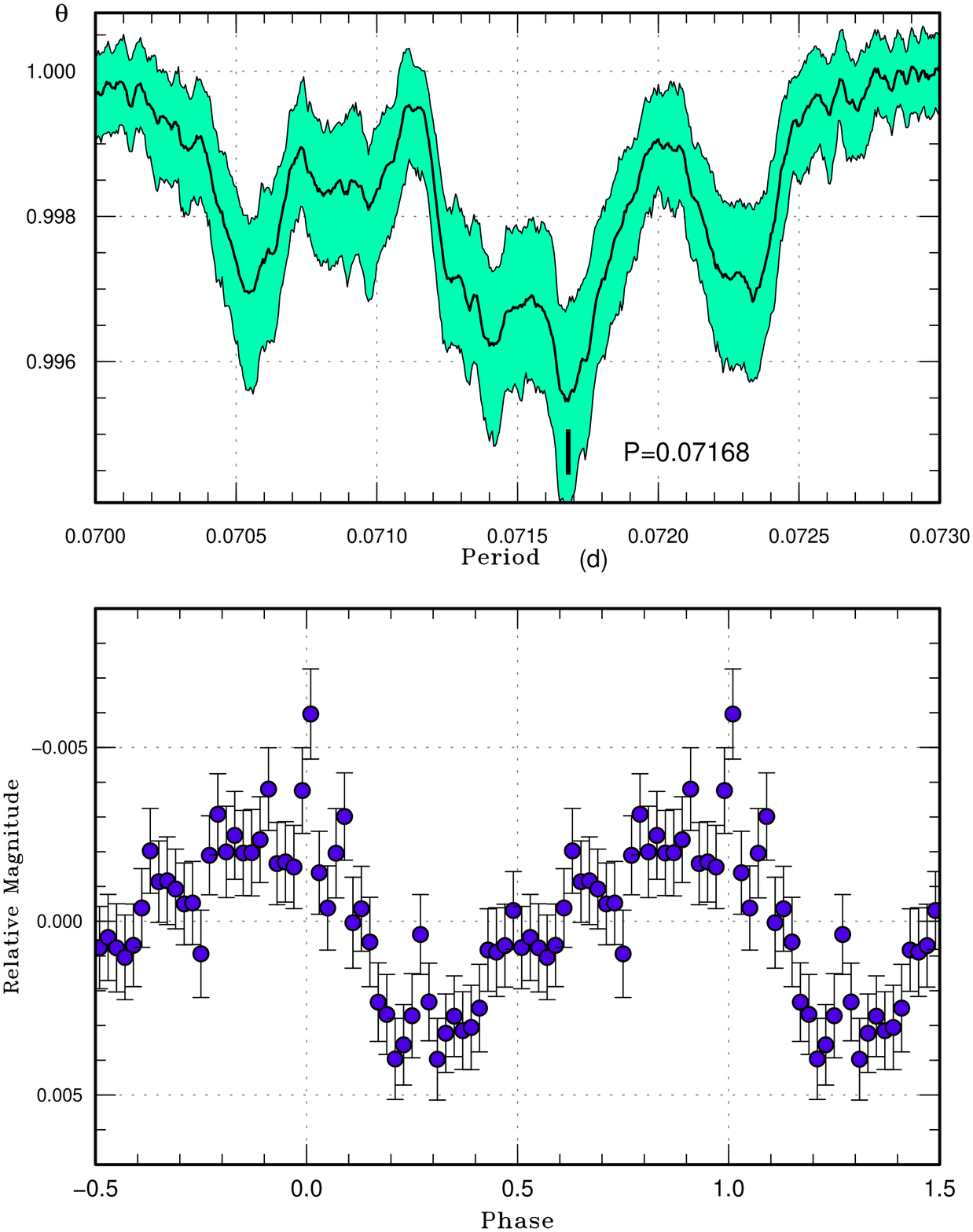}
  \end{center}
  \caption{Possible early superhumps in OT J184228 (2011).
     (Upper): PDM analysis.
     (Lower): Phase-averaged profile.}
  \label{fig:j1842eshpdm}
\end{figure}

\begin{table}
\caption{Superhump maxima of OT J184228 (2011).}\label{tab:j1842oc2011}
\begin{center}
\begin{tabular}{ccccc}
\hline
$E$ & max\commenta & error & $O-C$\commentb & $N$\commentc \\
\hline
0 & 55838.9637 & 0.0022 & $-$0.0366 & 172 \\
1 & 55839.0387 & 0.0005 & $-$0.0341 & 83 \\
17 & 55840.2225 & 0.0011 & $-$0.0097 & 164 \\
18 & 55840.2982 & 0.0009 & $-$0.0065 & 322 \\
19 & 55840.3668 & 0.0007 & $-$0.0104 & 270 \\
22 & 55840.5828 & 0.0017 & $-$0.0117 & 51 \\
31 & 55841.2434 & 0.0017 & $-$0.0033 & 104 \\
32 & 55841.3122 & 0.0009 & $-$0.0070 & 76 \\
33 & 55841.3884 & 0.0003 & $-$0.0032 & 211 \\
34 & 55841.4617 & 0.0008 & $-$0.0024 & 134 \\
35 & 55841.5395 & 0.0019 & 0.0029 & 91 \\
36 & 55841.6068 & 0.0005 & $-$0.0023 & 154 \\
37 & 55841.6791 & 0.0008 & $-$0.0024 & 113 \\
41 & 55841.9707 & 0.0011 & $-$0.0007 & 221 \\
42 & 55842.0431 & 0.0015 & $-$0.0007 & 355 \\
43 & 55842.1302 & 0.0046 & 0.0139 & 36 \\
45 & 55842.2669 & 0.0011 & 0.0057 & 76 \\
46 & 55842.3374 & 0.0019 & 0.0037 & 76 \\
49 & 55842.5624 & 0.0008 & 0.0113 & 118 \\
50 & 55842.6314 & 0.0006 & 0.0079 & 131 \\
56 & 55843.0700 & 0.0024 & 0.0117 & 87 \\
58 & 55843.1943 & 0.0071 & $-$0.0090 & 56 \\
59 & 55843.2815 & 0.0018 & 0.0058 & 76 \\
60 & 55843.3504 & 0.0022 & 0.0022 & 75 \\
63 & 55843.5736 & 0.0010 & 0.0080 & 164 \\
64 & 55843.6538 & 0.0017 & 0.0158 & 124 \\
65 & 55843.7276 & 0.0020 & 0.0171 & 65 \\
68 & 55843.9248 & 0.0013 & $-$0.0030 & 172 \\
69 & 55843.9854 & 0.0023 & $-$0.0149 & 165 \\
73 & 55844.3040 & 0.0008 & 0.0139 & 60 \\
74 & 55844.3727 & 0.0011 & 0.0100 & 74 \\
75 & 55844.4453 & 0.0010 & 0.0102 & 72 \\
76 & 55844.5252 & 0.0010 & 0.0176 & 47 \\
77 & 55844.5898 & 0.0011 & 0.0098 & 47 \\
82 & 55844.9495 & 0.0010 & 0.0072 & 114 \\
84 & 55845.0895 & 0.0020 & 0.0022 & 104 \\
87 & 55845.3107 & 0.0012 & 0.0060 & 73 \\
88 & 55845.3790 & 0.0015 & 0.0019 & 72 \\
89 & 55845.4512 & 0.0022 & 0.0016 & 64 \\
100 & 55846.2521 & 0.0009 & 0.0054 & 157 \\
101 & 55846.3275 & 0.0005 & 0.0083 & 254 \\
102 & 55846.4008 & 0.0008 & 0.0092 & 283 \\
103 & 55846.4745 & 0.0024 & 0.0103 & 46 \\
109 & 55846.9046 & 0.0013 & 0.0057 & 74 \\
114 & 55847.2699 & 0.0015 & 0.0087 & 226 \\
115 & 55847.3359 & 0.0017 & 0.0022 & 280 \\
128 & 55848.2806 & 0.0014 & 0.0049 & 70 \\
129 & 55848.3479 & 0.0012 & $-$0.0002 & 144 \\
130 & 55848.4222 & 0.0020 & 0.0015 & 145 \\
141 & 55849.2102 & 0.0164 & $-$0.0075 & 109 \\
142 & 55849.2933 & 0.0024 & 0.0030 & 257 \\
\hline
  \multicolumn{5}{l}{\commenta BJD$-$2400000.} \\
  \multicolumn{5}{l}{\commentb Against max $= 2455839.0003 + 0.072464 E$.} \\
  \multicolumn{5}{l}{\commentc Number of points used to determine the maximum.} \\
\end{tabular}
\end{center}
\end{table}

\addtocounter{table}{-1}
\begin{table}
\caption{Superhump maxima of OT J184228 (2011) (continued).}
\begin{center}
\begin{tabular}{ccccc}
\hline
$E$ & max\commenta & error & $O-C$\commentb & $N$\commentc \\
\hline
143 & 55849.3628 & 0.0013 & 0.0001 & 216 \\
144 & 55849.4368 & 0.0021 & 0.0016 & 141 \\
156 & 55850.3115 & 0.0015 & 0.0068 & 156 \\
157 & 55850.3759 & 0.0013 & $-$0.0013 & 138 \\
158 & 55850.4445 & 0.0018 & $-$0.0051 & 138 \\
165 & 55850.9518 & 0.0014 & $-$0.0051 & 150 \\
170 & 55851.3083 & 0.0032 & $-$0.0109 & 54 \\
171 & 55851.3874 & 0.0041 & $-$0.0043 & 71 \\
183 & 55852.2622 & 0.0018 & 0.0009 & 187 \\
184 & 55852.3280 & 0.0007 & $-$0.0057 & 271 \\
185 & 55852.4003 & 0.0010 & $-$0.0059 & 272 \\
186 & 55852.4717 & 0.0010 & $-$0.0069 & 126 \\
192 & 55852.9039 & 0.0011 & $-$0.0095 & 167 \\
193 & 55852.9774 & 0.0014 & $-$0.0085 & 251 \\
194 & 55853.0539 & 0.0018 & $-$0.0044 & 150 \\
198 & 55853.3424 & 0.0045 & $-$0.0058 & 65 \\
199 & 55853.4125 & 0.0026 & $-$0.0082 & 55 \\
198 & 55853.3424 & 0.0045 & $-$0.0058 & 65 \\
206 & 55853.9201 & 0.0018 & $-$0.0078 & 110 \\
226 & 55855.3771 & 0.0010 & $-$0.0001 & 56 \\
321 & 55862.2628 & 0.0015 & 0.0016 & 27 \\
377 & 55866.3262 & 0.0018 & 0.0070 & 19 \\
413 & 55868.9431 & 0.0010 & 0.0152 & 79 \\
428 & 55869.9909 & 0.0009 & $-$0.0241 & 91 \\
472 & 55873.2097 & 0.0053 & 0.0064 & 9 \\
\hline
  \multicolumn{5}{l}{\commenta BJD$-$2400000.} \\
  \multicolumn{5}{l}{\commentb Against max $= 2455839.0003 + 0.072464 E$.} \\
  \multicolumn{5}{l}{\commentc Number of points used to determine the maximum.} \\
\end{tabular}
\end{center}
\end{table}

\subsection{OT J210950.5$+$134840}\label{obj:j2109}

   This object (hereafter OT J210950) was discovered as a possible
nova (=PNV J21095047$+$1348396) by K. Itagaki as an 11.5 mag
(unfiltered CCD magnitude) object \citep{yam11j2109cbet2731}.
Although the initial discovery announcement suggested the absence of
the quiescent counterpart, independent examinations of plate
archives indicated the presence of an 18--19 mag counterpart
(S. Korotkiy, vsnet-alert 13342; E. Guido and G. Sostero,
\cite{yam11j2109cbet2731}, vsnet-alerr 13344).
It was already suggested to be a WZ Sge-type dwarf nova with
an amplitude exceeding 7 mag (vsnet-alert 13341).
D. Denisenko also noted the presence
of an M-dwarf having a common proper motion and the detection
of this object in GALEX UV data (vsnet-alert 13343).
Although only low-amplitude variations were detected soon
after the discovery, superhumps
appeared on May 30, 6~d after the discovery (vsnet-alert 13359).
The obtained period was not suggestive of an extreme
WZ Sge-type dwarf nova.  There was a hint of evolving (double wave)
superhumps on May 28 (vsnet-alert 13363).  The object was
spectroscopically confirmed to be a dwarf nova \citep{yam11j2109cbet2731}.

   The times of superhump maxima are listed in table
\ref{tab:j2109oc2011}.  Distinct stages of A--C are present.
Although the last two epochs were measured after the rapid fading,
the times of maxima were well on the extension of the timings
of stage C superhumps recorded before the rapid fading, and
we consider them as persisting stage C superhumps as we already
reported in earlier papers (\cite{Pdot}; \cite{Pdot2}; \cite{Pdot3}).
The resultant $P_{\rm dot}$ for the stage B was
$+8.5(0.6) \times 10^{-5}$, whose large value is consistent
with the idea that this object is not an extreme
WZ Sge-type dwarf nova.
No post-superoutburst rebrightening was recorded.
CRTS data\footnote{
$<$http://nesssi.cacr.caltech.edu/catalina/20110606/\\1106061121124182967p.html$>$
}
did not record a prior outburst, and the data indicated
that the object remained brighter than quiescence five months
after the outburst.  These features suggest that the outburst
frequency is low, and the presence of a long-fading tail looks like
those of WZ Sge-type dwarf novae (e.g. GW Lib, figure 33 in \cite{Pdot}).
The object may be a WZ Sge-type dwarf nova with non-extreme
properties, and showed a type-D superoutburst in terms of
the lack of a post-superoutburst rebrightening.
The $P_{\rm dot}$-$\epsilon$ relation (equation 6 in \cite{Pdot})
suggests $\epsilon$ of 2.6\%.

\begin{figure}
  \begin{center}
    \FigureFile(88mm,110mm){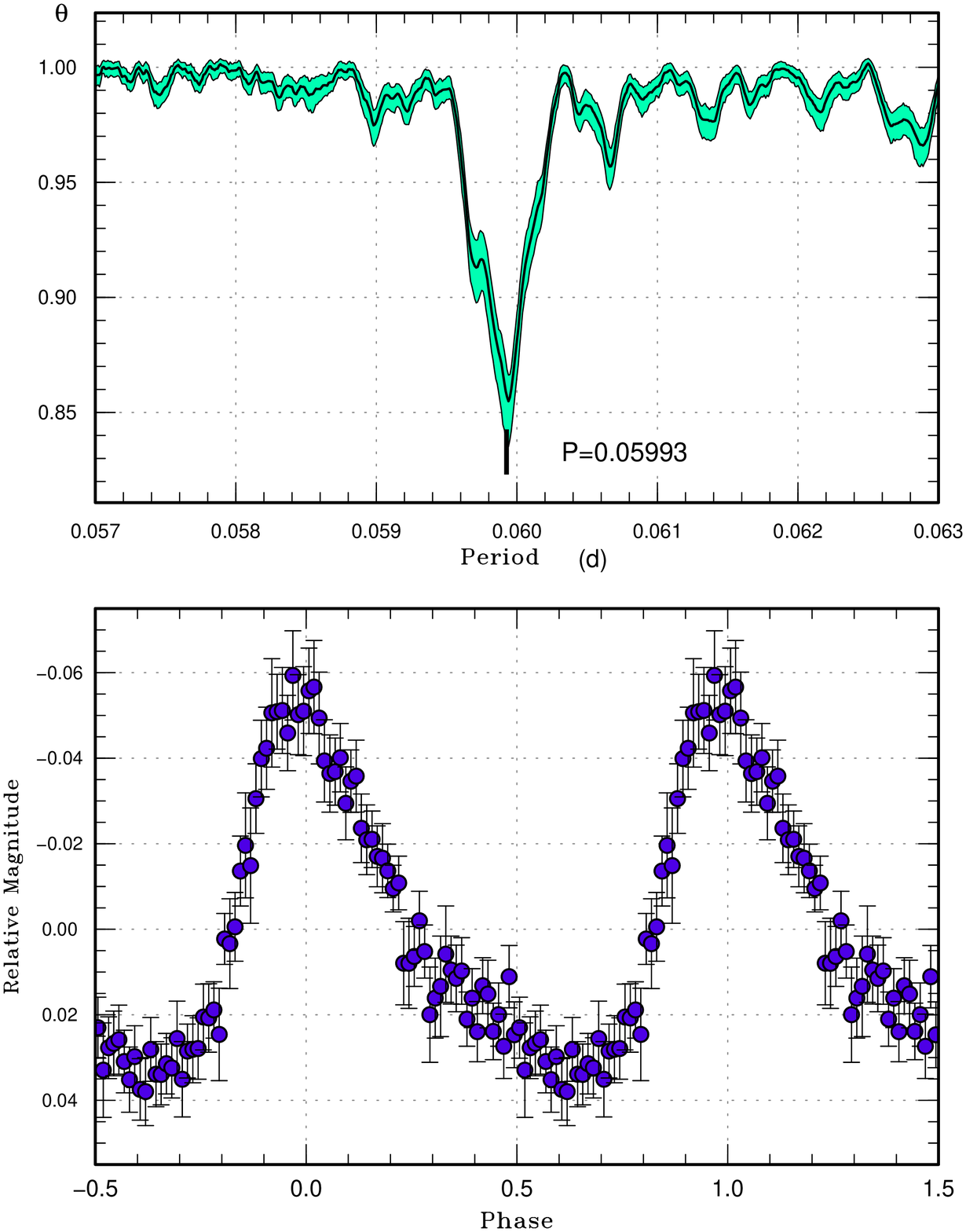}
  \end{center}
  \caption{Superhumps in OT J210950 (2011). (Upper): PDM analysis.
     (Lower): Phase-averaged profile.}
  \label{fig:j2109shpdm}
\end{figure}

   In the PDM analysis (figure \ref{fig:j2109shpdm}) there seems
to be a slightly enhanced signal shorter than $P_{\rm SH}$, we employed
lasso analysis to detect the possible $P_{\rm orb}$.
The obtained candidate period was 0.05865(1)~d, suggesting
$\epsilon$ for stage B superhumps of 2.4\%.  Although this period
is close to what was expected from the $P_{\rm dot}$-$\epsilon$ relation,
it needs to be tested by future observations.

\begin{table}
\caption{Superhump Maxima of OT J210950 (2011).}\label{tab:j2109oc2011}
\begin{center}
\begin{tabular}{ccccc}
\hline
$E$ & max\commenta & error & $O-C$\commentb & $N$\commentc \\
\hline
0 & 55710.5319 & 0.0009 & $-$0.0188 & 168 \\
1 & 55710.5897 & 0.0008 & $-$0.0211 & 168 \\
27 & 55712.1758 & 0.0006 & 0.0044 & 96 \\
28 & 55712.2356 & 0.0003 & 0.0042 & 92 \\
34 & 55712.5982 & 0.0002 & 0.0066 & 104 \\
37 & 55712.7804 & 0.0003 & 0.0088 & 87 \\
38 & 55712.8401 & 0.0002 & 0.0084 & 104 \\
39 & 55712.9005 & 0.0002 & 0.0088 & 122 \\
54 & 55713.7970 & 0.0002 & 0.0049 & 170 \\
55 & 55713.8564 & 0.0003 & 0.0043 & 87 \\
71 & 55714.8117 & 0.0004 & $-$0.0008 & 86 \\
72 & 55714.8706 & 0.0004 & $-$0.0019 & 86 \\
77 & 55715.1733 & 0.0008 & 0.0006 & 38 \\
78 & 55715.2326 & 0.0004 & $-$0.0001 & 48 \\
84 & 55715.5900 & 0.0003 & $-$0.0028 & 115 \\
87 & 55715.7693 & 0.0004 & $-$0.0036 & 61 \\
88 & 55715.8315 & 0.0004 & $-$0.0015 & 74 \\
89 & 55715.8904 & 0.0004 & $-$0.0026 & 75 \\
104 & 55716.7890 & 0.0004 & $-$0.0043 & 72 \\
105 & 55716.8497 & 0.0002 & $-$0.0036 & 146 \\
106 & 55716.9099 & 0.0004 & $-$0.0035 & 142 \\
117 & 55717.5703 & 0.0003 & $-$0.0033 & 168 \\
117 & 55717.5703 & 0.0003 & $-$0.0034 & 167 \\
121 & 55717.8098 & 0.0006 & $-$0.0040 & 106 \\
122 & 55717.8708 & 0.0008 & $-$0.0030 & 82 \\
128 & 55718.2312 & 0.0005 & $-$0.0027 & 101 \\
148 & 55719.4352 & 0.0008 & 0.0008 & 120 \\
165 & 55720.4667 & 0.0013 & 0.0119 & 122 \\
166 & 55720.5259 & 0.0010 & 0.0110 & 133 \\
187 & 55721.7886 & 0.0020 & 0.0132 & 68 \\
188 & 55721.8463 & 0.0010 & 0.0109 & 69 \\
220 & 55723.7585 & 0.0029 & 0.0023 & 35 \\
221 & 55723.8227 & 0.0008 & 0.0064 & 48 \\
222 & 55723.8818 & 0.0013 & 0.0055 & 49 \\
288 & 55727.8236 & 0.0018 & $-$0.0143 & 82 \\
289 & 55727.8802 & 0.0019 & $-$0.0178 & 93 \\
\hline
  \multicolumn{5}{l}{\commenta BJD$-$2400000.} \\
  \multicolumn{5}{l}{\commentb Against max $= 55710.5507 + 0.060025 E$.} \\
  \multicolumn{5}{l}{\commentc Number of points used to determine the maximum.} \\
\end{tabular}
\end{center}
\end{table}

\subsection{OT J214738.4$+$244553}\label{obj:j2147}

\begin{figure}
  \begin{center}
    \FigureFile(88mm,110mm){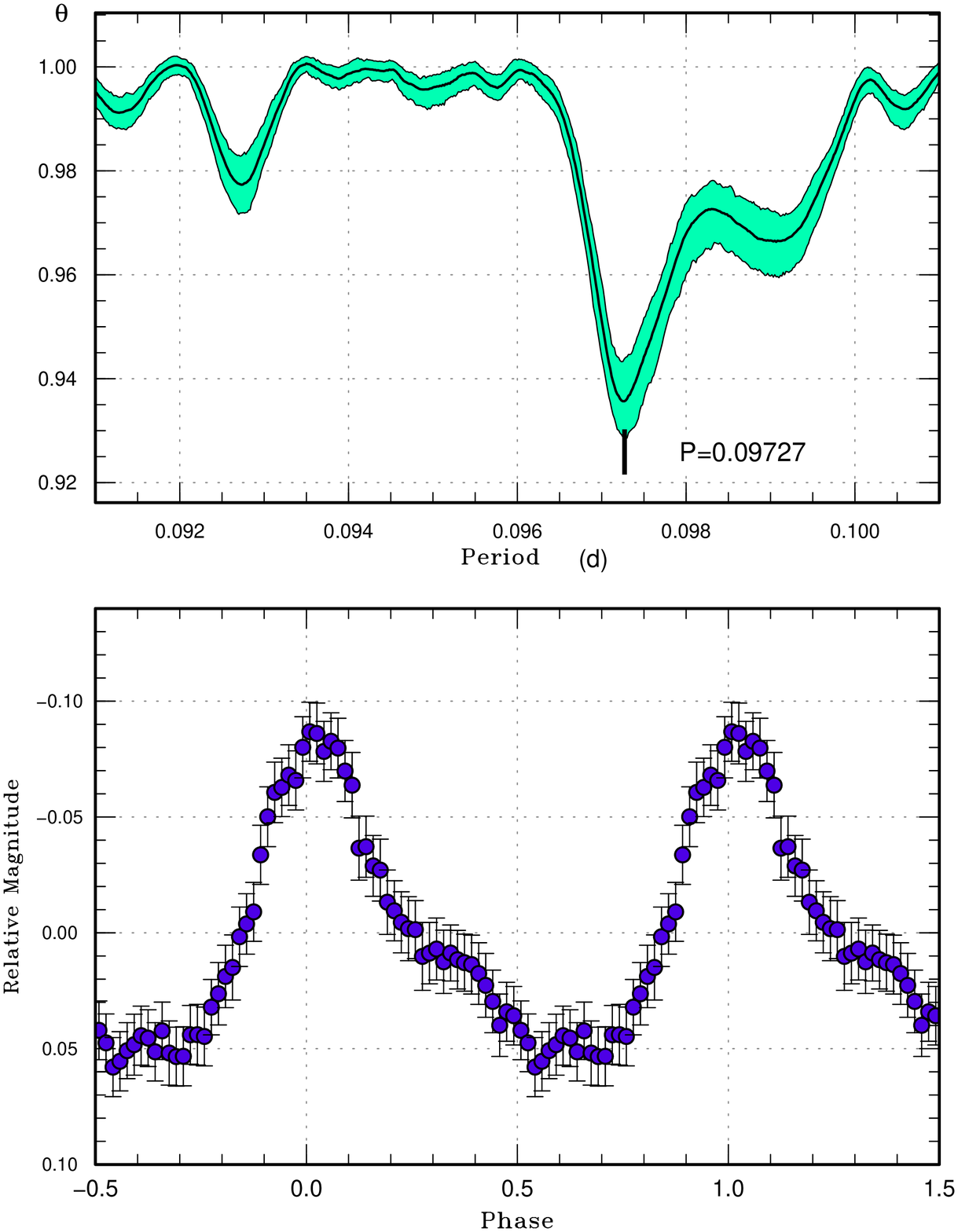}
  \end{center}
  \caption{Superhumps in OT J214738 (2011). (Upper): PDM analysis.
     (Lower): Phase-averaged profile.}
  \label{fig:j2147shpdm}
\end{figure}

   The object was detected as a transient (=CSS111004:214738$+$244554;
hereafter OT J214738) by CRTS on 2011 October 4.  ASAS-3 \citep{ASAS3}
recorded additional three outbursts.  Subsequent observations recorded
superhumps (vsnet-alert 13720, 13721; figure \ref{fig:j2147shpdm}).
The times of superhump maxima are listed in table \ref{tab:j2147oc2011}.
Although it is unclear whether we indeed observed the early stage
of the outburst, the superhumps for $E \le 20$ appears to be
stage A superhumps.  There was no clear transition to stage C,
and $P_{\rm dot}$ for $E \le 20$ was $+8.8(1.0) \times 10^{-5}$
anomalously high for a long-$P_{\rm SH}$ system.
The behavior resembles those of SDSS J170213 (subsection \ref{obj:j1702})
and GX Cas \citep{Pdot3}.
There is a strong signal at 0.09273(3)~d (see figure \ref{fig:j2147shpdm}), 
which we interpret as the orbital period.  Assuming this period,
the $\epsilon$ for the mean $P_{\rm SH}$ is 4.9\%.

\begin{table}
\caption{Superhump maxima of OT J214738 (2011).}\label{tab:j2147oc2011}
\begin{center}
\begin{tabular}{ccccc}
\hline
$E$ & max\commenta & error & $O-C$\commentb & $N$\commentc \\
\hline
0 & 55839.3213 & 0.0002 & $-$0.0289 & 244 \\
1 & 55839.4206 & 0.0002 & $-$0.0270 & 250 \\
10 & 55840.3196 & 0.0002 & $-$0.0038 & 253 \\
11 & 55840.4178 & 0.0003 & $-$0.0029 & 297 \\
12 & 55840.5161 & 0.0003 & $-$0.0019 & 252 \\
13 & 55840.6173 & 0.0005 & 0.0020 & 260 \\
14 & 55840.7175 & 0.0004 & 0.0049 & 140 \\
21 & 55841.4019 & 0.0002 & 0.0081 & 408 \\
22 & 55841.5001 & 0.0003 & 0.0090 & 237 \\
24 & 55841.6948 & 0.0003 & 0.0090 & 175 \\
25 & 55841.7914 & 0.0002 & 0.0083 & 186 \\
27 & 55841.9864 & 0.0002 & 0.0087 & 160 \\
28 & 55842.0864 & 0.0003 & 0.0114 & 88 \\
31 & 55842.3748 & 0.0005 & 0.0078 & 152 \\
32 & 55842.4717 & 0.0003 & 0.0074 & 327 \\
33 & 55842.5686 & 0.0003 & 0.0070 & 135 \\
34 & 55842.6645 & 0.0003 & 0.0056 & 237 \\
35 & 55842.7619 & 0.0006 & 0.0057 & 64 \\
44 & 55843.6347 & 0.0006 & 0.0027 & 60 \\
45 & 55843.7323 & 0.0004 & 0.0030 & 69 \\
51 & 55844.3131 & 0.0006 & $-$0.0001 & 110 \\
52 & 55844.4106 & 0.0007 & 0.0001 & 137 \\
53 & 55844.5068 & 0.0008 & $-$0.0011 & 96 \\
55 & 55844.7015 & 0.0002 & $-$0.0009 & 251 \\
56 & 55844.7976 & 0.0002 & $-$0.0022 & 212 \\
65 & 55845.6732 & 0.0006 & $-$0.0024 & 142 \\
66 & 55845.7686 & 0.0003 & $-$0.0043 & 186 \\
71 & 55846.2559 & 0.0004 & $-$0.0036 & 163 \\
72 & 55846.3539 & 0.0003 & $-$0.0029 & 381 \\
73 & 55846.4503 & 0.0003 & $-$0.0038 & 403 \\
75 & 55846.6462 & 0.0051 & $-$0.0026 & 94 \\
76 & 55846.7457 & 0.0004 & $-$0.0004 & 183 \\
82 & 55847.3285 & 0.0003 & $-$0.0015 & 180 \\
86 & 55847.7168 & 0.0004 & $-$0.0024 & 183 \\
87 & 55847.8098 & 0.0007 & $-$0.0067 & 166 \\
96 & 55848.6900 & 0.0007 & $-$0.0024 & 101 \\
97 & 55848.7888 & 0.0007 & $-$0.0009 & 102 \\
107 & 55849.7643 & 0.0017 & 0.0015 & 24 \\
\hline
  \multicolumn{5}{l}{\commenta BJD$-$2400000.} \\
  \multicolumn{5}{l}{\commentb Against max $= 2455839.3502 + 0.097313 E$.} \\
  \multicolumn{5}{l}{\commentc Number of points used to determine the maximum.} \\
\end{tabular}
\end{center}
\end{table}

\subsection{OT J215818.5$+$241925}\label{obj:j2158}

\begin{figure}
  \begin{center}
    \FigureFile(88mm,110mm){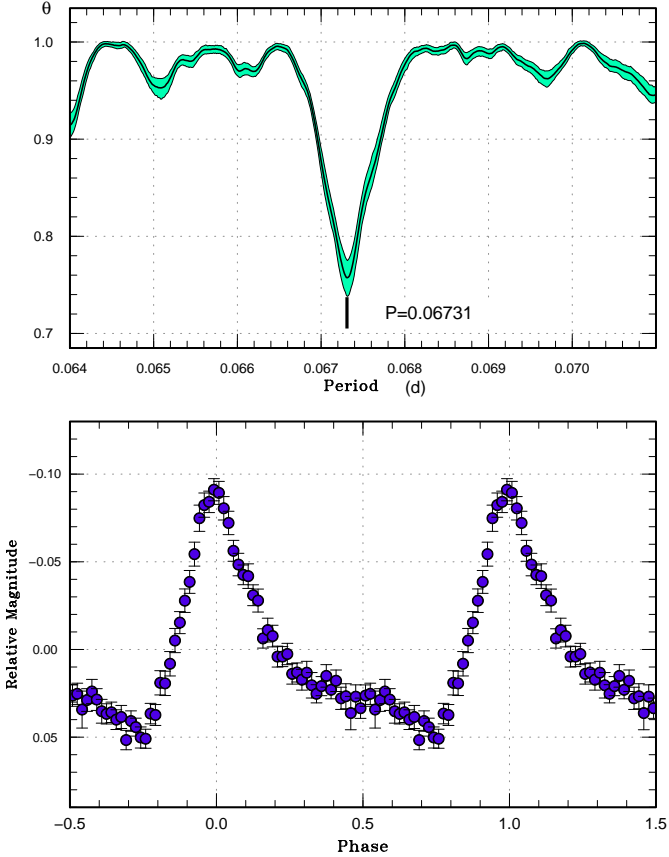}
  \end{center}
  \caption{Superhumps in OT J215818 (2011). (Upper): PDM analysis.
     (Lower): Phase-averaged profile.}
  \label{fig:j215818shpdm}
\end{figure}

   This object (hereafter OT J215818) is an object reported
by G. Sun and X. Gao to
Central Bureau for Astronomical Telegrams (CBAT) Transient Objects 
Confirmation Page (TOCP) originally suspected to be a nova
(=PNV J21581852$+$2419246).  Soon after the discovery, R. Koff\footnote{
$<$http://www.cbat.eps.harvard.edu/unconf/followups/\\J21581852$+$2419246.html$>$.
}
detected modulations similar to superhumps.  This finding was confirmed
by subsequent observations (vsnet-alert 13803, 13805, 13807;
figure \ref{fig:j215818shpdm}),
and the dwarf nova-type nature was confirmed.
The times of superhump maxima are listed in table \ref{tab:j2158oc2011}.
The $O-C$ diagram shows typical stage B and C superhumps.
The $P_{\rm dot}$ for stage B superhumps was not meaningfully
determined because the outburst was apparently observed only
during its late course.  \citet{she12j2158} reported a $P_{\rm dot}$
of 0.06728(21)~d using a slightly different data set and obtained
a similar pattern of $O-C$ variation with this analysis.
\citet{she12j2158} also reported the possible presence of
an orbital signal at 0.06606(35)~d.  Our analysis yielded only 
a very weak signal (see figure \ref{fig:j215818shpdm}), and
it appears to be still inconclusive.
Using lasso, we measured a period of 0.06607(5)~d assuming that
it is a real periodicity.

\begin{table}
\caption{Superhump maxima of OT J215818 (2011).}\label{tab:j2158oc2011}
\begin{center}
\begin{tabular}{ccccc}
\hline
$E$ & max\commenta & error & $O-C$\commentb & $N$\commentc \\
\hline
0 & 55863.2398 & 0.0003 & $-$0.0043 & 134 \\
1 & 55863.3048 & 0.0002 & $-$0.0064 & 231 \\
2 & 55863.3696 & 0.0004 & $-$0.0088 & 277 \\
3 & 55863.4404 & 0.0002 & $-$0.0051 & 438 \\
4 & 55863.5061 & 0.0003 & $-$0.0065 & 185 \\
17 & 55864.3810 & 0.0005 & $-$0.0039 & 78 \\
18 & 55864.4489 & 0.0002 & $-$0.0031 & 209 \\
19 & 55864.5162 & 0.0003 & $-$0.0029 & 201 \\
20 & 55864.5814 & 0.0006 & $-$0.0048 & 204 \\
21 & 55864.6512 & 0.0003 & $-$0.0021 & 318 \\
22 & 55864.7198 & 0.0004 & $-$0.0006 & 259 \\
23 & 55864.7865 & 0.0009 & $-$0.0010 & 99 \\
30 & 55865.2517 & 0.0023 & $-$0.0056 & 31 \\
31 & 55865.3245 & 0.0007 & 0.0002 & 112 \\
32 & 55865.3927 & 0.0004 & 0.0012 & 193 \\
33 & 55865.4587 & 0.0004 & 0.0002 & 195 \\
34 & 55865.5280 & 0.0005 & 0.0023 & 155 \\
36 & 55865.6653 & 0.0004 & 0.0054 & 175 \\
37 & 55865.7313 & 0.0004 & 0.0044 & 175 \\
40 & 55865.9351 & 0.0005 & 0.0068 & 119 \\
41 & 55865.9970 & 0.0009 & 0.0016 & 114 \\
50 & 55866.6093 & 0.0004 & 0.0100 & 179 \\
51 & 55866.6757 & 0.0003 & 0.0093 & 226 \\
52 & 55866.7436 & 0.0004 & 0.0101 & 176 \\
55 & 55866.9400 & 0.0008 & 0.0052 & 116 \\
56 & 55867.0100 & 0.0005 & 0.0081 & 74 \\
61 & 55867.3408 & 0.0008 & 0.0034 & 147 \\
62 & 55867.4088 & 0.0005 & 0.0042 & 193 \\
63 & 55867.4778 & 0.0006 & 0.0061 & 143 \\
80 & 55868.6185 & 0.0009 & 0.0061 & 91 \\
81 & 55868.6812 & 0.0010 & 0.0017 & 93 \\
125 & 55871.6216 & 0.0008 & $-$0.0105 & 127 \\
126 & 55871.6902 & 0.0008 & $-$0.0090 & 176 \\
127 & 55871.7544 & 0.0018 & $-$0.0119 & 148 \\
\hline
  \multicolumn{5}{l}{\commenta BJD$-$2400000.} \\
  \multicolumn{5}{l}{\commentb Against max $= 2455863.2441 + 0.067104 E$.} \\
  \multicolumn{5}{l}{\commentc Number of points used to determine the maximum.} \\
\end{tabular}
\end{center}
\end{table}

\subsection{OT J221232.0$+$160140}\label{obj:j2212}

\begin{figure}
  \begin{center}
    \FigureFile(88mm,110mm){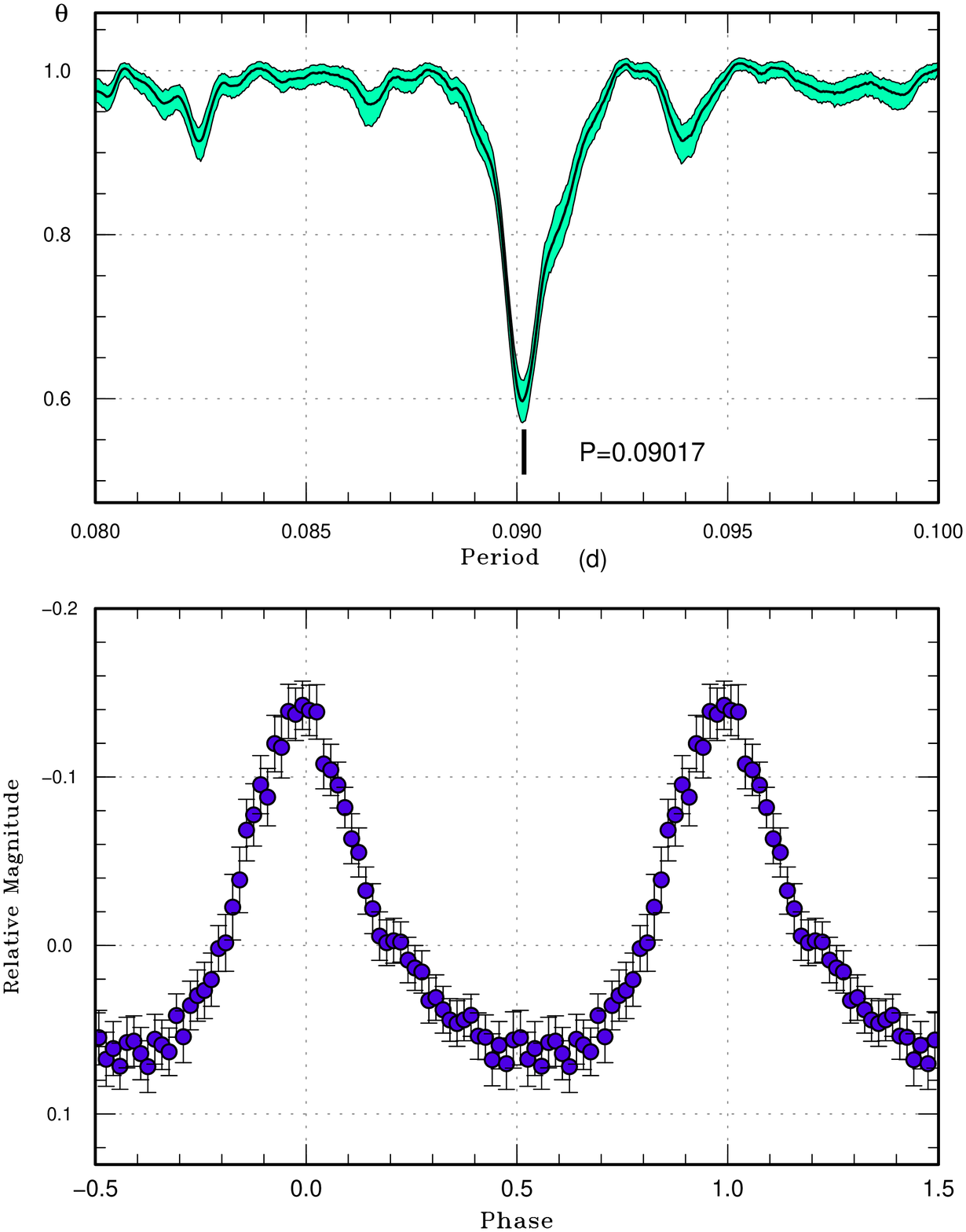}
  \end{center}
  \caption{Superhumps in OT J221232 (2011). (Upper): PDM analysis.
     (Lower): Phase-averaged profile.}
  \label{fig:j2212shpdm}
\end{figure}

   The object was detected as a transient (=CSS 090911:221232$+$160140;
hereafter OT J221232) by CRTS on 2009 September 11.
A bright outburst was detected on 2011 December 23 by E. Muyllaert
(BAAVSS alert 2804).  Subsequent observations confirmed
the presence of superhumps (vsnet-alert 14017, 14018, 14020;
figure \ref{fig:j2212shpdm}).
The times of superhump maxima are listed in table \ref{tab:j2212oc2011}.
The early stage of the outburst was not observed.
There was a likely stage B--C transition around $E=29$.

\begin{table}
\caption{Superhump maxima of OT J221232 (2011).}\label{tab:j2212oc2011}
\begin{center}
\begin{tabular}{ccccc}
\hline
$E$ & max\commenta & error & $O-C$\commentb & $N$\commentc \\
\hline
0 & 55921.6566 & 0.0004 & $-$0.0069 & 93 \\
6 & 55922.2042 & 0.0003 & $-$0.0000 & 116 \\
7 & 55922.2925 & 0.0004 & $-$0.0019 & 122 \\
11 & 55922.6537 & 0.0005 & $-$0.0012 & 91 \\
14 & 55922.9302 & 0.0018 & 0.0050 & 45 \\
17 & 55923.1942 & 0.0005 & $-$0.0015 & 48 \\
18 & 55923.2872 & 0.0005 & 0.0014 & 96 \\
22 & 55923.6480 & 0.0005 & 0.0018 & 91 \\
28 & 55924.1862 & 0.0023 & $-$0.0008 & 84 \\
29 & 55924.2797 & 0.0007 & 0.0026 & 221 \\
33 & 55924.6417 & 0.0008 & 0.0040 & 90 \\
55 & 55926.6221 & 0.0006 & 0.0017 & 82 \\
73 & 55928.2400 & 0.0004 & $-$0.0026 & 91 \\
106 & 55931.2153 & 0.0024 & $-$0.0016 & 30 \\
\hline
  \multicolumn{5}{l}{\commenta BJD$-$2400000.} \\
  \multicolumn{5}{l}{\commentb Against max $= 2455921.6635 + 0.090126 E$.} \\
  \multicolumn{5}{l}{\commentc Number of points used to determine the maximum.} \\
\end{tabular}
\end{center}
\end{table}

\subsection{OT J224736.4$+$250436}\label{obj:j2247}

\begin{figure}
  \begin{center}
    \FigureFile(88mm,110mm){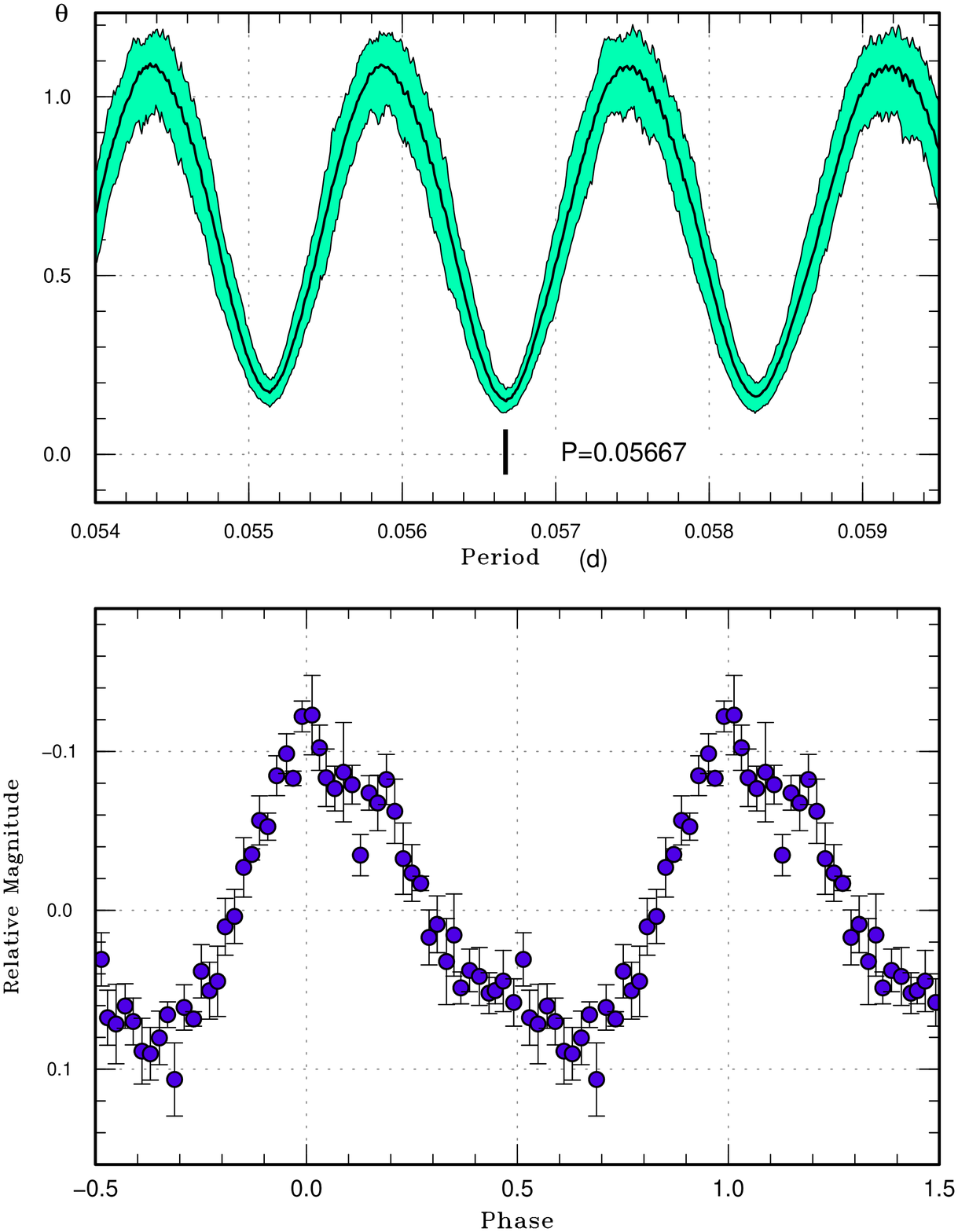}
  \end{center}
  \caption{Superhumps in OT J224736 (2012). (Upper): PDM analysis.
     (Lower): Phase-averaged profile.}
  \label{fig:j2247shpdm}
\end{figure}

   This object (=CSS120616:224736$+$250436, hereafter OT J224736) is 
a transient detected by CRTS on 2012 June 16.  There was a previous
outburst in 2006 November (CRTS data).  An analysis of the SDSS
color using the neural network \citep{kat12DNSDSS} suggested an object
below the period gap (vsnet-alert 14682).  Subsequent observation
detected superhumps (vsnet-alert 14684; figure \ref{fig:j2247shpdm}).
The times of superhump maxima are listed in table \ref{tab:j2247oc2012}.
Due to the gap in observation, the other 2-d aliases are still viable.
We used the period which gave the smallest residuals to superhump maxima
on individual nights.

\begin{table}
\caption{Superhump maxima of OT J224736 (2012).}\label{tab:j2247oc2012}
\begin{center}
\begin{tabular}{ccccc}
\hline
$E$ & max\commenta & error & $O-C$\commentb & $N$\commentc \\
\hline
0 & 56095.4674 & 0.0005 & $-$0.0005 & 27 \\
1 & 56095.5251 & 0.0004 & 0.0005 & 28 \\
36 & 56097.5076 & 0.0007 & $-$0.0006 & 30 \\
37 & 56097.5653 & 0.0005 & 0.0005 & 29 \\
\hline
  \multicolumn{5}{l}{\commenta BJD$-$2400000.} \\
  \multicolumn{5}{l}{\commentb Against max $= 2456095.4679 + 0.056673 E$.} \\
  \multicolumn{5}{l}{\commentc Number of points used to determine the maximum.} \\
\end{tabular}
\end{center}
\end{table}

\subsection{TCP J08461690$+$3115554}\label{obj:j0846}

   This object (hereafter TCP J084616) is a transient discovered
by T. Kryachko et al. on 2012 March 19.\footnote{
$<$http://www.cbat.eps.harvard.edu/unconf/followups/\\J08461690+3115554.html$>$.
}  The object was soon recognized as a deeply eclipsing SU UMa-type
dwarf nova (vsnet-alert 14347, 14348, 14351).
The object has a $g=21.8$ mag SDSS counterpart.
Using the MCMC method (appendix \ref{sec:app:mcmcecl}),
we obtained an eclipse ephemeris of
\begin{equation}
{\rm Min(BJD)} = 2456007.33870(6) + 0.091383(6) E
\label{equ:j0846ecl}.
\end{equation}

\begin{figure}
  \begin{center}
    \FigureFile(88mm,110mm){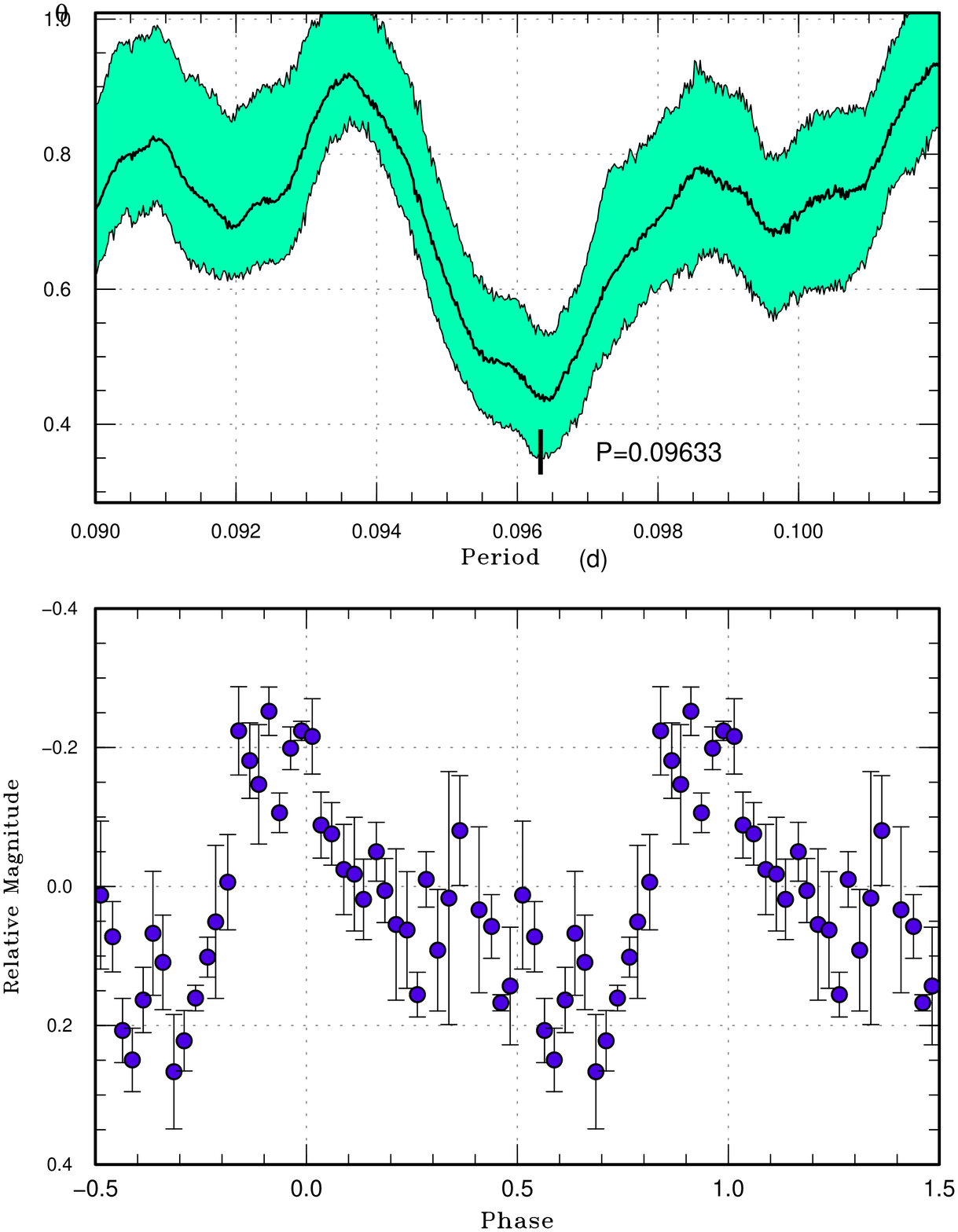}
  \end{center}
  \caption{Superhumps in TCP J084616 (2012). (Upper): PDM analysis.
     (Lower): Phase-averaged profile.}
  \label{fig:j0846shpdm}
\end{figure}

   The times of superhumps maxima are listed in table
\ref{tab:j0846oc2012}.  A PDM analysis (figure \ref{fig:j0846shpdm})
yielded a period of 0.09633(11)~d, giving $\epsilon$ = 5.4(1)\%.

\begin{table}
\caption{Superhump maxima of TCP J084616 (2012).}\label{tab:j0846oc2012}
\begin{center}
\begin{tabular}{cccccc}
\hline
$E$ & max\commenta & error & $O-C$\commentb & phase\commentc & $N$\commentd \\
\hline
0 & 56007.2977 & 0.0006 & $-$0.0009 & 0.35 & 32 \\
1 & 56007.3955 & 0.0014 & 0.0006 & 0.33 & 27 \\
11 & 56008.3626 & 0.0046 & 0.0046 & 0.62 & 35 \\
12 & 56008.4500 & 0.0028 & $-$0.0043 & 0.59 & 40 \\
\hline
  \multicolumn{6}{l}{\commenta BJD$-$2400000.} \\
  \multicolumn{6}{l}{\commentb Against max $= 2456007.2986 + 0.096303 E$.} \\
  \multicolumn{6}{l}{\commentc Orbital phase.} \\
  \multicolumn{6}{l}{\commentd Number of points used to determine the maximum.} \\
\end{tabular}
\end{center}
\end{table}

\subsection{TCP J23130812$+$2337018}\label{obj:j2313}\label{lastobj}

\begin{figure}
  \begin{center}
    \FigureFile(88mm,110mm){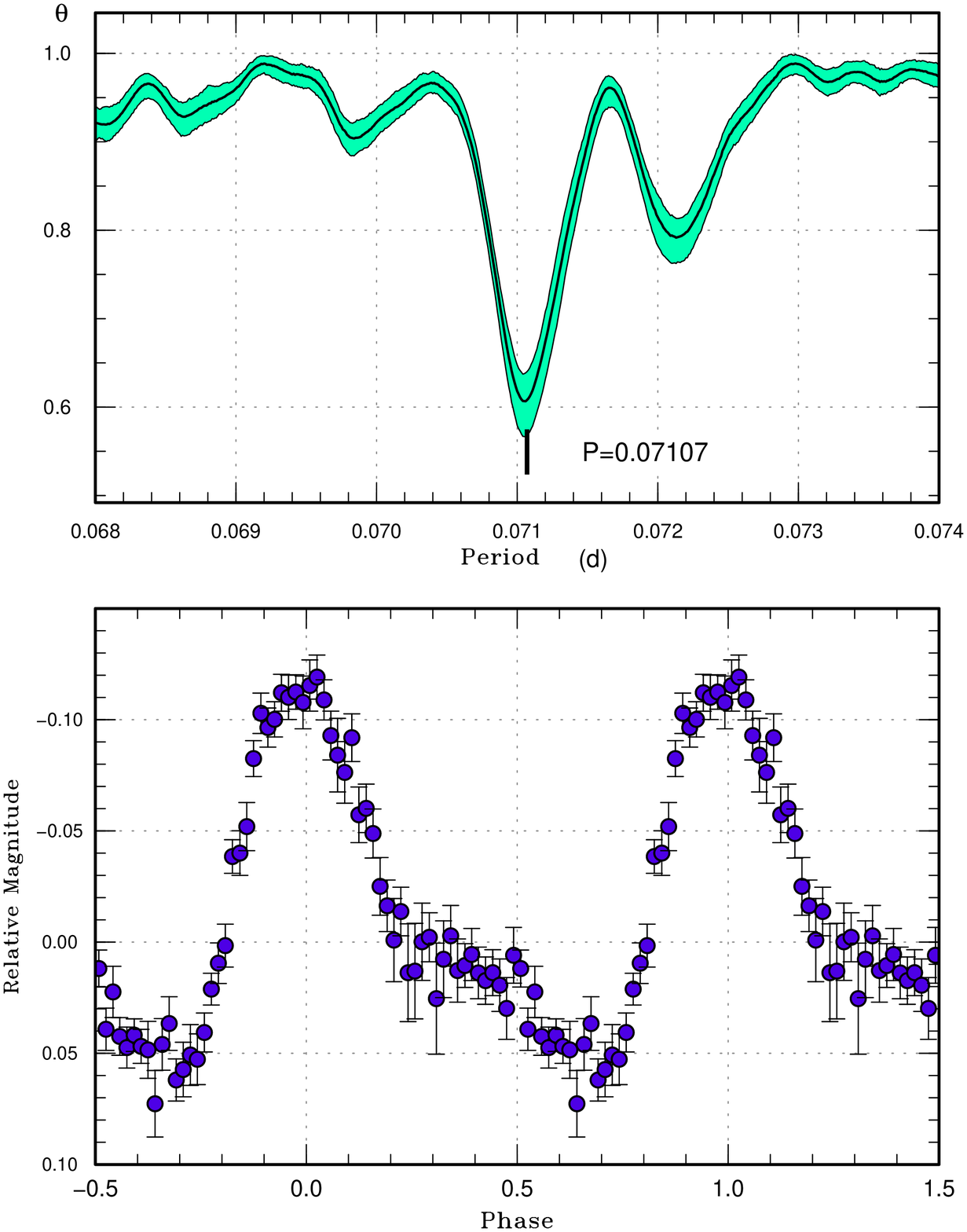}
  \end{center}
  \caption{Superhumps in TCP J231308 (2011). (Upper): PDM analysis.
     (Lower): Phase-averaged profile.}
  \label{fig:j2313shpdm}
\end{figure}

   This object (hereafter TCP J231308) was discovered by Itagaki and Kaneda
as a 14.3 mag (unfiltered CCD magnitude) object
(TCP J23130812$+$2337018).\footnote{
$<$http://www.cbat.eps.harvard.edu/unconf/followups/\\J23130812$+$2337018.html$>$.
}
According to ASAS-3 data, the object underwent a brighter ($V=13.4$)
outburst in 2005 August.  Subsequent observations confirmed the
presence of superhumps (vsnet-alert 13438; figure \ref{fig:j2313shpdm}).
The times of superhump maxima are listed in table
\ref{tab:j2313oc2011}.  The $O-C$ values clearly showed
a stage B--C transition.  The observation of stage B was not
sufficiently long to determine $P_{\rm dot}$.

\begin{table}
\caption{Superhump maxima of TCP J231308 (2011).}\label{tab:j2313oc2011}
\begin{center}
\begin{tabular}{ccccc}
\hline
$E$ & max\commenta & error & $O-C$\commentb & $N$\commentc \\
\hline
0 & 55732.5371 & 0.0003 & $-$0.0036 & 188 \\
1 & 55732.6077 & 0.0002 & $-$0.0041 & 212 \\
14 & 55733.5369 & 0.0003 & 0.0015 & 91 \\
15 & 55733.6066 & 0.0003 & 0.0001 & 59 \\
19 & 55733.8916 & 0.0003 & 0.0009 & 94 \\
23 & 55734.1796 & 0.0013 & 0.0047 & 176 \\
24 & 55734.2488 & 0.0014 & 0.0029 & 118 \\
28 & 55734.5318 & 0.0002 & 0.0017 & 78 \\
29 & 55734.6024 & 0.0004 & 0.0013 & 50 \\
51 & 55736.1610 & 0.0034 & $-$0.0031 & 74 \\
56 & 55736.5193 & 0.0003 & $-$0.0001 & 45 \\
57 & 55736.5888 & 0.0004 & $-$0.0017 & 44 \\
71 & 55737.5845 & 0.0003 & $-$0.0006 & 176 \\
85 & 55738.5799 & 0.0003 & 0.0001 & 199 \\
\hline
  \multicolumn{5}{l}{\commenta BJD$-$2400000.} \\
  \multicolumn{5}{l}{\commentb Against max $= 2455732.5407 + 0.071048 E$.} \\
  \multicolumn{5}{l}{\commentc Number of points used to determine the maximum.} \\
\end{tabular}
\end{center}
\end{table}

\section{Discussion}

\subsection{Period Derivatives during Stage B}\label{sec:stagebpdot}

\begin{figure}
  \begin{center}
    \FigureFile(88mm,77mm){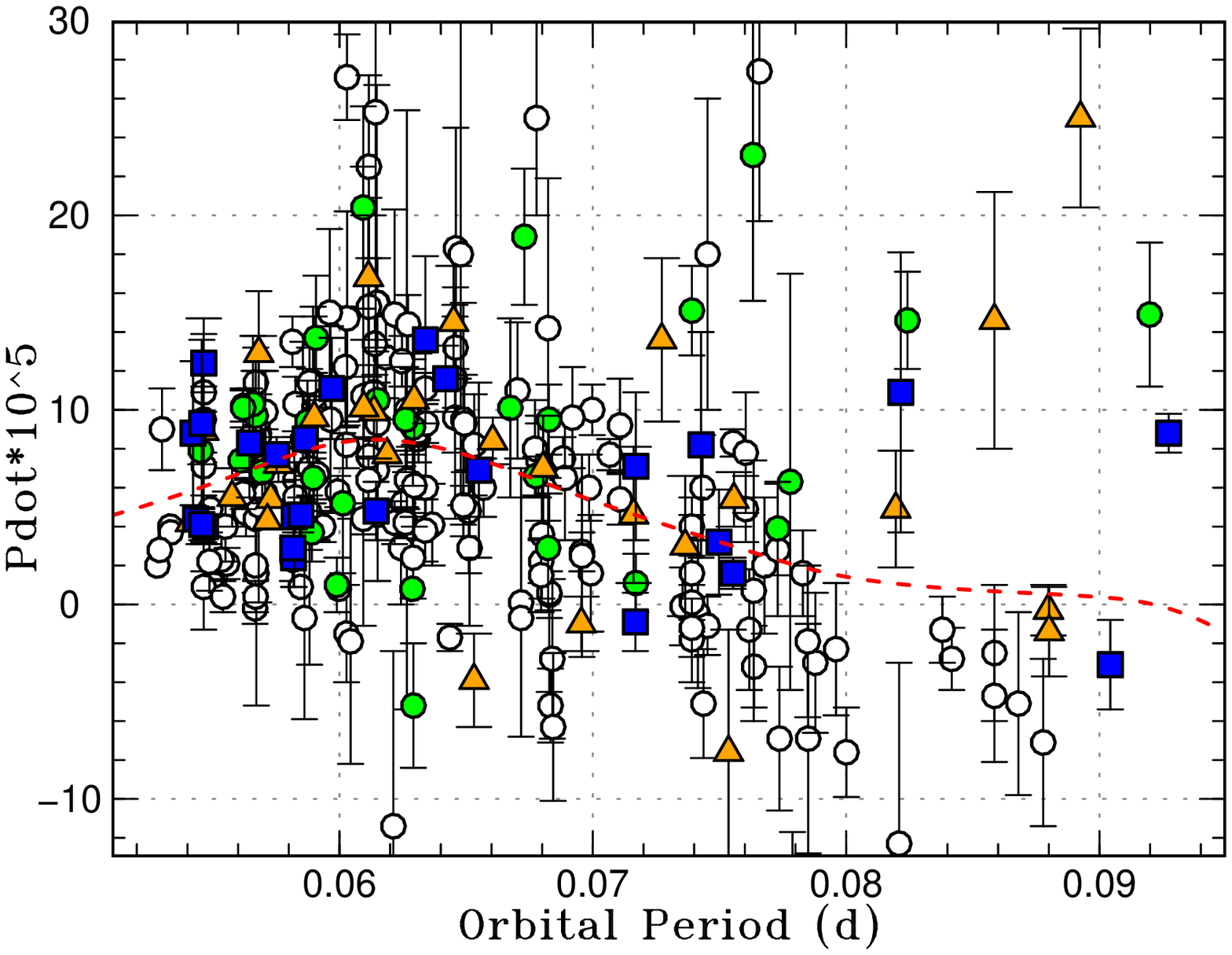}
  \end{center}
  \caption{$P_{\rm dot}$ for stage B versus $P_{\rm orb}$.
  Open circles, filled circles, filled triangles and filled squares
  represent samples in \citet{Pdot}, \citet{Pdot2}, \citet{Pdot3}
  and this paper, respectively.
  The curve represents the spline-smoothed global trend.
  }
  \label{fig:pdotporb4}
\end{figure}

   As in \citet{Pdot3}, we compared period derivatives during stage B
(figure \ref{fig:pdotporb4}).  The newly obtained $P_{\rm dot}$ for
object for $P_{\rm orb} < 0.076$~d followed the trend obtained
in the previous research.  There were also two $P_{\rm dot} > 0$
systems (SDSS J170213 and OT J145921) for $P_{\rm orb} > 0.080$~d,
as noted in \citet{Pdot3}.  Among them, SDSS J170213 showed
infrequent outbursts and indeed resembles EF Peg, the representative
$P_{\rm dot}\sim0$ object with a long $P_{\rm orb}$ (cf. \cite{Pdot}).
The other object, OT J145921, showed more frequent outbursts and
may resemble GX Cas, an unexpected object with $P_{\rm dot} > 0$
with typical SU UMa-type outburst behavior.  There may be two
classes of objects with $P_{\rm dot} > 0$ among SU UMa-type dwarf novae
with long $P_{\rm orb}$.

\subsection{Periods of Stage A Superhumps}

   Stage A superhumps recorded in the present study are listed
in table \ref{tab:pera}.  Although most of objects in this study
followed the trend obtained in the previous study, one object
(SDSS J170213) has a substantially smaller fractional excess for
stage A superhumps.  This may have been either due to small number
of observations (insufficient coverage for stage A), a systematic
effect by overlapping eclipses, or the unusual period evolution
of this object for this $P_{\rm orb}$ (cf. subsection
\ref{sec:stagebpdot}).  Although the object may also resemble 
short-$P_{\rm orb}$ objects in its small fractional excess 
for stage A superhumps, this needs to be confirmed by further
observations.

\begin{table}
\caption{Superhump Periods during Stage A}\label{tab:pera}
\begin{center}
\begin{tabular}{cccc}
\hline
Object & Year & period (d) & err \\
\hline
SV Ari & 2011 & 0.05575 & 0.00012 \\
VW Hyi & 2011 & 0.07770 & 0.00013 \\
BW Scl & 2011 & 0.05623 & 0.00012 \\
PU UMa & 2012 & 0.08382 & -- \\
SDSS J080303 & 2011 & 0.09540 & 0.00028 \\
SDSS J170213 & 2011 & 0.10605 & 0.00011 \\
OT J102842 & 2012 & 0.03844 & 0.00002 \\
OT J184228 & 2011 & 0.07287 & 0.00008 \\
OT J210950 & 2011 & 0.06087 & 0.00006 \\
OT J214738 & 2011 & 0.09928 & 0.00022 \\
\hline
\end{tabular}
\end{center}
\end{table}

\begin{figure}
  \begin{center}
    \FigureFile(88mm,77mm){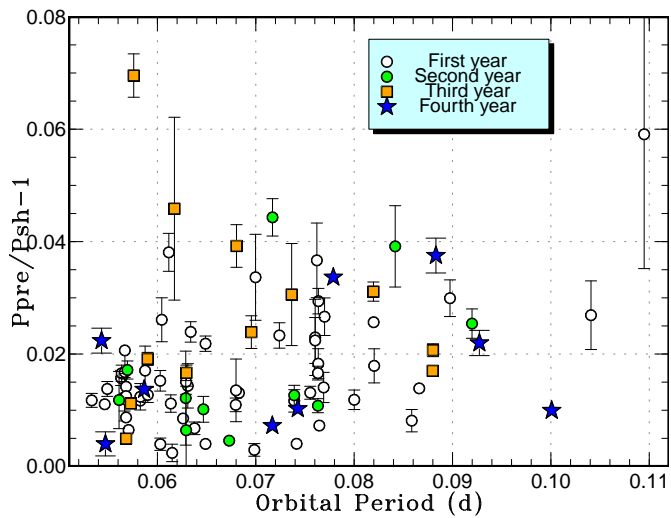}
  \end{center}
  \caption{Superhump periods during the stage A.  Superhumps in this stage
  has a period typically 1.0--1.5\% longer than the one during the stage B.
  There is a slight tendency of increasing fractional period excess for
  longer-$P_{\rm orb}$ systems.  The symbols for first, second, third
  and fourth years represent data in \citet{Pdot}, \citet{Pdot2},
  \citet{Pdot3} and this paper.
  }
  \label{fig:ppre3}
\end{figure}

\subsection{WZ Sge-Type Stars}\label{sec:wzsgestat}

   New WZ Sge-type dwarf novae and candidates are listed in
table \ref{tab:wztab}.  Among them PR Her, BW Scl and OT J184228 were
well characterized.  SV Ari was observed only in the late stage
of its superoutburst, and SDSS J220553 and OT J210950 were included
in this table due to their resemblance to WZ Sge-type objects
in the post-superoutburst behavior.  For OT J001952 and OT J055721
we have only very limited information and these objects were included
based on the large outburst amplitudes.

\begin{figure}
  \begin{center}
    \FigureFile(88mm,70mm){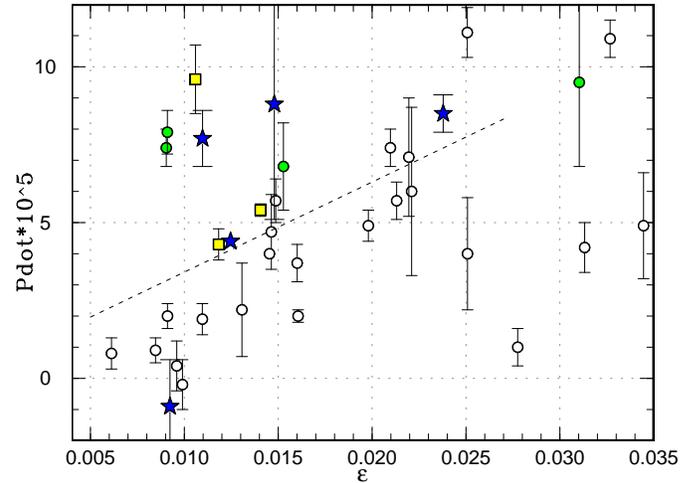}
  \end{center}
  \caption{$P_{\rm dot}$ versus $\epsilon$ for WZ Sge-type
  dwarf novae.  Open circles, filled circles, filled squares and filled stars
  represent outbursts reported in \citet{Pdot}, \citet{Pdot2}, \citet{Pdot3}
  and this paper, respectively.
  The dashed line represents a linear regression for points with
  $\epsilon < 0.026$ as in \citet{Pdot} figure 35.
  }
  \label{fig:wzpdoteps3}
\end{figure}

   The relation between $P_{\rm dot}$ versus $\epsilon$ for WZ Sge-type 
dwarf novae is shown in figure \ref{fig:wzpdoteps3}.  Although
there is a tendency of increasing $P_{\rm dot}$ for objects with
larger $\epsilon$ as stated in \citet{Pdot}, several objects lie
well above this relation, as discussed in \citet{Pdot3}.  Although we may
add two additional examples in the present study, these two objects,
SDSS J220553 and PR Her, were not very well sampled and it is not
certain whether these objects are outliers to this relation.
The reverse case OT J184228 is remarkable in that it showed almost
zero $P_{\rm dot}$.  This object was indeed unusual in its
``double plateau'' superoutburst.  The unusually small $P_{\rm dot}$
may be related to the unusual binary parameters (long $P_{\rm orb}$
and small expected $q$), and probably to its evolutionary stage
as a candidate period bouncer.

\begin{table*}
\caption{Parameters of WZ Sge-type superoutbursts.}\label{tab:wztab}
\begin{center}
\begin{tabular}{cccccccccccc}
\hline
Object & Year & $P_{\rm SH}$ & $P_{\rm orb}$ & $P_{\rm dot}$\commenta & err\commenta & $\epsilon$ & Type\commentb & $N_{\rm reb}$\commentc & delay\commentd & Max & Min \\
\hline
SV Ari & 2011 & 0.055524 & -- & 4.0 & 0.2 & -- & D & 0 & -- & ]15.0\commente & 22.1 \\
PR Her & 2011 & 0.055022 & 0.05422 & 8.8 & 3.7 & 0.015 & -- & -- & 13 & 12.8 & 21.0 \\
BW Scl & 2011 & 0.055000 & 0.054323 & 4.4 & 0.3 & 0.012 & D & 0 & 10 & 9.0 & 16.4 \\
SDSS J220553 & 2011 & 0.058151 & 0.05752 & 7.7 & 0.9 & 0.011 & -- & -- & -- & ]14.4 & 20.1 \\
OT J001952 & 2012 & 0.056770 & -- & -- & -- & -- & -- & -- & -- & ]15.6 & 21.5: \\
OT J055721 & 2011 & 0.059756 & -- & 4.6 & 0.9 & -- & C & 1 & -- & ]14.7 & 21.0: \\
OT J184228 & 2011 & 0.072342 & 0.07168 & $-$0.9 & 1.5 & 0.009 & B & ]2 & ]29 & ]11.8 & 20.6 \\
OT J210950 & 2011 & 0.060045 & 0.05865 & 8.5 & 0.6 & 0.024 & D & 0 & ]6 & ]11.5 & 18.7 \\
\hline
  \multicolumn{12}{l}{\commenta Unit $10^{-5}$.} \\
  \multicolumn{12}{l}{\commentb A: long-lasting rebrightening; B: multiple rebegitehnings; C: single rebrightening; D: no rebrightening.} \\
  \multicolumn{12}{l}{\commentc Number of rebrightenings.} \\
  \multicolumn{12}{l}{\commentd Days before ordinary superhumps appeared.} \\
  \multicolumn{12}{l}{\commente ``]'' represents the lower limit.} \\
\end{tabular}
\end{center}
\end{table*}

\begin{figure}
  \begin{center}
    \FigureFile(88mm,70mm){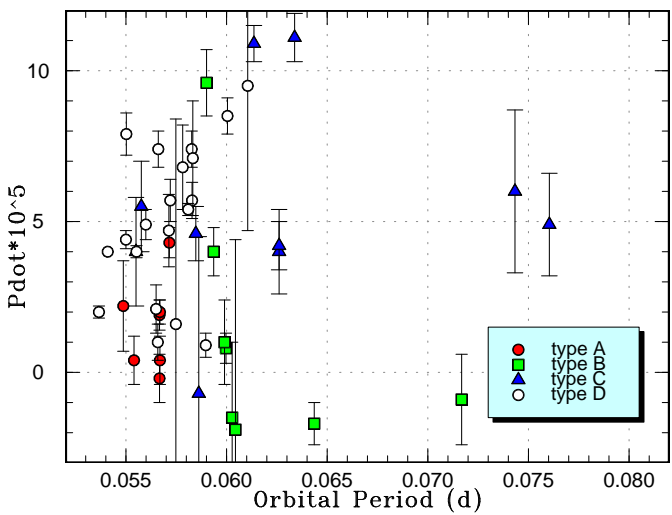}
  \end{center}
  \caption{$P_{\rm dot}$ versus $P_{\rm orb}$ for WZ Sge-type
  dwarf novae.  Symbols represent the type (cf. \cite{Pdot}) of outburst:
  type-A (filled circles), type-B (filled squares),
  type-C (filled triangles), type-D (open circles).
  }
  \label{fig:wzpdottype4}
\end{figure}

   Figure \ref{fig:wzpdottype4} indicates the updated relation
between $P_{\rm dot}$ and $P_{\rm orb}$ and its relation to
the type of post-superoutburst rebrightening phenomenon:
type-A (long-lasting post-outburst rebrightening), type-B (multiple
discrete rebrightenings), type-C (single rebrightening) and type-D
(no rebrightening), cf. \citet{Pdot}.   In the present study,
type-C and type-D superoutbursts followed the same trend as in the
past studies.  There is noteworthy presence of a new type-B
superoutburst (OT J184228) with an exceptionally long $P_{\rm orb}$ and
small $P_{\rm dot}$.  This presence seems to support the earlier
suggestion (\cite{Pdot}; \cite{Pdot3}) that type-B superoutbursts
are associated with low-$q$ systems and they are likely period bouncers.

\subsection{VW Hydri -- Revisiting the Prototype of SU UMa-Type Dwarf novae}\label{sec:vwhyidiscuss}

   In discussing superoutbursts and superhumps, we often refer
to historical observations of bright southern SU UMa-type dwarf
novae (VW Hyi, Z Cha and OY Car), from which
our early knowledge of phenomenology of superhumps was established.
These early findings were also summarized in textbooks such as
\citet{war95book}.  These early observations were, however, based
on photoelectric photometry and only limited parts of the entire
observations were published as figures, and these observations
are not accessible in electronic form.  This has been an obstacle
in comparing the results of modern-day CCD observations with
historical knowledge.  Although Kepler observations of
V344 Lyr and V1504 Cyg (\cite{Pdot3}; \cite{woo11v344lyr}) partly
filled this gap, a direct comparison in the ``prototype'' object
VW Hyi had been wanted.  We have fortunately been able to obtain
the entire course of the 2011 November--December superoutburst
and two subsequent normal outbursts and the intervening quiescent
period, although they were obtained only at a single observing
location and suffered from unavoidable gaps in coverage.
These data are available at the AAVSO database.

   The results of these observations (subsection \ref{obj:vwhyi})
can be summarized as follows:

\begin{enumerate}

\item The superoutburst started with a precursor outburst.

\item There was no hint of superhumps during the rising phase and
the early phase of the precursor outburst.

\item During the later stage of the precursor outburst,
superhumps started to grow and the object brightened.

\item The amplitudes of superhumps reached a maximum when the
object reached the maximum brightness.

\item The amplitudes of superhumps decreased during the
superoutburst plateau.

\item The global $P_{\rm dot}$ was negative.

\item Stage A was recognized during the evolving stage of superhumps.

\item There was likely stage B with an almost constant period, followed
by likely stage C with a shorter period.

\item The transition to stage B to C was smoother than in
short-$P_{\rm orb}$ systems.  This feature is similar to the ones 
in Kepler data for V344 Lyr and V1504 Cyg.

\item Superhumps having phases $\sim$0.5 offset appeared during the
late stage of the superoutburst.  These superhumps indeed appear to be
``traditional'' late superhumps, rather than a simple extension of
stage C superhumps as in short-$P_{\rm orb}$ systems.

\item The late superhumps persisted during the quiescent period
before the next normal outburst and they were still detected
even after this normal outburst.  The signal became undetectable
after the second normal outburst.  The situation was very similar
to the Kepler result for V344 Lyr.

\item In contrast to V344 Lyr, secondary maxima of superhumps
were not prominent.

\end{enumerate}

   We thus confirmed most of the classical superhump phenomenology
including the ``traditional'' late superhumps, whose presence
has been questioned in most of recently observed objects
(\cite{Pdot}; \cite{Pdot2}; \cite{Pdot3}).  Combined with the
knowledge with V344 Lyr, the traditional descriptions of the development
of superhumps and the appearance of late superhumps seem to be
common among relatively long-$P_{\rm orb}$ and high mass-transfer
systems, and VW Hyi can indeed be regarded as the prototype of
these systems.  Things, however, are somewhat different
in shorter-$P_{\rm orb}$ and lower mass-transfer systems
in the clear presence of stages B and C and no clear signature
of ``traditional'' late superhumps.  Although the systematics of 
superhumps in VW Hyi would be adequate for describing
long-$P_{\rm orb}$ systems, we regard it dangerous to describe
the phenomena seen in shorter-$P_{\rm orb}$ systems in the same
manner.  This will be particularly true for the term
``late superhumps'' which has been a major cause of confusion in
describing the stage C superhumps.  Although some authors refer
superhumps seen in the late stages of superoutburst as
late superhumps, this leads to a confusion since the original term
``late superhumps'' implies an $\sim$0.5 phase shift, while stage C
superhumps don't show such a phase shift.  This ambiguity in terminology
could lead to a confusion in interpreting the mechanism [see e.g.
\citet{hes92lateSH}, a work before the clarification between
``traditional'' superhumps and stage C superhumps, analyzed
stage C superhumps with a term of ``late superhumps'' and
an assumption of an $\sim$0.5 phase shift].  We propose that we
should not use the term ``late superhumps'' unless there is
a $\sim$0.5 phase shift.

   \citet{sch04vwhyimodel} recently compared the calculations
of the pure thermal-tidal instability (TTI) model and
the enhanced mass-transfer (EMT) model for VW Hyi.  Although both
models (TTI and EMT) well explain the many of observed
characteristics, they claimed the advantage of the EMT
model in that it can explain varieties in the observed
light curve of single systems such as VW Hyi.  Although we don't
aim to validate or invalidate their claim here, we need to be
specially careful in interpreting observations.  They referred to
varieties of light curves based on historical visual observations,
and these observations may have not been very sensitive to subtle signatures
of light curves.  For example, while the present superoutburst
showed a clear signature of a precursor, visual observations
of the same superoutburst reported to the AAVSO did not recognize 
this feature.  Considering that all six superoutbursts of V344 Lyr and
all six superoutbursts of V1504 Cyg showed precursors in Kepler data
\citep{can12v344lyr}, and considering that these light curves
are very similar to the present one of VW Hyi (figure \ref{fig:vwhyihumpall},
lower panel), we may postulate that precursors are more commonly 
present in superoutburst of these systems than assumed
in \citet{sch04vwhyimodel}, and that variations within the single system 
is less pronounced.
This stability of appearance of precursors might in turn favor
the pure TTI model, and it needs to be examined further.

\subsection{ER UMa Stars}\label{sec:erumastars}

   Early years from the discovery of ER UMa-type stars
(\cite{kat95eruma}; \cite{rob95eruma}; \cite{pat95v1159ori}),
it was not feasible to fully analyze period variations and $O-C$ 
diagrams in these systems (\cite{kat96erumaSH}; \cite{kat03erumaSH}).
Only recently superhumps in DI UMa (\cite{rut09diuma}; subsection
\ref{obj:diuma}) and RZ LMi \citep{ole08rzlmi} were systematically studied.
Although ER UMa showed positive superhumps at least until 2007
(our unpublished observations) and 2008 (AAVSO data),\footnote{
   We don't completely rule out that negative superhumps may have
   appeared during the later course of superoutbursts in 2007 and 2008,
   since we don't have data during the late course of the superoutbursts.
}
the object now predominantly shows negative superhumps even during
superoutbursts at least since 2011 \citep{ohs12eruma} and in quiescence
in 2008 \citep{kju10eruma} [there was also a possible signature of
negative superhumps in 1998 \citep{gao99erumaSH}], and it is
now impossible to follow the evolution of positive superhumps
during the entire course of a superoutburst of ER UMa as in
the 1990s.

   V1159 Ori, on the other hand, still shows positive superhumps
(subsection \ref{obj:v1159ori}), and it would be very desirable
to study this object in detail.  The recently recognized member
of this family, BK Lyn, now shows almost the same behavior
of ER UMa during its ``negative superhump'' (present) state.

   We also studied RZ LMi and found some evidence of period
variation (subsection \ref{obj:rzlmi}).  We also suggested
a candidate orbital period from photometry, and this needs to be
tested by further observations.

   We list currently known ER UMa stars and their
periods in table \ref{tab:erumalist}.  The orbital periods were
taken from \citet{tho97erumav1159ori} (ER UMa, V1159 Ori),
\citet{tho02gwlibv844herdiuma} (DI UMa), \citet{rin96bklyn}
(BK Lyn).  The superhump periods were from \citet{kat95eruma}
and \citet{ohs12eruma} (ER UMa), \citet{rut09diuma} (DI UMa),
\citet{ole04ixdra} (IX Dra), and this work (V1159 Ori, RZ LMi, BK Lyn).
Although \citet{ole04ixdra} suggested a possible orbital period,
we did not include it because it is less likely to detect an orbital
signal having a period close to the superhump period (as they
claimed) from such a limited coverage.  Since the behavior of
IX Dra is very similar to that of ER UMa in the 1990s
\citep{ish01ixdra}, we would expect an $\epsilon$ similar to
ER UMa.  A search for the definite orbital period is still needed.
It would be also interesting to see whether this object
currently shows positive or negative superhumps.

\begin{table}
\caption{List of ER UMa-type stars}\label{tab:erumalist}
\begin{center}
\begin{tabular}{cccc}
\hline
Object & $P_{\rm orb}$ (d) & $P_{\rm SH}$ (d)\commenta & State\commentb \\
\hline
ER UMa (1995--2008) & 0.06366  & 0.0657 & $+$ \\
ER UMa (2011--)     &          & 0.06226 & $-$ \\
V1159 Ori           & 0.062178 & 0.0643 & $+$ \\
RZ LMi              & (0.059053) & 0.05944 & $+$ \\
DI UMa              & 0.054567 & 0.05531 & $+$ \\
IX Dra              &    --    & 0.06697 & $+$ \\
BK Lyn (2005--)     & 0.07498  & 0.07279 & $-$ \\
\hline
  \multicolumn{4}{l}{\commenta The periods of dominant periodicities are given.} \\
  \multicolumn{4}{l}{\commentb positive ($+$) and negative ($-$) superhumps.} \\
\end{tabular}
\end{center}
\end{table}

\subsection{Superoutbursts of AM CVn Stars}\label{sec:amcvnstars}

   We analyzed the superhumps in a dwarf nova (CR Boo) belonging to
AM CVn-type objects (subsection \ref{obj:crboo}).  CR Boo recently showed
a regular pattern of superoutbursts similar to that of ER UMa 
(as noted in \cite{kat00crboo}).  We recorded
a stage B--C transition similar to hydrogen-rich SU UMa-type
dwarf novae.  This is the first indication that superhumps 
in helium dwarf novae evolve in a similar way to hydrogen-rich 
SU UMa-type dwarf novae.  Since this pattern of stage B--C
transition was also recorded in the black-hole X-ray transient
KV UMa \citep{Pdot}, the presence of stages B and C appear to be
ubiquitous to all low-$q$ systems.  Theoretical studies for
the origin of the superhump stages are required.
Although the early stages of the superoutburst was not observed,
SDSS J172102 underwent a superoutburst followed by a short
post-superoutburst rebrightening resembling that of short-$P_{\rm orb}$
hydrogen-rich SU UMa-type dwarf novae.  This also strengthens
the similarity of phenomenology between helium-rich and hydrogen-rich
SU UMa-type dwarf novae.

   The peculiar object SBS 1108$+$574, a hydrogen-rich CV below 
the period minimum, also showed distinct stages B and C as
in ordinary short-$P_{\rm orb}$ SU UMa-type dwarf novae
(subsection \ref{obj:sbs1108}).  Although the system parameters
of this object is similar to those of AM CVn-type stars,
the superoutburst was much longer than those of AM CVn-type
superoutbursts and there were no ``dip''-like fadings during
the superoutburst plateau (cf. \cite{kot12amcvnoutburst}).
Such a difference in the behavior may be a result from the
different properties of between pure-helium and hydrogen-rich
accretion disks and warrants further study.

\subsection{Double Periodic Superhumps?}

   In the present paper, we encountered three unusual objects
(CC Scl, MASTER J072948, OT J173516) which showed superoutbursts 
similar to other SU UMa-type dwarf novae but with only low-amplitude and
rather irregular superhumps.  The light curves of these systems
appear to be expressed by a combination of closely separated
two periods.  In CC Scl, these periods are almost certainly
$P_{\rm orb}$ and positive superhumps, while the situation for
MASTER J072948 and OT J173516 is unclear: either positive superhumps 
with an unusual $\epsilon$ or negative superhumps.  Although these
objects comprise only a minority of known SU UMa-type dwarf novae,
there may have been ``overlooked'' systems since the amplitudes of
variations are very small.  We cannot explain
the unusual behavior in these systems.  If negative superhumps
(or a tilted disk) were excited as in the present-day ER UMa, 
the coexistence of both signals of $P_{\rm orb}$ and
negative superhumps and the rather irregular waveform may be 
easier to reconcile.

\section{Summary}

   We studied the characteristics of superhumps for 86
SU UMa-type dwarf novae whose superoutbursts were mainly observed 
during the 2011--2012 season.  Most of the new data for systems with 
short orbital periods basically confirmed the earlier findings.

   Among WZ Sge-type dwarf novae, BW Scl showed an $O-C$ variation
similar to other WZ Sge-type dwarf novae such as V455 And and GW Lib,
and this pattern of period variation appears to be common among
WZ Sge-type dwarf novae with shortest orbital periods.
The WZ Sge-type object OT J184228.1$+$483742
showed an unusual pattern of double outbursts composed of an outburst
with early superhumps and another with ordinary superhumps, separated
by a temporary fading.  We propose an interpretation that a very small 
growth rate of the 3:1 resonance due to an extremely low mass-ratio 
led to a quenching of the superoutburst before ordinary superhumps 
appeared.  We suspect that this object is a good candidate for
a period bouncer.

   We studied VW Hyi during its superoutburst in 2011 November--December
and subsequent two normal outbursts.  We confirmed the presence of
``traditional'' late superhumps with a $\sim$0.5 phase shift.
These late superhumps persisted until the second next normal outburst.
The behavior was very similar to the results of Kepler observations
of V344 Lyr and it is likely these late superhumps seem to be
common among relatively long-$P_{\rm orb}$ and high mass-transfer
systems.

   We extended our research to the analysis of positive and negative
superhumps in ER UMa-type dwarf novae, and found that the current
V1159 Ori shows positive superhumps similar to ER UMa in the 1990s.
In two extreme ER UMa stars (DI UMa and RZ LMi), there is an indication
of positive period derivatives.   We identified likely orbital
periods for these objects, and both objects likely have small
mass ratios.  The recently identified ER UMa-type object BK Lyn
has been in dwarf nova-type state at least since 2005, and its
current variation is dominated by negative superhumps as in ER UMa
at least since 2011.

   We further examined superhumps in AM CVn-type objects, and for the
first time established the pattern of period variations very similar
to short-period hydrogen-rich SU UMa-type dwarf novae, and these
objects are indeed a helium analogue of hydrogen-rich SU UMa-type
dwarf novae.

   We also studied a peculiar object SBS 1108$+$574, a hydrogen-rich
dwarf nova below the period minimum, and showed a very similar pattern
of period variations to those of short-period SU UMa-type dwarf novae.
We detected a likely orbital period in this system and estimated
the mass ratio to be $q = 0.06$.  This finding suggests that this 
secondary is a somewhat evolved star whose hydrogen envelope was 
mostly stripped during the mass-exchange.

   We identified a new group of SU UMa-type dwarf novae (CC Scl,
MASTER J072948 and OT J173516.9$+$154708) with low-amplitude superhumps 
with complex profiles.
The complex profile in CC Scl is likely a result of combination of 
orbital humps and positive superhumps.  The cases for MASTER J072948
and OT J173516.9$+$154708 are less clear and the second signal may be 
negative superhumps.

\medskip

This work was supported by the Grant-in-Aid for the Global COE Program
``The Next Generation of Physics, Spun from Universality and Emergence"
from the Ministry of Education, Culture, Sports, Science and Technology
(MEXT) of Japan.
The authors are grateful to observers of VSNET Collaboration and
VSOLJ observers who supplied vital data.
We acknowledge with thanks the variable star
observations from the AAVSO International Database contributed by
observers worldwide and used in this research.
This work is deeply indebted to outburst detections and announcement
by a number of variable star observers worldwide, including participants of
CVNET, BAA VSS alert and AVSON networks. 
We are grateful to the Catalina Real-time Transient Survey
team for making their real-time
detection of transient objects available to the public.

\appendix
\section{MCMC Analysis of Eclipses}\label{sec:app:mcmcecl}

   The KW method is widely used to determine the mid-eclipse times
of eclipsing binaries.  Although this method is useful for
densely sampled light curves, it is difficult to determine
the period of eclipsing binaries with sparse samples.
This is particularly the case for TCP J084616 in which
each eclipse was observed with low time-resolution and with
large photometric errors due to the faintness of the object.
In such cases, a usual approach to measure the mid-eclipse
times by the KW method and then make a linear regression does not
give a good result.  We solved this problem by extending
the application of the MCMC analysis introduced in \citet{Pdot2}.

\begin{figure}
  \begin{center}
    \FigureFile(88mm,50mm){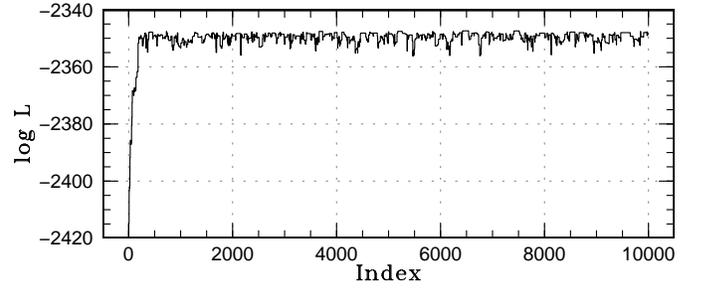}
  \end{center}
  \caption{MCMC analysis of TCP J084616: behavior of likelihood.}
  \label{fig:j0846mcmc}
\end{figure}

\begin{figure}
  \begin{center}
    \FigureFile(70mm,30mm){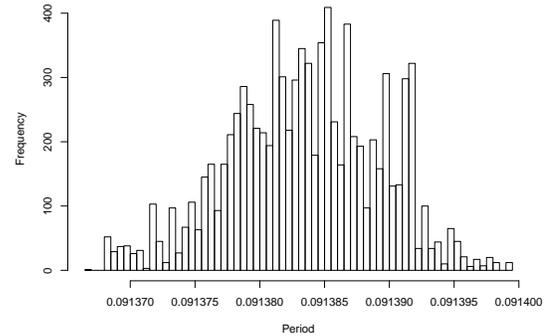}
  \end{center}
  \caption{Posterior probablistic function of $P$ of TCP J084616.}
  \label{fig:j0846perhist}
\end{figure}

\begin{figure}
  \begin{center}
    \FigureFile(88mm,70mm){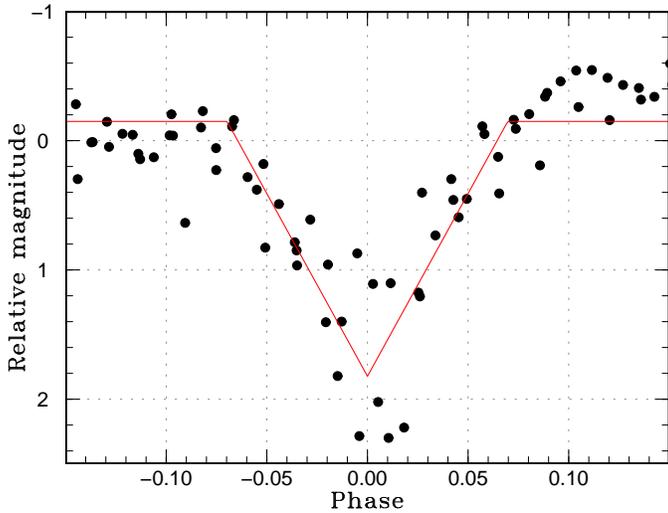}
  \end{center}
  \caption{Best-fit model for TCP J084616.  Points are observations
     and line is the model.}
  \label{fig:j0846fit}
\end{figure}

In this problem, $D = \{y_{\rm obs}(t_i)\}$ are the observed magnitudes
(corrected for trends if necessary) for the epochs $\{t_i\}$,
parameter space is $\theta = \{P,E_0,a,b,c\}$
defined by the model ($y_{\rm model}$):
\begin{eqnarray}
\phi_i = & 1/2 - |(t_i-E_0)/P \bmod 1 - 1/2| & \nonumber \\
y_{\rm model}(\phi_i) = & b+c(a-\phi_i) & (\phi_i < a) \nonumber \\
y_{\rm model}(\phi_i) = & b & (\phi_i \ge a), \nonumber \\
\end{eqnarray}
where $P$ and $E_0$ are the period and epoch, respectively, and
$\bmod 1$ means the fractional part.
and other parameters define the shape of the light curve.
Assuming that $\epsilon_i = y_{\rm obs}(t_i) - y_{\rm model}(t_i)$
follows a normal distribution
$N(0,\sigma_i^2)$, the likelihood function can be written as
\begin{equation}
\mathcal{L}(\theta) = \prod_i \frac{1}{\sqrt{2\pi\sigma_i^2}}
\exp{\biggl[-\frac{\{y_{\rm obs}(t_i)-y_{\rm model}(t_i)\}^2}{2\sigma_i^2} \biggr]},
\end{equation}
and apply MCMC algorithm to this $\mathcal{L}(\theta)$.
A sample of results is shown in figures \ref{fig:j0846mcmc},
\ref{fig:j0846perhist} and \ref{fig:j0846fit}.  We used
the resultant $P$ and $E_0$ in subsection \ref{obj:j0846}.
This method is also applicable to usual determination of 
minima of eclipsing binaries by appropriately defining the shape
of the light curve.  This method is advantageous to classical
KW method in its plain formulation, robustness of the solution
for noisy data, and easy incorporation of errors in individual
measurements.

\begin{thebibliography}{}

\bibitem[Abbott et~al.(1997)]{abb97bwscl}
  Abbott, T. M.~C., Fleming, T.~A., \& Pasquini, L.\ 1997, \aap, 318, 134

\bibitem[Augusteijn et~al.(1996)]{aug96v485cen}
  Augusteijn, T., van~der Hooft, F., de Jong, J.~A., \& van Paradijs, J.\ 1996,
  \aap, 311, 889

\bibitem[Augusteijn et~al.(1993)]{aug93v485cen}
  Augusteijn, T., van Kerkwijk, M.~H., \& van Paradijs, J.\ 1993, \aap, 267,
  L55

\bibitem[Augusteijn, Wisotzki(1997)]{aug97bwscl}
  Augusteijn, T., \& Wisotzki, L.\ 1997, \aap, 324, L57

\bibitem[{Balanutsa} et~al.(2012a)]{bal12j1743atel4022}
  {Balanutsa}, P., {et~al.}\ 2012a, \ATel, 4022

\bibitem[{Balanutsa} et~al.(2012b)]{bal12j1822atel4084}
  {Balanutsa}, P., {et~al.}\ 2012b, \ATel, 4084

\bibitem[{Balanutsa} et~al.(2012c)]{bal12j0729atl3935}
  {Balanutsa}, P., {et~al.}\ 2012c, \ATel, 3935

\bibitem[{Baraffe} et~al.(2003)]{bar03BDmodel}
  {Baraffe}, I., {Chabrier}, G., {Barman}, T.~S., {Allard}, F., \&
  {Hauschildt}, P.~H.\ 2003, \aap, 402, 701

\bibitem[{Boyd} et~al.(2006)]{boy06j1702}
  {Boyd}, D., {Oksanen}, A., \& {Henden}, A.\ 2006, \JBAA, 116, 187

\bibitem[{Breedt} et~al.(2012)]{bre12j1122}
  {Breedt}, E., {G{\"a}nsicke}, B.~T., {Marsh}, T.~R., {Steeghs}, D., {Drake},
  A.~J., \& {Copperwheat}, C.~M.\ 2012, \mnras, 425, 2548

\bibitem[{Cannizzo} et~al.(2012)]{can12v344lyr}
  {Cannizzo}, J.~K., {Smale}, A.~P., {Wood}, M.~A., {Still}, M.~D., \&
  {Howell}, S.~B.\ 2012, \apj, 747, 117

\bibitem[{Cannizzo} et~al.(2010)]{can10v344lyr}
  {Cannizzo}, J.~K., {Still}, M.~D., {Howell}, S.~B., {Wood}, M.~A., \&
  {Smale}, A.~P.\ 2010, \apj, 725, 1393

\bibitem[{Chabrier}, {Baraffe}(1997)]{cha97lowmassstar}
  {Chabrier}, G., \& {Baraffe}, I.\ 1997, \aap, 327, 1039

\bibitem[Chen et~al.(2001)]{che01ECCV}
  Chen, A., O'Donoghue, D., Stobie, R.~S., Kilkenny, D., \& Warner, B.\ 2001,
  \mnras, 325, 89

\bibitem[{Denisenko}, {Sokolovsky}(2011)]{den11ROSATCVs}
  {Denisenko}, D.~V., \& {Sokolovsky}, K.~V.\ 2011, \AstL, 37, 91

\bibitem[{Dillon} et~al.(2008)]{dil08SDSSCV}
  {Dillon}, M., {et~al.}\ 2008, \mnras, 386, 1568

\bibitem[Duerbeck(1987)]{due87novaatlas}
  Duerbeck, H.~W.\ 1987, \ssr, 45, 1

\bibitem[Gao et~al.(1999)]{gao99erumaSH}
  Gao, W., Li, Z., Wu, X., Zhang, Z., \& Li, Y.\ 1999, \apjl, 527, L55

\bibitem[{Garnavich} et~al.(2012)]{gar12sbs1108atel4112}
  {Garnavich}, P., {Littlefield}, C., {Marion}, G.~H., {Irwin}, J., {Kirshner},
  R.~P., \& {Vinko}, J.\ 2012, \ATel, 4112

\bibitem[{Gessner}(1966)]{ges66VS8}
  {Gessner}, H.\ 1966, \VeSon, 7, 61

\bibitem[{Hambsch}(2012)]{ham12ROADaavso}
  {Hambsch}, F.-J.\ 2012, \JAVSO, 40, 1003

\bibitem[Harvey et~al.(1995)]{har95v503cyg}
  Harvey, D., Skillman, D.~R., Patterson, J., \& Ringwald, F.~A.\ 1995, \pasp,
  107, 551

\bibitem[Hessman, Hopp(1990)]{hes90gd552}
  Hessman, F.~V., \& Hopp, U.\ 1990, \aap, 228, 387

\bibitem[Hessman et~al.(1992)]{hes92lateSH}
  Hessman, F.~V., Mantel, K.-H., Barwig, H., \& Schoembs, R.\ 1992, \aap, 263,
  147

\bibitem[{Hoffmeister}(1949)]{hof49newvar}
  {Hoffmeister}, C.\ 1949, \ErgAN, 12, 12

\bibitem[{Hoffmeister}(1951)]{hof49prher}
  {Hoffmeister}, C.\ 1951, \ErgAN, 12, 14

\bibitem[{Hoffmeister}(1968)]{hof68an290277}
  {Hoffmeister}, C.\ 1968, \an, 290, 277

\bibitem[Howell et~al.(1993)]{how93efpeg}
  Howell, S.~B., Schmidt, R., DeYoung, J.~A., Fried, R., Schmeer, P., \& Gritz,
  L.\ 1993, \pasp, 105, 579

\bibitem[{Imada} et~al.(2006)]{ima06j0137}
  {Imada}, A., {et~al.}\ 2006, \pasj, 58, 143

\bibitem[{Imada} et~al.(2008a)]{ima08fltractcv0549}
  {Imada}, A., {Kato}, T., {Monard}, L.~A.~G.~B., {Stubbings}, R., {Uemura},
  M., {Ishioka}, R., \& {Nogami}, D.\ 2008a, \pasj, 60, 267

\bibitem[{Imada} et~al.(2008b)]{ima08egaqr}
  {Imada}, A., {et~al.}\ 2008b, \pasj, 60, 1151

\bibitem[{Imada} et~al.(2009)]{ima09nsv4838}
  {Imada}, A., {et~al.}\ 2009, \pasj, 61, 535

\bibitem[Ishioka et~al.(2001a)]{ish01j2315}
  Ishioka, R., Kato, T., Matsumoto, K., Uemura, M., Iwamatsu, H., \& Stubbings,
  R.\ 2001a, \ibvs, 5023

\bibitem[Ishioka et~al.(2002)]{ish02htcam}
  Ishioka, R., {et~al.}\ 2002, \pasj, 54, 581

\bibitem[Ishioka et~al.(2001b)]{ish01ixdra}
  Ishioka, R., Kato, T., Uemura, M., Iwamatsu, H., Matsumoto, K., Martin,
  B.~E., Billings, G.~W., \& Novak, R.\ 2001b, \pasj, 53, L51

\bibitem[{Ishioka} et~al.(2007)]{ish07CVIR}
  {Ishioka}, R., {Sekiguchi}, K., \& {Maehara}, H.\ 2007, \pasj, 59, 929

\bibitem[Kato(2001)]{kat01v1159ori}
  Kato, T.\ 2001, \pasj, 53, L17

\bibitem[Kato(2002a)]{kat02wzsgeESH}
  Kato, T.\ 2002a, \pasj, 54, L11

\bibitem[Kato(2002b)]{kat02efpeg}
  Kato, T.\ 2002b, \pasj, 54, 87

\bibitem[Kato et~al.(2000a)]{kat00ssumi}
  Kato, T., Hanson, G., Poyner, G., Muyllaert, E., Reszelski, M., \& Dubovsky,
  P.~A.\ 2000a, \ibvs, 4932

\bibitem[{Kato} et~al.(2009)]{Pdot}
  {Kato}, T., {et~al.}\ 2009, \pasj, 61, S395

\bibitem[Kato et~al.(2002)]{kat02v503cyg}
  Kato, T., Ishioka, R., \& Uemura, M.\ 2002, \pasj, 54, 1029

\bibitem[Kato, Kunjaya(1995)]{kat95eruma}
  Kato, T., \& Kunjaya, C.\ 1995, \pasj, 47, 163

\bibitem[{Kato} et~al.(2012a)]{Pdot3}
  {Kato}, T., {et~al.}\ 2012a, \pasj, 64, 21

\bibitem[{Kato} et~al.(2012b)]{kat12DNSDSS}
  {Kato}, T., {Maehara}, H., \& {Uemura}, M.\ 2012b, \pasj, 64, 62

\bibitem[{Kato} et~al.(2010)]{Pdot2}
  {Kato}, T., {et~al.}\ 2010, \pasj, 62, 1525

\bibitem[Kato et~al.(2000b)]{kat00crboo}
  Kato, T., Nogami, D., Baba, H., Hanson, G., \& Poyner, G.\ 2000b, \mnras,
  315, 140

\bibitem[Kato et~al.(1996a)]{kat96alcom}
  Kato, T., Nogami, D., Baba, H., Matsumoto, K., Arimoto, J., Tanabe, K., \&
  Ishikawa, K.\ 1996a, \pasj, 48, L21

\bibitem[Kato et~al.(1996b)]{kat96erumaSH}
  Kato, T., Nogami, D., \& Masuda, S.\ 1996b, \pasj, 48, L5

\bibitem[Kato et~al.(2003a)]{kat03erumaSH}
  Kato, T., Nogami, D., \& Masuda, S.\ 2003a, \pasj, 55, L7

\bibitem[Kato et~al.(2003b)]{kat03hodel}
  Kato, T., Nogami, D., Moilanen, M., \& Yamaoka, H.\ 2003b, \pasj, 55, 989

\bibitem[Kato et~al.(2004a)]{kat04v803cen}
  Kato, T., Stubbings, R., Monard, B., Butterworth, N.~D., Bolt, G., \&
  Richards, T.\ 2004a, \pasj, 56, S89

\bibitem[{Kato}, {Uemura}(2012)]{kat12perlasso}
  {Kato}, T., \& {Uemura}, M.\ 2012, \pasj, 64, 122

\bibitem[Kato et~al.(2004b)]{VSNET}
  Kato, T., Uemura, M., Ishioka, R., Nogami, D., Kunjaya, C., Baba, H., \&
  Yamaoka, H.\ 2004b, \pasj, 56, S1

\bibitem[{Kemp} et~al.(2012)]{kem12bklynsass}
  {Kemp}, J., {et~al.}\ 2012, \SASS, 31, 7

\bibitem[{Kjurkchieva}, {Marchev}(2010)]{kju10eruma}
  {Kjurkchieva}, D., \& {Marchev}, D.\ 2010, \POBeo, 90, 147

\bibitem[{Kotko} et~al.(2012)]{kot12amcvnoutburst}
  {Kotko}, I., {Lasota}, J.-P., {Dubus}, G., \& {Hameury}, J.-M.\ 2012, \aap,
  544, A13

\bibitem[{Kwee}, {van Woerden}(1956)]{KWmethod}
  {Kwee}, K.~K., \& {van Woerden}, H.\ 1956, \bain, 12, 327

\bibitem[{Levitan} et~al.(2011)]{lev11j0719}
  {Levitan}, D., {et~al.}\ 2011, \apj, 739, 68

\bibitem[{Littlefair} et~al.(2006)]{lit06j1702}
  {Littlefair}, S.~P., {Dhillon}, V.~S., {Marsh}, T.~R., \& {G{\"a}nsicke},
  B.~T.\ 2006, \mnras, 371, 1435

\bibitem[{Markarian}, {Stepanian}(1983)]{mar83SBS1}
  {Markarian}, B.~E., \& {Stepanian}, D.~A.\ 1983, \Afz, 19, 639

\bibitem[Mennickent et~al.(1999)]{men99tucrt}
  Mennickent, R.~E., Patterson, J., O'Donoghue, D., Unda, E., Harvey, D.,
  Vanmuster, T., \& Bolt, G.\ 1999, \apss, 262, 1

\bibitem[{Nakano} et~al.(2011)]{nak11j1842cbet2818}
  {Nakano}, S., {Nishimura}, H., {Noguchi}, T., \& {Munari}, U.\ 2011, \CBET,
  2818, 1

\bibitem[Nogami et~al.(1995)]{nog95v1159ori}
  Nogami, D., Kato, T., Masuda, S., \& Hirata, R.\ 1995, \ibvs, 4155

\bibitem[{Nogami} et~al.(2004)]{nog04v406hya}
  {Nogami}, D., {Monard}, B., {Retter}, A., {Liu}, A., {Uemura}, M., {Ishioka},
  R., {Imada}, A., \& {Kato}, T.\ 2004, \pasj, 56, L39

\bibitem[{Ohshima} et~al.(2012)]{ohs12eruma}
  {Ohshima}, T., {et~al.}\ 2012, \pasj, 64, L3

\bibitem[{Ohshima} et~al.(2011)]{ohs11qzvir}
  {Ohshima}, T., {et~al.}\ 2011, \pasj, \submitted

\bibitem[Olech(1997)]{ole97v485cen}
  Olech, A.\ 1997, \actaa, 47, 281

\bibitem[{Olech} et~al.(2006)]{ole06ssumi}
  {Olech}, A., {Mularczyk}, K., {K{\c e}dzierski}, P., {Z{\l}oczewski}, K.,
  {Wi{\'s}niewski}, M., \& {Szaruga}, K.\ 2006, \aap, 452, 933

\bibitem[{Olech} et~al.(2008)]{ole08rzlmi}
  {Olech}, A., {Wisniewski}, M., {Zloczewski}, K., {Cook}, L.~M., {Mularczyk},
  K., \& {Kedzierski}, P.\ 2008, \actaa, 58, 131

\bibitem[{Olech} et~al.(2004)]{ole04ixdra}
  {Olech}, A., {Zloczewski}, K., {Mularczyk}, K., {Kedzierski}, P.,
  {Wisniewski}, M., \& {Stachowski}, G.\ 2004, \actaa, 54, 57

\bibitem[{Pastukhova}(1988)]{pas88nyher}
  {Pastukhova}, E.~N.\ 1988, \ATsir, 1534, 17

\bibitem[Patterson et~al.(1995)]{pat95v1159ori}
  Patterson, J., Jablonski, F., Koen, C., O'Donoghue, D., \& Skillman, D.~R.\
  1995, \pasp, 107, 1183

\bibitem[Patterson et~al.(1997)]{pat97crboo}
  Patterson, J., {et~al.}\ 1997, \pasp, 109, 1100

\bibitem[{Pavlenko} et~al.(2010a)]{pav10mndraproc}
  {Pavlenko}, E., {et~al.}\ 2010a, in AIP Conference Proceedings, 17th European
  White Dwarf Workshop, ed. {K.~Werner \& T.~Rauch} (\PublisherAIP), p.~320

\bibitem[{Pavlenko} et~al.(2012a)]{pav12j1051atel3889}
  {Pavlenko}, E., {Kato}, T., {Pit}, N., {Baklanov}, A., {Antonyuk}, K., \&
  {Stein}, W.\ 2012a, \ATel, 3889, 1

\bibitem[{Pavlenko} et~al.(2012b)]{pav12v503cyg}
  {Pavlenko}, E.~P., {Samsonov}, D.~A., {Antonyuk}, O.~I., {Andreev}, M.~V.,
  {Baklanov}, A.~V., \& {Sosnovskij}, A.~A.\ 2012b, \Ap, 55, 494

\bibitem[{Pavlenko} et~al.(2010b)]{pav10mndra}
  {Pavlenko}, E.~P., {et~al.}\ 2010b, \ARep, 54, 6

\bibitem[{Pojma\'nski}(2002)]{ASAS3}
  {Pojma\'nski}, G.\ 2002, \actaa, 52, 397

\bibitem[Provencal et~al.(1997)]{pro97crboo}
  Provencal, J.~L., {et~al.}\ 1997, \apj, 480, 383

\bibitem[{Ramsay} et~al.(2011)]{ram11amcvn}
  {Ramsay}, G., {Barclay}, T., {Steeghs}, D., {Wheatley}, P.~J., {Hakala}, P.,
  {Kotko}, I., \& {Rosen}, S.\ 2011, \mnras, 419, 2836

\bibitem[{Ramsay} et~al.(2012)]{ram12amcvnLC}
  {Ramsay}, G., {Barclay}, T., {Steeghs}, D., {Wheatley}, P.~J., {Hakala}, P.,
  {Kotko}, I., \& {Rosen}, S.\ 2012, \mnras, 419, 2836

\bibitem[{Rau} et~al.(2010)]{rau10HeDN}
  {Rau}, A., {Roelofs}, G.~H.~A., {Groot}, P.~J., {Marsh}, T.~R., {Nelemans},
  G., {Steeghs}, D., {Salvato}, M., \& {Kasliwal}, M.~M.\ 2010, \apj, 708, 456

\bibitem[{Richards} et~al.(2009)]{ric09SDSSQSOcand}
  {Richards}, G.~T., {et~al.}\ 2009, \apjs, 180, 67

\bibitem[Richter(1990)]{ric90gd552}
  Richter, G.~A.\ 1990, \MitVS, 12, 59

\bibitem[Ringwald et~al.(1996)]{rin96bklyn}
  Ringwald, F.~A., Thorstensen, J.~R., Honeycutt, R.~K., \& Robertson, J.~W.\
  1996, \mnras, 278, 125

\bibitem[{Robertson} et~al.(1998)]{rob98oldnovaproc}
  {Robertson}, J.~W., {Honeycutt}, R.~K., {Hillwig}, T., \& {Jurcevic}, J.\
  1998, in \ASPConf{137}{Wild Stars in the Old West}, ed. S. Howell, E.
  Kuulkers, \& C. Woodward (\PublisherASP), p.~469

\bibitem[Robertson et~al.(2000)]{rob00oldnova}
  Robertson, J.~W., Honeycutt, R.~K., Hillwig, T., Jurcevic, J.~S., \& Henden,
  A.~A.\ 2000, \aj, 119, 1365

\bibitem[Robertson et~al.(1995)]{rob95eruma}
  Robertson, J.~W., Honeycutt, R.~K., \& Turner, G.~W.\ 1995, \pasp, 107, 443

\bibitem[Rutkowski et~al.(2009)]{rut09diuma}
  Rutkowski, A., Olech, A., {Wi\'{s}niewski}, M., Pietrukowicz, P., Pala, J.,
  \& Poleski, R.\ 2009, \aap, 497, 437

\bibitem[Schoembs, Vogt(1980)]{sch80vwhyi}
  Schoembs, R., \& Vogt, N.\ 1980, \aap, 91, 25

\bibitem[{Schreiber} et~al.(2004)]{sch04vwhyimodel}
  {Schreiber}, M.~R., {Hameury}, J.-M., \& {Lasota}, J.-P.\ 2004, \aap, 427,
  621

\bibitem[Schwope et~al.(2000)]{sch00RASSID}
  Schwope, A., {et~al.}\ 2000, \an, 321, 1

\bibitem[{Shears} et~al.(2012a)]{she12puuma}
  {Shears}, J., {Hambsch}, F.-J., {Littlefield}, C., {Miller}, I., {Morelle},
  E., {Pickard}, R., {Pietz}, J., \& {Sabo}, R.\ 2012a, \JBAA,
  \inpress\arxiv{1209.4062}

\bibitem[{Shears} et~al.(2012b)]{she12j2158}
  {Shears}, J., {et~al.}\ 2012b, \JBAA, \inpress\arxiv{1205.0898}

\bibitem[Skillman, Patterson(1993)]{ski93bklyn}
  Skillman, D.~R., \& Patterson, J.\ 1993, \apj, 417, 298

\bibitem[{Sokolovsky} et~al.(2012)]{sol12j1051atel3849}
  {Sokolovsky}, K.~V., {Baryshev}, K.~O., \& {Korotkiy}, S.~A.\ 2012, \ATel,
  3849

\bibitem[{Solheim}(2010)]{sol10amcvnreview}
  {Solheim}, {J.-E.}\ 2010, \pasp, 122, 1133

\bibitem[{Southworth} et~al.(2008)]{sou08CVperiod}
  {Southworth}, J., {et~al.}\ 2008, \mnras, 391, 591

\bibitem[{Southworth} et~al.(2007)]{sou07SDSSCV2}
  {Southworth}, J., {Marsh}, T.~R., {G{\"a}nsicke}, B.~T., {Aungwerojwit}, A.,
  {Hakala}, P., {de Martino}, D., \& {Lehto}, H.\ 2007, \mnras, 382, 1145

\bibitem[Stellingwerf(1978)]{PDM}
  Stellingwerf, R.~F.\ 1978, \apj, 224, 953

\bibitem[{Szkody} et~al.(2003)]{szk03SDSSCV2}
  {Szkody}, P., {et~al.}\ 2003, \aj, 126, 1499

\bibitem[{Szkody} et~al.(2006)]{szk06SDSSCV5}
  {Szkody}, P., {et~al.}\ 2006, \aj, 131, 973

\bibitem[{Szkody} et~al.(2005)]{szk05SDSSCV4}
  {Szkody}, P., {et~al.}\ 2005, \aj, 129, 2386

\bibitem[{Szkody} et~al.(2007)]{szk07CVWDpuls}
  {Szkody}, P., {et~al.}\ 2007, \apj, 658, 1188

\bibitem[{Tappert} et~al.(2004)]{tap04CTCV}
  {Tappert}, C., {Augusteijn}, T., \& {Maza}, J.\ 2004, \mnras, 354, 321

\bibitem[Thorstensen et~al.(2002a)]{tho02j2329}
  Thorstensen, J.~R., Fenton, W.~H., Patterson, J.~O., Kemp, J., Krajci, T., \&
  Baraffe, I.\ 2002a, \apjl, 567, L49

\bibitem[Thorstensen et~al.(2002b)]{tho02gwlibv844herdiuma}
  Thorstensen, J.~R., Patterson, J.~O., Kemp, J., \& Vennes, S.\ 2002b, \pasp,
  114, 1108

\bibitem[Thorstensen et~al.(1996)]{tho96Porb}
  Thorstensen, J.~R., Patterson, J.~O., Shambrook, A., \& Thomas, G.\ 1996,
  \pasp, 108, 73

\bibitem[{Thorstensen}, {Skinner}(2012)]{tho12CRTSCVs}
  {Thorstensen}, J.~R., \& {Skinner}, J.~N.\ 2012, \aj, 144, 81

\bibitem[Thorstensen et~al.(1997)]{tho97erumav1159ori}
  Thorstensen, J.~R., Taylor, C.~J., Becker, C.~M., \& Remillard, R.~A.\ 1997,
  \pasp, 109, 477

\bibitem[{Tiurina} et~al.(2012)]{tiu12j1051atel3845}
  {Tiurina}, N., {et~al.}\ 2012, \ATel, 3845

\bibitem[Tsugawa, Osaki(1997)]{tsu97amcvn}
  Tsugawa, M., \& Osaki, Y.\ 1997, \pasj, 49, 75

\bibitem[Uemura et~al.(2002)]{uem02j2329letter}
  Uemura, M., {et~al.}\ 2002, \pasj, 54, L15

\bibitem[Uemura et~al.(2001)]{uem01v725aql}
  Uemura, M., Kato, T., Pavlenko, E., Baklanov, A., \& Pietz, J.\ 2001, \pasj,
  53, 539

\bibitem[{Uemura} et~al.(2005)]{uem05tvcrv}
  {Uemura}, M., {et~al.}\ 2005, \aap, 432, 261

\bibitem[{Unda-Sanzana} et~al.(2008)]{und08gd552}
  {Unda-Sanzana}, E., {et~al.}\ 2008, \mnras, 388, 889

\bibitem[{Uthas} et~al.(2012)]{uth12j1457bwscl}
  {Uthas}, H., {et~al.}\ 2012, \mnras, 420, 379

\bibitem[Vogt(1983)]{vog83lateSH}
  Vogt, N.\ 1983, \aap, 118, 95

\bibitem[Warner(1995)]{war95book}
  Warner, B.\ 1995, Cataclysmic Variable Stars (\PublisherCambridge)

\bibitem[{Warner}, {Woudt}(2004)]{war04CVnewZZproc}
  {Warner}, B., \& {Woudt}, P.~A.\ 2004, in \ASPConf{310}{IAU Colloq. 193:
  Variable Stars in the Local Group}, ed. {D.~W.~Kurtz \& K.~R.~Pollard}
  (\PublisherASP), p.~382

\bibitem[{Wils} et~al.(2010)]{wil10newCVs}
  {Wils}, P., {G{\"a}nsicke}, B.~T., {Drake}, A.~J., \& {Southworth}, J.\ 2010,
  \mnras, 402, 436

\bibitem[Wolf, Wolf(1905)]{wol05svari}
  Wolf, M., \& Wolf, G.\ 1905, \an, 169, 415

\bibitem[{Wood} et~al.(2011)]{woo11v344lyr}
  {Wood}, M.~A., {Still}, M.~D., {Howell}, S.~B., {Cannizzo}, J.~K., \&
  {Smale}, A.~P.\ 2011, \apj, 741, 105

\bibitem[{Woudt} et~al.(2012)]{wou12ccscl}
  {Woudt}, P.~A., {et~al.}\ 2012, \mnras, 427, 1004

\bibitem[{Yamaoka} et~al.(2011)]{yam11j2109cbet2731}
  {Yamaoka}, H., {et~al.}\ 2011, \CBET, 2731, 1

\end{thebibliography}
\end{document}